%% file: main.tex
\documentclass[a4paper,11pt]{article}     

\include{python_preamble}
\usepackage{jheppub}
\usepackage     [utf8]                  {inputenc}
\usepackage     [T1]                    {fontenc}
\usepackage                             {color}
\usepackage                             {amsmath}
\usepackage                             {amssymb}
\usepackage{bbold}
\usepackage                             {graphicx}
\usepackage     [english]               {babel}
\usepackage                             {hyperref}   
\usepackage{amssymb}
\usepackage{mathtools}
\usepackage{color}
\usepackage{multirow}
\usepackage{listings}
\usepackage{natbib}
\usepackage{tabularx}
\usepackage{tikz-feynman}
\usepackage{tikz}
\usetikzlibrary{shapes,arrows}
\usepackage{caption}
\usepackage{subcaption}
\usepackage{graphicx}

\usepackage[edges]{forest}
\definecolor{foldercolor}{RGB}{124,166,198}

\tikzset{pics/folder/.style={code={%
    \node[inner sep=0pt, minimum size=#1](-foldericon){};
    \node[folder style, inner sep=0pt, minimum width=0.3*#1, minimum height=0.6*#1, above right, xshift=0.05*#1] at (-foldericon.west){};
    \node[folder style, inner sep=0pt, minimum size=#1] at (-foldericon.center){};}
    },
    pics/folder/.default={20pt},
    folder style/.style={draw=foldercolor!80!black,top color=foldercolor!40,bottom color=foldercolor}
}

\forestset{is file/.style={edge path'/.expanded={%
        ([xshift=\forestregister{folder indent}]!u.parent anchor) |- (.child anchor)},
        inner sep=1pt},
    this folder size/.style={edge path'/.expanded={%
        ([xshift=\forestregister{folder indent}]!u.parent anchor) |- (.child anchor) pic[solid]{folder=#1}}, inner xsep=0.6*#1},
    folder tree indent/.style={before computing xy={l=#1}},
    folder icons/.style={folder, this folder size=#1, folder tree indent=3*#1},
    folder icons/.default={12pt},
}

\newcommand{\0}{$0\nu\beta\beta$}

\newcommand{\str}[1]{\PY{l+s+s2}{\PYZdq{}}\PY{l+s+s2}{#1}\PY{l+s+s2}{\PYZdq{}}}
\newcommand{\bool}[1]{\PY{k+kc}{#1}}
\newcommand{\Or}{\mathcal{O}}
\newcommand{\nudobe}{\texttt{$\nu$DoBe} }
\newcommand{\UILink}{https://nudobe.streamlit.app/}
\newcommand{\GitHubLink}{https://github.com/OScholer/nudobe}
\renewcommand{\vec}[1]{{\mathbf{#1}}}

\title{\nudobe - A Python Tool for Neutrinoless Double Beta Decay}

\author[a]{Oliver Scholer,\,}
\emailAdd{scholer@mpi-hd.mpg.de}

\author[b,c]{Jordy de Vries,\,}
\emailAdd{j.devries4@uva.nl}

\author[d,e]{and Luk\'{a}\v{s} Gr\'{a}f}
\emailAdd{lukas.graf@berkeley.edu}

\affiliation[a]{Max-Planck-Institut f{\"u}r Kernphysik, Saupfercheckweg 1, 69117 Heidelberg, Germany}
\affiliation[b]{Institute for Theoretical Physics Amsterdam and Delta Institute for Theoretical Physics, University of Amsterdam, Science Park 904, 1098 XH Amsterdam, The Netherlands}
\affiliation[c]{Nikhef, Theory Group, Science Park 105, 1098 XG, Amsterdam, The Netherlands}
\affiliation[d]{Department of Physics, University of California, Berkeley, California 94720, USA}
\affiliation[e]{Department of Physics, University of California, San Diego, La Jolla, CA 92093-0319, USA}

\date{September 2020}
\abstract{We present $\nu$DoBe, a Python tool for the computation of neutrinoless double beta decay ($0\nu\beta\beta$) rates in terms of lepton-number-violating operators in the Standard Model Effective Field Theory (SMEFT). The tool can be used for automated calculations of \0 rates, electron spectra and angular correlations for all isotopes of experimental interest, for lepton-number-violating operators up to and including dimension 9. The tool takes care of renormalization-group running to lower energies and  provides the matching to the low-energy effective field theory and, at lower scales, to a chiral effective field theory description of \0 rates. The user can specify different sets of nuclear matrix elements from various many-body methods and hadronic low-energy constants. The tool can be used to quickly generate analytical and numerical expressions for \0 rates and to generate a large variety of plots. In this work, we provide examples of possible use along with a detailed code documentation. The code can be accessed through:
\\
\href{\GitHubLink}{GitHub}: \GitHubLink
\\
\href{\UILink}{Online User-Interface}: \UILink
}
\begin{document}
\preprint{
\hfill N3AS-23-010
}

\maketitle

\section{Introduction}
Neutrinoless double beta decay ($0\nu\beta\beta$) searches are the most sensitive probe of the violation of lepton number (LNV). The process takes place in a nucleus and converts two neutrons to two protons and two electrons, but no outgoing anti-neutrinos. So far only upper limits have been set on $0\nu\beta\beta$ rates~\citep{Umehara:2008ru,GERDA:2020xhi,CUPID-0:2018rcs,NEMO-3:2009fxe,CUPID:2020aow,Danevich:2016eot,Arnaboldi:2002te,CUORE:2019yfd,EXO-200:2017vqi,KamLAND-Zen:2022tow,NEMO:2008kpp}, but next-generation experiments hope to make a first detection by probing $0\nu\beta\beta$ lifetimes reaching $10^{27}$-$10^{28}$ years~\citep{2020CupidProspects,2017LEGEND200,legendcollaboration2021legend1000,nEXO:2021ujk,SNO:2021xpa}. A nonzero signal would have profound implications. It would imply lepton number is violated by two units. Following the famous black-box theorem~\citep{PhysRevD.25.2951} a positive observation of \0 would indicate that neutrinos are indeed Majorana mass eigenstates. Additionally, it would give important clues towards the mechanism of the neutrino mass and also be a strong hint for leptogenesis scenarios to explain the absence of anti-matter in our universe. Detailed reviews about \0 can be found in Refs.~\citep{2019review_Werner,Agostini:2022zub,Cirigliano:2022oqy}.

While tremendous experimental effort is going towards the first $0\nu\beta\beta$ detection, we must keep in mind that it is a complicated process involving particle, hadronic, nuclear, and atomic physics. The interpretation of a signal (or lack thereof) requires care. First of all, even if a nonzero decay rate is measured this does not immediately point towards the underlying source of LNV. Typically, $0\nu\beta\beta$ experiments are interpreted in terms of the effective neutrinos mass, $m_{\beta\beta}$, that enters through the exchange of light Majorana neutrinos (either active or potential sterile neutrinos). In a broad class of BSM models there can be competing mechanisms from the exchange of heavy particles such as heavy right-handed neutrinos or double charged scalars~\citep{Deppisch:2012nb,Li:2020flq}. Studies of \0-decay in multiple isotopes can be utilized to identify or reject certain types of BSM models~\citep{Deppisch:2006hb,Gehman:2007qg,Graf:2022lhj,Agostini:2022bjh}.

As $0\nu\beta\beta$ decay is a low-energy process, the typical Q value of the reaction is a few MeV, it can be efficiently described through the use of effective field theory (EFT) techniques. In particular, EFTs provide a useful method to systematically categorize and order the possible LNV mechanisms~\citep{Pas:1999fc,Pas:2000vn,Prezeau:2003xn,Graf:2018ozy, Cirigliano:2017djv,Cirigliano_2018}. If the origin of LNV happens at scales well above the scale of  $0\nu\beta\beta$ decay, this allows for a model-independent description. 
EFT techniques are also very powerful in the description of the associated hadronic and nuclear physics that is required to connect data to the underlying LNV mechanism that is typically written at the level of elementary particles.

\begin{figure}[t]
    \centering
    \includegraphics[width=0.9\textwidth]{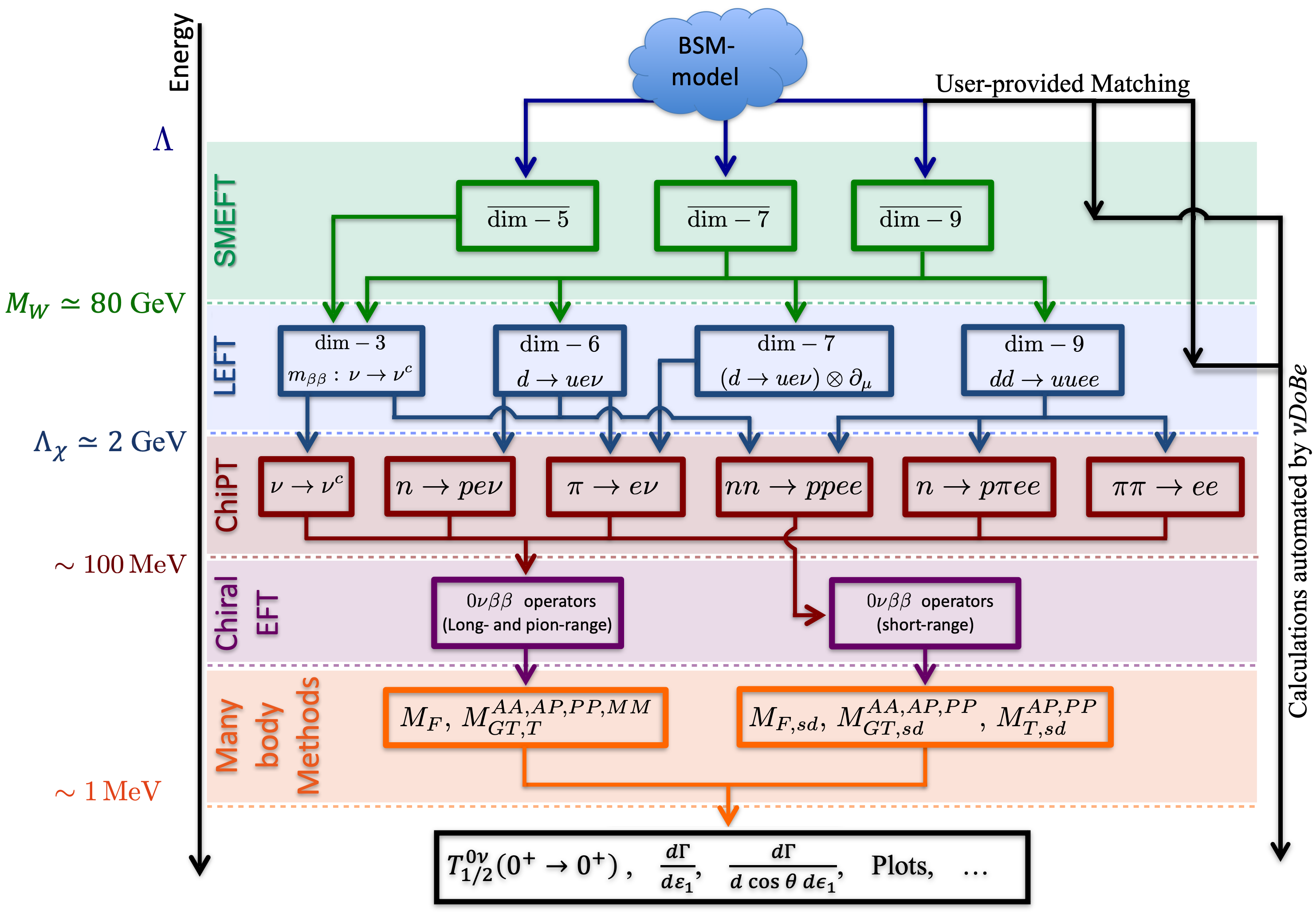}
    \caption{The EFT approach applied in \nudobe. Given a certain lepton number violating BSM-model, all \0 observables of interest can be calculated by following a subsequent chain of EFTs. In \nudobe the user is required to provide the matching conditions at either SMEFT or LEFT level. The calculations of \0 observables, plots, etc. are then automated by \nudobe. Figure modified from Ref.~\citep{Cirigliano_2018}.}
    \label{fig:eftChain}
\end{figure}

$\nu$DoBe is based on an end-to-end EFT framework, from particle to nuclear physics, that has been developed in Refs.~\cite{Cirigliano:2017djv,Cirigliano_2018} and is illustrated in Fig.~\ref{fig:eftChain}. Assuming that LNV originates at scales well above the electroweak scale\footnote{This excludes the possibility of light sterile neutrinos which require a modified framework \cite{Dekens:2020ttz}. $\nu$DoBe cannot (yet) handle such scenarios.}, $v\simeq 246$ GeV, possible LNV mechanisms can be described by local effective LNV operators among Standard Model fields. The resulting EFT, often called the Standard Model Effective Field Theory (SMEFT), contains an infinite number of operators that can be ordered by their dimension. The higher the dimension of the operator, the more suppressed its low-energy effects are by additional powers of $Q/\Lambda$ where $Q$ is a low-energy scale and $\Lambda \gg v$ the scale where the SMEFT operators are generated. LNV operators appear with odd dimension \cite{Kobach:2016ami} and $\nu$DoBe includes operators with dimension up-to-and-including dimension nine \cite{Weinberg:1979sa,Lehman:2014jma,Liao:2020jmn}.

The SMEFT operators are then evolved from $\Lambda$ to the electroweak scale, where heavy Standard Model fields (W, Z, Higgs, top) are integrated out and the theory is matched to an EFT that is often-called the low-energy EFT (LEFT). The LEFT operators are then evolved to the GeV scale and matched to hadronic LNV operators by using chiral EFT which provides an expansion in $m_\pi/\Lambda_\chi$ where $m_\pi$ denotes the pion mass, around 100 MeV, and  $\Lambda_\chi \sim 2$ GeV, the chiral-symmetry-breaking scale. The chiral perturbation theory Lagrangian is used to calculate the so-called neutrino transition operator that mitigates $nn\rightarrow pp + ee$ transitions. This step is non-trivial as strong interactions among nucleons lead to deviations from naive dimensional analysis expectations typically used to power count different contributions \cite{Cirigliano:2018hja,Cirigliano:2019vdj}. The neutrino transition operator is then inserted into nuclear wave functions to obtain the relevant nuclear matrix elements (NMEs) that can be combined with atomic calculations to obtain $0\nu\beta\beta$ decay rates.

The EFT approach calculates $0\nu\beta\beta$ decay rates in a systematic expansion \mbox{$(v/\Lambda)^\alpha (\Lambda_\chi/v)^\beta (m_\pi/\Lambda_\chi)^\gamma$}, where $\alpha$, $\beta$, and $\gamma$ are exponents that depend on the LNV mechanism under consideration. The decay rate is expressed in terms of phase space factors (PSFs), NMEs, low-energy constants (LECs), evolution factors, and the Wilson coefficient of the SMEFT operators, as outlined in detail in Refs.~\cite{Cirigliano:2017djv,Cirigliano_2018}. The actual procedure is, however, somewhat challenging and, frankly, tedious for the expert and non-expert alike because of the large number of steps and the sizeable number of QCD, nuclear, and atomic matrix elements. $\nu$DoBE is constructed to automate this procedure: the user chooses which LNV SMEFT operators appear at the high-energy scale $\Lambda$ (or alternatively which LEFT operators at the electroweak scale), and the tool presents the resulting $0\nu\beta\beta$ decay observables. In this way, any high-scale model of LNV can be quickly analyzed. 

In this work we describe $\nu$DoBe (Neutrinoless DOuble-BEta calculator) in detail and give some explicit examples of its use. Before doing so let us list the most important applications of $\nu$DoBe:
\\
\begin{itemize}
    \item An automatic calculation of $0\nu\beta\beta$ decay rates and electron kinematics of all isotopes of experimental interest in terms of LNV SMEFT operators up to and including dimension nine. Alternatively, the decay rate can be expressed in terms of LNV LEFT operators. 
    \item The tool computes differential decays rates based on state-of-the-art information regarding renormalization-group factors, hadronic and nuclear matrix elements, and atomic phase space factors. 
    \item It is possible to choose various sets of NMEs (based on different many-body nuclear-structure methods) to assess the theoretical uncertainty of the predictions. Users can define their own set of NMEs if necessary.
    \item The expression for $0\nu\beta\beta$-decay rates includes recently identified short-range contributions associated to hard-neutrino-exchange processes \cite{Cirigliano:2018hja,Cirigliano:2019vdj}. The code uses current best estimates for the associated LECs but different values can be specified.
    \item $\nu$DoBe can be used to generate interesting plots of half-lives or various different distributions as function of neutrino masses or LNV Wilson coefficients. This can be used to quickly analyse specific BSM scenarios.
    \item For a given BSM model or a set of Wilson coefficients \nudobe can generate analytical expressions of the decay half-life and output them in latex (or html) form.
    \item {Given an experimental bound on the half-life of certain isotope, \nudobe can be used to extract limits on the effective Majorana neutrino mass or higher-dimensional LNV LEFT or SMEFT operators.}
\end{itemize}

\section{Physics Formalism - The \0 Master Formula Framework}
The effects of LNV physics appearing at some high-energy scale $\Lambda\gg M_W$ on low-energy phenomena such as \0 can be described by a chain of EFTs. On the high-energy end of this chain we employ the Standard Model EFT (SMEFT)~\citep{Grzadkowski:2010es,Henning:2015alf,Lehman:2014jma,Liao:2019tep,Li:2020gnx,Murphy:2020rsh,Li:2020xlh,Liao:2020jmn} describing the effects of new physics above the electroweak scale. At the scale of electroweak symmetry breaking SMEFT is matched onto the low-energy EFT (LEFT)~\citep{Jenkins:2017jig,Jenkins:2017dyc,Dekens:2019ept,Liao:2020zyx,Li:2020tsi} respecting the $SU(3)_C\times U(1)_{em}$ gauge group. At lower scales we match subsequently onto lepton-extended chiral EFT to incorporate the non-perturbative nature of QCD below $\Lambda_\chi\sim 2\,\mathrm{GeV}$. $\nu$DoBe follows the procedure developed in Refs.~\citep{Cirigliano:2017djv,Cirigliano_2018} which is illustrated in Fig.~\ref{fig:eftChain}. Below, we present the relevant expressions and how they are used in $\nu$DoBe and refer to \citep{Cirigliano:2017djv,Cirigliano_2018} for more details. 

The \0 half-life equation can be written in very short-hand notation
\begin{align}
    T^{-1}_{1/2} = g_A^4\sum_k G_{0k} |\mathcal{A}_k(\{C_i\})|^2,
\end{align}
where conventionally $g_A^4$, with $g_A \simeq 1.27$, is factored out. The $G_{0k}$ denote atomic phase space factors. Most of the physics is captured by the so-called subamplitudes $\mathcal{A}_k(\{C_i\})$ which depend on NMEs, LECs, and the Wilson coefficients of LNV higher-dimensional operators labeled by $\{C_i\}$. The full expressions of the subamplitudes are rather lengthy and can be found in Ref.~\citep{Cirigliano_2018}, but we provide some details in appendices. Specifically, Appendix \ref{app:operator_list} gives the full lists of SMEFT and LEFT operators included in $\nu$DoBe. The matching of SMEFT to LEFT is summarized in Appendix~\ref{app:operator_matching} where the matching of dim-9 LNV SMEFT operators to LEFT is performed for the first time. Appendices \ref{app:PSFs} and \ref{app:NMEs} discuss the phase space factors and NMEs, respectively. 

\section{Examples of $\nu$DoBe applications}
We now present three analyses revisited from the literature and re-performed with $\nu$DoBe to illustrate the practical use of the tool. We begin by analyzing the KamLAND-ZEN upper limit on the \0 decay of \textsuperscript{136}Xe and the implications for the effective neutrino Majorana mass. We then turn to the analysis of \0 in specific beyond-the-Standard Model models including the minimal left-right symmetric extension to the Standard Model as well as possible realizations of leptoquarks. 

\subsection{An analysis of recent KamLAND-Zen limits}
\subsubsection{The Light-Neutrino-Exchange Mechanism}
Recently, the KamLAND-Zen experiment obtained a new lower limit on the half-life in \textsuperscript{136}Xe of $2.3\times10^{26}\;\mathrm{yr}$ \citep{KamLAND-Zen:2022tow}. We will analyze the resulting limits on the effective Majorana mass $m_{\beta\beta}$ using $\nu$DoBe. In particular, we will apply different sets of NMEs and study the potential impact of the recently introduced short-range contributions originating from hard neutrino-exchange~\citep{Cirigliano:2018hja,Cirigliano:2019vdj} which were not included in Ref.~\citep{KamLAND-Zen:2022tow}. We briefly describe the necessary inputs to the code here while a Jupyter notebook with all details is accessible on \href{\GitHubLink}{GitHub}. 

\begin{table}[t!]
    \centering
    \begin{tabular}{l c c c c c}
    \hline\hline
                                         & $M_\nu$ & $M_{F,sd}$                    & $M_{F,sd}/M_\nu$ 
                                         & $m_{\beta\beta}$ [meV]
                                         & $\Tilde{m}_{\beta\beta}$ [meV]\\
        \hline
        QRPA~\citep{Hyvarinen:2015bda}*  & $3.009$  & $-61.8\frac{m_em_p}{m_\pi^2}$ & $-0.51$ & 57 & 27          \\
        QRPA~\citep{Terasaki:2020ndc}    & $3.384$ & $-$                           & $-$ & 51 & 35              \\
        QRPA~\citep{Simkovic:2013qiy}    & $2.460$ & $-$                           & $-$ & 70 & 48              \\
        QRPA~\citep{Mustonen:2013zu}     & $1.89$  & $-$                           & $-$ & 91 & 63              \\
        QRPA~\citep{Fang:2018tui}        & $1.18$  & $-28.8\frac{m_em_p}{m_\pi^2}$ & $-0.60$ & 146 & 62      \\
        EDF~\citep{LopezVaquero:2013yji} & $4.773$ & $-$                           & $-$ & 36 & 25              \\
        EDF~\citep{Yao:2014uta}          & $4.32$  & $-$                             & $-$ & 40 & 28            \\
        EDF~\citep{Rodriguez:2010mn}     & $4.20$  & $-$                           & $-$ & 41 & 28              \\
        IBM2~\citep{Deppisch:2020ztt}*   & $3.387$ & $-29.8\frac{m_em_p}{m_\pi^2}$ & $-0.22$ & 51 & 34      \\
        IBM2~\citep{Barea:2015kwa}       & $3.05$ & $-29.7\frac{m_em_p}{m_\pi^2}$ & $-0.24$ & 57 & 37           \\
        SM~\citep{Coraggio:2020hwx}      & $2.39$ & $-$                           & $-$ & 72 & 50               \\
        SM~\citep{Neacsu:2014bia}        & $1.76$ & $-$                           & $-$ & 98 & 68               \\
        SM~\citep{Menendez:2008jp}       & $1.77$ & $-$                           & $-$ & 98 & 67               \\
        SM~\citep{Menendez:2017fdf}*     & $2.45$ & $-52\frac{m_em_p}{m_\pi^2}$   & $-0.52$ & 71 & 32               \\
    \hline\hline
    \end{tabular}
    \caption{Different nuclear matrix elements in \textsuperscript{136}Xe. The NMEs are given assuming $g_A\sim 1.271$. Appropriate rescaling has been applied to the IBM2 NMEs. The sets labeled with an asterisk (*) are pre-installed in \nudobe. In the columns labeled with $m_{\beta\beta}$ and $\Tilde{m}_{\beta\beta}$ we show the resulting limits on the effective Majorana mass for each set of NMEs without ($m_{\beta\beta}$) and including ($\Tilde{m}_{\beta\beta}$) the short-range contribution proportional to $g_\nu^{NN}$.}
    \label{tab:KamLANDNMEs}
\end{table}

\begin{figure}[t!]
    \centering
    \includegraphics[width=0.495\textwidth]{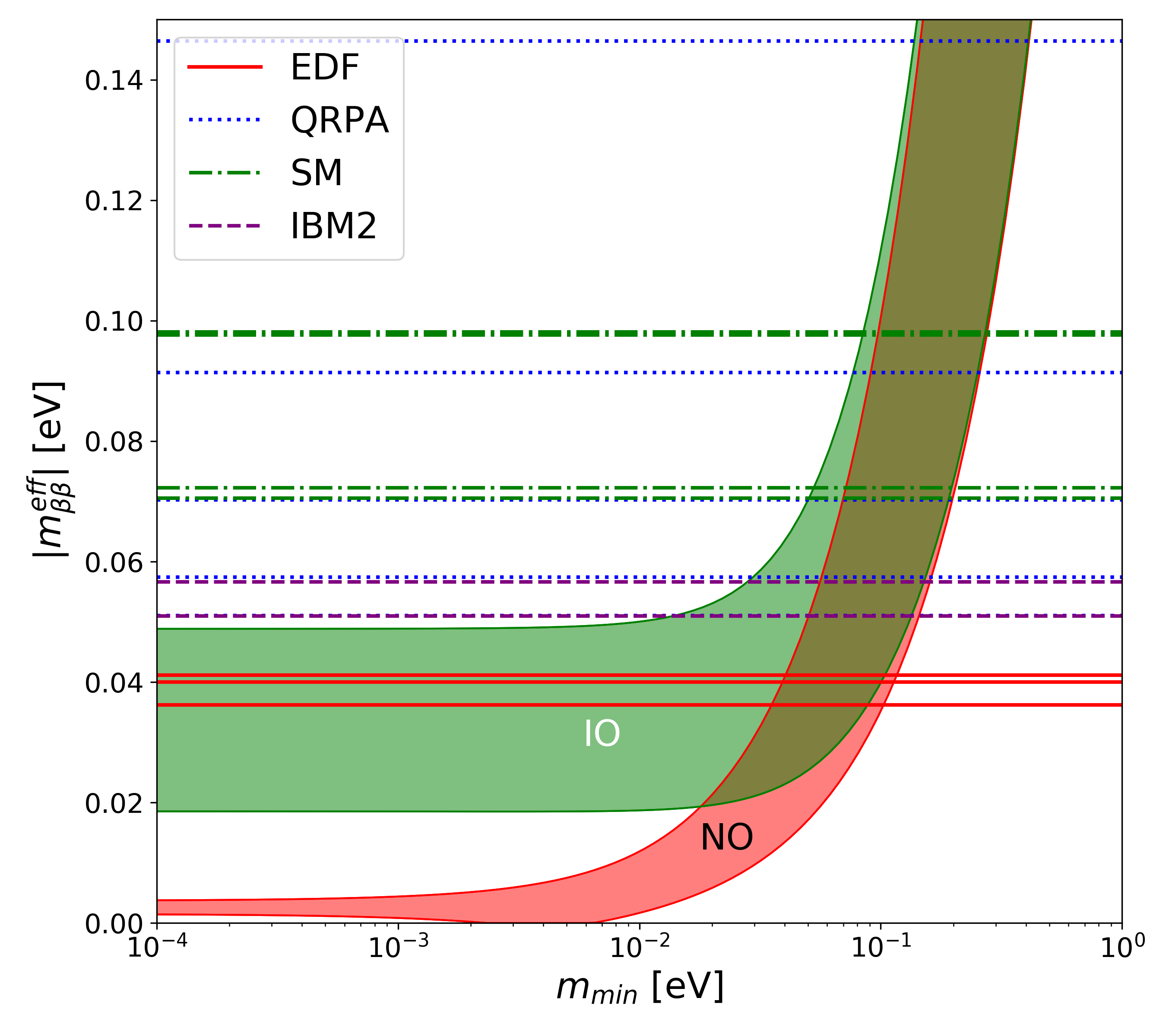}
    \includegraphics[width=0.495\textwidth]{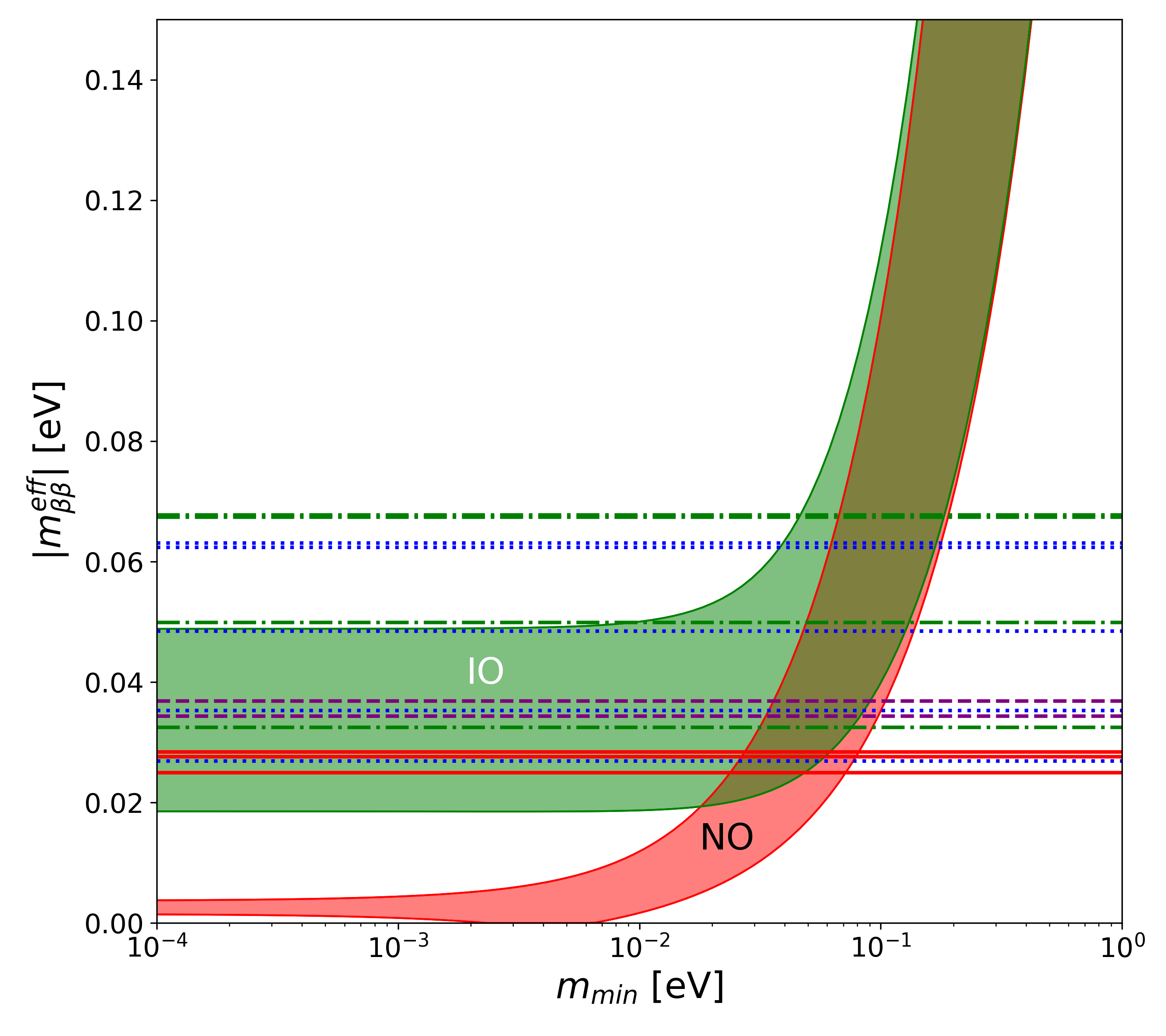}
    \includegraphics[width=0.495\textwidth]{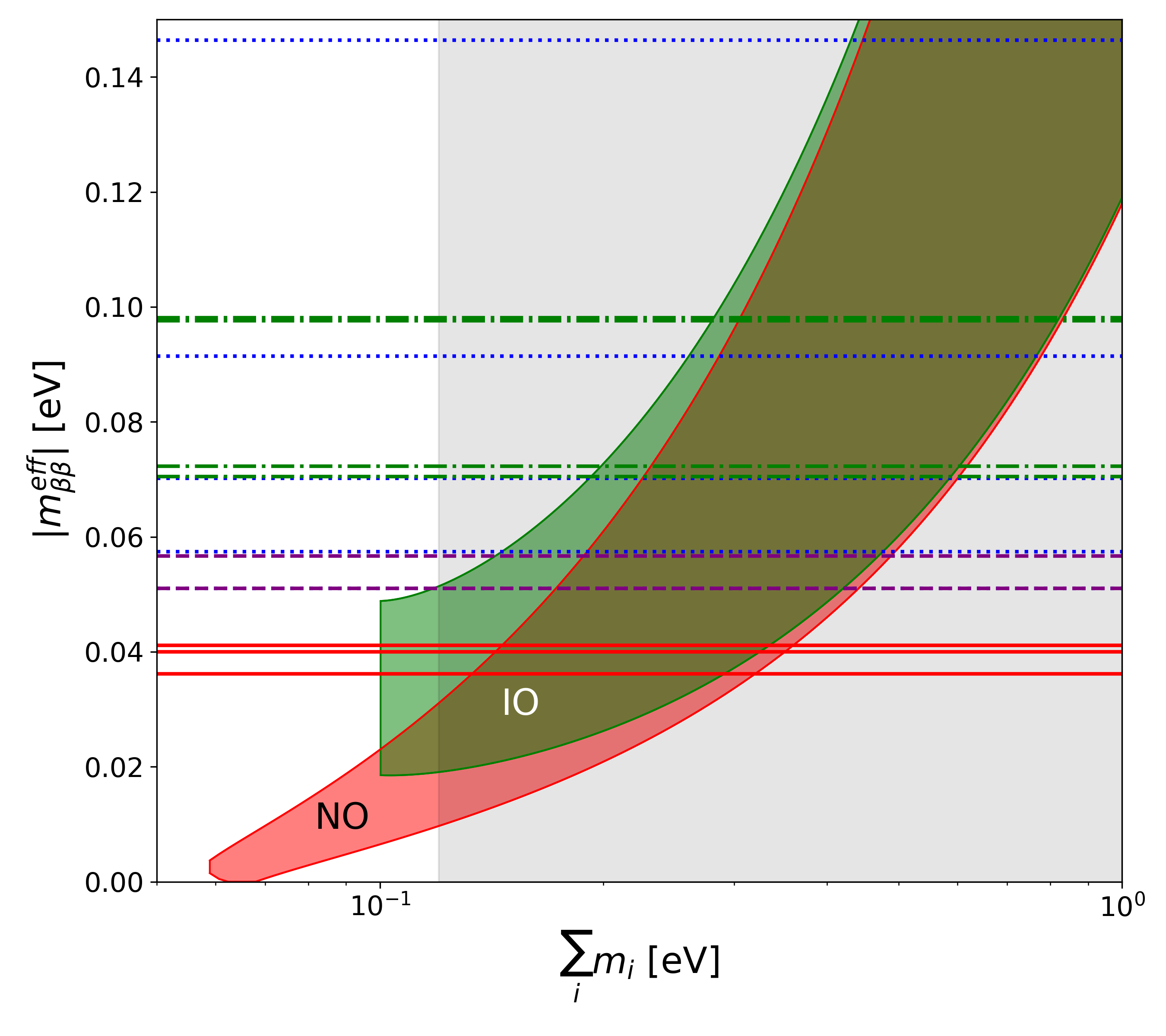}
    \includegraphics[width=0.495\textwidth]{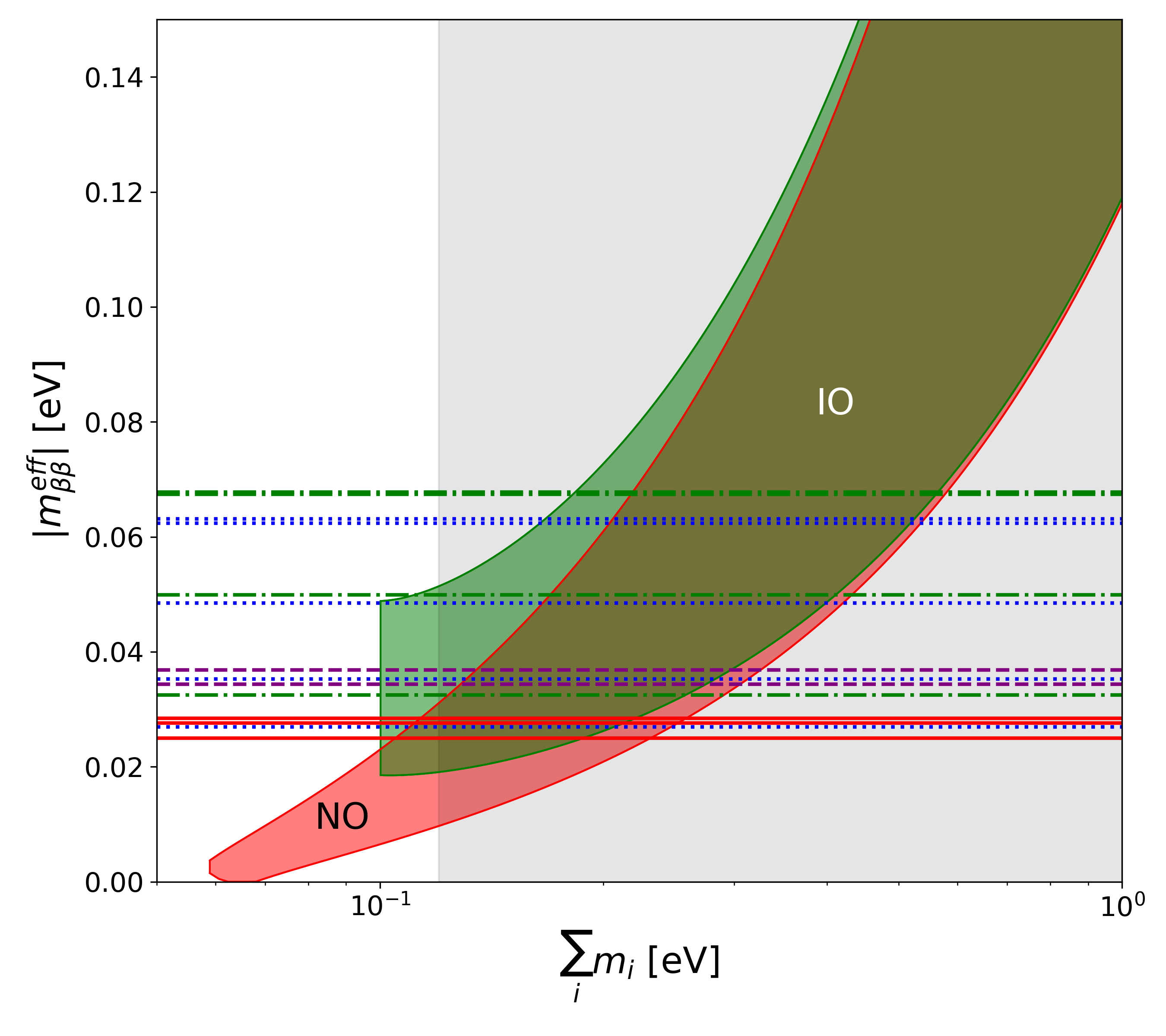}
    \caption{Limits on the effective Majorana mass $m_{\beta\beta}$ obtained from the recent half-life limit by the KamLAND-Zen experiment~\citep{KamLAND-Zen:2022tow}. On the left, the limits are shown without including the short-range contribution proportional to $g_\nu^{NN}$ as in~\citep{KamLAND-Zen:2022tow}, while on the right we do include it (see text for details). In the lower plots we show the current limit on the sum of the neutrino masses $\sum_i m_i\lesssim 0.12\,\mathrm{eV}$~\citep{Workman:2022ynf} as a gray band. The allowed parameter regions for normal (NO) and inverse (IO) mass ordering are displayed as red and green bands, respectively. They are obtained by variation of the unknown Majorana phases utilizing the \protect\hyperlink{model.plot_m_bb}{\texttt{plot\_m\_eff()}} function of \nudobe.}\label{fig:NME_mbb_limits}
\end{figure}
If only the standard mass mechanism is considered, the half-life is expressed as
\begin{align}
    T_{1/2}^{-1} = & g_A^4\left|\frac{m_{\beta\beta}}{m_e}\right|^2G_{01}V_{ud}^4\left|\left(-\frac{1}{g_A^2}M_F + M_{GT} + M_{T} + 2\frac{m_\pi^2 g_\nu^{NN}}{g_A^2}M_{F,sd}\right)\right|^2\,,
\end{align}
where the various $M_i$ denote NMEs, while $g_\nu^{NN}$ is an LEC associated with hard-neutrino exchange. To replicate the results from~\citep{KamLAND-Zen:2022tow} we apply the same value for the phase space factor $V_{ud}^4G_{01}(^{136}\mathrm{Xe}) = 1.458\times10^{-14}\,\mathrm{y}^{-1}$ (the factor $V_{ud}^4$ is included because of a slightly different definition of $G_{0k}$). $\nu$DoBe has the NMEs split into various components, but most literature is based on the effective combination
\begin{equation}
M_\nu = -\frac{1}{g_A}M_F + M_{GT} + M_T\,,
\end{equation}
while setting $g_\nu^{NN}=0$. In Table~\ref{tab:KamLANDNMEs} we show the different sets of NMEs studied here.\footnote{For some NME calculation the magnetic factor has been calculated incorrectly through $g_M = \kappa_1 g_V$, where $\kappa_1$ is the isovector anomalous magnetic moment of the nucleon, instead of the correct $g_M = (1+\kappa_1) g_V$. For more details see the appendices or Refs.~\cite{Cirigliano:2017djv,Cirigliano_2018}. If the individual NMEs $M_{GT,T}^{MM}$ are given explicitly in the original publication, we correct $M_\nu$ appropriately. In case that both CD Bonn and Argonne potentials are given we take the values corresponding to the CD Bonn potential.} 

In general, to apply \nudobe we need the individual NMEs $M_{F, GT, T, Fsd}$. However, often only the combined NME $M_\nu = -\frac{1}{g_A}M_F + M_{GT} + M_T$ is given in the literature. Within \nudobe we can work around this by simply defining $M_{GT}=M_\nu$ while putting all other NMEs (including $M_{F,sd})$ to zero. By utilizing the internal \hyperlink{model.get_limits}{\texttt{get\_limits()}} function we then arrive at the limits on $m_{\beta\beta}$ in the range $[36,146]\,\mathrm{meV}$. The individual limits on $m_{\beta\beta}$ derived from each NME method are listed in Table~\ref{tab:KamLANDNMEs} and the left column of Figure~\ref{fig:NME_mbb_limits}. Whenever NMEs calculated using both CD Bonn and Argonne nuclear potentials are available we choose to show the CD-Bonn ones. If, instead, the Argonne potential NMEs are considered the upper limit moves to $156\,\mathrm{meV}$ as in~\citep{KamLAND-Zen:2022tow}.
\begin{table}[t!]
\begin{tabularx}{\textwidth}{c c c c c c c c c c c c c}
\hline\hline
$m_{\beta\beta}$ & $C_{S}^{(6)}$ & $C_{T}^{(6)}$ & $C_{VL}^{(6)}$ & $C_{VR}^{(6)}$ & $C_{V}^{(7)}$ & $C_{S1}^{(9)}$ & $C_{S2}^{(9)}$ & $C_{S3}^{(9)}$ & $C_{S4}^{(9)}$ & $C_{S5}^{(9)}$ & $C_{V}^{(9)}$  & $C_{\Tilde{V}}^{(9)}$
\\
\hline
$m_{\beta\beta}$ & $C_{SL}^{(6)}$ & $C_{T}^{(6)}$ & $C_{VL}^{(6)}$ & $C_{VR}^{(6)}$ & $C_{VL}^{(7)}$ & $C_{1L}^{(9)}$ & $C_{2L}^{(9)}$ & $C_{3L}^{(9)}$ & $C_{4L}^{(9)}$ & $C_{5L}^{(9)}$ & $C_{6}^{(9)}$  & $C_{7}^{(9)}$
\\
  & $C_{SR}^{(6)}$ &   &   &   & $C_{VR}^{(7)}$ & $C_{1R}^{(9)}$ & $C_{2R}^{(9)}$ & $C_{3R}^{(9)}$ & $C_{4R}^{(9)}$ & $C_{5R}^{(9)}$ & $C_{8}^{(9)}$  & $C_{9}^{(9)}$
\\
  &   &   &   &   &   & ${C_{1L}^{(9)}}'$ & ${C_{2L}^{(9)}}'$ & ${C_{3L}^{(9)}}'$ &   &   & ${C_{6}^{(9)}}'$  & ${C_{7}^{(9)}}'$
\\
  &   &   &   &   &   & ${C_{1R}^{(9)}}'$ & ${C_{2R}^{(9)}}'$ & ${C_{3R}^{(9)}}'$ &   &   & ${C_{8}^{(9)}}'$  & ${C_{9}^{(9)}}'$
\\
\hline\hline
\end{tabularx}\caption{Groups of LEFT operators that result in the same half-lives.}\label{tab:operator_groups}
\end{table}

Now we want to study the effect of the additional short-range contribution proportional to $g_\nu^{NN}$ which has not been considered in~\citep{KamLAND-Zen:2022tow}. For those NME calculations where the corresponding short-range NME $M_{F,sd}$ is available we simply add this contribution. However, for many NMEs no short-range contributions have been calculated. Here, we take a very simplistic approach by approximating the short-range NME $M_{F,sd}$ from the ratios $M_{F,sd}/M_\nu$ found in those NME calculations with $M_{F,sd}$ available. In \textsuperscript{136}Xe we find that $|M_{F,sd}/M_\nu|>0.2$, hence, we take $M_{F,sd}=-\frac{1}{5}M_\nu$ as a lower limit approximation of the resulting half-life (Note, again, that $g_\nu^{NN}$ is negative and, hence, the contribution of a negative $M_{F,sd}$ is constructive). In this approximation we arrive at limits on $m_{\beta\beta}$ in the range $[25, 68]\,\mathrm{meV}$ which are more stringent, illustrating the importance of fixing the short-distance contributions. The limits from all different NMEs with short-range contributions are shown in the right column of Figure~\ref{fig:NME_mbb_limits} and Table~\ref{tab:KamLANDNMEs}. 

We stress that these limits are only indicative as the short-distance contributions must be calculated for each nuclear many-body method consistently with the long-distance contributions, see Sect.~(5.3.5) for a more detailed discussion. This has recently been done in Refs.~\cite{Cirigliano:2020dmx,Wirth:2021pij}. 
\subsubsection{Higher Dimensional Mechanisms}
\begin{figure}[t!]
    \centering
    \includegraphics[width=\textwidth]{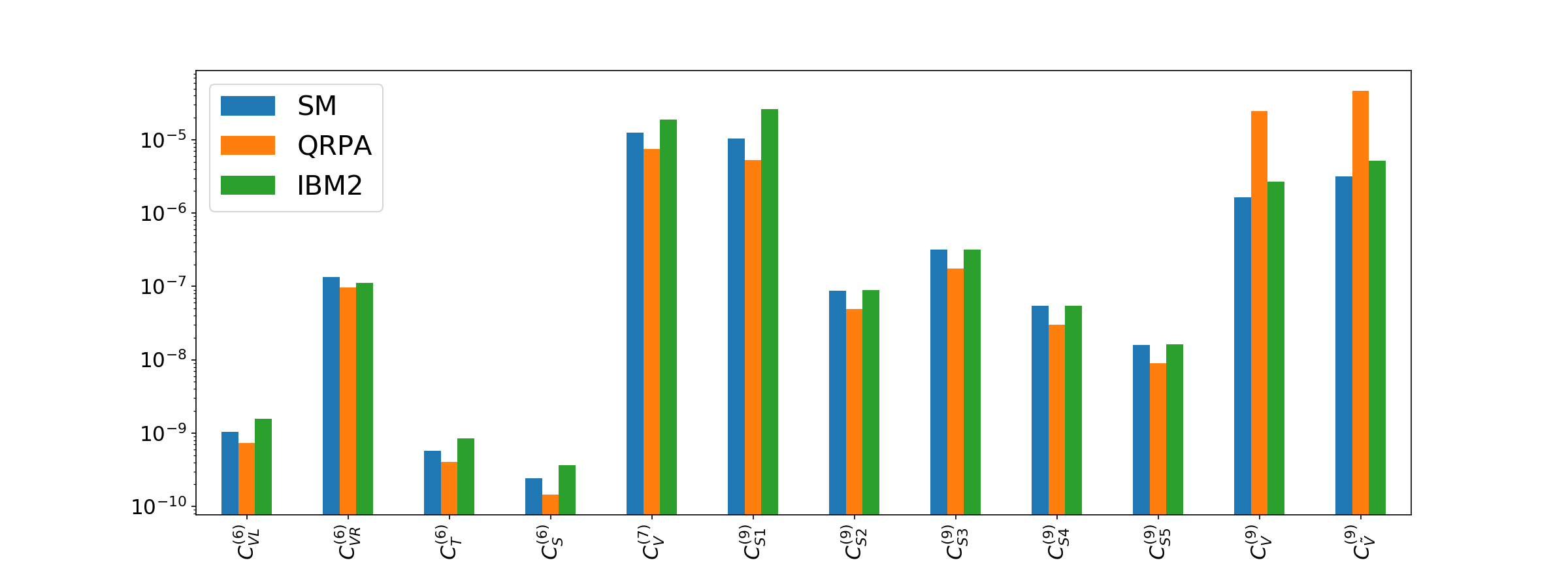}
    \caption{Limits on the higher dimensional ($d\geq 6$) LEFT operators obtained from the recent limit on the half-life of \textsuperscript{136}Xe by the KamLAND-Zen experiment in combination with different NMEs.}
    \label{fig:nonStandardLimits}
\end{figure}

\noindent
\textbf{LEFT:}
We can use \nudobe to put limits on the higher-dimensional LNV operators. Considering operators from the low-energy EFT we can put these into 13 different groups of operators that result in the same half-lives (see Table~\ref{tab:LEFT_operators} in Appendix~\ref{app:operator_list}) because of parity conservation in QCD. For convenience we summarize these operator groups here again in Table~\ref{tab:operator_groups}.
We require the corresponding NMEs to calculate the limits on the higher dimensional ($d\geq 6$) operators of interest, which means we can only use a few NME calculation methods namely those from the shell model~\cite{Menendez:2017fdf}, QRPA~\cite{Hyvarinen:2015bda} and IBM2~\cite{Deppisch:2020ztt}. 
The resulting limits on the higher dimensional operators are given in Figure~\ref{fig:nonStandardLimits}. The limits are extracted at the SMEFT~$\rightarrow$~LEFT matching scale $m_W$ and acquire a broad range of values from $\sim10^{-10}$ (for $C_S^{(6)}$) to  $\sim10^{-5}$ (for $C_V^{(7)}$, $C_{S1}^{(9)}$, $C_V^{(9)}$, and $C_{\tilde V}^{(9)}$), thus illustrating the fact that different operators can have very different $0\nu\beta\beta$ efficiencies. 

We can estimate the scale of new physics $\Lambda$ through
\begin{align}
    \Lambda_i\simeq \frac{v}{C_i^{1/(d-4)}}\,
\end{align}
with $d\in[6,7,9]$ being the operator's LEFT dimension and the Higgs vacuum expectation value (vev) $v\simeq 246\,\mathrm{GeV}$. This translates to limits on the scale of new physics in the range of $\sim 2\,\mathrm{TeV} \,$ -- $\,20\,\mathrm{PeV}$. 

\begin{figure}[t]
    \centering
    \includegraphics[width=\textwidth]{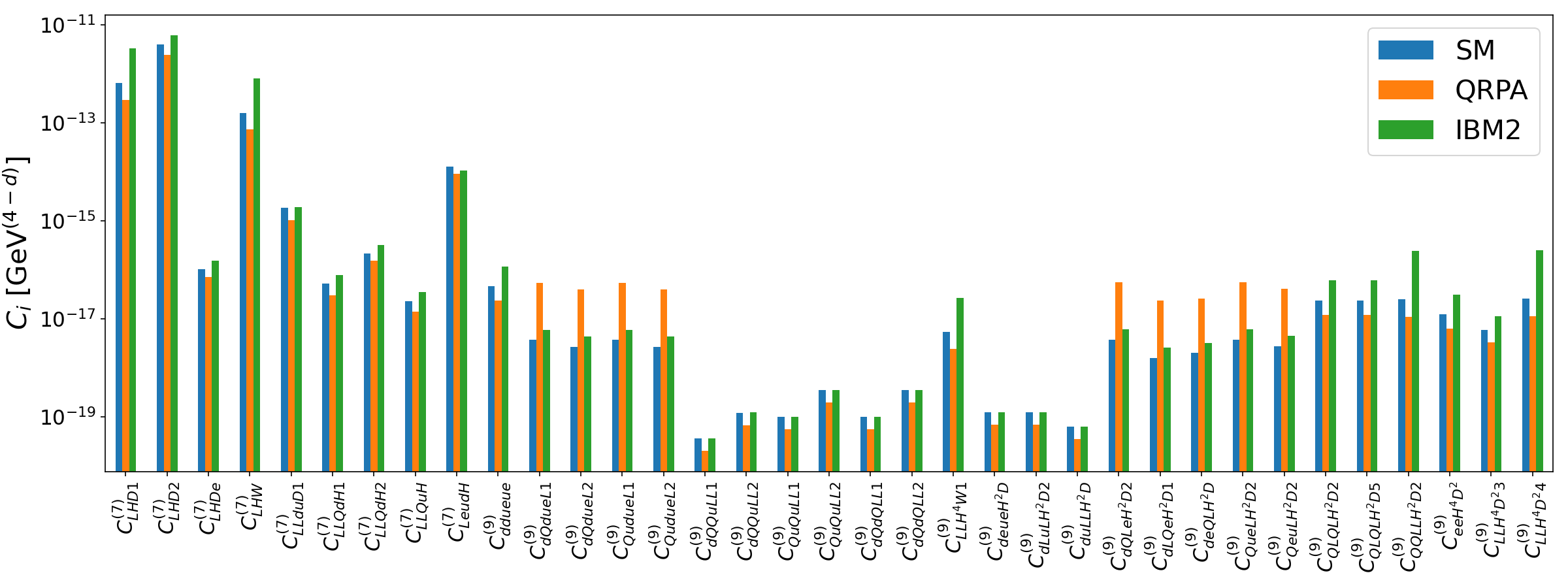}
    \caption{Limits on the higher dimensional ($d\geq 7$) SMEFT operators obtained from the recent KamLAND-Zen results assuming different NMEs.}
    \label{fig:KamLandOperatorLimitsSMEFT}
\end{figure}

\noindent{\textbf{SMEFT:}}
Alternatively, we can use the new limit on the half-life in \textsuperscript{136}Xe from KamLAND-Zen to provide limits on the dimensionfull Wilson coefficients of the SMEFT operators. These limits are given in Fig.~\ref{fig:KamLandOperatorLimitsSMEFT} and correspond to scales in the range of $\sim 1\,\mathrm{TeV}$ -- $400\,\mathrm{TeV}$
For more details, we refer to the provided Jupyter notebook on \href{\GitHubLink}{GitHub} as well as Section~\ref{sec:opertor_limits}.

\subsection{The Minimal Left-Right Symmetric Model}
The minimal left-right symmetric model (mLRSM)~\citep{PhysRevD.10.275,PhysRevD.11.2558,PhysRevD.12.1502,Duka_2000} is a well-studied extension to the Standard Model. It enlarges the Standard Model gauge group by adding the right-handed $SU(2)$; hence, the mLRSM gauge group reads $SU(3)_C\times SU(2)_L \times SU(2)_R \times U(1)_{B-L}$. At the same time, this additional symmetry requires the existence of new fermionic and bosonic degrees of freedom. Typically, the mLRSM incorporates two scalar triplets $\Delta_L\in(1,3,1,2)$ and $\Delta_R\in(1,1,3,2)$ and a scalar bidoublet $\Phi\in(1,2,2*,0)$. Fermions are grouped into left- and right-handed $SU(2)_{L,R}$ doublets requiring the introduction of right-handed neutrinos $\nu_R$
\begin{align}
    L_L &= \left(\begin{matrix}\nu_L\\e_L\end{matrix}\right)\in\left(1, 2, 1, -1\right),\qquad Q_L= \left(\begin{matrix}u_L\\d_L\end{matrix}\right)\in\left(3, 2, 1, 1/3\right),\\
    L_R &= \left(\begin{matrix}\nu_R\\e_R\end{matrix}\right)\in\left(1, 1, 2, -1\right),\qquad Q_R= \left(\begin{matrix}u_R\\d_R\end{matrix}\right)\in\left(3, 1, 2, 1/3\right)\;.
\end{align}
In this section, we revisit the mLRSM as studied in~\citep{Cirigliano_2018,Graf:2022lhj} (conventions are equal to~\citep{Graf:2022lhj}) utilizing \nudobe and the shell model NMEs~\cite{Menendez:2017fdf} provided therein.
Lepton number violation in the mLRSM is introduced via Yukawa interactions,
\begin{align}
\begin{split}
     \mathcal{L}_{y} =& \sum_{ij} \bigg[
     Y^l_{ij}{\overline{L_L}}_i\Phi L_{R,j} + \Tilde{Y}^l_{ij}\overline{L_L}_{i}\Tilde{\Phi} L_{R,j} + Y^L_{ij}L^T_{L,i}Ci\tau_2\Delta_L L_{L,j} + {Y^R_{ij}}^\dagger L^T_{R,i}Ci\tau_2\Delta_R L_{R,j}
     \bigg] + \text{h.c.}\label{eq:mLRSMyukawas}
\end{split}
\end{align}
When the neutral components of the scalar multiplets acquire non-zero vacuum expectation values,
\begin{align}
    \left<\Phi\right> = \frac{1}{\sqrt{2}}\left(\begin{matrix}\kappa & 0 \\ 0 & \kappa'e^{i\alpha}\end{matrix}\right),\qquad \left<\Delta_L\right> = \frac{1}{\sqrt{2}}\left(\begin{matrix}0 & 0 \\ v_Le^{i\theta_L} & 0\end{matrix}\right),\qquad \left<\Delta_R\right> = \frac{1}{\sqrt{2}}\left(\begin{matrix}0 & 0 \\ v_R & 0\end{matrix}\right),
\end{align}
they give rise to neutrino mass matrices
\begin{align}
    \begin{split}
    M^\nu_{D,ij} &= \frac{1}{\sqrt{2}}\left[Y^l_{ij}\kappa + \Tilde{Y}^l_{ij}\kappa'\exp{-i\alpha}\right],\quad\; \\
    {M^\nu_{L,ij}}^\dagger &= \sqrt{2}Y^L_{ij}v_L\exp{i\theta_L},\qquad\qquad\quad\;\;\; M^\nu_{R,ij}=\sqrt{2}Y^R_{ij}v_R\;.
\end{split}
\end{align}
The ratio of the bidoublet's vevs $\xi = \frac{{\kappa'}}{\kappa}$ describes the mixing between the left-handed and right-handed $W$-bosons.
When the right-handed triplet gains a non-zero vev, the mLRSM gauge group is broken down to the Standard Model gauge structure. Matching the mLRSM onto the SMEFT results in the following LNV Lagrangian,
\begin{align}
\begin{split}
    \mathcal{L}_{\Delta L=2}=&C^{(5)}\left(\left(L^TCi\tau_2 H\right)\left(\Tilde{H}^\dagger L\right)\right) \\
    &+ \left(L^T\gamma^\mu e_R\right)i\tau_2 H\bigg[C^{(7)}_{Leu\overline{d}\Phi}\overline{d_R}\gamma_\mu u_R +C^{(7)}_{L\Phi De} H^Ti\tau_2(D_\mu\Phi_{SM}) \bigg]\\
    &+\overline{e_R}e_R^c\bigg[C^{(9)}_{eeud}\overline{u_R}\gamma^\mu d_R\overline{u_R}\gamma_\mu d_R + C^{(9)}_{ee\Phi ud}\overline{u_R}\gamma^\mu d_R\left(\left[iD_\mu H\right]^\dagger\Tilde{H}\right) 
    \\
    &\qquad\qquad+ C^{(9)}_{ee\Phi D}\left(\left[iD_\mu H\right]^\dagger\Tilde{H}\right)^2\bigg],
\end{split}
\end{align}
where $H$ is the Standard Model Higgs with the vev $v=\sqrt{\kappa^2 + {\kappa'}^2}$ and the SMEFT Wilson coefficients are given by
\begin{align}
\begin{split}
    C^{(5)}\;\;\;\;\,&=\frac{1}{v^2}\left({M_D^{\nu}}^T{M_R^\nu}^{-1}M_D^\nu-M_L^\nu\right)\,,\\
    C^{(7)}_{Leu\overline{d}\Phi}&=\frac{\sqrt{2}}{v}\frac{1}{v_R^2}\left(V_R^{ud}\right)^*\left(M_D^{\nu T}{M_R^\nu}^{-1}\right)_{ee}\,,\qquad\qquad\quad
    C^{(7)}_{L\Phi De}\,=\frac{2i\xi\exp{i\alpha}}{\left(1+\xi^2\right){V_R^{ud}}^*}C^{(7)}_{Leu\overline{d}\Phi}\,,\\
    C^{(9)}_{eeud}\;\;\,&=-\frac{1}{2v_R^4}{V_R^{ud}}^2\left[\left({M_R^\nu}^\dagger\right)^{-1}+\frac{2}{m^2_{\Delta_R}}M_R^\nu\right]\,,\qquad
    C^{(9)}_{ee\Phi ud}\,=-4\frac{\xi\exp{-i\alpha}}{\left(1+\xi^2\right)V_R^{ud}}C^{(9)}_{eeud}\,,\\
    C^{(9)}_{ee\Phi D}\;&=4\frac{\xi^2\exp{-2i\alpha}}{\left(1+\xi^2\right)^2 {V_R^{ud}}^2}C^{(9)}_{eeud}\,.
\end{split}
\end{align}
The Wilson coefficients are completely described by fixing the vevs of the triplet scalars $v_{L,R}$, the heavy right-handed triplet mass $m_{\Delta_R}$, the heavy neutrino masses $m_{\nu_{R,i}}, i=1...3$, the lightest active neutrino mass $m_{\nu_\mathrm{min}}$, the neutrino mixing matrices $U_{L,R}$, the complex vev phases $\theta_L$ and $\alpha$ as well as the left-right mixing parameter $\xi$. 

\begin{figure}[t!]
    \centering
    \includegraphics[width=0.495\textwidth]{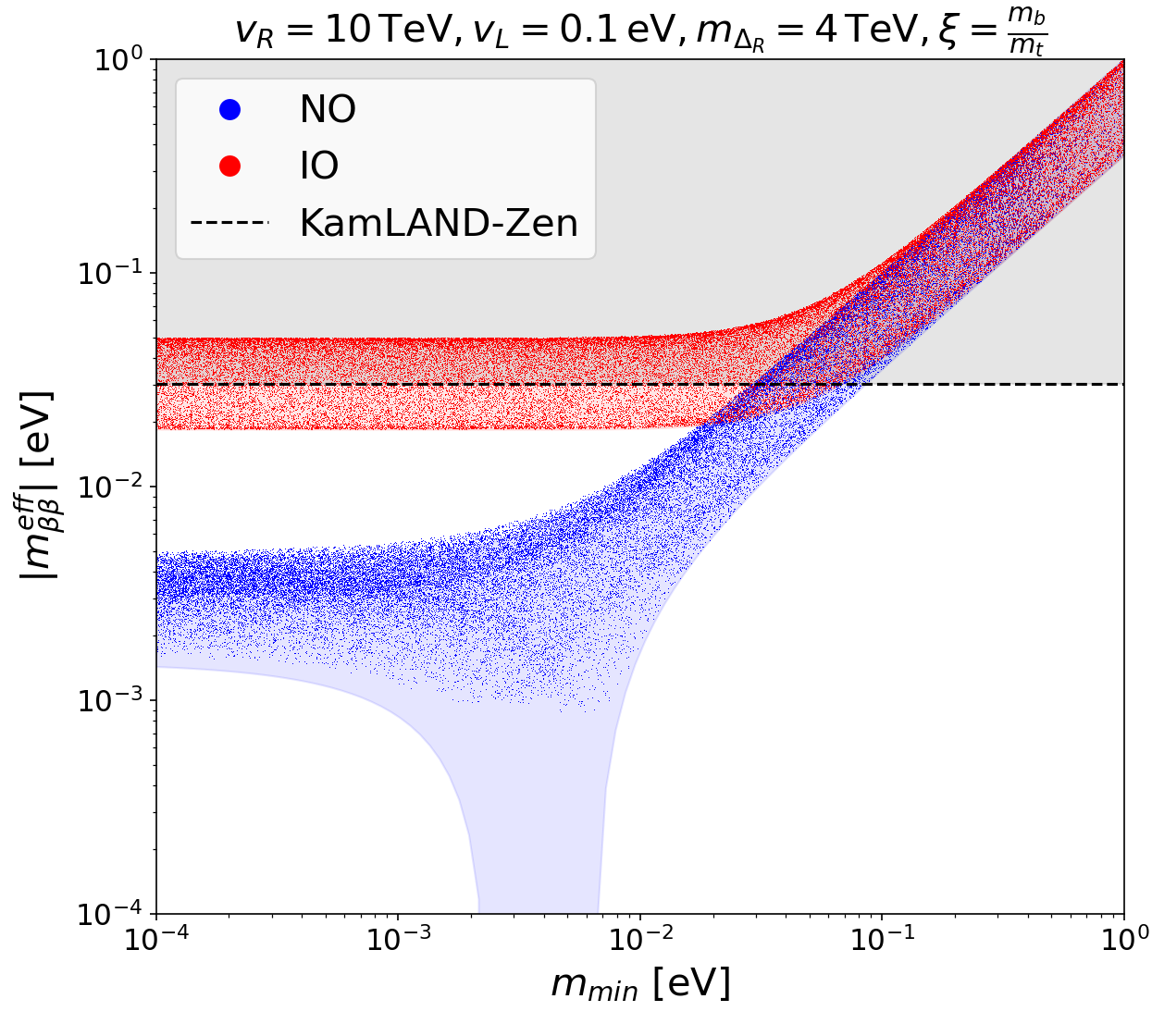}
    \includegraphics[width=0.495\textwidth]{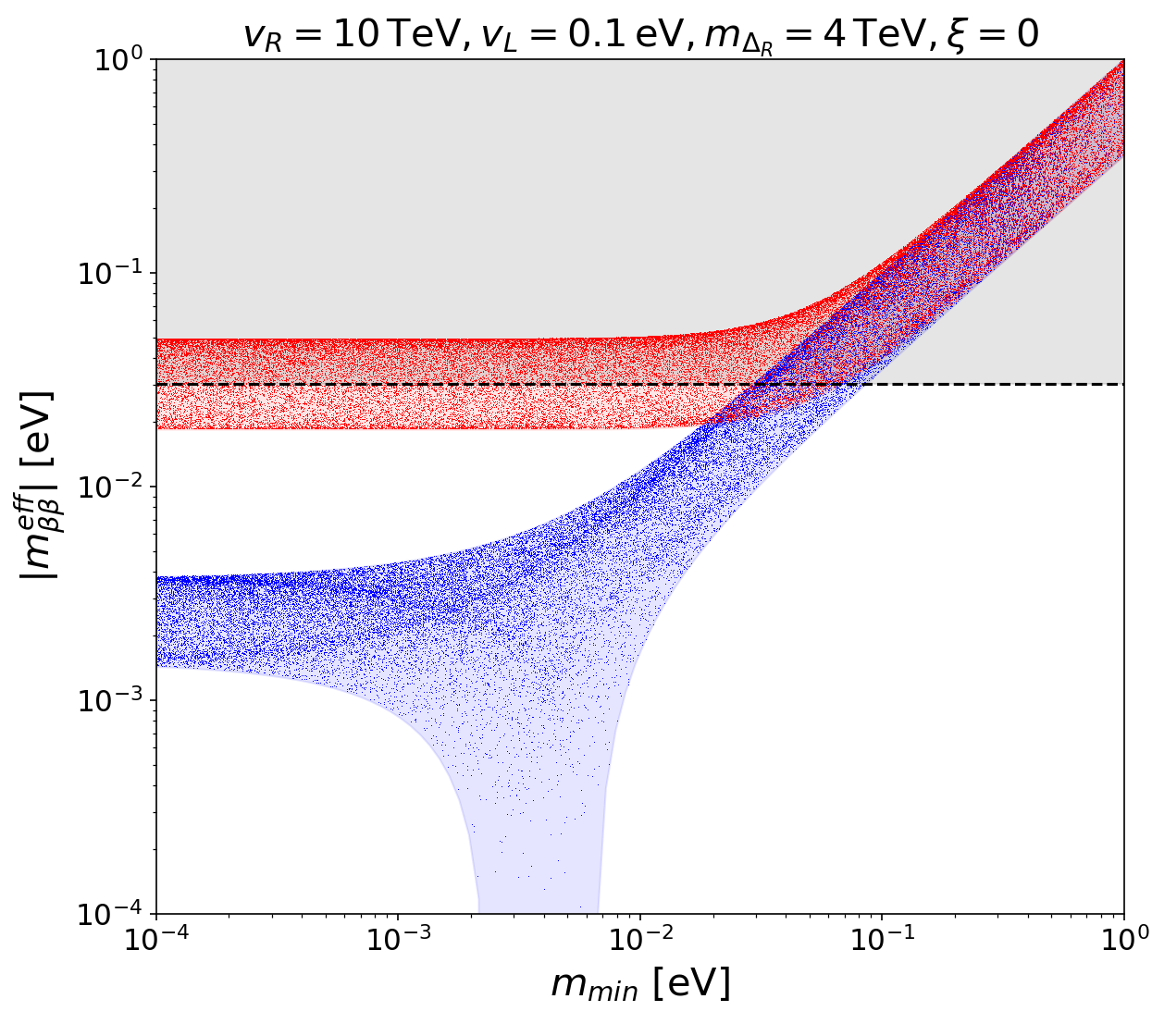}
    \includegraphics[width=0.495\textwidth]{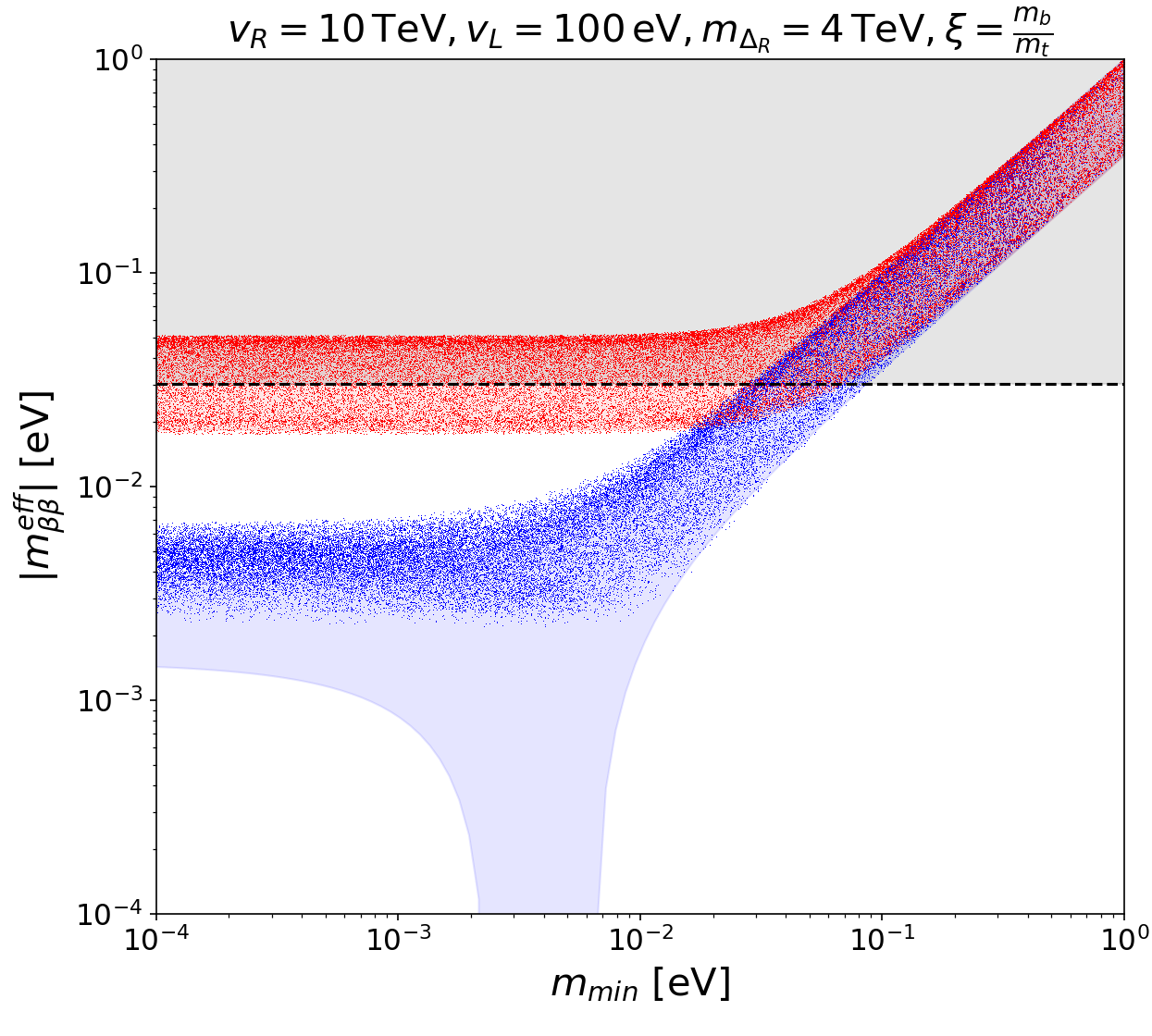}
    \includegraphics[width=0.495\textwidth]{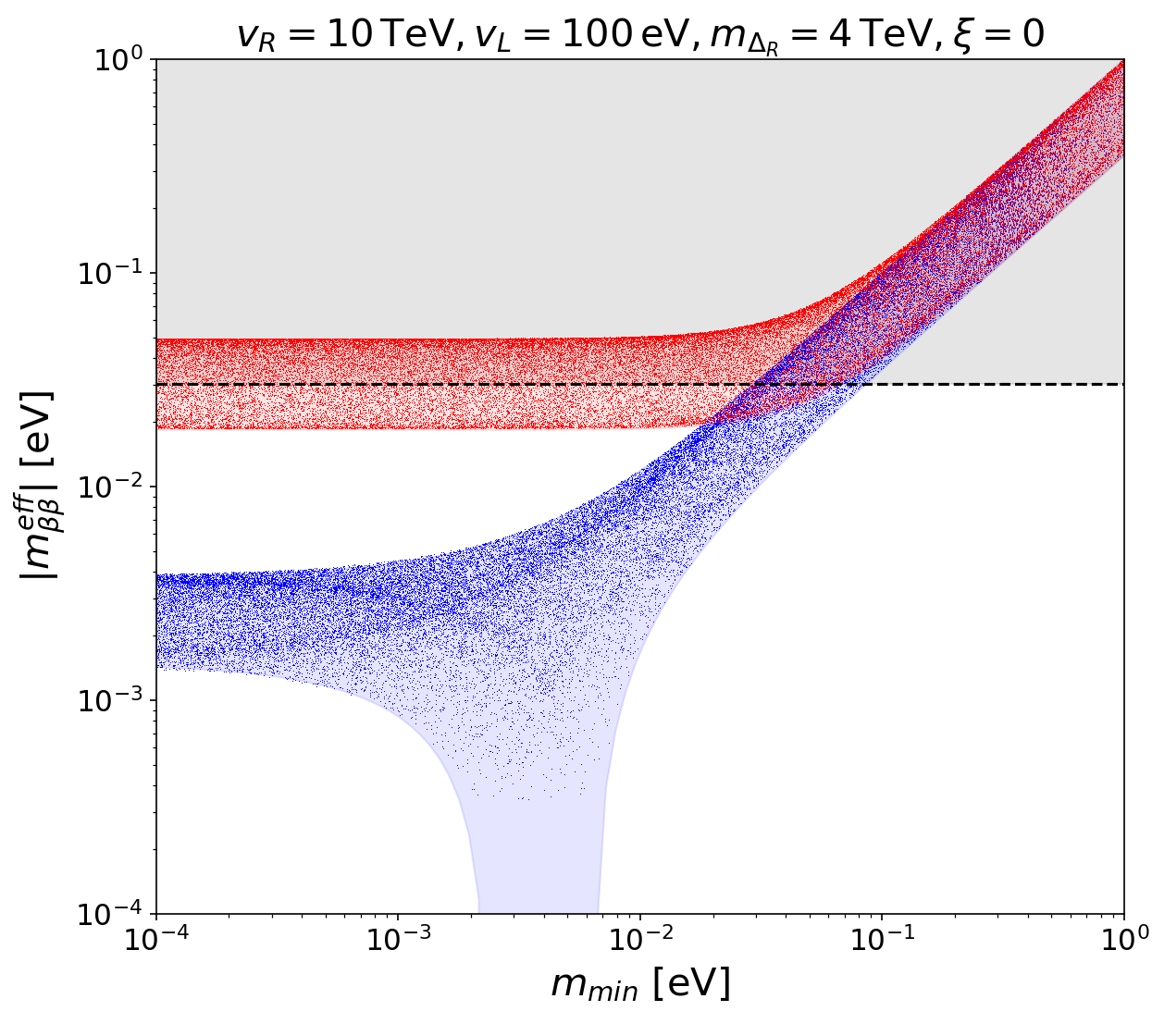}
    \caption{The effective Majorana mass $m^{eff}_{\beta\beta}$ in dependence on the minimal light neutrino mass for the four parameter settings listed in~\eqref{eq:modelsmLRSM}. The shaded areas show the Majorana mass $m_{\beta\beta}$ arising from the light-neutrino-exchange mechanism alone, while the scattered points show the resulting effective mass for the different mLRSM realizations when varying the unknown Majorana phases, $\theta_L$ and $\alpha$. Both the normal (blue) and the inverted (red) mass orderings are displayed. The gray area shows the current limit given by KamLAND-Zen.}
    \label{fig:mLRSMmodels}
\end{figure}

To apply \nudobe, we need to translate these Wilson coefficients to the correct operator basis (see Appendix~\ref{app:operator_list}) via
\begin{align}
    \begin{split}
        &C_{LH}^{(5)}       = \left(C^{(5)}\right)_{ee}\;,\\
        &C_{LeudH}^{(7)}    = C_{Leu\overline{d}\Phi}^{(7)}\;,\\
        &C_{LHDe}^{(7)}     = C_{L\Phi De}^{(7)}\;,\\
        &C_{ddueue}^{(9)}   = 4 \left(C_{eeud}^{(9)}\right)^*\;,\\
        &C_{deueH^2D}^{(9)} = -2 \left(C_{ee\Phi ud}^{(9)}\right)^*\;,\\
        &C_{eeH^4D}^{(9)}   = - \left(C_{ee\Phi D}^{(9)}\right)^*\;.\\
    \end{split}
\end{align}
\begin{figure}[t!]
    \centering
    \includegraphics[width=0.495\textwidth]{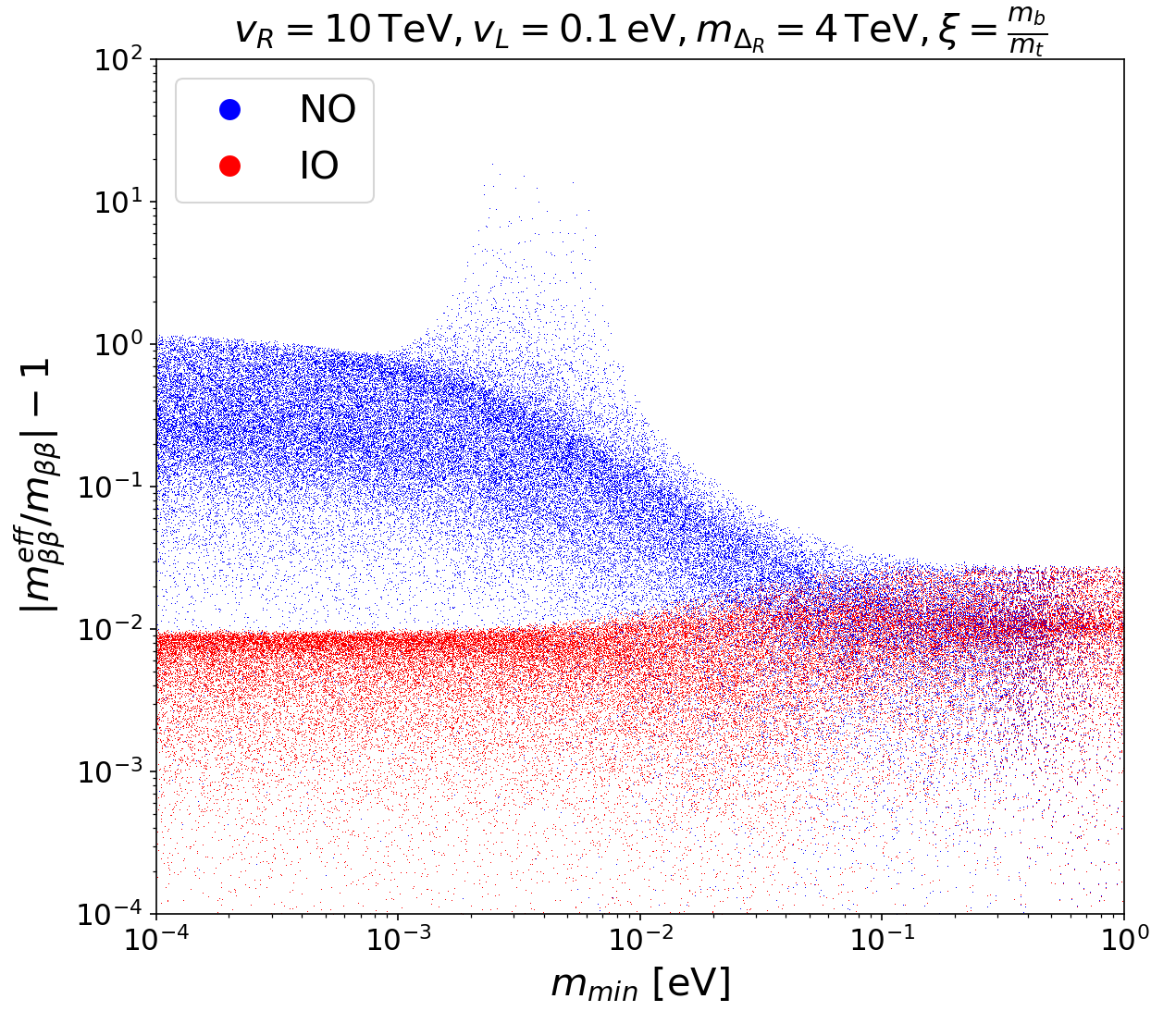}
    \includegraphics[width=0.495\textwidth]{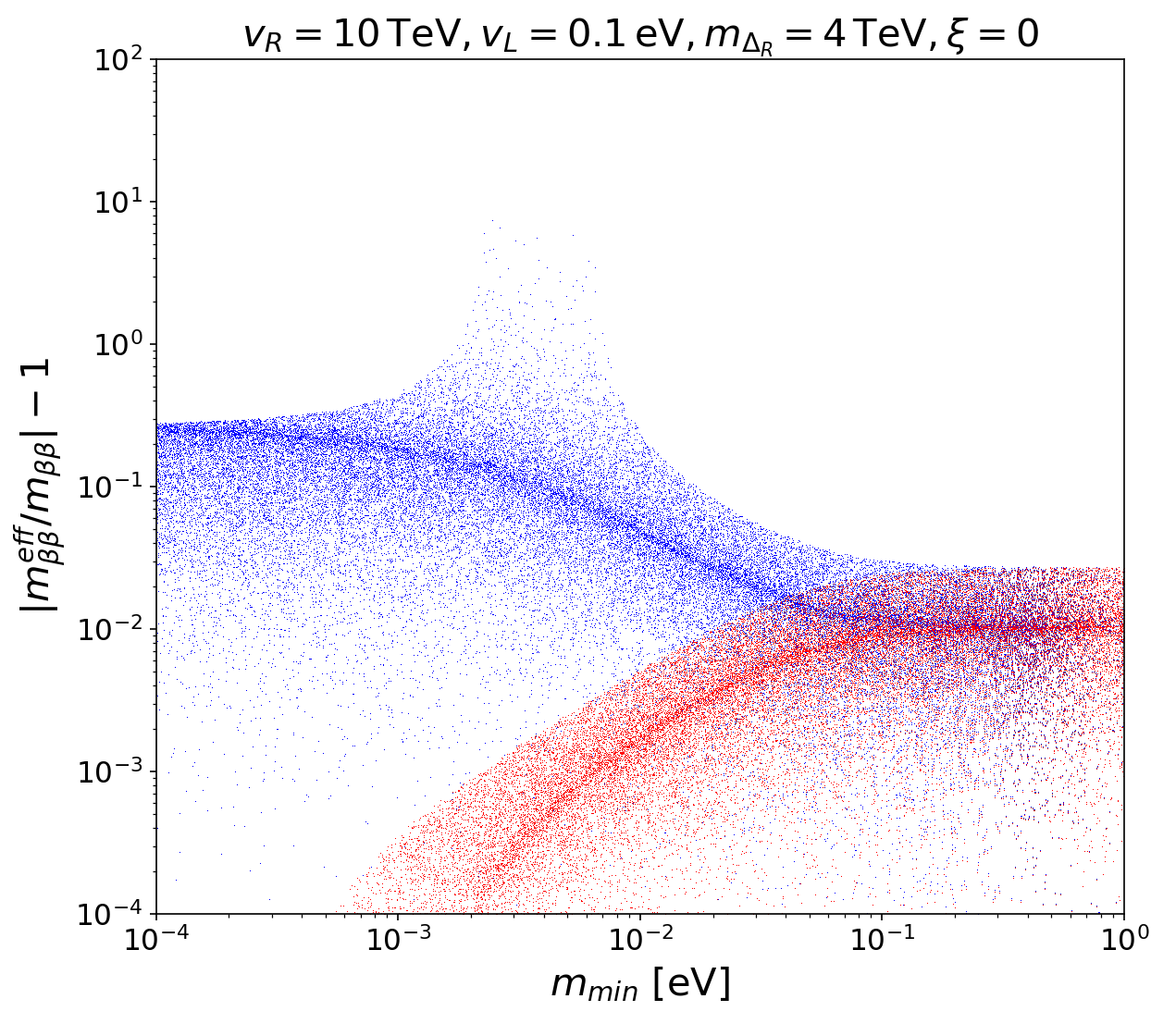}
    \includegraphics[width=0.495\textwidth]{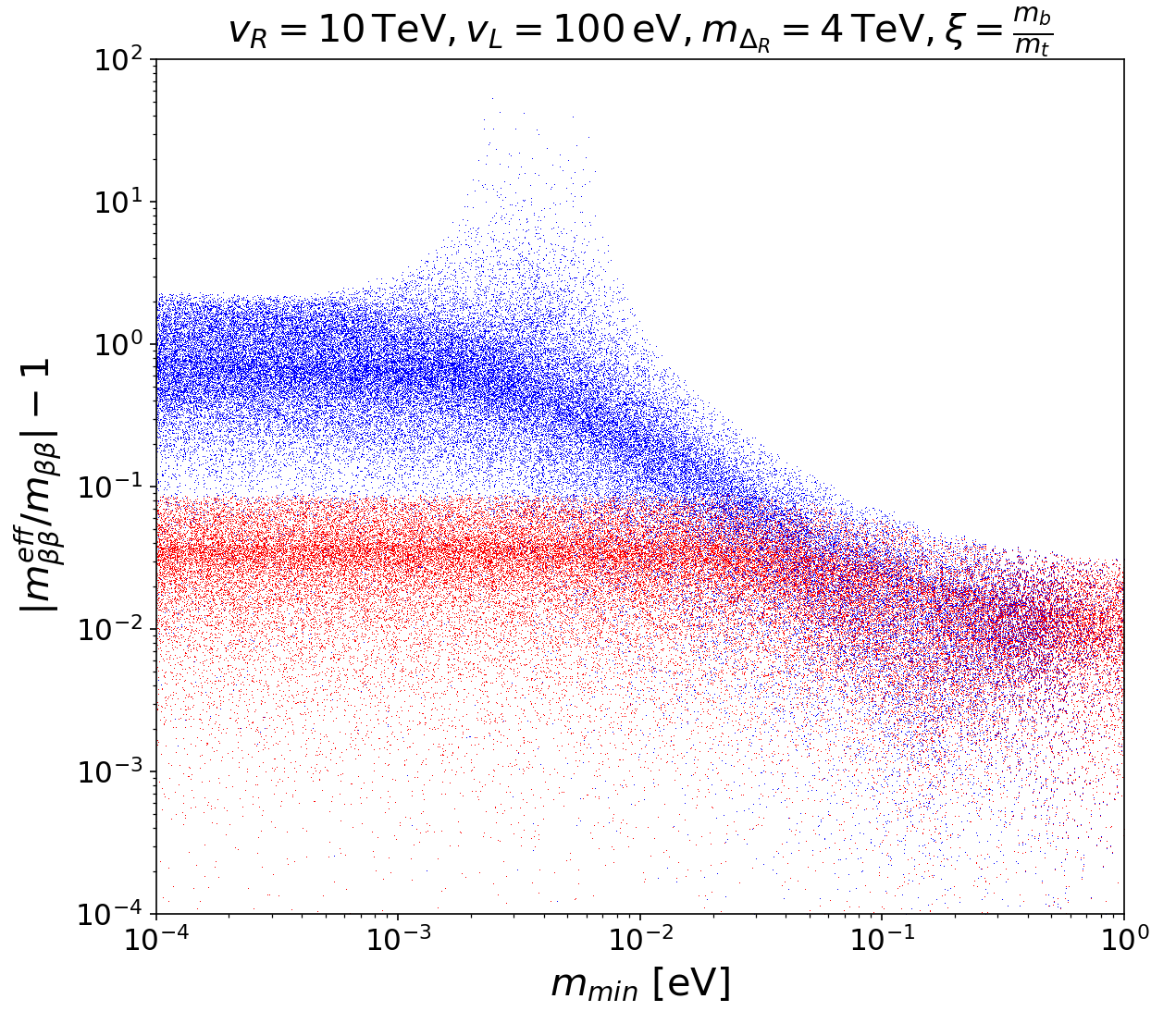}
    \includegraphics[width=0.495\textwidth]{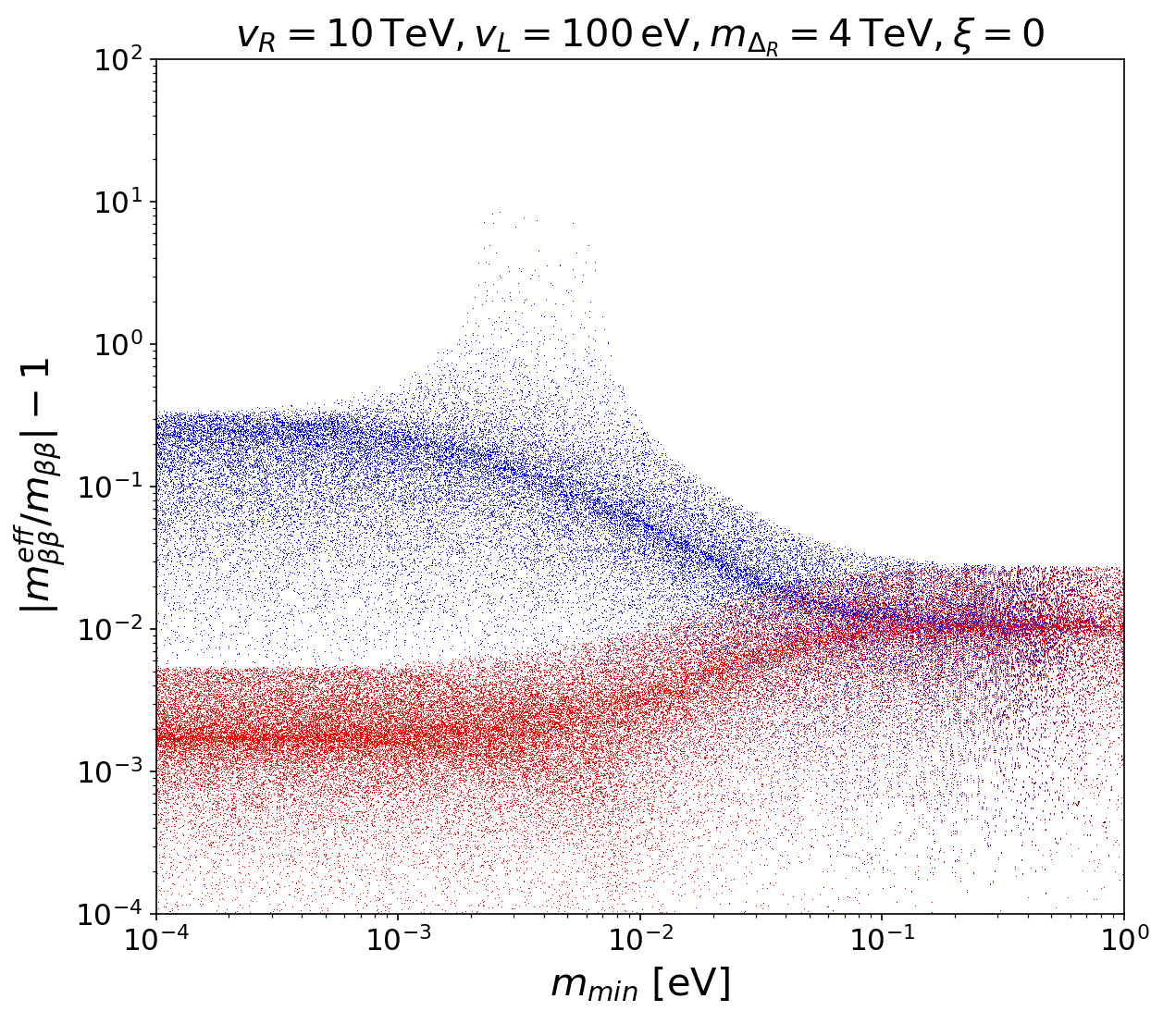}
    \caption{Similar to Figure~\ref{fig:mLRSMmodels} but normalized to the standard mass mechanism induced by $m_{\beta\beta}$.}
    \label{fig:mLRSMmodels_normalized}
\end{figure}
With the help of \nudobe we can use these Wilson coefficients to generate a \hyperlink{nudobe.EFT.SMEFT}{\texttt{SMEFT} model class} to study different parametric scenarios of mLRSM realizations in Python. As an example, we revisit four different realizations of the mLRSM studied in Ref.~\citep{Cirigliano_2018}, given by
\begin{align}
\begin{split}
    &m_{{\nu}_{R1}} = 10\,\textrm{TeV}\;,\quad m_{{\nu}_{R2}}= 12\,\textrm{TeV}\;,\quad m_{{\nu}_{R3}} = 13\,\textrm{TeV}\\
    &m_{\Delta_R} = 4\,\textrm{TeV}\,,\quad v_R=10\,\textrm{TeV}\,,\quad V_{ud}^R=V_{ud}^L\,,\quad U_R = U_L\,,
\end{split}
\end{align}
and by setting the left-handed triplets vev $v_L$ and the ratio of the bidoublet vevs $\xi$ to one of the following values
\begin{align}
\begin{split}
    \mathrm{Model\:1:}&\quad v_L = 0.1\,\mathrm{eV}\,,\quad \xi = \frac{m_b}{m_t}\,,\\
    \mathrm{Model\:2:}&\quad v_L = 0.1\,\mathrm{eV}\,,\quad \xi = 0\,,\\
    \mathrm{Model\:3:}&\quad v_L = 100\,\mathrm{eV}\,,\quad \xi = \frac{m_b}{m_t}\,,\\
    \mathrm{Model\:4:}&\quad v_L = 100\,\mathrm{eV}\,,\quad \xi = 0\,.\label{eq:modelsmLRSM}
\end{split}
\end{align}

In Figure~\ref{fig:mLRSMmodels} we show the evolution of the effective neutrino mass parameter,
\begin{align}
    m^{eff}_{\beta\beta} = \frac{m_e}{g_A^2 V_{ud}^2\mathcal{M}_3^{(\nu)}G_{01}^{1/2}} T_{1/2}^{-1/2},
\end{align}
where $\mathrm{M}_3^{(\nu)}$ is the NME for the light-neutrino-exchange mechanism (L$\nu$EM)(see~\citep{Cirigliano_2018}). Figure~\ref{fig:mLRSMmodels_normalized} shows similar information but now we normalized to $m_{\beta\beta}$, the effective mass in the L$\nu$EM. The scattered points show variations of the unknown Majorana phases in the neutrino mixing matrix as well as $\theta_L$ and $\alpha$. We observe that in all four parameter settings the case of inverted mass ordering is hardly influenced by the additional higher dimensional interactions introduced in the mLRSM. This would change for smaller values of the left-right symmetry breaking scale.
On the contrary, for normal mass ordering the higher dimensional $d\geq 7$ contributions do increase the expected decay rate when compared to the L$\nu$EM. In addition, the case of normal mass ordering does no longer show the typical funnel. 

We refrain from a more detailed analysis of the parameter space as our main goal is to illustrate how $\nu$DoBE can be readily applied for $0\nu\beta\beta$ analyses of complicated models. Again, a detailed Jupyter notebook is accessible via \href{\GitHubLink}{GitHub}.

\subsection{A Leptoquark Mechanism}
\begin{figure*}[h]
    \centering
    \begin{tikzpicture}
      \begin{feynman}
        \vertex (n){\(d\)};
        \vertex[right=1.7cm of n](nnuLQ);
        \vertex[right=1.5cm of nnuLQ](peLQ);
        \vertex[right=1.5cm of peLQ](pemiddle);
        \vertex[above=0.5cm of pemiddle](p){\(u\)};
        \vertex[below=0.5cm of pemiddle](e){\(e^-\)};
        \vertex[below=2cm of nnuLQ](wnue);
        \vertex[right=3cm of wnue] (e2){\(e^-\)};
        \vertex[below=1cm of wnue](npw);
        \vertex[left=1.5cm of npw](n2){\(d\)};
        \vertex[right=3cm of npw](p2){\(u\)};
        \diagram* {
          (n) -- (nnuLQ) -- [edge label=\(\nu\)] (wnue),
          (n2) --  (npw) -- (p2),
          (nnuLQ) -- [boson, edge label=\(V(S)\)] (peLQ) -- (p),
          (peLQ) -- (e),
          (wnue) -- [boson, edge label=\(W\)] (npw),
          (wnue) -- (e2),
          };
          
        \vertex[right=7cm of n](n3){\(d\)};
        \vertex[right=2.45cm of n3](neLQ);
        \vertex[right=2.25cm of neLQ](e3){\(e^-\)};
        \vertex[below=1cm of neLQ](pnu);
        \vertex[right=2.25cm of pnu](p3){\(u\)};
        \vertex[below=1cm of pnu](wnue2);
        \vertex[right=2.25cm of wnue2](e4){\(e^-\)};
        \vertex[below=1cm of wnue2](npw2);
        \vertex[left=2.25cm of npw2](n4){\(d\)};
        \vertex[right=2.25cm of npw2](p4){\(u\)};
        \diagram* {
        (n3) -- (neLQ) -- (e3),
        (neLQ) -- [boson, edge label=\(V(S)\)] (pnu) -- (p3),
        (pnu) -- (wnue2) -- (e4),
        (wnue2) -- [boson, edge label=\(W\)] (npw2) -- (p4),
        (n4) -- (npw2),
          };
      \end{feynman}
    \end{tikzpicture}
    \caption{Feynman diagrams of the vector $(V)$ and scalar $(S)$ leptoquark interactions contributing to \0.}
    \label{fig:feynman_LQ}
\end{figure*}
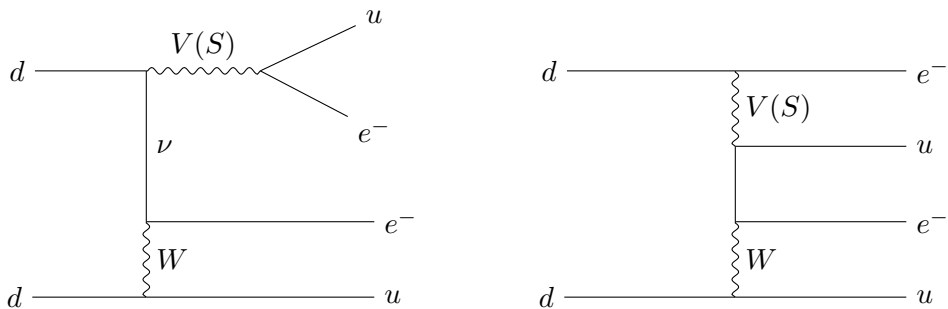
In this section we revisit the \0 decay induced by new leptoquark (LQ) interactions first studied in Ref.~\citep{Hirsch:1996ye}. Assuming the Standard Model without right-handed neutrinos, one can add up to 10 scalar or vector LQs with non-trivial couplings to the Standard Model at tree level. These are summarized in Table~\ref{tab:leptoquarks}.
LQs with $Q=\pm 1/3$ and $Q=\pm 2/3$ then generate new decay channels for \0 decay different from the usually considered L$\nu$EM with the corresponding Feynman diagrams displayed in Figure~\ref{fig:feynman_LQ}. The matching procedure involves a minor subtlety when compared to the previous case study of the mLRSM as the LNV interactions relevant for \0 decay are generated when electroweak symmetry breaking introduces non-diagonal mixing between different LQs. The analysis is more straightforward if one matches the BSM model directly onto LEFT instead of SMEFT.

After doing so, the parts of the low-energy Lagrangian relevant for \0 decay are given by~\citep{Hirsch:1996ye}
\begin{align}
\begin{split}
    \mathcal{L}_{LQ} =& \left[\overline{e}P_L\nu^c\right]\left\{\frac{\epsilon_S}{M^2_S}\left[\overline{u}P_Rd\right] + \frac{\epsilon_V}{M^2_V}\left[\overline{u}P_Ld\right]\right\} \\
    -& \left[\overline{e}\gamma^\mu P_L\nu^c\right]\bigg\{\left(\frac{\alpha_S^R}{M_S^2}+\frac{\alpha_V^R}{M_V^2}\right)\left[\overline{u}\gamma_\mu P_Rd\right] - \sqrt{2}\left(\frac{\alpha_S^L}{M_S^2}+\frac{\alpha_V^L}{M_V^2}\right)\left[\overline{u}\gamma_\mu P_Ld\right]\bigg\} + \text{h.c.}.
    \label{eq:low_energy_LQ_lagrangian}
\end{split}
\end{align}
Hence, we obtain the following LEFT Wilson coefficients
\begin{align}
\begin{split}
    C^{(6)}_{SL} &= \frac{v^2}{M_V^2}\epsilon_V\,,
    \\
    C^{(6)}_{SR} &= \frac{v^2}{M_S^2}\epsilon_S,
    \\C^{(6)}_{VL} &= \sqrt{2}v^2\left(\frac{\alpha_{S}^L}{M_S^2}+\frac{\alpha_{V}^L}{M_V^2}\right)
    \,,
    \\
    C^{(6)}_{VR} &= -v^2\left(\frac{\alpha_{S}^R}{M_S^2}+\frac{\alpha_{V}^R}{M_V^2}\right)\,.
\end{split}
\end{align}
\begin{table}[t!]
    \centering
    \begin{tabular}{l|c c c c}
    \hline\hline
        LQ $(\Omega)$ & $SU(3)_C$ & $SU(2)_L$ & $U(1)_Y$ & Q \\
         \hline
        $S_0$ & 3 & 1 & -2/3 & -1/3 \\
        $\Tilde{S}_0$ & 3 & 1 & -8/3 & -4/3 \\
        $S_{1/2}$ & $\overline{3}$ & 2 & -7/3 & $\left(-2/3, -5/3\right)$ \\
        $\Tilde{S}_{1/2}$ & $\overline{3}$ & 2 & -1/3 & $\left(1/3, -2/3\right)$ \\
        $S_1$ & 3 & 3 & -2/3 & $\left(2/3, -1/3, -4/3\right)$ \\
        $V_0$ & $\overline{3}$ & 1 & -4/3 & -2/3 \\
        $\Tilde{V}_0$ & $\overline{3}$ & 1 & -10/3 & -5/3 \\
        $V_{1/2}$ & 3 & 2 & -5/3 & $\left(-1/3, -4/3\right)$ \\
        $\Tilde{V}_{1/2}$ & 3 & 2 & 1/3 & $\left(2/3, -1/3\right)$ \\
        $V_1$ & $\overline{3}$ & 3 & -4/3 & $\left(1/3, -2/3, -5/3\right)$\\
        \hline\hline
    \end{tabular}
    \caption{List of possible scalar and vector leptoquarks and their transformation properties under the Standard Model symmetries~\citep{Hirsch:1996ye}.}
    \label{tab:leptoquarks}
\end{table}

Now we can use the \hyperlink{functions.limits_LEFT}{\texttt{get\_limits\_LEFT()}} function of \nudobe to evaluate the limits on the parameters $\epsilon_{i}, \alpha_{i}^{S,V}, i\in[L,R]$ imposed by the recent KamLAND-Zen results when assuming only one parameter at a time to be non-vanishing. In doing so, using the IBM2 NMEs~\cite{Deppisch:2020ztt} provided with \nudobe we find that
\begin{align}
\begin{split}
    \epsilon_i &\lesssim 5.80\times10^{-9}\left(\frac{M_i}{1\,\mathrm{TeV}}\right)^2\,,\\
    \alpha_i^L &\lesssim 1.81\times10^{-8}\left(\frac{M_i}{1\,\mathrm{TeV}}\right)^2\,,\\
    \alpha_i^R &\lesssim 1.86\times10^{-6}\left(\frac{M_i}{1\,\mathrm{TeV}}\right)^2\,.
\end{split}
\end{align}

In comparison with the limits obtained in the original work \cite{Hirsch:1996ye} the above bounds are somewhat more stringent, which is mainly due the improved KamLAND-Zen results~\cite{KamLAND-Zen:2022tow}. This aspect is partially compensated by using a different set of NMES and PSFs reflecting the improved status of theoretical calculations. A more detailed comparison with Ref.~\citep{Hirsch:1996ye} can be found in the provided Jupyter notebook.

\section{Conclusion \& Outlook}

Neutrinoless double beta decay experiments provide the most stringent tests of lepton number violation. The experimental prospects are excellent and it will play an important role in the resolution of the neutrino mass puzzle and the search for beyond-the-Standard-Model physics in general. 

That being said,  $0\nu\beta\beta$ is a complicated process involving particle, nuclear, and atomic physics and its interpretation requires a great care. The main goal of \nudobe is to help the community with these difficulties. Being a low-energy process, effective field theory methods can be used efficiently to describe $0\nu\beta\beta$ for a large class of LNV scenarios stemming from possibly rich physics living at UV scales. \nudobe computes $0\nu\beta\beta$ (differential) decay rates in terms of Wilson coefficients of LNV operators constructed within the Standard Model Effective Field Theory (SMEFT). It uses the state-of-the-art renormalization-group evolution factors and QCD, nuclear, and atomic matrix elements and comes with a large number of built-in plotting and analysis tools. Naturally, \nudobe can help to easily extract limits on the $m_{\beta\beta}$ parameter associated with the L$\nu$EM, but also on all the other beyond-the-Standard-Model Wilson coefficients of interest, imposed by current and upcoming experimental results. We hope it will be used by the community to quickly and accurately analyze $0\nu\beta\beta$ predictions and constraints related to beyond-the-Standard-Model models involving lepton number violation. We encourage users of \nudobe to share their BSM analyses with the community by adding a Jupyter notebook to \nudobe's \href{\GitHubLink}{GitHub}.

In the upcoming updates we aim to extend \nudobe in several directions, some of which are work in progress. Most importantly, we will include the possibility of relatively light sterile neutrinos, which means addition of effective operators arising in the SMEFT framework extended by three right-handed singlet neutrinos, so called $\nu$SMEFT. Several works have computed the impact of light sterile neutrinos on \0 observables, but the analysis is complicated and essentially done on a case-by-case basis \cite{Blennow:2010th,Barea:2015zfa,Asaka:2016zib,Dekens:2020ttz, Li:2020flq,deVries:2022nyh,Dekens:2023iyc}. An automated tool would be very helpful in this regard. In addition to the currently available approximate calculations of the phase-space factors described in Appendix~\ref{app:PSFs} we plan to add the option of employing the exact numerical solutions for the radial electron wave functions using a broader variety of nuclear potentials and including other subtle effects such as the electron screening along with and beyond the treatments adopted in Refs.~\cite{Kotila:2012zza,Stoica:2013lka}. Last but not least, we envision to incorporate predictions for and constraints from other lepton-number-violating processes, such as kaon decays.

\subsection{Online Tool}
Finally, for quick analyses we created an online \href{\UILink}{user-interface} using Streamlit. The online user-interface is aimed towards delivering an easy and fast to use possibility for the most relevant use-cases like, e.g., calculating the decay observables (half-life, spectra, angular correlation) from a given model or studying limits on the different Wilson coefficients given half-life limits from experiments.

\section*{Acknowledgements}
The authors are grateful to Vaisakh Plakkot, Jacob Spisak and Nele Volmer for testing the code and to Wouter Dekens and Xiao-Dong Ma for helpful discussions and clarifications of encountered discrepancies. 
LG acknowledges support from the National Science Foundation, Grant PHY-1630782, and the Heising-Simons Foundation, Grant 2017-228. JdV acknowledges support from the Dutch Research Council (NWO) in the form of a VIDI grant. 

\newpage
\section{Documentation: \nudobe - A Python Tool for Neutrinoless Double Beta Decay}
In the following we provide a detailed description of all the different functionalities of \nudobe including various explicit code examples.
\subsection{List of Functions and Classes}
\begin{enumerate}
    \item {nudobe.EFT:}
    \begin{enumerate}
        \item \hyperlink{nudobe.EFT.SMEFT}{\texttt{EFT.SMEFT}}
        \item \hyperlink{nudobe.EFT.LEFT}{\texttt{EFT.LEFT}}
    \end{enumerate}
    \item {nudobe.EFT.model: (model = SMEFT or LEFT)}
    \begin{enumerate}
        \item \hyperlink{model.run}{\texttt{model.run}}
        \item \hyperlink{model.half_lives}{\texttt{model.half\_lives}}
        \item \hyperlink{model.t_half}{\texttt{model.t\_half}}
        \item \hyperlink{model.generate_formula}{\texttt{model.generate\_formula}}
        \item \hyperlink{model.generate_matrix}{\texttt{model.generate\_matrix}}
        \item \hyperlink{model.plot_WC_variation}{\texttt{model.plot\_WC\_variation}}
        \item \hyperlink{model.plot_t_half}{\texttt{model.plot\_t\_half}}
        \item \hyperlink{model.plot_t_half_inv}{\texttt{model.plot\_t\_half\_inv}}
        \item \hyperlink{model.plot_m_bb}{\texttt{model.plot\_m\_eff}}
        \item \hyperlink{model.plot_WC_variation_scatter}{\texttt{model.plot\_WC\_variation\_scatter}}
        \item \hyperlink{model.plot_t_half_scatter}{\texttt{model.plot\_t\_half\_scatter}}
        \item \hyperlink{model.plot_t_half_inv_scatter}{\texttt{model.plot\_t\_half\_inv\_scatter}}
        \item \hyperlink{model.plot_m_bb_scatter}{\texttt{model.plot\_m\_bb\_scatter}}
        \item \hyperlink{model.ratios}{\texttt{model.ratios}}
        \item \hyperlink{model.plot_ratios}{\texttt{model.plot\_ratios}}
        \item \hyperlink{model.spectrum}{\texttt{model.spectrum}}
        \item \hyperlink{model.angular_corr}{\texttt{model.angular\_corr}}
        \item \hyperlink{model.plot_spec}{\texttt{model.plot\_spec}}
        \item \hyperlink{model.plot_corr}{\texttt{model.plot\_corr}}
        \item \hyperlink{model.get_limits}{\texttt{model.get\_limits}}
    \end{enumerate}
    \item {nudobe.functions:}
    \begin{enumerate}
        \item \hyperlink{functions.generate_formula}{\texttt{functions.generate\_formula}}
        \item \hyperlink{functions.generate_matrix}{\texttt{functions.generate\_matrix}}
        \item \hyperlink{functions.limits_LEFT}{\texttt{functions.get\_limits\_LEFT}}
        \item \hyperlink{functions.limits_SMEFT}{\texttt{functions.get\_limits\_SMEFT}}
        \item \hyperlink{functions.get_contours}{\texttt{functions.get\_contours}}
    \end{enumerate}
    \item {nudobe.plots:}
    \begin{enumerate}
        \item \hyperlink{plots.limits_LEFT}{\texttt{plots.limits\_LEFT}}
        \item \hyperlink{plots.limits_SMEFT}{\texttt{plots.limits\_SMEFT}}
        \item \hyperlink{plots.contours}{\texttt{plots.contours}}
    \end{enumerate}
\end{enumerate}

\subsection{Installation}
%
%
\subsubsection{Requirements}
Using \nudobe requires the following packages to be installed (the versions we used are given in [brackets], though more recent version should work, too.):
\begin{enumerate}
    \item \texttt{NumPy}~\citep{numpy} [v. 1.19.2]
    \item \texttt{Pandas}~\citep{pandas_software, pandas_paper} [v. 1.1.3]
    \item \texttt{Matplotlib}~\citep{matplotlib} [v. 3.3.2]
    \item \texttt{SciPy}~\citep{scipy} [v. 1.5.2]
    \item \texttt{mpmath}~\citep{mpmath} [v. 1.1.0] (only necessary if both PSF schemes are required)
\end{enumerate}
Additionally, Python 3 is required while 3.6 or higher is recommended. If you use Python 2 the code will give wrong results! Besides internal usage of pandas, \nudobe generally uses pandas' \href{https://pandas.pydata.org/docs/reference/api/pandas.DataFrame.html}{DataFrame} class to output results in a table format. Pandas DataFrames provide a convenient framework for further analyses similar to numpy with additional features like, e.g., latex exports or plotting functions.

To ensure that there are no conflicts with other third-party python modules we recommend setting up a dedicated virtual environment when using \nudobe.
\subsubsection{Setting up \nudobe}
To use \nudobe simply download the code from \href{\GitHubLink}{GitHub} and copy it into your project's main directory.
Your projects directory structure should look something like this:
\begin{figure}[h!]
    \centering
        \begin{forest}
        for tree={font=\sffamily, grow'=0,
        folder indent=.9em, folder icons,
        edge=densely dotted}
        [MyProjectDirectory
          [MyProject.py, is file]
          [MyProjectNotebook.ipynb, is file]
          [nudobe,
              [\_\_init\_\_.py, is file]
              [ExampleNotebooks]
              [NMEs]
              [PSFs]
              [src]]
        ]
        \end{forest}
    \caption{An example of how you should include \nudobe in your project. The \texttt{src} folder contains all the \nudobe sub-modules, the \texttt{NMEs} and \texttt{PSFs} folders contain the different sets of phase-space factors and nuclear matrix elements stored as \texttt{.csv} files and the \texttt{ExampleNotebooks} folder contains some \href{https://jupyter.org}{Jupyter notebook} examples of how to use \nudobe.}
    \label{fig:directory_tree}
\end{figure}
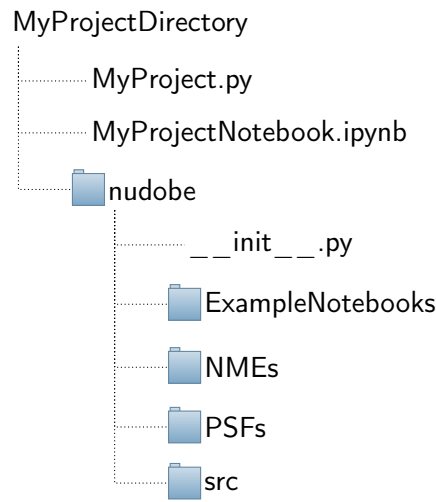

%
%
\subsection{Parameters and Constants}
\subsubsection{Physical Constants}
Physical constants such as particles masses and mixing angles are defined in the \texttt{constants.py} file and can be changed if that is required.
\subsubsection{Units}
Generally, all dimensionful parameters in \nudobe are defined in GeV, i.e., we set $\mathrm{GeV}=1$, numerically. The only exception to this is the minimal neutrino mass in plotting functions which is taken in eV. To make it easier to not get confused with units, the numerical values of different energy units can be accessed in the \texttt{nudobe.constants} module.
    \begin{tcolorbox}[breakable, size=fbox, boxrule=1pt, pad at break*=1mm,colback=cellbackground, colframe=cellborder]
\prompt{In}{incolor}{1}{\boxspacing}
\begin{Verbatim}[commandchars=\\\{\}]
\PY{k+kn}{from} \PY{n+nn}{nudobe} \PY{k+kn}{import} \PY{n}{constants}
\PY{k+kn}{from} \PY{n+nn}{constants} \PY{k+kn}{import} \PY{o}{*}
\end{Verbatim}
\end{tcolorbox}

    \begin{tcolorbox}[breakable, size=fbox, boxrule=1pt, pad at break*=1mm,colback=cellbackground, colframe=cellborder]
\prompt{In}{incolor}{2}{\boxspacing}
\begin{Verbatim}[commandchars=\\\{\}]
\PY{c+c1}{\PYZsh{}We can define a scale Lambda = 100TeV directly as}
\PY{n}{Lambda} \PY{o}{=} \PY{l+m+mf}{1e+5}
\PY{n+nb}{print}\PY{p}{(}\PY{n}{Lambda}\PY{p}{)}

\PY{c+c1}{\PYZsh{}or by using the TeV parameter defined in constants}
\PY{n}{Lambda} \PY{o}{=} \PY{l+m+mi}{100}\PY{o}{*}\PY{n}{TeV}
\PY{n+nb}{print}\PY{p}{(}\PY{n}{Lambda}\PY{p}{)}
\end{Verbatim}
\end{tcolorbox}

    \begin{Verbatim}[commandchars=\\\{\}]
100000.0
100000.0
    \end{Verbatim}
\nudobe includes parameters for the energy units of \texttt{TeV, GeV, MeV, keV, eV} and \texttt{meV}.
\subsubsection{Nuclear Matrix Elements}
\0 decay rates depend on a set of nuclear matrix elements that involve complicated nuclear structure calculations. As of today, the NMEs are computed within different nuclear many-body methods and the results, unfortunately, tend to differ. $\nu$DoBe comes with three sets of NMEs that one can choose from; the \textit{interacting boson model 2} (\str{IBM2})~\citep{Deppisch:2020ztt}, the \textit{quasi random-phase approximation} (\str{QRPA})~\citep{Hyvarinen:2015bda} and the \textit{shell model} (\str{SM})~\citep{Menendez:2017fdf}. These particular sets of NMEs are chosen because within the corresponding method all NMEs required to analyze \0 from dimension-9 SMEFT operators have been computed. The corresponding files are stored in the \texttt{NMEs/} folder as \texttt{.csv} files. 
Other NME approximation methods can be studied by, simply, adding a new \texttt{.csv} file to the \texttt{NMEs/} folder. If you do so, please make sure to follow the definitions of the NMEs described in appendix~\ref{app:NMEs}.
\\\\
Within \nudobe there are two possibilities to set NME methods:\\\\
1. You can set the method right at the start when initiating a model as will be shown in the following section. Every calculation you will do within this model will then use the corresponding NMEs. \\\\
2. Alternatively, most functions have a parameter called \texttt{method} that will reset the NMEs temporarily when calling the function (see the functions definitions).
\\\\
Additionally, within the model classes the NMEs can be accessed and changed in the \texttt{model.NMEs} dictionary.
\subsubsection{Phase-Space Factors}
\nudobe includes two approximation schemes for the calculation of PSFs and electron wave functions~\citep{Stefanik:2015twa}. The PSF-scheme is defined when initiating a model class via the \texttt{PSF\_scheme} parameter. Scheme \str{A} includes PSFs calculated from approximate wave functions in a uniform charge distribution while scheme \str{B} gives PSFs calculated exactly assuming a point-like nucleus.
The Phase-Space factors for the approximation schemes A and B are stored in the \texttt{PSFs/} folder as \texttt{PSFs\_A.csv} and \texttt{PSFs\_B.csv}. 
Again, similar to the NMEs you can use custom PSFs by replacing entries in the \texttt{.csv} files accordingly. Note, however, that this will only change the overall magnitude of the PSFs while the spectra and angular correlations are still calculated from the electron wave functions defined in the \texttt{PSFs.py} file and by the choice of the PSF-scheme. Additional methods for calculating the electron wave functions will be added in a future update. Note, however, that while different approximation schemes do have a noticeable effect on the magnitude of the PSFs, the general shape of the electron spectra and angular correlation is not expected to be influenced largely by a different choice of wave functions~\citep{Stefanik:2015twa}. Within a model class, you can also access and change the PSFs by replacing the corresponding entries in the \texttt{model.PSFpanda} DataFrame.
\subsubsection{Low Energy Constants}
The choice of low energy constants (LECs) strongly influences the resulting half-lives. The values of the known LECs as well as order of magnitude estimates have been summarized in~\citep{Cirigliano_2018} and are shown in Table~\ref{Tab:LECs} for convenience. 

The preset values of the LECs are defined in the \texttt{constants.py} file. Global changes of the LECs are best taken care of in this file. Locally, within the model classes we have implemented two generic settings that can be applied when doing calculations.\\\\
\textbf{1.} In the first setting we use the known LECs and set the unknown LECs to null except keeping $g_{6,7}^{NN}=g_V^{\pi N}=\Tilde{g}_V^{\pi N}=1$ such that we still allow for short-range vector operators in the LEFT framework to contribute to the overall amplitude. This setting can be applied by calling \texttt{unknown\_LECs = \bool{False}} when initiating a model.\\\\
\textbf{2.} In the second setting we take the unknown LECs to be equal to their positive order of magnitude estimate. This option can be chosen by applying \texttt{unknown\_LECs = \bool{True}}.
\begin{table}
\center
\begin{tabular}{|c|cc||c|cc|}
\hline
 \multicolumn{3}{|c||}{ $n\rightarrow pe\nu$, $\pi \rightarrow e \nu$ } &  \multicolumn{3}{c|}{$\pi \pi \rightarrow e e$} \\
 \hline
 $g_A$ & $1.271\pm 0.002$ & \cite{ParticleDataGroup:2016lqr}      & $g^{\pi\pi}_{1}$   		& $  0.36 \pm 0.02 $             & \cite{Nicholson:2018mwc}  \\
 $g_S$ & $0.97\pm 0.13$ & \cite{Bhattacharya:2016zcn} & $g^{\pi\pi}_{2}$   		& $  2.0  \pm 0.2 $  \, GeV$^2$  & \cite{Nicholson:2018mwc}  \\
 $g_M$ &  $ 4.7$ &  \cite{ParticleDataGroup:2016lqr}              & $g^{\pi\pi}_{3}$ 	        & $ -(0.62 \pm 0.06)$  \, GeV$^2$  & \cite{Nicholson:2018mwc}  \\
 $g_T$ & $0.99\pm 0.06$ & \cite{Bhattacharya:2016zcn} & $g^{\pi\pi}_{4}$   		& $ -(1.9  \pm 0.2)$   \, GeV$^2$  & \cite{Nicholson:2018mwc}\\  
 $|g'_{T}|$ & $\mathcal O(1)$  	&		      & $g^{\pi\pi}_{5}$ 		& $ -(8.0  \pm 0.6)$   \, GeV$^2$  & \cite{Nicholson:2018mwc}  \\
 $B$      &    $2.7$~GeV       &     & $|g^{\pi\pi}_{ \rm T}|$     & $\mathcal O(1)$  & \\ 
 \hline
  \multicolumn{3}{|c||}{$n \rightarrow p\pi ee$} & \multicolumn{3}{c|}{$nn\rightarrow pp\, ee$}   
 \\
  \hline 
   $|g^{\pi N}_{1} |$       & $\mathcal{O}(1)$ &    & $|g^{N N}_1|$         & $\mathcal{O}(1)$ &  \\  
   $|g^{\pi N}_{6,7,8,9}|$ & $\mathcal{O}(1)$ &    & $|g_{6,7}^{N N}|$ & $\mathcal{O}(1)$ & \\
     $|g^{\pi N}_{\rm VL}|$ & $\mathcal{O}(1)$ &    &$ |g^{NN}_{\rm VL}|$ & $ \mathcal{O}(1) $ &     \\ 
           $|g^{\pi N}_{\rm T}|$ 			   & 		$ \Or(1)$     &    &
        $ |g_{\rm T}^{NN}|$ &$\Or(1)$&   
  \\ 
			   & 		      &    & $ |g^{NN}_{\nu}|$ & $ -92.9\,\mathrm{GeV}^{-2} \pm 50\%$ &   \citep{Cirigliano:2020dmx,Cirigliano:2021qko,Wirth:2021pij}  \\ 
			   &&&$ |g_{VL,VR}^{E,m_e}|$ &$\Or(1)$&\\
			   			   &&&            $ |g^{NN}_{2,3,4,5}|$ & $ \mathcal{O}((4\pi)^2) $ &  
  \\\hline
\end{tabular}
\caption{The low-energy constants as used in \nudobe. The table is taken from~\citep{Cirigliano_2018} with an updated $g_\nu^{NN}$. When varying LECs \nudobe uses a 50\% uncertainty for $g_\nu^{NN}$. Additionally, the unknown LECs are varied within their order of magnitude estimates (i.e. in the range $\pm [1/\sqrt{10}, \sqrt{10}] \times \mathcal{O}(|g|)$ while all other LECs are kept constant.
}\label{Tab:LECs}
\end{table}
\\\\
You can also manually adjust each LEC. For each initiated model class, the LECs are stored within a dictionary. You can simply change the values for each LEC by replacing the corresponding values in the \texttt{model.LEC} dict.

We stress that the value of short-distance LECs, such as $g_\nu^{NN}$ are only meaningful within a given renormalization scheme that also affects the corresponding nuclear forces and thus the long- and short-distance NMEs. Refs.~\citep{Cirigliano:2020dmx,Cirigliano:2021qko} outlined a strategy how this can be achieved. The idea is that the amplitude of the process $nn \rightarrow pp + ee$, $A_\nu$, is observable and should therefore be regulator independent. While the process itself cannot be measured in any practical way, the amplitude has been computed in~\citep{Cirigliano:2020dmx,Cirigliano:2021qko} by relating it to momentum integral of a known kernel (proportional to the neutrino propagator) multiplied by the generalized forward Compton scattering amplitude $nnW^+ \rightarrow pp W^-$. The low- and high-energy regime of this integral can be described model independently through, respectively, chiral EFT and the operator product expansion for perturbative QCD. The full amplitude is then obtained by interpolating between the two regimes using appropriate form factors for single-nucleon and nucleon-nucleon interactions.  Once $A_\nu$ is obtained at some kinematic point, it becomes possible to fix $g_\nu^{NN}$ for an EFT of nucleon-nucleon interactions using any regularization scheme. Crucially, the value of $g_\nu^{NN}$ is not fixed but depends on the applied EFT and the associated regularization scheme, however, the observable amplitude $A_\nu$ is always correctly described. This procedure was followed in Ref.~\citep{Wirth:2021pij} which applied chiral EFT nucleon-nucleon interactions to extract $g_\nu^{NN}$ and then performed ab initio computations of the NMEs related to $0\nu\beta\beta$ of \textsuperscript{48}Ca. The value of $g_\nu^{NN}$ in the table is the value taken  from Ref.~\citep{Wirth:2021pij}. However, strictly speaking this value is not compatible with the NME sets of $\nu$DoBe for heavier isotopes which were obtained with different many-body methods. For that reason we have assigned a $50\%$ uncertainty on the value of $g_\nu^{NN}$. 

%
%
\subsection{Setting Up a Model}
The goal of the tool is to directly connect BSM models that contain additional LNV sources to $0\nu\beta\beta$ decay rates and electron-kinematics measured in experiments. The main task of the user (``you'', from now on) is to provide the matching relations between the specific BSM model to the LNV SMEFT or LEFT operators. Currently, the tool contains all $\Delta L=2$ operators relevant for \0 involving first-generations quarks and leptons up-to-and-including dim-9 operators (all $\Delta L=2$ SMEFT operators have odd dimension \cite{Kobach:2016ami}). The basis of dim-5, dim-7, and dim-9 SMEFT operators has been derived in Refs.~\cite{Weinberg:1979sa,Lehman:2014jma,Liao:2020jmn,Li:2020xlh} and we follow the notation of these references. A full list of the relevant operators is given in Appendix \ref{app:operator_list}.

%
%
\subsubsection{EFT Model Classes}
The code allows you to set up an EFT model consisting of LNV operators that trigger $0\nu\beta\beta$ decay. You can set up models either above the scale of electroweak symmetry breaking (EWSB) as a SMEFT model or below EWSB as a LEFT model. The most important classes are\\\\
\begin{tabularx}{\textwidth}{llX}
    \hline\hline\\
    \multicolumn{3}{l}{\hypertarget{nudobe.EFT.SMEFT}{\PY{n}{nudobe.EFT}\PY{o}{.}\PY{n}{SMEFT}\PY{p}{(}\PY{n}{WC}\PY{p}{,} \PY{n}{scale} \PY{o}{=} \PY{n}{m\PYZus{}W}\PY{p}{,} \PY{n}{name} \PY{o}{=} \PY{k+kc}{None}\PY{p}{,} \PY{n}{unknown\PYZus{}LECs} \PY{o}{=} \PY{k+kc}{False}\PY{p}{,}}}\\
    \multicolumn{3}{l}{\qquad\quad\PY{n}{method} \PY{o}{=} \PY{l+s+s2}{\PYZdq{}}\PY{l+s+s2}{IBM2}\PY{l+s+s2}{\PYZdq{}}\PY{p}{,} \PY{n}{PSF\PYZus{}scheme} \PY{o}{=} \PY{l+s+s2}{\PYZdq{}}\PY{l+s+s2}{A}\PY{l+s+s2}{\PYZdq{}}\PY{p}{)}}\\\\
    \multicolumn{3}{l}{\hypertarget{nudobe.EFT.LEFT}{\PY{n}{nudobe.EFT}\PY{o}{.}\PY{n}{LEFT}\PY{p}{(}\PY{n}{WC}\PY{p}{,} \PY{n}{name} \PY{o}{=} \PY{k+kc}{None}\PY{p}{,} \PY{n}{unknown\PYZus{}LECs} \PY{o}{=} \PY{k+kc}{False}\PY{p}{,} \PY{n}{method} \PY{o}{=} \PY{l+s+s2}{\PYZdq{}}\PY{l+s+s2}{IBM2}\PY{l+s+s2}{\PYZdq{}}\PY{p}{,} }}\\
    \multicolumn{3}{l}{\qquad\quad\PY{n}{PSF\PYZus{}scheme} \PY{o}{=} \PY{l+s+s2}{\PYZdq{}}\PY{l+s+s2}{A}\PY{l+s+s2}{\PYZdq{}}\PY{p}{)}}
    \\\\
    \hline\\
    \textbf{Parameter} & \textbf{Type} & \textbf{Description}\\\\\hline\\
    \textbf{WC} & dictionay & Defines the Wilson coefficients as \{\str{WCname1} : WCvalue1\, ..., \str{WCnameN} : WCvalueN\}. You only need to define the non-vanishing WCs here.
    \\\\
    \textbf{scale} & float & Optional - Sets the scale of new physics $\Lambda$ for the WCs. Needs to be set larger than or equal to $80$ i.e. higher than the mass of the $W$ boson. If \texttt{scale}>\texttt{m\_W} \nudobe will automatically run the provided WCs down to \texttt{m\_W} when initiating a model.
    \\\\
    \textbf{name} & string & Optional - Defines a name for the model. This will show up in plots.
\end{tabularx}
\begin{tabularx}{\textwidth}{llX}
    \textbf{method} & string & Optional - Sets the NME calculation method. You can choose from \str{IBM2}, \str{SM} and \str{QRPA}. The preset value is \str{IBM2}.
    \\\\
    \textbf{unknown\_LECs} & bool & Optional - If set to \bool{True} the unknown LECs will be set to their NDA estimates (see Table~\ref{Tab:LECs}). If set to \bool{False} the unknown LECs will be turned off i.e. set to 0.
    \\\\
    \textbf{PSF\_scheme} & string & Optional - Choose PSFs and electron wave functions - \str{A}: approximate solution to a uniform charge distribution. \str{B}: exact solution to a point-like charge
    \\\\\hline\hline
\end{tabularx}
\\\\\\
Both are mostly identical in what they can do. In fact, the SMEFT class generally does all its calculations by matching the SMEFT Wilson coefficients onto LEFT to internally generate a LEFT class which then performs the desired calculations. See section~\ref{sec:WilsonCoefficients} and Appendix~\ref{app:operator_list} for the different Wilson coefficients. The SMEFT$\rightarrow$LEFT matching is provided in Appendix~\ref{app:operator_matching}.
%
%
\subsubsection{Importing the Model Classes}
To get started you first need to import the EFT classes which allow you to set up a model in either SMEFT or LEFT:
    \begin{tcolorbox}[breakable, size=fbox, boxrule=1pt, pad at break*=1mm,colback=cellbackground, colframe=cellborder]
\prompt{In}{incolor}{ }{\boxspacing}
\begin{Verbatim}[commandchars=\\\{\}]
\PY{k+kn}{from} \PY{n+nn}{nudobe} \PY{k+kn}{import} \PY{n}{EFT}
\end{Verbatim}
\end{tcolorbox}
%
%
\subsubsection{Initiating a Model}
You can set up a model by defining the corresponding non-vanishing Wilson coefficients in a dictionary style.
    \begin{tcolorbox}[breakable, size=fbox, boxrule=1pt, pad at break*=1mm,colback=cellbackground, colframe=cellborder]
\prompt{In}{incolor}{ }{\boxspacing}
\begin{Verbatim}[commandchars=\\\{\}]
\PY{c+c1}{\PYZsh{}define Wilson coefficients}
\PY{n}{Lambda} \PY{o}{=} \PY{l+m+mf}{1e+16}
\PY{n}{SMEFT\PYZus{}WCs} \PY{o}{=} \PY{p}{\PYZob{}}\PY{l+s+s2}{\PYZdq{}}\PY{l+s+s2}{LH(5)}\PY{l+s+s2}{\PYZdq{}} \PY{p}{:} \PY{l+m+mi}{1}\PY{o}{/}\PY{n}{Lambda}\PY{p}{,}
             \PY{l+s+s2}{\PYZdq{}}\PY{l+s+s2}{LH(7)}\PY{l+s+s2}{\PYZdq{}} \PY{p}{:} \PY{l+m+mf}{0.5}\PY{o}{/}\PY{n}{Lambda}\PY{o}{*}\PY{o}{*}\PY{l+m+mi}{3}\PY{p}{,}
             \PY{o}{.}
             \PY{o}{.}
             \PY{o}{.}
            \PY{p}{\PYZcb{}}
\PY{n}{LEFT\PYZus{}WCs} \PY{o}{=} \PY{p}{\PYZob{}}\PY{l+s+s2}{\PYZdq{}}\PY{l+s+s2}{m\PYZus{}bb}\PY{l+s+s2}{\PYZdq{}}  \PY{p}{:} \PY{l+m+mf}{1e\PYZhy{}9}\PY{p}{,} \PY{c+c1}{\PYZsh{}Remember that 1eV = 1e-9GeV}
            \PY{l+s+s2}{\PYZdq{}}\PY{l+s+s2}{SL(6)}\PY{l+s+s2}{\PYZdq{}} \PY{p}{:} \PY{l+m+mf}{1e\PYZhy{}6}\PY{p}{,}
            \PY{o}{.}
            \PY{o}{.}
            \PY{o}{.}
           \PY{p}{\PYZcb{}}
\end{Verbatim}
\end{tcolorbox}
Afterwards you can initiate a model with the specific choice of the Wilson coefficients.
When initiating LEFT models you have to set the WCs at the matching scale $m_W$.
The LEFT model class will then automatically run the Wilson coefficients to the scale relevant for neutrinoless double-$\beta$ decay and save them internally to the parameter \texttt{model.WC} while the original input WCs are stored to the \texttt{model.WC\_input} parameter. SMEFT models can be defined at an arbitrary scale $\Lambda\geq m_W$. See Section~\ref{sec:RGE_running} for more details on the relevant scales and RGE evolution.

All LEFT Wilson coefficients except for $m_{\beta\beta}$ are dimensionless while the SMEFT WCs are of mass dimension $4-d$, where $d$ is the operators dimension. Masses and scales are, generally, taken in GeV. That
is, $\mathrm{m\_bb = 1e-}9$ corresponds to $m_{\beta\beta} = 1\,$eV.
You can initiate the models by typing:

    \begin{tcolorbox}[breakable, size=fbox, boxrule=1pt, pad at break*=1mm,colback=cellbackground, colframe=cellborder]
\prompt{In}{incolor}{ }{\boxspacing}
\begin{Verbatim}[commandchars=\\\{\}]
\PY{c+c1}{\PYZsh{}generate models}
\PY{n}{SMEFT\PYZus{}model} \PY{o}{=} \PY{n}{EFT}\PY{o}{.}\PY{n}{SMEFT}\PY{p}{(}\PY{n}{SMEFT\PYZus{}WCs}\PY{p}{,} \PY{n}{scale} \PY{o}{=} \PY{n}{Lambda}\PY{p}{)}

\PY{n}{LEFT\PYZus{}model} \PY{o}{=} \PY{n}{EFT}\PY{o}{.}\PY{n}{LEFT}\PY{p}{(}\PY{n}{LEFT\PYZus{}WCs}\PY{p}{)}
\end{Verbatim}
\end{tcolorbox}
If you know you only want to work within SMEFT or LEFT or you do not want to put the ``EFT.'' in front each time you initiate a new model class, you can also directly import the corresponding classes via
    \begin{tcolorbox}[breakable, size=fbox, boxrule=1pt, pad at break*=1mm,colback=cellbackground, colframe=cellborder]
\prompt{In}{incolor}{ }{\boxspacing}
\begin{Verbatim}[commandchars=\\\{\}]
\PY{k+kn}{from} \PY{n+nn}{nudobe} \PY{k+kn}{import} \PY{n}{EFT}
\PY{k+kn}{from} \PY{n+nn}{EFT} \PY{k+kn}{import} \PY{n}{LEFT, SMEFT}
\PY{o}{.}
\PY{o}{.}
\PY{o}{.}
\PY{n}{SMEFT\PYZus{}model} \PY{o}{=} \PY{n}{SMEFT}\PY{p}{(}\PY{n}{SMEFT\PYZus{}WCs}\PY{p}{,} \PY{n}{scale} \PY{o}{=} \PY{l+m+mf}{1e+16}\PY{p}{)}
\PY{n}{LEFT\PYZus{}model}  \PY{o}{=} \PY{n}{LEFT}\PY{p}{(}\PY{n}{LEFT\PYZus{}WCs}\PY{p}{)}
\end{Verbatim}
\end{tcolorbox}
After having set up a model class you can call the Wilson coefficients at $m_W$ (SMEFT) or $\Lambda_\chi$ (LEFT) via e.g.
    \begin{tcolorbox}[breakable, size=fbox, boxrule=1pt, pad at break*=1mm,colback=cellbackground, colframe=cellborder]
\prompt{In}{incolor}{ }{\boxspacing}
\begin{Verbatim}[commandchars=\\\{\}]
\PY{n}{SMEFT\_model}\PY{o}{.}\PY{n}{WC}
\end{Verbatim}
\end{tcolorbox}
\subsubsection{List of Wilson coefficients}\label{sec:WilsonCoefficients}
The relevant Wilson coefficients under consideration are given in Tables~\ref{tab:LEFT_operators} (LEFT) and \ref{tab:SMEFT_operators_SR} (SMEFT). Within \texttt{nudobe} they are defined as dictionaries. The dictionary key for each WC is given in the column ``Code Label'' in each table.
After importing the EFT module you can access the WC dictionaries directly via
    \begin{tcolorbox}[breakable, size=fbox, boxrule=1pt, pad at break*=1mm,colback=cellbackground, colframe=cellborder]
\prompt{In}{incolor}{ }{\boxspacing}
\begin{Verbatim}[commandchars=\\\{\}]
\PY{k+kn}{from} \PY{n+nn}{nudobe} \PY{k+kn}{import} \PY{n}{EFT}
\end{Verbatim}
\end{tcolorbox}

    \begin{tcolorbox}[breakable, size=fbox, boxrule=1pt, pad at break*=1mm,colback=cellbackground, colframe=cellborder]
\prompt{In}{incolor}{ }{\boxspacing}
\begin{Verbatim}[commandchars=\\\{\}]
\PY{n}{EFT}\PY{o}{.}\PY{n}{LEFT\PYZus{}WCs}
\end{Verbatim}
\end{tcolorbox}

    \begin{tcolorbox}[breakable, size=fbox, boxrule=1pt, pad at break*=1mm,colback=cellbackground, colframe=cellborder]
\prompt{In}{incolor}{ }{\boxspacing}
\begin{Verbatim}[commandchars=\\\{\}]
\PY{n}{EFT}\PY{o}{.}\PY{n}{SMEFT\PYZus{}WCs}
\end{Verbatim}
\end{tcolorbox}
Alternatively, the list of WCs are also stored in the \texttt{nudobe.constants} module and can be accessed there.
\subsubsection{Example: Light neutrino exchange mechanism in LEFT}
You can generate a model class that represents the $0\nu\beta\beta$-decay induced by the exchange of light Majorana neutrinos with $m_{\beta\beta} = 1\,$eV using NMEs calculated in the IBM2 framework via
\begin{tcolorbox}[breakable, size=fbox, boxrule=1pt, pad at break*=1mm,colback=cellbackground, colframe=cellborder]
\prompt{In}{incolor}{1}{\boxspacing}
\begin{Verbatim}[commandchars=\\\{\}]
\PY{k+kn}{from} \PY{n+nn}{nudobe} \PY{k+kn}{import} \PY{n}{EFT}
\PY{k+kn}{from} \PY{n+nn}{EFT} \PY{k+kn}{import} \PY{n}{LEFT}

\PY{n}{mass\PYZus{}mechanism} \PY{o}{=} \PY{n}{LEFT}\PY{p}{(}\PY{p}{\PYZob{}}\PY{l+s+s2}{\PYZdq{}}\PY{l+s+s2}{m\PYZus{}bb}\PY{l+s+s2}{\PYZdq{}} \PY{p}{:} \PY{l+m+mf}{1e\PYZhy{}9}\PY{p}{\PYZcb{}}\PY{p}{,} \PY{n}{method} \PY{o}{=} \PY{l+s+s2}{\PYZdq{}}\PY{l+s+s2}{IBM2}\PY{l+s+s2}{\PYZdq{}}\PY{p}{)}
\end{Verbatim}
\end{tcolorbox}

\subsubsection{RGE running}\label{sec:RGE_running}
When initiating either a LEFT or a SMEFT model the different Wilson coefficients should generally be defined at the scale $\Lambda = m_W$. 
\nudobe will then take care of the running and matching procedure down to the chiral EFT scale $\Lambda_\chi = 2\,\mathrm{GeV}$. The different RGEs are defined in the \texttt{RGE.py} file and the running of the LEFT WCs between $m_W$ and $\Lambda_\chi$ is calculated and stored as a matrix when importing the \texttt{EFT} module. If you want to define WCs at a different scale than $m_W$ you can run them down to $m_W$ (SMEFT) or $\Lambda_\chi$ (LEFT) - or generally any other scale - by using the \texttt{RGE.run()} function that is also built into the model classes. Note that the matching scale SMEFT$\rightarrow$LEFT is $m_W$.\\\\
\begin{tabularx}{\textwidth}{llX}
    \hline\hline\\
    \multicolumn{3}{l}{\hypertarget{model.run}{\PY{n}{nudobe}\PY{o}{.}\PY{n}{EFT}\PY{o}{.}\PY{n}{LEFT}\PY{o}{.}\PY{n}{run}\PY{p}{(}\PY{n}{WC} \PY{o}{=} \PY{k+kc}{None}\PY{p}{,} \PY{n}{initial\PYZus{}scale} \PY{o}{=} \PY{n}{m\PYZus{}W}\PY{p}{,} \PY{n}{final\PYZus{}scale} \PY{o}{=} \PY{n}{lambda\PYZus{}chi}\PY{p}{,}}}\\
    \multicolumn{3}{l}{\qquad\quad\PY{n}{inplace} \PY{o}{=} \PY{k+kc}{False}\PY{p}{)}}\\\\
    \multicolumn{3}{l}{
    \PY{n}{nudobe}\PY{o}{.}\PY{n}{EFT}\PY{o}{.}\PY{n}{SMEFT}\PY{o}{.}\PY{n}{run}\PY{p}{(}\PY{n}{WC} \PY{o}{=} \PY{k+kc}{None}\PY{p}{,} \PY{n}{initial\PYZus{}scale} \PY{o}{=} \PY{k+kc}{None}\PY{p}{,} \PY{n}{final\PYZus{}scale} \PY{o}{=} \PY{n}{m\PYZus{}W}\PY{p}{,}}\\
    \multicolumn{3}{l}{\qquad\quad\PY{n}{inplace} \PY{o}{=} \PY{k+kc}{False}\PY{p}{)}}
    \\\\
    \hline\\
    \textbf{Parameter} & \textbf{Type} & \textbf{Description}\\\\\hline\\
    \textbf{WC} & dictionay & Optional - Defines the Wilson coefficients as \{\str{WCname1} : WCvalue1\, ..., \str{WCnameN} : WCvalueN\}. You only need to define the non-vanishing WCs here.
    \\\\
    \textbf{initial\_scale} & float & Optional - Scale the WCs are defined at initially.
    \\\\
    \textbf{final\_scale} & float & Optional - Scale the WCs should be evolved to.
\end{tabularx}
\begin{tabularx}{\textwidth}{llX}
    \textbf{inplace} & bool & Optional - If \bool{True} the resulting WCs after running will replace the models WCs.
    \\\\\hline\hline
\end{tabularx}
\\\\\\
The RGE running of the LEFT operators is given in~\citep{Cirigliano_2018} (Eqs. 14-16) while the RGEs for the dimension 7 SMEFT operators can be found in~\citep{Liao:2019tep} (Eqs. 21-24). 
The RGEs of the SMEFT dimension 9 operators will be included in a future update. Currently, \nudobe does not evolve the SMEFT dimension 9 operators when using the \texttt{run()} function. Therefore, for reasons of consistency, we recommend defining SMEFT models directly at the matching scale $m_W$ if both dimension 7 and dimension 9 operators are to be studied. Alternatively, you can use the SMEFT models \texttt{.run()} function to run the dimension 7 SMEFT operators from any arbitrary scale to the LEFT matching scale.
\\\\
\noindent\textit{\textbf{Example: RGE evolution of a specific SMEFT model.}}
    \begin{tcolorbox}[breakable, size=fbox, boxrule=1pt, pad at break*=1mm,colback=cellbackground, colframe=cellborder]
\prompt{In}{incolor}{1}{\boxspacing}
\begin{Verbatim}[commandchars=\\\{\}]
\PY{k+kn}{from} \PY{n+nn}{nudobe} \PY{k+kn}{import} \PY{n}{EFT}\PY{p}{,} \PY{n}{constants}
\PY{k+kn}{from} \PY{n+nn}{EFT} \PY{k+kn}{import} \PY{n}{SMEFT}
\PY{k+kn}{from} \PY{n+nn}{constants} \PY{k+kn}{import} \PY{o}{*}
\end{Verbatim}
\end{tcolorbox}
After importing the SMEFT class we define a model with two non-vanishing dimension 7 operators. To show the impact of the RGE running we first define the model class without setting a scale explicitly such that the SMEFT class assumes the scale is $m_W$. This time we use the unit parameters defined in the \texttt{constants} module to define scales and masses.
    \begin{tcolorbox}[breakable, size=fbox, boxrule=1pt, pad at break*=1mm,colback=cellbackground, colframe=cellborder]
\prompt{In}{incolor}{2}{\boxspacing}
\begin{Verbatim}[commandchars=\\\{\}]
\PY{c+c1}{\PYZsh{}Scale of new physics @ 50TeV}
\PY{n}{Lambda} \PY{o}{=} \PY{l+m+mi}{50}\PY{o}{*}\PY{n}{TeV}

\PY{c+c1}{\PYZsh{}Generate SMEFT model}
\PY{n}{model} \PY{o}{=} \PY{n}{SMEFT}\PY{p}{(}\PY{p}{\PYZob{}}\PY{l+s+s2}{\PYZdq{}}\PY{l+s+s2}{LeudH(7)}\PY{l+s+s2}{\PYZdq{}} \PY{p}{:} \PY{l+m+mf}{0.05}\PY{o}{/}\PY{n}{Lambda}\PY{o}{*}\PY{o}{*}\PY{l+m+mi}{3}\PY{p}{,} \PY{l+s+s2}{\PYZdq{}}\PY{l+s+s2}{LHD1(7)}\PY{l+s+s2}{\PYZdq{}} \PY{p}{:} \PY{l+m+mi}{5}\PY{o}{/}\PY{n}{Lambda}\PY{o}{*}\PY{o}{*}\PY{l+m+mi}{3}\PY{p}{\PYZcb{}}\PY{p}{)}
\end{Verbatim}
\end{tcolorbox}

    \begin{tcolorbox}[breakable, size=fbox, boxrule=1pt, pad at break*=1mm,colback=cellbackground, colframe=cellborder]
\prompt{In}{incolor}{3}{\boxspacing}
\begin{Verbatim}[commandchars=\\\{\}]
\PY{c+c1}{\PYZsh{}Non\PYZhy{}vanishing SMEFT WCs before RGE running}
\PY{k}{for} \PY{n}{WC} \PY{o+ow}{in} \PY{n}{model}\PY{o}{.}\PY{n}{WC}\PY{p}{:}
    \PY{k}{if} \PY{n}{model}\PY{o}{.}\PY{n}{WC}\PY{p}{[}\PY{n}{WC}\PY{p}{]} \PY{o}{!=} \PY{l+m+mi}{0}\PY{p}{:}
        \PY{n+nb}{print}\PY{p}{(}\PY{n}{WC}\PY{p}{,} \PY{n}{model}\PY{o}{.}\PY{n}{WC}\PY{p}{[}\PY{n}{WC}\PY{p}{]}\PY{p}{)}
\end{Verbatim}
\end{tcolorbox}
\begin{tcolorbox}[breakable, size=fbox, boxrule=.5pt, pad at break*=1mm, opacityfill=0]
\prompt{Out}{outcolor}{3}{\boxspacing}
    \begin{Verbatim}[commandchars=\\\{\}]
LHD1(7) 4e-14
LeudH(7) 4.0000000000000004e-16
    \end{Verbatim}
\end{tcolorbox}
We now calculate the half-life in \textsuperscript{136}Xe before running the SMEFT WCs from $50\,$TeV down to $m_W$
    \begin{tcolorbox}[breakable, size=fbox, boxrule=1pt, pad at break*=1mm,colback=cellbackground, colframe=cellborder]
\prompt{In}{incolor}{4}{\boxspacing}
\begin{Verbatim}[commandchars=\\\{\}]
\PY{c+c1}{\PYZsh{}Half\PYZhy{}Life in 136Xe}
\PY{n}{model}\PY{o}{.}\PY{n}{t\PYZus{}half}\PY{p}{(}\PY{l+s+s2}{\PYZdq{}}\PY{l+s+s2}{136Xe}\PY{l+s+s2}{\PYZdq{}}\PY{p}{)}
\end{Verbatim}
\end{tcolorbox}
            \begin{tcolorbox}[breakable, size=fbox, boxrule=.5pt, pad at break*=1mm, opacityfill=0]
\prompt{Out}{outcolor}{4}{\boxspacing}
\begin{Verbatim}[commandchars=\\\{\}]
1.7039062688609342e+29
\end{Verbatim}
\end{tcolorbox}
Now we use the \texttt{run()} function of the SMEFT class to run the WCs down to $m_W$ and recalculate the half-life
    \begin{tcolorbox}[breakable, size=fbox, boxrule=1pt, pad at break*=1mm,colback=cellbackground, colframe=cellborder]
\prompt{In}{incolor}{5}{\boxspacing}
\begin{Verbatim}[commandchars=\\\{\}]
\PY{c+c1}{\PYZsh{}Run the WCs down to the matching scale m\PYZus{}W}
\PY{n}{model}\PY{o}{.}\PY{n}{run}\PY{p}{(}\PY{n}{initial\PYZus{}scale} \PY{o}{=} \PY{n}{Lambda}\PY{p}{,} \PY{n}{inplace} \PY{o}{=} \PY{k+kc}{True}\PY{p}{)}

\PY{c+c1}{\PYZsh{}Recalculate Half\PYZhy{}Life}
\PY{n}{model}\PY{o}{.}\PY{n}{t\PYZus{}half}\PY{p}{(}\PY{l+s+s2}{\PYZdq{}}\PY{l+s+s2}{136Xe}\PY{l+s+s2}{\PYZdq{}}\PY{p}{)}
\end{Verbatim}
\end{tcolorbox}

            \begin{tcolorbox}[breakable, size=fbox, boxrule=.5pt, pad at break*=1mm, opacityfill=0]
\prompt{Out}{outcolor}{5}{\boxspacing}
\begin{Verbatim}[commandchars=\\\{\}]
1.7973515786734228e+29
\end{Verbatim}
\end{tcolorbox}
        
To see how the WCs change when running them from $50\,$TeV down to $m_W$ we can print them again
    \begin{tcolorbox}[breakable, size=fbox, boxrule=1pt, pad at break*=1mm,colback=cellbackground, colframe=cellborder]
\prompt{In}{incolor}{6}{\boxspacing}
\begin{Verbatim}[commandchars=\\\{\}]
\PY{c+c1}{\PYZsh{}Non\PYZhy{}vanishing SMEFT WCs after RGE running}
\PY{k}{for} \PY{n}{WC} \PY{o+ow}{in} \PY{n}{model}\PY{o}{.}\PY{n}{WC}\PY{p}{:}
    \PY{k}{if} \PY{n}{model}\PY{o}{.}\PY{n}{WC}\PY{p}{[}\PY{n}{WC}\PY{p}{]} \PY{o}{!=} \PY{l+m+mi}{0}\PY{p}{:}
        \PY{n+nb}{print}\PY{p}{(}\PY{n}{WC}\PY{p}{,} \PY{n}{model}\PY{o}{.}\PY{n}{WC}\PY{p}{[}\PY{n}{WC}\PY{p}{]}\PY{p}{)}
\end{Verbatim}
\end{tcolorbox}

    \begin{Verbatim}[commandchars=\\\{\}]
LHD1(7)  3.105078059915025e-14
LHD2(7)  4.0238127608753095e-15
LHW(7)  -3.3074147823810323e-16
LeudH(7) 3.915696874279481e-16
    \end{Verbatim}

%
%
\subsection{Getting the Half-Life}\label{get-half-lives}
There are two ways you can get the half-lives from a specific model. To get the half-lives in all isotopes that are included in the chosen NME method you can simply run
\\\\
\begin{tabularx}{\textwidth}{llX}
    \hline\hline\\
    \multicolumn{3}{l}{\hypertarget{model.half_lives}{\PY{n}{model}\PY{o}{.}\PY{n}{half\PYZus{}lives}\PY{p}{(}\PY{n}{WC} \PY{o}{=} \PY{k+kc}{None}\PY{p}{,} \PY{n}{method} \PY{o}{=} \PY{k+kc}{None}\PY{p}{,} \PY{n}{vary\PYZus{}LECs} \PY{o}{=} \PY{k+kc}{False}\PY{p}{,} \PY{n}{n\PYZus{}points} \PY{o}{=} \PY{l+m+mi}{1000}\PY{p}{)}}}
    \\\\
    \hline\\
    \textbf{Parameter} & \textbf{Type} & \textbf{Description}\\\\\hline\\
    \textbf{WC} & dictionary & Optional - Defines the Wilson coefficients as \{\str{WCname1} : WCvalue1\, ..., \str{WCnameN} : WCvalueN\}. If \bool{None} the models WCs will be used.
    \\\\
    \textbf{method} & string & Optional - Sets the NME calculation method. You can choose from \str{IBM2}, \str{SM} and \str{QRPA}. If \bool{None} the models method will be used.
    \\\\
    \textbf{vary\_LECs} & bool & Optional - If set to \bool{True} the unknown LECs will be varied within their NDA estimates (see Table~\ref{Tab:LECs}). If set to \bool{False} the unknown LECs will stay fixed.
\end{tabularx}
\begin{tabularx}{\textwidth}{llX}
    \textbf{n\_points} & integer & Optional - Number of variations.
    \\\\\hline\hline
\end{tabularx}
\\\\\\
with \texttt{model} initiated via either \texttt{nudobe.EFT.SMEFT} or \texttt{nudobe.EFT.SMEFT}.
This will output a pandas DataFrame containing the half-lives of all isotopes included. It works for both SMEFT and LEFT models. If you only want to get the half-life for a single isotope you can run the function\\\\
\begin{tabularx}{\textwidth}{llX}
    \hline\hline\\
    \multicolumn{3}{l}{\hypertarget{model.t_half}{\PY{n}{model}\PY{o}{.}\PY{n}{t\PYZus{}half}\PY{p}{(}\PY{n}{isotope}\PY{p}{,} \PY{n}{WC} \PY{o}{=} \PY{k+kc}{None}\PY{p}{,} \PY{n}{method} \PY{o}{=} \PY{k+kc}{None}\PY{p}{)}}}
    \\\\
    \hline\\
    \textbf{Parameter} & \textbf{Type} & \textbf{Description}\\\\\hline\\
    \textbf{isotope} & string & Optional - Defines the isotope to be studied.
    \\\\
    \textbf{WC} & dictionary & Optional - Defines the Wilson coefficients as \{\str{WCname1} : WCvalue1\, ..., \str{WCnameN} : WCvalueN\}. If \bool{None} the models WCs will be used.
    \\\\
    \textbf{method} & string & Optional - Sets the NME calculation method. You can choose from \str{IBM2}, \str{SM} and \str{QRPA}. If \bool{None} the models method will be used.
    \\\\\hline\hline
\end{tabularx}
\\\\\\
\textit{\textbf{Example: Light neutrino exchange mechanism in LEFT.}} \\
As an example we study once more the standard mass mechanism induced by the exchange of light Majorana neutrinos. We want to find the expected half-lives in the case of an effective Majorana mass $m_{\beta\beta} = 100\,$meV:
    \begin{tcolorbox}[breakable, size=fbox, boxrule=1pt, pad at break*=1mm,colback=cellbackground, colframe=cellborder]
\prompt{In}{incolor}{1}{\boxspacing}
\begin{Verbatim}[commandchars=\\\{\}]
\PY{k+kn}{from} \PY{n+nn}{nudobe} \PY{k+kn}{import} \PY{n}{EFT}\PY{p}{,} \PY{n}{constants}
\PY{k+kn}{from} \PY{n+nn}{EFT} \PY{k+kn}{import} \PY{n}{LEFT}
\PY{k+kn}{from} \PY{n+nn}{constants} \PY{k+kn}{import} \PY{o}{*}
\end{Verbatim}
\end{tcolorbox}
The resulting range of values can be used as an estimate for the theoretical uncertainty of the predictions arising from QCD matrix elements (that is, the LECs). 

    \begin{tcolorbox}[breakable, size=fbox, boxrule=1pt, pad at break*=1mm,colback=cellbackground, colframe=cellborder]
\prompt{In}{incolor}{2}{\boxspacing}
\begin{Verbatim}[commandchars=\\\{\}]
\PY{c+c1}{\PYZsh{}Define Wilson coefficients for the standard mass mechanism}
\PY{c+c1}{\PYZsh{}with m\PYZus{}bb = 100meV}
\PY{n}{WC} \PY{o}{=} \PY{p}{\PYZob{}}\PY{l+s+s2}{\PYZdq{}}\PY{l+s+s2}{m\PYZus{}bb}\PY{l+s+s2}{\PYZdq{}} \PY{p}{:} \PY{l+m+mf}{100}\PY{o}{*}\PY{n}{meV}\PY{p}{\PYZcb{}}

\PY{c+c1}{\PYZsh{}NME method}
\PY{n}{method} \PY{o}{=} \PY{l+s+s2}{\PYZdq{}}\PY{l+s+s2}{IBM2}\PY{l+s+s2}{\PYZdq{}}

\PY{c+c1}{\PYZsh{}initiate model}
\PY{n}{mass\PYZus{}mechanism} \PY{o}{=} \PY{n}{LEFT}\PY{p}{(}\PY{n}{WC}\PY{p}{,} \PY{n}{method} \PY{o}{=} \PY{n}{method}\PY{p}{)}
\end{Verbatim}
\end{tcolorbox}

    \begin{tcolorbox}[breakable, size=fbox, boxrule=1pt, pad at break*=1mm,colback=cellbackground, colframe=cellborder]
\prompt{In}{incolor}{3}{\boxspacing}
\begin{Verbatim}[commandchars=\\\{\}]
\PY{c+c1}{\PYZsh{}half\PYZhy{}life for 76Ge}
\PY{n}{mass\PYZus{}mechanism}\PY{o}{.}\PY{n}{t\PYZus{}half}\PY{p}{(}\PY{l+s+s2}{\PYZdq{}}\PY{l+s+s2}{76Ge}\PY{l+s+s2}{\PYZdq{}}\PY{p}{)}
\end{Verbatim}
\end{tcolorbox}

            \begin{tcolorbox}[breakable, size=fbox, boxrule=.5pt, pad at break*=1mm, opacityfill=0]
\prompt{Out}{outcolor}{3}{\boxspacing}
\begin{Verbatim}[commandchars=\\\{\}]
4.754252877417845e+25
\end{Verbatim}
\end{tcolorbox}
        
    \begin{tcolorbox}[breakable, size=fbox, boxrule=1pt, pad at break*=1mm,colback=cellbackground, colframe=cellborder]
\prompt{In}{incolor}{4}{\boxspacing}
\begin{Verbatim}[commandchars=\\\{\}]
\PY{c+c1}{\PYZsh{}half\PYZhy{}lives in all isotopes}
\PY{n}{mass\PYZus{}mechanism}\PY{o}{.}\PY{n}{half\PYZus{}lives}\PY{p}{(}\PY{p}{)}
\end{Verbatim}
\end{tcolorbox}

            \begin{tcolorbox}[breakable, size=fbox, boxrule=.5pt, pad at break*=1mm, opacityfill=0]
\prompt{Out}{outcolor}{4}{\boxspacing}
\begin{Verbatim}[commandchars=\\\{\}]
           76Ge          82Se        ...              232Th          238U
0  4.754253e+25  1.578798e+25        ...       9.476641e+24  2.565784e+24
\end{Verbatim}
\end{tcolorbox}
    \begin{tcolorbox}[breakable, size=fbox, boxrule=1pt, pad at break*=1mm,colback=cellbackground, colframe=cellborder]
\prompt{In}{incolor}{5}{\boxspacing}
\begin{Verbatim}[commandchars=\\\{\}]
\PY{c+c1}{\PYZsh{}half\PYZhy{}lives in all isotopes with varied LECs}
\PY{n}{mass\PYZus{}mechanism}\PY{o}{.}\PY{n}{half\PYZus{}lives}\PY{p}{(}\PY{n}{vary\PYZus{}LECs} \PY{o}{=} \PY{k+kc}{True}\PY{p}{,} \PY{n}{n\PYZus{}points} \PY{o}{=} \PY{l+m+mi}{100}\PY{p}{)}
\end{Verbatim}
\end{tcolorbox}

            \begin{tcolorbox}[breakable, size=fbox, boxrule=.5pt, pad at break*=1mm, opacityfill=0]
\prompt{Out}{outcolor}{5}{\boxspacing}
\begin{Verbatim}[commandchars=\\\{\}]
            76Ge          82Se           {\ldots}         232Th          238U 
0   4.108211e+25  1.360687e+25           {\ldots}  7.861367e+24  2.130089e+24
1   3.791413e+25  1.254053e+25           {\ldots}  7.105329e+24  1.925981e+24
2   5.139999e+25  1.709430e+25           {\ldots}  1.048842e+25  2.838449e+24
...          {\ldots}           {\ldots}           {\ldots}           {\ldots}           {\ldots}
97  5.193175e+25  1.727460e+25           {\ldots}  1.063069e+25  2.876775e+24
98  3.979882e+25  1.317465e+25           {\ldots}  7.552259e+24  2.046653e+24
99  5.260296e+25  1.750227e+25           {\ldots}  1.081124e+25  2.925407e+24

[100 rows x 18 columns]
\end{Verbatim}
\end{tcolorbox}

%
%
\subsection{Decay-Rate Formula}
In case you want to get an analytical expression of the decay rate in terms of the different Wilson coefficients in your model you can use the function \texttt{generate\_formula()}
\\\\
\begin{tabularx}{\textwidth}{llX}
    \hline\hline\\\multicolumn{3}{l}{\hypertarget{functions.generate_formula}{\PY{n}{nudobe}\PY{o}{.}\PY{n}{functions}\PY{o}{.}\PY{n}{generate\PYZus{}formula}\PY{p}{(}\PY{n}{WC}\PY{p}{,} \PY{n}{isotope} \PY{o}{=} \str{136Xe}\PY{p}{,} \PY{n}{method} \PY{o}{=} \str{IBM2}\PY{p}{,}}}\\
    \multicolumn{3}{l}{\qquad\quad\PY{n}{decimal} \PY{o}{=} \PY{l+m+mi}{2}\PY{p}{,} \PY{n}{output} \PY{o}{=} \PY{l+s+s2}{\PYZdq{}}\PY{l+s+s2}{latex}\PY{l+s+s2}{\PYZdq{}}\PY{p}{,} \PY{n}{unknown\PYZus{}LECs} \PY{o}{=} \PY{k+kc}{False}, \PY{n}{PSF\PYZus{}scheme} \PY{o}{=}\PY{l+s+s2}{\PYZdq{}}\PY{l+s+s2}{A}\PY{l+s+s2}{\PYZdq{}}\PY{p}{)}}
    \\\\
    \multicolumn{3}{l}{\hypertarget{model.generate_formula}{
    \PY{n}{model}\PY{o}{.}\PY{n}{generate\PYZus{}formula}\PY{p}{(}\PY{n}{isotope}\PY{p}{,} \PY{n}{WC} \PY{o}{=} \PY{k+kc}{None}\PY{p}{,} \PY{n}{method} \PY{o}{=} \PY{k+kc}{None}\PY{p}{,} \PY{n}{decimal} \PY{o}{=} \PY{l+m+mi}{2}\PY{p}{,}}}\\
    \multicolumn{3}{l}{\qquad\quad\PY{n}{output} \PY{o}{=} \PY{l+s+s2}{\PYZdq{}}\PY{l+s+s2}{latex}\PY{l+s+s2}{\PYZdq{}}\PY{p}{,} \PY{n}{unknown\PYZus{}LECs} \PY{o}{=} \PY{k+kc}{None}\PY{p}{,} \PY{n}{PSF\PYZus{}scheme} \PY{o}{=} \PY{k+kc}{None}\PY{p}{)}}
    \\\\
    \hline\\
\end{tabularx}
\begin{tabularx}{\textwidth}{llX}
    \textbf{Parameter} & \textbf{Type} & \textbf{Description}\\\\\hline\\
    \textbf{isotope} & string & Defines the isotope to be studied.
    \\\\
    \textbf{WC} & list & Optional - List of non-zero Wilson coefficients that should contribute to the half-life [\str{WCname1} \, ..., \str{WCnameN}]. If \bool{None} the models non-zero WCs will be used.
    \\\\
    \textbf{method} & string & Optional - Sets the NME calculation method. You can choose from \str{IBM2}, \str{SM} and \str{QRPA}. If called via the \texttt{functions} module the preset value is \str{IBM2}. If called via a \texttt{model} class he models method will be used if \bool{None}.
    \\\\
    \textbf{decimal} & integer & Optional - Sets the numbers of decimals used in rounding.
    \\\\
    \textbf{output} & string & Optional - Get output in \str{latex} or \str{html} format.
    \\\\
    \textbf{unknown\_LECs} & bool & Optional - If set to \bool{True} the unknown LECs will be set to their NDA estimates (see Table~\ref{Tab:LECs}). If set to \bool{False} the unknown LECs will be turned off i.e. set to 0.
    \\\\
    \textbf{PSF\_scheme} & string & Optional - Choose PSFs and electron wave functions - \str{A}: approximate solution to a uniform charge distribution. \str{B}: exact solution to a point-like charge
    \\\\\hline\hline
\end{tabularx}
\\\\\\
to generate an expression for the decay rate
\begin{align}
    T_{1/2}^{-1} = \sum_{ij} M_{ij} C_i^\dagger C_j
\end{align}
either in latex or html format. This equation represents a matrix equation of the form
\begin{align}
    T_{1/2}^{-1} = C^\dagger M C\,
\end{align}
and you can obtain the matrix coefficients from 
\\\\
\begin{tabularx}{\textwidth}{llX}
    \hline\hline\\
    \multicolumn{3}{l}{\hypertarget{functions.generate_matrix}{\PY{n}{nudobe}\PY{o}{.}\PY{n}{functions}\PY{o}{.}\PY{n}{generate\PYZus{}matrix}\PY{p}{(}\PY{n}{WC}\PY{p}{,} \PY{n}{isotope} \PY{o}{=} \str{136Xe}\PY{p}{,} \PY{n}{method} \PY{o}{=} \str{IBM2}\PY{p}{,}}}\\
    \multicolumn{3}{l}{\qquad\quad\PY{n}{unknown\PYZus{}LECs} \PY{o}{=} \PY{k+kc}{False}, \PY{n}{PSF\PYZus{}scheme} \PY{o}{=}\PY{l+s+s2}{\PYZdq{}}\PY{l+s+s2}{A}\PY{l+s+s2}{\PYZdq{}}\PY{p}{)}}
    \\\\
    \multicolumn{3}{l}{\hypertarget{model.generate_matrix}{\PY{n}{model}\PY{o}{.}\PY{n}{generate\PYZus{}matrix}\PY{p}{(}\PY{n}{isotope}\PY{p}{,} \PY{n}{WC} \PY{o}{=} \PY{k+kc}{None}\PY{p}{,} \PY{n}{method} \PY{o}{=} \PY{k+kc}{None}\PY{p}{,}}}\\
    \multicolumn{3}{l}{\qquad\quad\PY{n}{unknown\PYZus{}LECs} \PY{o}{=} \PY{k+kc}{False}\PY{p}{,} \PY{n}{PSF\PYZus{}scheme} \PY{o}{=} \PY{k+kc}{None}\PY{p}{)}}
    \\\\
    \hline
\end{tabularx}
\begin{tabularx}{\textwidth}{llX}
    \textbf{Parameter} & \textbf{Type} & \textbf{Description}\\\\\hline\\
    \textbf{isotope} & string & Defines the isotope to be studied.
    \\\\
    \textbf{WC} & list & Optional - List of non-zero Wilson coefficients that should contribute to the half-life [\str{WCname1} \, ..., \str{WCnameN}]. If \bool{None} the models non-zero WCs will be used.
\end{tabularx}
\begin{tabularx}{\textwidth}{llX}
    \textbf{method} & string & Optional - Sets the NME calculation method. You can choose from \str{IBM2}, \str{SM} and \str{QRPA}. If called via the \texttt{functions} module the preset value is \str{IBM2}. If called via a \texttt{model} class he models method will be used if \bool{None}.
    \\\\
    \textbf{unknown\_LECs} & bool & Optional - If set to \bool{True} the unknown LECs will be set to their NDA estimates (see Table~\ref{Tab:LECs}). If set to \bool{False} the unknown LECs will be turned off i.e. set to 0.
    \\\\
    \textbf{PSF\_scheme} & string & Optional - Choose PSFs and electron wave functions - \str{A}: approximate solution to a uniform charge distribution. \str{B}: exact solution to a point-like charge
    \\\\\hline\hline
\end{tabularx}
\\\\\\
Both functions will only consider those Wilson coefficients which are non-zero in your model.\\\\
\textit{\textbf{Example: The light-neutrino-exchange mechanism with an additional LNV right-handed current}}\\
Let's assume a LEFT model with two lepton number violating terms in the Lagrangian
\begin{align}
    \mathcal{L}_{\Delta L=2} = -\frac{1}{2}m_{\beta\beta}\Big[\overline{\nu_{L,e}^C}\nu_{L,e}\Big] + \frac{1}{v^2}C_\mathrm{VR}^{(6)}\Big[\overline{u_R}\gamma^\mu d_R\Big]\Big[\overline{e_R}\gamma_\mu \nu_{L,e}^C\Big]
\end{align}
    \begin{tcolorbox}[breakable, size=fbox, boxrule=1pt, pad at break*=1mm,colback=cellbackground, colframe=cellborder]
\prompt{In}{incolor}{1}{\boxspacing}
\begin{Verbatim}[commandchars=\\\{\}]
\PY{k+kn}{import} \PY{n+nn}{numpy} \PY{k}{as} \PY{n+nn}{np}
\PY{k+kn}{from} \PY{n+nn}{nudobe} \PY{k+kn}{import} \PY{n}{EFT}\PY{p}{,} \PY{n}{constants}
\PY{k+kn}{from} \PY{n+nn}{EFT} \PY{k+kn}{import} \PY{n}{LEFT}
\PY{k+kn}{from} \PY{n+nn}{constants} \PY{k+kn}{import} \PY{o}{*}
\end{Verbatim}
\end{tcolorbox}
%
    \begin{tcolorbox}[breakable, size=fbox, boxrule=1pt, pad at break*=1mm,colback=cellbackground, colframe=cellborder]
\prompt{In}{incolor}{2}{\boxspacing}
\begin{Verbatim}[commandchars=\\\{\}]
\PY{c+c1}{\PYZsh{}define Wilson coefficients and model}
\PY{n}{WC} \PY{o}{=} \PY{p}{\PYZob{}}\PY{l+s+s2}{\PYZdq{}}\PY{l+s+s2}{m\PYZus{}bb}\PY{l+s+s2}{\PYZdq{}} \PY{p}{:} \PY{l+m+mi}{100}\PY{o}{*}\PY{n}{meV}\PY{p}{,}
      \PY{l+s+s2}{\PYZdq{}}\PY{l+s+s2}{VR(6)}\PY{l+s+s2}{\PYZdq{}} \PY{p}{:} \PY{l+m+mf}{1e\PYZhy{}7}
     \PY{p}{\PYZcb{}}

\PY{n}{model} \PY{o}{=} \PY{n}{LEFT}\PY{p}{(}\PY{n}{WC}\PY{p}{,} \PY{n}{method} \PY{o}{=} \PY{l+s+s2}{\PYZdq{}}\PY{l+s+s2}{IBM2}\PY{l+s+s2}{\PYZdq{}}\PY{p}{)}
\end{Verbatim}
\end{tcolorbox}

    \begin{tcolorbox}[breakable, size=fbox, boxrule=1pt, pad at break*=1mm,colback=cellbackground, colframe=cellborder]
\prompt{In}{incolor}{3}{\boxspacing}
\begin{Verbatim}[commandchars=\\\{\}]
\PY{c+c1}{\PYZsh{}get the decay rate R in 76Ge}
\PY{n}{R} \PY{o}{=} \PY{l+m+mi}{1}\PY{o}{/}\PY{n}{model}\PY{o}{.}\PY{n}{t\PYZus{}half}\PY{p}{(}\PY{l+s+s2}{\PYZdq{}}\PY{l+s+s2}{76Ge}\PY{l+s+s2}{\PYZdq{}}\PY{p}{)}
\PY{n}{R}
\end{Verbatim}
\end{tcolorbox}

            \begin{tcolorbox}[breakable, size=fbox, boxrule=.5pt, pad at break*=1mm, opacityfill=0]
\prompt{Out}{outcolor}{3}{\boxspacing}
\begin{Verbatim}[commandchars=\\\{\}]
2.418831433432257e-26
\end{Verbatim}
\end{tcolorbox}

    \begin{tcolorbox}[breakable, size=fbox, boxrule=1pt, pad at break*=1mm,colback=cellbackground, colframe=cellborder]
\prompt{In}{incolor}{4}{\boxspacing}
\begin{Verbatim}[commandchars=\\\{\}]
\PY{c+c1}{\PYZsh{}get decay rate formula for 76Ge in latex form}
\PY{n+nb}{print}\PY{p}{(}\PY{n}{model}\PY{o}{.}\PY{n}{generate\PYZus{}formula}\PY{p}{(}\PY{l+s+s2}{\PYZdq{}}\PY{l+s+s2}{76Ge}\PY{l+s+s2}{\PYZdq{}}\PY{p}{)}\PY{p}{)}
\end{Verbatim}
\end{tcolorbox}

    \begin{Verbatim}[commandchars=\\\{\}]
\$T\_\{1/2\}\^{}\{-1\} = +2.1\textbackslash{}times
10\^{}\{-6\}\textbackslash{}left|\textbackslash{}frac\{m\_\{\textbackslash{}beta\textbackslash{}beta\}\}\{1\textbackslash{}mathrm\{GeV\}\}\textbackslash{}right|\^{}2+1.07\textbackslash{}times
10\^{}\{-13\}\textbackslash{}left|C\_\{VR\}\^{}\{(6)\}\textbackslash{}right|\^{}2+2.09\textbackslash{}times 10\^{}\{-10\}
\textbackslash{}mathrm\{Re\}\textbackslash{}left[\textbackslash{}frac\{m\_\{\textbackslash{}beta\textbackslash{}beta\}\}\{1\textbackslash{}mathrm\{GeV\}\}
(\{C\_\{VR\}\^{}\{(6)\}\})\^{}*\textbackslash{}right]\$
    \end{Verbatim}
\quad\\
which if we put it into latex reads
\begin{align}
    T_{1/2}^{-1} = +2.1\times 10^{-6}\left|\frac{m_{\beta\beta}}{1\mathrm{GeV}}\right|^2+1.07\times 10^{-13}\left|C_{VR}^{(6)}\right|^2+2.09\times 10^{-10} \mathrm{Re}\left[\frac{m_{\beta\beta}}{1\mathrm{GeV}}({C_{VR}^{(6)}})^*\right]
\end{align}
    \begin{tcolorbox}[breakable, size=fbox, boxrule=1pt, pad at break*=1mm,colback=cellbackground, colframe=cellborder]
\prompt{In}{incolor}{5}{\boxspacing}
\begin{Verbatim}[commandchars=\\\{\}]
\PY{c+c1}{\PYZsh{}get the decay rate matrix M}
\PY{n}{M} \PY{o}{=} \PY{n}{model}\PY{o}{.}\PY{n}{generate\PYZus{}matrix}\PY{p}{(}\PY{l+s+s2}{\PYZdq{}}\PY{l+s+s2}{76Ge}\PY{l+s+s2}{\PYZdq{}}\PY{p}{)}
\PY{n}{M}
\end{Verbatim}
\end{tcolorbox}

            \begin{tcolorbox}[breakable, size=fbox, boxrule=.5pt, pad at break*=1mm, opacityfill=0]
\prompt{Out}{outcolor}{5}{\boxspacing}
\begin{Verbatim}[commandchars=\\\{\}]
array([[2.10337991e-06, 1.04362734e-10],
       [1.04362734e-10, 1.06726053e-13]])
\end{Verbatim}
\end{tcolorbox}

    \begin{tcolorbox}[breakable, size=fbox, boxrule=1pt, pad at break*=1mm,colback=cellbackground, colframe=cellborder]
\prompt{In}{incolor}{6}{\boxspacing}
\begin{Verbatim}[commandchars=\\\{\}]
\PY{c+c1}{\PYZsh{}get the decay rate from the matrix M}
\PY{n}{C}  \PY{o}{=} \PY{n}{np}\PY{o}{.}\PY{n}{array}\PY{p}{(}\PY{n+nb}{list}\PY{p}{(}\PY{n}{WC}\PY{o}{.}\PY{n}{values}\PY{p}{(}\PY{p}{)}\PY{p}{)}\PY{p}{)}
\PY{n}{R2} \PY{o}{=} {C}\PY{o}{.}\PY{n}{T}\PY{o}{.}\PY{n}{conj}\PY{p}{(}\PY{p}{)} \PY{n+nd}{@} M \PY{n+nd}{@} C
\PY{n}{R2}
\end{Verbatim}
\end{tcolorbox}

            \begin{tcolorbox}[breakable, size=fbox, boxrule=.5pt, pad at break*=1mm, opacityfill=0]
\prompt{Out}{outcolor}{6}{\boxspacing}
\begin{Verbatim}[commandchars=\\\{\}]
2.418831433432201e-26
\end{Verbatim}
\end{tcolorbox}
Comparing the results on the decay rate $T_{1/2}^{-1}$ $\mathrm{R}$ and $\mathrm{R2}$ from the functions $\mathrm{model.half\_lives()}$ and $\mathrm{model.generate\_matrix()}$, respectively, we see that they are equal up to some small accumulated rounding errors.
    \begin{tcolorbox}[breakable, size=fbox, boxrule=1pt, pad at break*=1mm,colback=cellbackground, colframe=cellborder]
\prompt{In}{incolor}{7}{\boxspacing}
\begin{Verbatim}[commandchars=\\\{\}]
\PY{p}{(}\PY{n}{R}\PY{o}{\PYZhy{}}\PY{n}{R2}\PY{p}{)}\PY{o}{/}\PY{n}{R}
\end{Verbatim}
\end{tcolorbox}
    \begin{tcolorbox}[breakable, size=fbox, boxrule=.5pt, pad at break*=1mm, opacityfill=0]
\prompt{Out}{outcolor}{7}{\boxspacing}
\begin{Verbatim}[commandchars=\\\{\}]
2.313607087198588e-14
\end{Verbatim}
\end{tcolorbox}
%
%
\subsection{Plot half-lives - varying a single Wilson coefficient}
%
%
\subsubsection{Line-Plots}
You can plot the half-lives $T_{1/2}$ (\texttt{model.plot\_t\_half(...)}), the decay-rate $T_{1/2}^{-1}$ (\texttt{model.plot\_t\_half\_inv(...)}) or the effective Majorana mass $m_\mathrm{eff}$ (\texttt{model.plot\_m\_eff(...)}) as a line-plot while varying a single Wilson coefficient via the functions:
\\\\
\begin{tabularx}{\textwidth}{llX}
    \hline\hline\\
    \multicolumn{3}{l}{\hypertarget{model.plot_WC_variation}{\PY{n}{model}\PY{o}{.}\PY{n}{plot\PYZus{}WC\PYZus{}variation}\PY{p}{(}\PY{n}{xaxis} \PY{o}{=} \PY{l+s+s2}{\PYZdq{}}\PY{l+s+s2}{m\PYZus{}min}\PY{l+s+s2}{\PYZdq{}}\PY{p}{,} \PY{n}{yaxis} \PY{o}{=} \PY{l+s+s2}{\PYZdq{}}\PY{l+s+s2}{t}\PY{l+s+s2}{\PYZdq{}}\PY{p}{,} \PY{n}{isotope} \PY{o}{=} \PY{l+s+s2}{\PYZdq{}}\PY{l+s+s2}{76Ge}\PY{l+s+s2}{\PYZdq{}}\PY{p}{,}}}
    \\
    \multicolumn{3}{l}{\qquad\quad\PY{n}{x\PYZus{}min} \PY{o}{=} \PY{l+m+mf}{1e\PYZhy{}4}\PY{p}{,} \PY{n}{x\PYZus{}max} \PY{o}{=} \PY{l+m+mf}{1e+0}\PY{p}{,} \PY{n}{y\PYZus{}min} \PY{o}{=} \PY{k+kc}{None}\PY{p}{,} \PY{n}{y\PYZus{}max} \PY{o}{=} \PY{k+kc}{None}\PY{p}{,} \PY{n}{xscale} \PY{o}{=} \PY{l+s+s2}{\PYZdq{}}\PY{l+s+s2}{log}\PY{l+s+s2}{\PYZdq{}}\PY{p}{,}}
    \\
    \multicolumn{3}{l}{\qquad\quad\PY{n}{yscale} \PY{o}{=} \PY{l+s+s2}{\PYZdq{}}\PY{l+s+s2}{log}\PY{l+s+s2}{\PYZdq{}}\PY{p}{,} \PY{n}{n\PYZus{}points} \PY{o}{=} \PY{l+m+mi}{100}\PY{p}{,} \PY{n}{cosmo} \PY{o}{=} \PY{k+kc}{False}\PY{p}{,} \PY{n}{m\PYZus{}cosmo} \PY{o}{=} \PY{l+m+mf}{0.15}\PY{p}{,} \PY{n}{limits} \PY{o}{=} \PY{k+kc}{None}\PY{p}{,}}\\
    \multicolumn{3}{l}{\qquad\quad\PY{n}{experiments} \PY{o}{=} \PY{k+kc}{None}\PY{p}{,} \PY{n}{ordering} \PY{o}{=} \PY{l+s+s2}{\PYZdq{}}\PY{l+s+s2}{both}\PY{l+s+s2}{\PYZdq{}}\PY{p}{,} \PY{n}{dcp} \PY{o}{=} \PY{l+m+mf}{1.36}\PY{p}{,} \PY{n}{numerical\PYZus{}method} \PY{o}{=} \PY{l+s+s2}{\PYZdq{}}\PY{l+s+s2}{Powell}\PY{l+s+s2}{\PYZdq{}}\PY{p}{,}}
    \\
    \multicolumn{3}{l}{\qquad\quad\PY{n}{show\PYZus{}mbb} \PY{o}{=} \PY{k+kc}{False}\PY{p}{,} \PY{n}{normalize} \PY{o}{=} \PY{k+kc}{False}\PY{p}{,} \PY{n}{colorNO} \PY{o}{=} \PY{l+s+s2}{\PYZdq{}}\PY{l+s+s2}{b}\PY{l+s+s2}{\PYZdq{}}\PY{p}{,} \PY{n}{colorIO} \PY{o}{=} \PY{l+s+s2}{\PYZdq{}}\PY{l+s+s2}{r}\PY{l+s+s2}{\PYZdq{}}\PY{p}{,}}
    \\
    \multicolumn{3}{l}{\qquad\quad\PY{n}{legend} \PY{o}{=} \PY{k+kc}{True}\PY{p}{,} \PY{n}{labelNO} \PY{o}{=} \PY{k+kc}{None}\PY{p}{,} \PY{n}{labelIO} \PY{o}{=} \PY{k+kc}{None}\PY{p}{,} \PY{n}{autolabel} \PY{o}{=} \PY{k+kc}{True}\PY{p}{,}}\\
    \multicolumn{3}{l}{\qquad\quad\PY{n}{alpha\PYZus{}plot} \PY{o}{=} \PY{l+m+mf}{0.5}\PY{p}{,} \PY{n}{alpha\PYZus{}mass} \PY{o}{=} \PY{l+m+mf}{0.05}\PY{p}{,} \PY{n}{alpha\PYZus{}cosmo} \PY{o}{=} \PY{l+m+mf}{0.1}\PY{p}{,} \PY{n}{vary\PYZus{}phases} \PY{o}{=} \PY{k+kc}{True}\PY{p}{,}}
    \\
    \multicolumn{3}{l}{\qquad\quad\PY{n}{alpha} \PY{o}{=} \PY{p}{[}\PY{l+m+mi}{0}\PY{p}{,}\PY{l+m+mi}{0}\PY{p}{]}\PY{p}{,} \PY{n}{savefig} \PY{o}{=} \PY{k+kc}{False}\PY{p}{,} \PY{n}{file} \PY{o}{=} \PY{l+s+s2}{\PYZdq{}}\PY{l+s+s2}{variation.png}\PY{l+s+s2}{\PYZdq{}}\PY{p}{,} \PY{n}{dpi} \PY{o}{=} \PY{l+m+mi}{300}\PY{p}{)}}
    \\\\
    \multicolumn{3}{l}{\hypertarget{model.plot_t_half}{\PY{n}{model}\PY{o}{.}\PY{n}{plot\PYZus{}t\PYZus{}half}\PY{p}{(}\PY{p}{)} \PY{o}{=} \PY{n}{model}\PY{o}{.}\PY{n}{plot\PYZus{}WC\PYZus{}variation}\PY{p}{(}\PY{n}{yaxis} \PY{o}{=}\PY{l+s+s2}{\PYZdq{}}\PY{l+s+s2}{t}\PY{l+s+s2}{\PYZdq{}}\PY{p}{,} ...\PY{p}{)}}}
    \\\\
    \multicolumn{3}{l}{\hypertarget{model.plot_t_half_inv}{\PY{n}{model}\PY{o}{.}\PY{n}{plot\PYZus{}t\PYZus{}half\PYZus{}inv}\PY{p}{(}\PY{p}{)} \PY{o}{=} \PY{n}{model}\PY{o}{.}\PY{n}{plot\PYZus{}WC\PYZus{}variation}\PY{p}{(}\PY{n}{yaxis} \PY{o}{=}\PY{l+s+s2}{\PYZdq{}}\PY{l+s+s2}{1/t}\PY{l+s+s2}{\PYZdq{}}\PY{p}{,} ...\PY{p}{)}}}
    \\\\
    \multicolumn{3}{l}{\hypertarget{model.plot_m_bb}{\PY{n}{model}\PY{o}{.}\PY{n}{plot\PYZus{}m\PYZus{}eff}\PY{p}{(}\PY{p}{)} \PY{o}{=} \PY{n}{model}\PY{o}{.}\PY{n}{plot\PYZus{}WC\PYZus{}variation}\PY{p}{(}\PY{n}{yaxis} \PY{o}{=}\PY{l+s+s2}{\PYZdq{}}\PY{l+s+s2}{m\_eff}\PY{l+s+s2}{\PYZdq{}}\PY{p}{,} ...\PY{p}{)}}}
    \\\\
    \hline\\
    \textbf{Parameter} & \textbf{Type} & \textbf{Description}\\\\\hline\\
    \textbf{xaxis} & string & Optional - Wilson coefficient varied on the x-axis.
    \\\\
    \textbf{yaxis} & string & Optional - Choose from \str{t}, \str{1/t}, \str{m\_eff} to get the half-life, inverse half-life or effective neutrino mass.
    \\\\
    \textbf{isotope} & string & Optional - Defines the isotope to be studied.
    \\\\
    \textbf{x\_min} & float & Optional - Minimal x value.
    \\\\
    \textbf{x\_max} & float & Optional - Maximal x value.
    \\\\
    \textbf{y\_min} & float & Optional - Minimal y value.
    \\\\
    \textbf{y\_max} & float & Optional - Maximal y value.
    \\\\
    \textbf{xscale} & string & Optional - Sets scaling of x-axis e.g. \str{log} or \str{lin}.
    \\\\
    \textbf{yscale} & string & Optional - Sets scaling of y-axis e.g. \str{log} or \str{lin}.
    \\\\
    \textbf{n\_points} & integer & Optional - Number of datapoints on the x-axis.
\end{tabularx}
\begin{tabularx}{\textwidth}{llX}
    \textbf{cosmo} & bool & Optional - If \bool{True} show cosmology limit.
    \\\\
    \textbf{m\_cosmo} & integer & Optional - Limit from cosmology.
    \\\\
    \textbf{limits} & dictionary & Optional - Plots limits from experiments \{Name : \{limit, color, linestyle, linewidth, alpha, fill, label\}\}. The Isotope is assumed to be the same as the isotope parameter for the plot.
    \newpage
    \\\\
    \textbf{experiments} & dictionary & Optional - The same as limits but instead of giving the y-axis limit you set the half-life limit.
    \\\\
    \textbf{ordering} & string & Optional - \str{both}, \str{NO} or \str{IO} sets neutrino mass ordering.
    \\\\
    \textbf{dcp} & float & Optional - Dirac CP phase.
    \\\\
    \textbf{numerical\_method} & string & Optional - Optimization method. See scipy.optimize.minimize
    \\\\
    \textbf{show\_mbb} & bool & Optional - If \bool{True} plot mass mechanism for comparison.
    \\\\
    \textbf{normalize} & float & Optional - If \bool{True} normalize to mass mechanism.
    \\\\
    \textbf{colorNO} & string & Optional - Set color for NO plot.
    \\\\
    \textbf{colorIO} & string & Optional - Set color for IO plot.
    \\\\
    \textbf{legend} & bool & Optional - If \bool{True} a legend is shown.
    \\\\
    \textbf{labelNO} & string & Optional - Legend label of NO plot.
    \\\\
    \textbf{labelIO} & string & Optional - Legend label of IO plot.
    \\\\
    \textbf{autolabel} & bool & Optional - If \bool{True}, legend labels are generated automatically.
    \\\\
    \textbf{alpha\_plot} & string & Optional - Set alpha for filled areas.
    \\\\
    \textbf{alpha\_mass} & string & Optional - Set alpha for mass mechanism if \texttt{show\_mbb=}\bool{True}.
    \\\\
    \textbf{alpha\_cosmo} & string & Optional - Set alpha for mass mechanism if cosmo=\bool{True}.
    \\\\
    \textbf{vary\_phases} & string & Optional - If \bool{True}, vary the unknown phases of
the x-axis WC.
\end{tabularx}
\begin{tabularx}{\textwidth}{llX}
    \textbf{alpha} & float & Optional - 2-entries float array that defines the unknown Majorana phases or float that defines the complex WC phase. Only necessary if phases are not varied on the x-axis WC.
    \\\\
    \textbf{savefig} & bool & Optional - If \bool{True} save figure as file
    \\\\
    \textbf{file} & string & Optional - Filename to save figure to.
    \\\\
    \textbf{dpi} & float & Optional - sets the resolution in dots per inch when saving the figure.
    \\\\\hline\hline
\end{tabularx}
\\\\\\
These functions will plot the allowed y-axis range of values by minimizing and maximizing the half-life (effective mass) with respect  the phase of the Wilson coefficient given in the \texttt{xaxis} parameter while varying the absolute value in the range given by the parameters \texttt{x\_min} and \texttt{x\_max}. You can turn off the variation of the unknown phases by setting \texttt{vary\_phases=}\bool{False}. All other Wilson coefficients will be kept constant as well as the unknown LECs. Note that this assumes that all Wilson coefficients are independent from each other. See the provided Jupyter notebook on the mLRSM for an example of how to approach studying a model with interdependent Wilson coefficients.
\\\\
\textit{\textbf{Example: The light-neutrino-exchange mechanism with an additional LNV right-handed current once more}}\\
Let us again consider the previous example of an LNV right-handed current appearing in addition to the effective neutrino Majorana mass. First we plot the half-life for the L$\nu$EM only and save it to \texttt{t\_half\_mass.png} by typing 

    \begin{tcolorbox}[breakable, size=fbox, boxrule=1pt, pad at break*=1mm,colback=cellbackground, colframe=cellborder]
\prompt{In}{incolor}{1}{\boxspacing}
\begin{Verbatim}[commandchars=\\\{\}]
\PY{k+kn}{from} \PY{n+nn}{nudobe} \PY{k+kn}{import} \PY{n}{EFT}
\PY{k+kn}{from} \PY{n+nn}{EFT} \PY{k+kn}{import} \PY{n}{LEFT}
\end{Verbatim}
\end{tcolorbox}

    \begin{tcolorbox}[breakable, size=fbox, boxrule=1pt, pad at break*=1mm,colback=cellbackground, colframe=cellborder]
\prompt{In}{incolor}{2}{\boxspacing}
\begin{Verbatim}[commandchars=\\\{\}]
\PY{n}{fig} \PY{o}{=} \PY{n}{LEFT}\PY{p}{(}\PY{p}{\PYZob{}}\PY{p}{\PYZcb{}}\PY{p}{,} \PY{n}{method} \PY{o}{=} \PY{l+s+s2}{\PYZdq{}}\PY{l+s+s2}{SM}\PY{l+s+s2}{\PYZdq{}}\PY{p}{)}\PY{o}{.}\PY{n}{plot\PYZus{}t\PYZus{}half}\PY{p}{(}\PY{n}{savefig} \PY{o}{=} \PY{k+kc}{True}\PY{p}{,} 
                                          \PY{n}{file}    \PY{o}{=} \PY{l+s+s2}{\PYZdq{}}\PY{l+s+s2}{t\PYZus{}half\PYZus{}mass.png}\PY{l+s+s2}{\PYZdq{}}\PY{p}{)}
\end{Verbatim}
\end{tcolorbox}

    \begin{center}
    \adjustimage{max size={0.75\linewidth}{0.75\paperheight}}{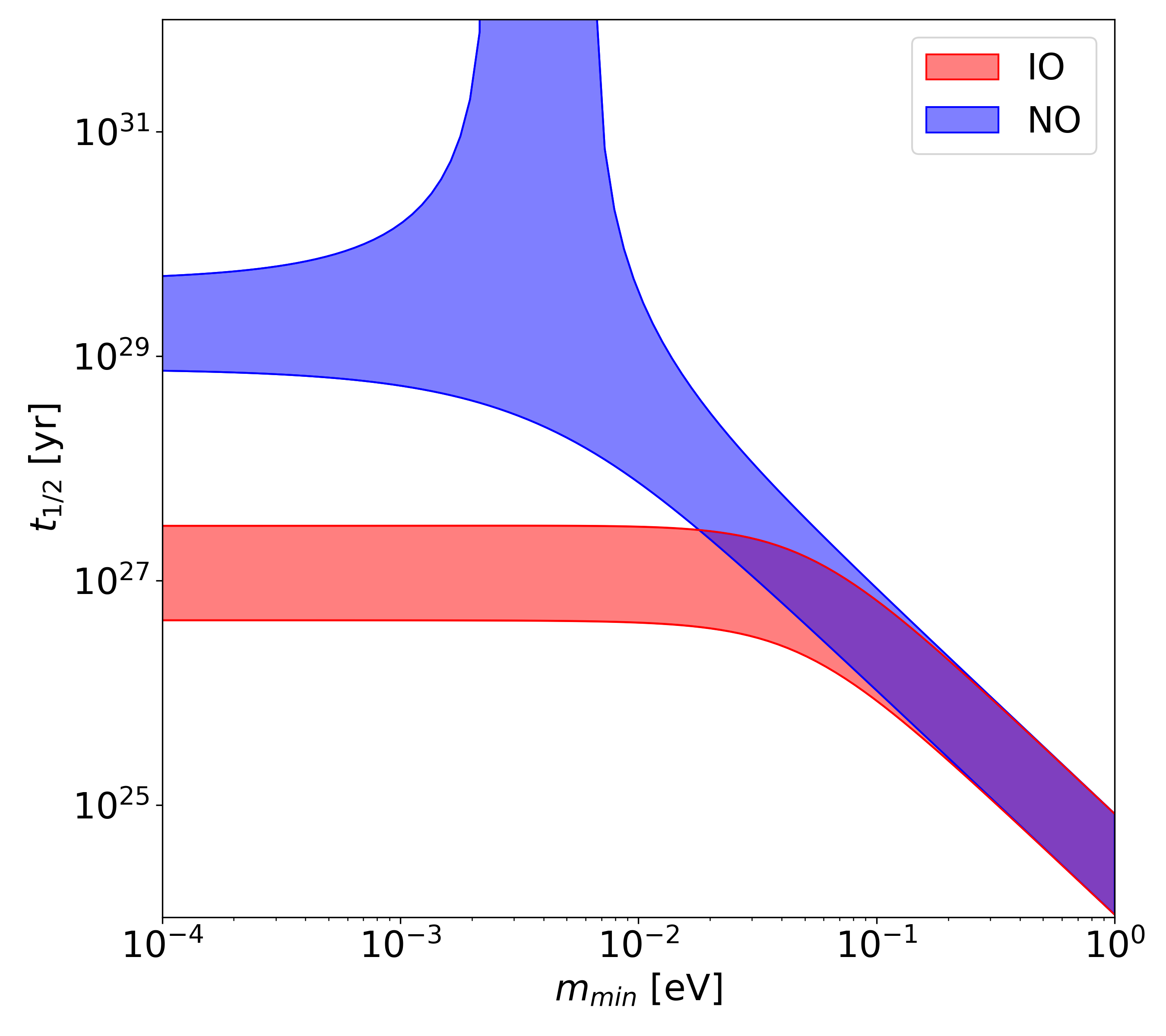}
    \end{center}
    { \hspace*{\fill} \\}
This time we chose NMEs generated in the shell model (\str{SM}), including the $g_\nu^{NN}$ contribution. Then we want to plot the half-lives for our model of interest which we first have to initialize via
    \begin{tcolorbox}[breakable, size=fbox, boxrule=1pt, pad at break*=1mm,colback=cellbackground, colframe=cellborder]
\prompt{In}{incolor}{3}{\boxspacing}
\begin{Verbatim}[commandchars=\\\{\}]
\PY{n}{WC} \PY{o}{=} \PY{p}{\PYZob{}}\PY{l+s+s2}{\PYZdq{}}\PY{l+s+s2}{m\PYZus{}bb}\PY{l+s+s2}{\PYZdq{}} \PY{p}{:} \PY{l+m+mf}{0.1e\PYZhy{}9}\PY{p}{,}
      \PY{l+s+s2}{\PYZdq{}}\PY{l+s+s2}{VR(6)}\PY{l+s+s2}{\PYZdq{}} \PY{p}{:} \PY{l+m+mf}{1e\PYZhy{}7}
     \PY{p}{\PYZcb{}}
\PY{n}{model} \PY{o}{=} \PY{n}{LEFT}\PY{p}{(}\PY{n}{WC}\PY{p}{,} \PY{n}{method} \PY{o}{=} \PY{l+s+s2}{\PYZdq{}}\PY{l+s+s2}{SM}\PY{l+s+s2}{\PYZdq{}}\PY{p}{)}
\end{Verbatim}
\end{tcolorbox}
In our half-life figure we want to show the current experimental limit from GERDA~\citep{GERDA:2020xhi} in yellow. Hence, we choose \textsuperscript{76}Ge as the isotope and define GERDA in the parameter \texttt{experiments}. We plot the standard mass mechanism alongside by putting \texttt{show\_mbb=}\bool{True}. Finally, we save the generated figure as a \texttt{.png} file to \texttt{t\_half\_model.png}
    \begin{tcolorbox}[breakable, size=fbox, boxrule=1pt, pad at break*=1mm,colback=cellbackground, colframe=cellborder]
\prompt{In}{incolor}{4}{\boxspacing}
\begin{Verbatim}[commandchars=\\\{\}]
\PY{n}{isotope} \PY{o}{=} \PY{l+s+s2}{\PYZdq{}}\PY{l+s+s2}{76Ge}\PY{l+s+s2}{\PYZdq{}}
\PY{n}{experiments} \PY{o}{=} \PY{p}{\PYZob{}}\PY{l+s+s2}{\PYZdq{}}\PY{l+s+s2}{GERDA}\PY{l+s+s2}{\PYZdq{}} \PY{p}{:} \PY{p}{\PYZob{}}\PY{l+s+s2}{\PYZdq{}}\PY{l+s+s2}{limit}\PY{l+s+s2}{\PYZdq{}}     \PY{p}{:} \PY{l+m+mf}{1.8e+26}\PY{p}{,} 
                          \PY{l+s+s2}{\PYZdq{}}\PY{l+s+s2}{color}\PY{l+s+s2}{\PYZdq{}}     \PY{p}{:} \PY{l+s+s2}{\PYZdq{}}\PY{l+s+s2}{y}\PY{l+s+s2}{\PYZdq{}}\PY{p}{,} 
                          \PY{l+s+s2}{\PYZdq{}}\PY{l+s+s2}{linestyle}\PY{l+s+s2}{\PYZdq{}} \PY{p}{:} \PY{l+s+s2}{\PYZdq{}}\PY{l+s+s2}{\PYZhy{}}\PY{l+s+s2}{\PYZdq{}}\PY{p}{,} 
                          \PY{l+s+s2}{\PYZdq{}}\PY{l+s+s2}{linewidth}\PY{l+s+s2}{\PYZdq{}} \PY{p}{:} \PY{l+m+mf}{1}\PY{p}{,} 
                          \PY{l+s+s2}{\PYZdq{}}\PY{l+s+s2}{alpha}\PY{l+s+s2}{\PYZdq{}}     \PY{p}{:} \PY{l+m+mf}{0.5}\PY{p}{,} 
                          \PY{l+s+s2}{\PYZdq{}}\PY{l+s+s2}{fill}\PY{l+s+s2}{\PYZdq{}}      \PY{p}{:} \PY{k+kc}{True}\PY{p}{,} 
                          \PY{l+s+s2}{\PYZdq{}}\PY{l+s+s2}{label}\PY{l+s+s2}{\PYZdq{}}     \PY{p}{:} \PY{l+s+s2}{\PYZdq{}}\PY{l+s+s2}{GERDA}\PY{l+s+s2}{\PYZdq{}}\PY{p}{\PYZcb{}}\PY{p}{\PYZcb{}}

\PY{n}{fig} \PY{o}{=} \PY{n}{model}\PY{o}{.}\PY{n}{plot\PYZus{}t\PYZus{}half}\PY{p}{(}\PY{n}{isotope} \PY{o}{=} \PY{n}{isotope}\PY{p}{,} \PY{n}{experiments} \PY{o}{=} \PY{n}{experiments}\PY{p}{,}
                        \PY{n}{show\PYZus{}mbb} \PY{o}{=} \PY{k+kc}{True}\PY{p}{,} \PY{n}{y\PYZus{}min} \PY{o}{=} \PY{l+m+mf}{1e+24}\PY{p}{,} \PY{n}{y\PYZus{}max} \PY{o}{=} \PY{l+m+mf}{1e+31}\PY{p}{,}
                        \PY{n}{savefig} \PY{o}{=} \PY{k+kc}{True}\PY{p}{,} \PY{n}{file} \PY{o}{=} \PY{l+s+s2}{\PYZdq{}}\PY{l+s+s2}{t\PYZus{}half\PYZus{}model.png}\PY{l+s+s2}{\PYZdq{}}\PY{p}{)}
\end{Verbatim}
\end{tcolorbox}

    \begin{center}
    \adjustimage{max size={0.75\linewidth}{0.75\paperheight}}{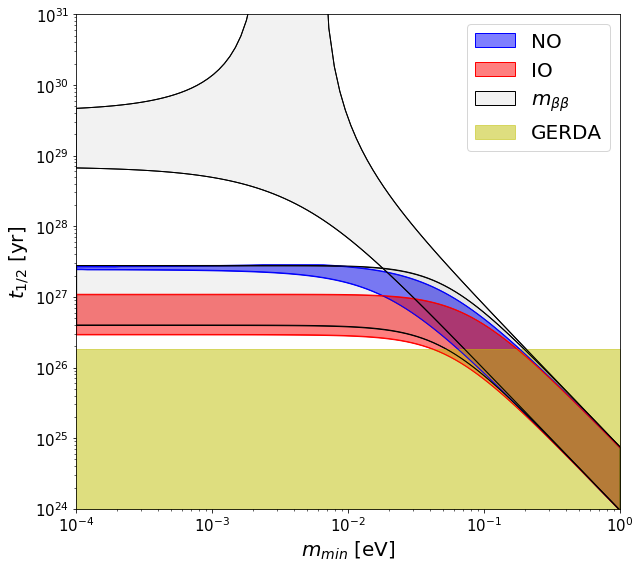}
    \end{center}
    { \hspace*{\fill} \\}

    The gray area now corresponds to the plot above with just $m_{\beta\beta}$, i.e., it shows the expected half-life from the L$\nu$EM. We see that due to the new LNV operator, the \0 rates are significantly enhanced for the normal ordering for small $m_{\text{min}}$ but this effect is less pronounced in the inverted ordering. The yellow area shows the regions of the parameter space excluded by the GERDA experiment~\citep{GERDA:2020xhi}.
    
Additionally, instead of plotting the mass mechanism alongside our BSM model we can normalize the half-life generated by our model to the L$\nu$EM and plot it via
    \begin{tcolorbox}[breakable, size=fbox, boxrule=1pt, pad at break*=1mm,colback=cellbackground, colframe=cellborder]
\prompt{In}{incolor}{5}{\boxspacing}
\begin{Verbatim}[commandchars=\\\{\}]
\PY{n}{model}\PY{o}{.}\PY{n}{plot\PYZus{}t\PYZus{}half}\PY{p}{(}\PY{n}{isotope}   \PY{o}{=} \PY{n}{isotope}\PY{p}{,}
                  \PY{n}{normalize} \PY{o}{=} \PY{k+kc}{True}\PY{p}{,} 
                  \PY{n}{savefig}   \PY{o}{=} \PY{k+kc}{True}\PY{p}{,} 
                  \PY{n}{file}      \PY{o}{=} \PY{l+s+s2}{\PYZdq{}}\PY{l+s+s2}{t\PYZus{}half\PYZus{}normalized.png}\PY{l+s+s2}{\PYZdq{}}\PY{p}{)}
\end{Verbatim}
\end{tcolorbox}
    
    \begin{center}
    \adjustimage{max size={0.75\linewidth}{0.75\paperheight}}{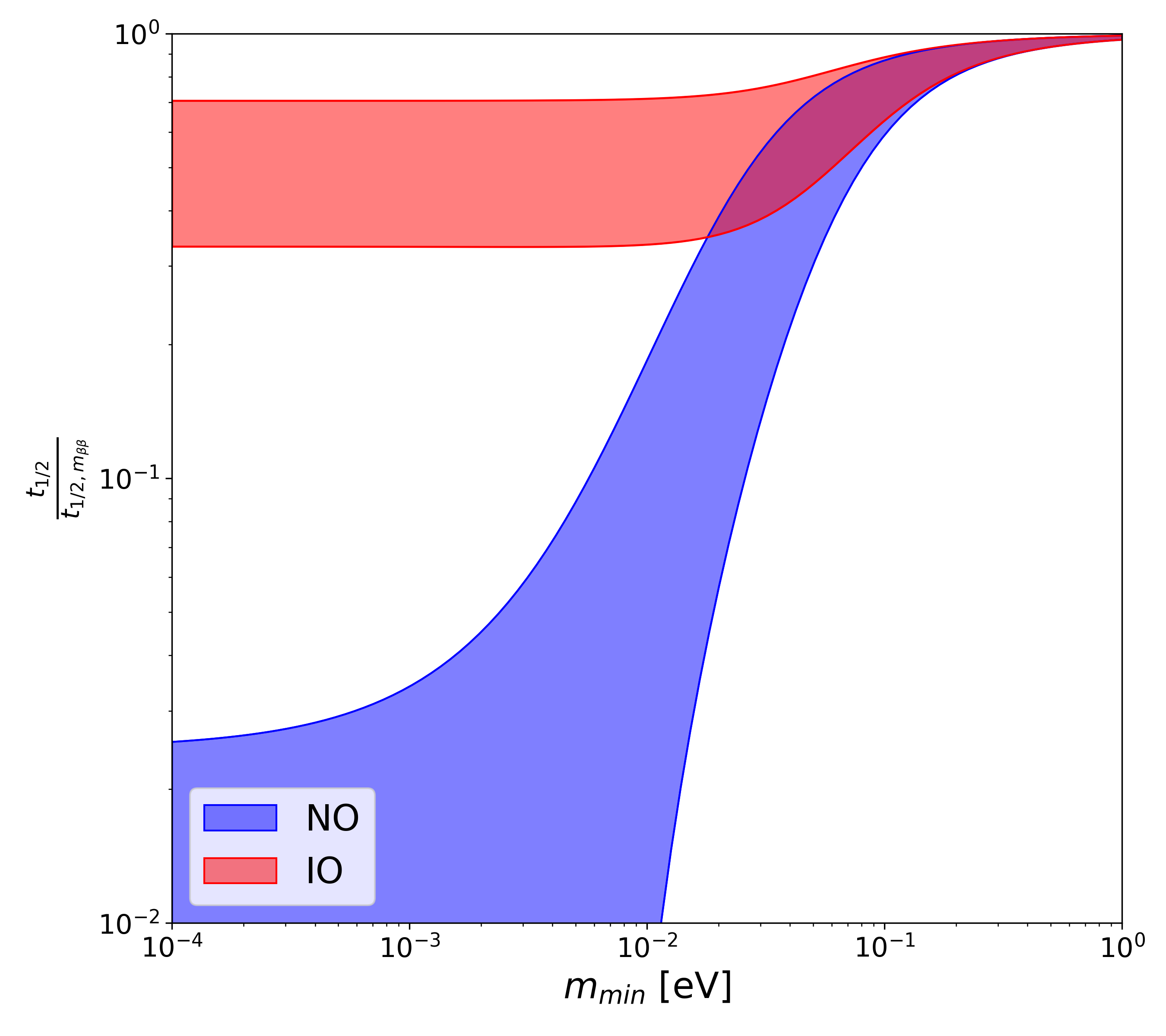}
    \end{center}
    { \hspace*{\fill} \\}

From these plots we can easily see that while the case of inverse mass hierarchy is almost unaffected by the additional higher dimensional contribution, the normal mass hierarchy case results in significantly higher decay rates for minimal neutrino masses below $0.1\,$eV.
%
%
\subsubsection{Scatter Plots}
The above plots show all allowed values when varying the unknown phase(s) of the Wilson coefficient on the x-axis. If, instead, you want to additionally vary the unknown LECs, get a graphical representation of the probability distribution when doing a variation, or simply want a different design you can generate scatter plots through
\\\\
\begin{tabularx}{\textwidth}{llX}
    \hline\hline\\
    \multicolumn{3}{l}{\hypertarget{model.plot_WC_variation_scatter}{\PY{n}{plot\PYZus{}WC\PYZus{}variation\PYZus{}scatter}\PY{p}{(}\PY{n}{xaxis} \PY{o}{=} \PY{l+s+s2}{\PYZdq{}}\PY{l+s+s2}{m\PYZus{}min}\PY{l+s+s2}{\PYZdq{}}\PY{p}{,} \PY{n}{yaxis} \PY{o}{=} \PY{l+s+s2}{\PYZdq{}}\PY{l+s+s2}{t}\PY{l+s+s2}{\PYZdq{}}\PY{p}{,} \PY{n}{isotope} \PY{o}{=} \PY{l+s+s2}{\PYZdq{}}\PY{l+s+s2}{76Ge}\PY{l+s+s2}{\PYZdq{}}\PY{p}{,}}}
    \\
    \multicolumn{3}{l}{\qquad\quad\PY{n}{vary\PYZus{}phases} \PY{o}{=} \PY{k+kc}{True}\PY{p}{,} \PY{n}{vary\PYZus{}LECs} \PY{o}{=} \PY{k+kc}{False}\PY{p}{,} \PY{n}{alpha} \PY{o}{=} \PY{p}{[}\PY{l+m+mi}{0}\PY{p}{,}\PY{l+m+mi}{0}\PY{p}{]}\PY{p}{,} \PY{n}{n\PYZus{}points} \PY{o}{=} \PY{l+m+mi}{10000}\PY{p}{,}}
    \\
    \multicolumn{3}{l}{\qquad\quad\PY{n}{markersize} \PY{o}{=} \PY{l+m+mf}{0.5}\PY{p}{,} \PY{n}{x\PYZus{}min} \PY{o}{=} \PY{l+m+mf}{1e\PYZhy{}4}\PY{p}{,} \PY{n}{x\PYZus{}max} \PY{o}{=} \PY{l+m+mi}{1}\PY{p}{,} \PY{n}{y\PYZus{}min} \PY{o}{=} \PY{k+kc}{None}\PY{p}{,} \PY{n}{y\PYZus{}max} \PY{o}{=} \PY{k+kc}{None}\PY{p}{,}}
    \\
    \multicolumn{3}{l}{\qquad\quad\PY{n}{xscale} \PY{o}{=} \PY{l+s+s2}{\PYZdq{}}\PY{l+s+s2}{log}\PY{l+s+s2}{\PYZdq{}}\PY{p}{,} \PY{n}{yscale} \PY{o}{=} \PY{l+s+s2}{\PYZdq{}}\PY{l+s+s2}{log}\PY{l+s+s2}{\PYZdq{}}\PY{p}{,} \PY{n}{limits} \PY{o}{=} \PY{k+kc}{None}\PY{p}{,} \PY{n}{experiments} \PY{o}{=} \PY{k+kc}{None}\PY{p}{,}}
    \\
    \multicolumn{3}{l}{\qquad\quad\PY{n}{ordering} \PY{o}{=} \PY{l+s+s2}{\PYZdq{}}\PY{l+s+s2}{both}\PY{l+s+s2}{\PYZdq{}}\PY{p}{,} \PY{n}{dcp} \PY{o}{=} \PY{l+m+mf}{1.36}\PY{p}{,} \PY{n}{show\PYZus{}mbb} \PY{o}{=} \PY{k+kc}{False}\PY{p}{,} \PY{n}{normalize} \PY{o}{=} \PY{k+kc}{False}\PY{p}{,}}
    \\
    \multicolumn{3}{l}{\qquad\quad\PY{n}{cosmo} \PY{o}{=} \PY{k+kc}{False}\PY{p}{,} \PY{n}{m\PYZus{}cosmo} \PY{o}{=} \PY{l+m+mf}{0.15}\PY{p}{,} \PY{n}{colorNO} \PY{o}{=} \PY{l+s+s2}{\PYZdq{}}\PY{l+s+s2}{b}\PY{l+s+s2}{\PYZdq{}}\PY{p}{,} \PY{n}{colorIO} \PY{o}{=} \PY{l+s+s2}{\PYZdq{}}\PY{l+s+s2}{r}\PY{l+s+s2}{\PYZdq{}}\PY{p}{,} \PY{n}{legend} \PY{o}{=} \PY{k+kc}{True}\PY{p}{,}}
    \\
    \multicolumn{3}{l}{\qquad\quad\PY{n}{labelNO} \PY{o}{=} \PY{k+kc}{None}\PY{p}{,} \PY{n}{labelIO} \PY{o}{=} \PY{k+kc}{None}\PY{p}{,} \PY{n}{autolabel} \PY{o}{=} \PY{k+kc}{True}\PY{p}{,} \PY{n}{alpha\PYZus{}plot} \PY{o}{=} \PY{l+m+mi}{1}\PY{p}{,}}
    \\
    \multicolumn{3}{l}{\qquad\quad\PY{n}{alpha\PYZus{}mass} \PY{o}{=} \PY{l+m+mf}{0.05}\PY{p}{,} \PY{n}{alpha\PYZus{}cosmo} \PY{o}{=} \PY{l+m+mf}{0.1}\PY{p}{,} \PY{n}{savefig} \PY{o}{=} \PY{k+kc}{False}\PY{p}{,} \PY{n}{file} \PY{o}{=} \PY{l+s+s2}{\PYZdq{}}\PY{l+s+s2}{var\PYZus{}scat.png}\PY{l+s+s2}{\PYZdq{}}\PY{p}{,}}
    \\
    \multicolumn{3}{l}{\qquad\quad \PY{n}{dpi} \PY{o}{=} \PY{l+m+mi}{300}\PY{p}{)}}
    \\\\
    \multicolumn{3}{l}{\hypertarget{model.plot_t_half_scatter}{\PY{n}{model}\PY{o}{.}\PY{n}{plot\PYZus{}t\PYZus{}half\PYZus{}scatter}\PY{p}{(}...\PY{p}{)} \PY{o}{=} \PY{n}{model}\PY{o}{.}\PY{n}{plot\PYZus{}WC\PYZus{}variation}\PY{p}{(}\PY{n}{yaxis} \PY{o}{=}\PY{l+s+s2}{\PYZdq{}}\PY{l+s+s2}{m\_eff}\PY{l+s+s2}{\PYZdq{}}\PY{p}{,} ...\PY{p}{)}}}
    \\\\
    \multicolumn{3}{l}{\hypertarget{model.plot_t_half_inv_scatter}{\PY{n}{model}\PY{o}{.}\PY{n}{plot\PYZus{}t\PYZus{}half\PYZus{}inv\PYZus{}scatter}\PY{p}{(}...\PY{p}{)} \PY{o}{=} \PY{n}{model}\PY{o}{.}\PY{n}{plot\PYZus{}WC\PYZus{}variation}\PY{p}{(}\PY{n}{yaxis} \PY{o}{=}\PY{l+s+s2}{\PYZdq{}}\PY{l+s+s2}{m\_eff}\PY{l+s+s2}{\PYZdq{}}\PY{p}{,} ...\PY{p}{)}}}
    \\\\
    \multicolumn{3}{l}{\hypertarget{model.plot_m_bb_scatter}{\PY{n}{model}\PY{o}{.}\PY{n}{plot\PYZus{}m\PYZus{}eff\PYZus{}scatter}\PY{p}{(}...\PY{p}{)} \PY{o}{=} \PY{n}{model}\PY{o}{.}\PY{n}{plot\PYZus{}WC\PYZus{}variation}\PY{p}{(}\PY{n}{yaxis} \PY{o}{=}\PY{l+s+s2}{\PYZdq{}}\PY{l+s+s2}{m\_eff}\PY{l+s+s2}{\PYZdq{}}\PY{p}{,} ...\PY{p}{)}}}
    \\\\
    \hline\\    
    \textbf{Parameter} & \textbf{Type} & \textbf{Description}\\\\\hline\\
    \textbf{xaxis} & string & Optional - Wilson coefficient varied on the x-axis.
    \\\\
    \textbf{yaxis} & string & Optional - Choose from \str{t}, \str{1/t}, \str{m\_eff} to get the half-life, inverse half-life or effective neutrino mass.
    \\\\
    \textbf{isotope} & string & Optional - Defines the isotope to be studied.
    \\\\
    \textbf{vary\_phases} & string & Optional - If \bool{True}, vary the unknown phases of
the x-axis WC.
    \\\\
    \textbf{vary\_LECs} & string & Optional - If \bool{True}, vary the unknown LECs within
their NDA estimate (see Table~\ref{Tab:LECs}).
    \\\\
    \textbf{alpha} & float & Optional - 2-entries float array that defines the unknown Majorana phases or float that defines the complex WC phase. Only necessary if phases are not varied on the x-axis WC.
    \\\\
    \textbf{n\_points} & integer & Optional - Number of scattered datapoints.
    \\\\
    \textbf{markersize} & float & Optional - Markersize of scattered points.
    \\\\
    \textbf{x\_min} & float & Optional - Minimal x value.
    \\\\
    \textbf{x\_max} & float & Optional - Maximal x value.
\end{tabularx}
\begin{tabularx}{\textwidth}{llX}
    \textbf{y\_min} & float & Optional - Minimal y value.
    \\\\
    \textbf{y\_max} & float & Optional - Maximal y value.
    \\\\
    \textbf{xscale} & string & Optional - Sets scaling of x-axis e.g. \str{log} or \str{lin}.
    \\\\
    \textbf{yscale} & string & Optional - Sets scaling of y-axis e.g. \str{log} or \str{lin}.
    \\\\
    \textbf{limits} & dictionary & Optional - Plots limits from experiments \{Name : \{limit, color, linestyle, linewidth, alpha, fill, label\}\}. The Isotope is assumed to be the same as the isotope parameter for the plot.
    \newpage
    \\\\
    \textbf{experiments} & dictionary & Optional - The same as limits but instead of giving the y-axis limit you set the half-life limit.
    \\\\
    \textbf{ordering} & string & Optional - ``both'', ``NO'' or ``IO'' sets neutrino mass ordering.
    \\\\
    \textbf{dcp} & float & Optional - Dirac CP phase.
    \\\\
    \textbf{numerical\_method} & string & Optional - Optimization method. See scipy.optimize.minimize
    \\\\
    \textbf{show\_mbb} & bool & Optional - If \bool{True} plot mass mechanism for comparison.
    \\\\
    \textbf{normalize} & float & Optional - If \bool{True} normalize to mass mechanism.
    \\\\
    \textbf{cosmo} & bool & Optional - If \bool{True} show cosmology limit.
    \\\\
    \textbf{m\_cosmo} & integer & Optional - Limit from cosmology.
    \\\\
    \textbf{colorNO} & string & Optional - Set color for NO plot.
    \\\\
    \textbf{colorIO} & string & Optional - Set color for IO plot.
    \\\\
    \textbf{legend} & bool & Optional - If \bool{True} a legend is shown.
    \\\\
    \textbf{labelNO} & string & Optional - Legend label of NO plot.
    \\\\
    \textbf{labelIO} & string & Optional - Legend label of IO plot.
\end{tabularx}
\begin{tabularx}{\textwidth}{llX}
    \textbf{autolabel} & bool & Optional - If \bool{True}, legend labels are generated automatically.
    \\\\
    \textbf{alpha\_plot} & string & Optional - Set alpha for filled areas.
    \\\\
    \textbf{alpha\_mass} & string & Optional - Set alpha for mass mechanism if \texttt{show\_mbb=}\bool{True}.
    \\\\
    \textbf{alpha\_cosmo} & string & Optional - Set alpha for mass mechanism if cosmo=\bool{True}.
    \\\\
    \textbf{savefig} & bool & Optional - If \bool{True} save figure as file
    \\\\
    \textbf{file} & string & Optional - Filename to save figure to.
    \\\\
    \textbf{dpi} & float & Optional - sets the resolution in dots per inch when saving the figure.
    \\\\\hline\hline
\end{tabularx}
\\\\\\
\textit{\textbf{Example: light-neutrino-exchange mechanism with variation of $g_\nu^{NN}$}}\\
Again, we import the module and define the model by setting the NME method and WCs
    \begin{tcolorbox}[breakable, size=fbox, boxrule=1pt, pad at break*=1mm,colback=cellbackground, colframe=cellborder]
\prompt{In}{incolor}{1}{\boxspacing}
\begin{Verbatim}[commandchars=\\\{\}]
\PY{k+kn}{from} \PY{n+nn}{nudobe} \PY{k+kn}{import} \PY{n}{EFT}
\PY{k+kn}{from} \PY{n+nn}{EFT} \PY{k+kn}{import} \PY{n}{LEFT}
\end{Verbatim}
\end{tcolorbox}

    \begin{tcolorbox}[breakable, size=fbox, boxrule=1pt, pad at break*=1mm,colback=cellbackground, colframe=cellborder]
\prompt{In}{incolor}{2}{\boxspacing}
\begin{Verbatim}[commandchars=\\\{\}]
\PY{n}{WC} \PY{o}{=} \PY{p}{\PYZob{}}\PY{l+s+s2}{\PYZdq{}}\PY{l+s+s2}{m\PYZus{}bb}\PY{l+s+s2}{\PYZdq{}} \PY{p}{:} \PY{l+m+mf}{0.1e\PYZhy{}9}\PY{p}{\PYZcb{}}
\PY{n}{method} \PY{o}{=} \PY{l+s+s2}{\PYZdq{}}\PY{l+s+s2}{SM}\PY{l+s+s2}{\PYZdq{}}

\PY{n}{model} \PY{o}{=} \PY{n}{LEFT}\PY{p}{(}\PY{n}{WC}\PY{p}{,} \PY{n}{method} \PY{o}{=} \PY{n}{method}\PY{p}{)}
\end{Verbatim}
\end{tcolorbox}
We can generate a scatter-plot comparing the standard mass mechanism with and without variation of the unknown LEC $g_\nu^{NN}$ in a range of $\pm50\%$ around the central value
    \begin{tcolorbox}[breakable, size=fbox, boxrule=1pt, pad at break*=1mm,colback=cellbackground, colframe=cellborder]
\prompt{In}{incolor}{3}{\boxspacing}
\begin{Verbatim}[commandchars=\\\{\}]
\PY{n}{model}\PY{o}{.}\PY{n}{plot\PYZus{}t\PYZus{}half\PYZus{}scatter}\PY{p}{(}\PY{n}{vary\PYZus{}LECs} \PY{o}{=} \PY{k+kc}{True}\PY{p}{,} \PY{n}{show\PYZus{}mbb} \PY{o}{=} \PY{k+kc}{True}\PY{p}{,}
                          \PY{n}{savefig} \PY{o}{=} \PY{k+kc}{True}\PY{p}{,} \PY{n}{file} \PY{o}{=} \PY{l+s+s2}{\PYZdq{}}\PY{l+s+s2}{mbb\PYZus{}scatter.png}\PY{l+s+s2}{\PYZdq{}}\PY{p}{,} 
                          \PY{n}{dpi} \PY{o}{=} \PY{l+m+mi}{300}\PY{p}{)}
\end{Verbatim}
\end{tcolorbox}

    \begin{center}
    \adjustimage{max size={0.75\linewidth}{0.75\paperheight}}{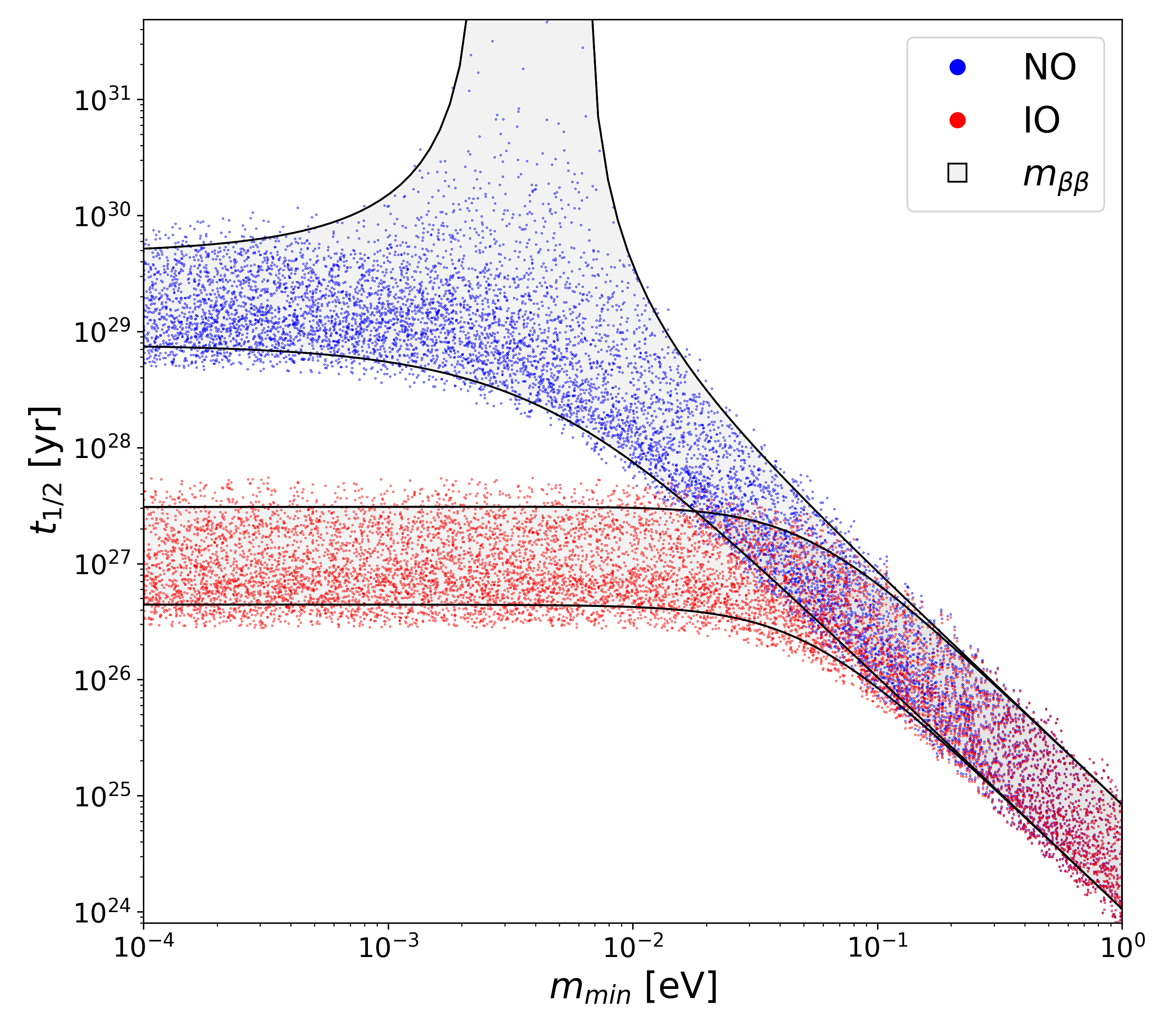}
    \end{center}
    { \hspace*{\fill} \\}

In the above plot, the scattered points show the resulting half-lives when varying both the Majorana phases as well as $g_\nu^{NN}$ while the shaded regions show the allowed half-life region when only the Majorana phases are varied.
%
%
\subsection{Half-life Ratios}
To distinguish different models among each other it can be interesting to look at the ratios of half-lives compared to a reference isotope~\cite{Graf:2022lhj}, e.g. \textsuperscript{76}Ge,
\begin{align}
    R = \frac{T_{1/2}(^AZ)}{T_{1/2}(^{76}\mathrm{Ge})}.
\end{align}
You can return the ratios in all isotopes of interest in a tabular form by using
\\\\
\begin{tabularx}{\textwidth}{llX}
    \hline\hline\\
    \multicolumn{3}{l}{\hypertarget{model.ratios}{\PY{n}{model}\PY{o}{.}\PY{n}{ratios}\PY{p}{(}\PY{n}{reference\PYZus{}isotope} \PY{o}{=} \PY{l+s+s2}{\PYZdq{}}\PY{l+s+s2}{76Ge}\PY{l+s+s2}{\PYZdq{}}\PY{p}{,} \PY{n}{normalized} \PY{o}{=} \PY{k+kc}{True}\PY{p}{,} \PY{n}{WC} \PY{o}{=} \PY{k+kc}{None}\PY{p}{,}}}
    \\
    \multicolumn{3}{l}{\qquad\quad\PY{n}{method}\PY{o}{=}\PY{k+kc}{None}\PY{p}{,} \PY{n}{vary\PYZus{}LECs} \PY{o}{=} \PY{k+kc}{False}\PY{p}{,} \PY{n}{n\PYZus{}points} \PY{o}{=} \PY{l+m+mi}{100}\PY{p}{)}}
    \\\\
    \hline\\
    \textbf{Parameter} & \textbf{Type} & \textbf{Description}\\\\\hline\\
    \textbf{reference\_isotope} & string & Optional - Sets the reference isotope for the half-life ratios.
    \\\\
    \textbf{normalized} & bool & Optional - If \bool{True} the ratios will be normalized to the standard mass mechanism $R/R_{m_{\beta\beta}}$.
    \\\\
    \textbf{WC} & dictionary & Optional - Defines the Wilson coefficients as \{\str{WCname1} : WCvalue1\, ..., \str{WCnameN} : WCvalueN\}. If \bool{None} the models WCs will be used.
    \\\\
    \textbf{method} & string & Optional - Sets the NME calculation method. You can choose from \str{IBM2}, \str{SM} and \str{QRPA}. If \bool{None} the models method will be used.
    \\\\
    \textbf{vary\_LECs} & bool & Optional - If set to \bool{True} the unknown LECs will be varied within their NDA estimates (see Table~\ref{Tab:LECs}). If set to \bool{False} the unknown LECs will stay fixed
    \\\\
    \textbf{n\_points} & integer & Optional - Number of variations.
    \\\\\hline\hline
\end{tabularx}
\\\\\\
which will output a pandas DataFrame containing the half-life ratios. You can depict the ratio $R$ or, if you set \texttt{normalized=}\bool{True}, $R/R_{m_{\beta\beta}}$. Additionally, you can generate a scatter plot of these ratios through
\\\\
\begin{tabularx}{\textwidth}{llX}
    \hline\hline\\
    \multicolumn{3}{l}{\hypertarget{model.plot_ratios}{\PY{n}{model}\PY{o}{.}\PY{n}{plot\PYZus{}ratios}\PY{p}{(}\PY{n}{reference\PYZus{}isotope} \PY{o}{=} \PY{l+s+s2}{\PYZdq{}}\PY{l+s+s2}{76Ge}\PY{l+s+s2}{\PYZdq{}}\PY{p}{,} \PY{n}{normalized} \PY{o}{=} \PY{k+kc}{True}\PY{p}{,} \PY{n}{WC} \PY{o}{=} \PY{k+kc}{None}\PY{p}{,}}}
    \\
    \multicolumn{3}{l}{\qquad\quad\PY{n}{method}\PY{o}{=}\PY{k+kc}{None}\PY{p}{,} \PY{n}{vary\PYZus{}LECs} \PY{o}{=} \PY{k+kc}{False}\PY{p}{,} \PY{n}{n\PYZus{}points} \PY{o}{=} \PY{l+m+mi}{100}\PY{p}{,} \PY{n}{color} \PY{o}{=} \PY{l+s+s2}{\PYZdq{}}\PY{l+s+s2}{b}\PY{l+s+s2}{\PYZdq{}}\PY{p}{,}  \PY{n}{alpha} \PY{o}{=} \PY{l+m+mi}{0.25}\PY{p}{,}}
    \\
    \multicolumn{3}{l}{\qquad\quad\PY{n}{addgrid} \PY{o}{=} \PY{k+kc}{True}\PY{p}{,} \PY{n}{savefig} \PY{o}{=} \PY{k+kc}{False}\PY{p}{,} \PY{n}{file} \PY{o}{=} \PY{l+s+s2}{\PYZdq{}}\PY{l+s+s2}{ratios.png}\PY{l+s+s2}{\PYZdq{}}\PY{p}{,} \PY{n}{n\PYZus{}dpi} \PY{o}{=} \PY{l+m+mi}{300}\PY{p}{)}}
    \\\\
    \hline\\
    \textbf{Parameter} & \textbf{Type} & \textbf{Description}\\\\\hline\\
    \textbf{reference\_isotope} & string & Optional - Sets the reference isotope for the half-life ratios.
    \\\\
    \textbf{normalized} & bool & Optional - If \bool{True} the ratios will be normalized to the standard mass mechanism $R/R_{m_{\beta\beta}}$.
    \\\\
    \textbf{WC} & dictionary & Optional - Defines the Wilson coefficients as \{\str{WCname1} : WCvalue1\, ..., \str{WCnameN} : WCvalueN\}. If \bool{None} the models WCs will be used.
    \\\\
    \textbf{method} & string & Optional - Sets the NME calculation method. You can choose from \str{IBM2}, \str{SM} and \str{QRPA}. If \bool{None} the models method will be used.
    \\\\
    \textbf{vary\_LECs} & bool & Optional - If set to \bool{True} the unknown LECs will be varied within their NDA estimates. If set to \bool{False} the unknown LECs will stay fixed
    \\\\
    \textbf{n\_points} & integer & Optional - Number of variations.
    \\\\
    \textbf{color} & string or list & Optional - Sets the color of the plot either in RGB or via string. See Matplotlib for details.
    \\\\
    \textbf{alpha} & float & Optional - Sets the alpha value for the variational points.
    \\\\
    \textbf{addgrid} & bool & Optional - If \bool{True} a grid is plotted.
    \\\\
    \textbf{savefig} & bool & Optional - If \bool{True} save figure as file
    \\\\
    \textbf{file} & string & Optional - Filename to save figure to.
    \\\\
    \textbf{dpi} & float & Optional - sets the resolution in dots per inch when saving the figure.
    \\\\\hline\hline
\end{tabularx}
\\\\\\
\textit{\textbf{Example: a LNV right-handed current (aka the $\lambda$-mechanism)}}
    
    \begin{tcolorbox}[breakable, size=fbox, boxrule=1pt, pad at break*=1mm,colback=cellbackground, colframe=cellborder]
\prompt{In}{incolor}{1}{\boxspacing}
\begin{Verbatim}[commandchars=\\\{\}]
\PY{k+kn}{from} \PY{n+nn}{nudobe} \PY{k+kn}{import} \PY{n}{EFT}
\PY{k+kn}{from} \PY{n+nn}{EFT} \PY{k+kn}{import} \PY{n}{LEFT}
\end{Verbatim}
\end{tcolorbox}

    \begin{tcolorbox}[breakable, size=fbox, boxrule=1pt, pad at break*=1mm,colback=cellbackground, colframe=cellborder]
\prompt{In}{incolor}{2}{\boxspacing}
\begin{Verbatim}[commandchars=\\\{\}]
\PY{c+c1}{\PYZsh{}Set WCs}
\PY{n}{WC} \PY{o}{=} \PY{p}{\PYZob{}}\PY{l+s+s2}{\PYZdq{}}\PY{l+s+s2}{VR(6)}\PY{l+s+s2}{\PYZdq{}}\PY{p}{:} \PY{l+m+mf}{1e\PYZhy{}7}\PY{p}{\PYZcb{}}

\PY{c+c1}{\PYZsh{}Initiate model}
\PY{n}{model} \PY{o}{=} \PY{n}{LEFT}\PY{p}{(}\PY{n}{WC}\PY{p}{,} \PY{n}{method} \PY{o}{=} \PY{l+s+s2}{\PYZdq{}}\PY{l+s+s2}{SM}\PY{l+s+s2}{\PYZdq{}}\PY{p}{)}
\end{Verbatim}
\end{tcolorbox}

    \begin{tcolorbox}[breakable, size=fbox, boxrule=1pt, pad at break*=1mm,colback=cellbackground, colframe=cellborder]
\prompt{In}{incolor}{3}{\boxspacing}
\begin{Verbatim}[commandchars=\\\{\}]
\PY{c+c1}{\PYZsh{}Define reference isotope}
\PY{n}{reference\PYZus{}isotope} \PY{o}{=} \PY{l+s+s2}{\PYZdq{}}\PY{l+s+s2}{76Ge}\PY{l+s+s2}{\PYZdq{}}
\end{Verbatim}
\end{tcolorbox}

    \begin{tcolorbox}[breakable, size=fbox, boxrule=1pt, pad at break*=1mm,colback=cellbackground, colframe=cellborder]
\prompt{In}{incolor}{4}{\boxspacing}
\begin{Verbatim}[commandchars=\\\{\}]
\PY{c+c1}{\PYZsh{}Calculate half\PYZhy{}life ratios R/R\PYZus{}mbb}
\PY{n}{ratios} \PY{o}{=} \PY{n}{model}\PY{o}{.}\PY{n}{ratios}\PY{p}{(}\PY{n}{reference\PYZus{}isotope}\PY{p}{)}
\PY{n}{ratios}
\end{Verbatim}
\end{tcolorbox}

            \begin{tcolorbox}[breakable, size=fbox, boxrule=.5pt, pad at break*=1mm, opacityfill=0]
\prompt{Out}{outcolor}{4}{\boxspacing}
\begin{Verbatim}[commandchars=\\\{\}]
       48Ca  76Ge    82Se     124Sn     130Te    136Xe
0  0.313404   1.0  0.4864  0.842622  0.721081  0.73654
\end{Verbatim}
\end{tcolorbox}
        
    \begin{tcolorbox}[breakable, size=fbox, boxrule=1pt, pad at break*=1mm,colback=cellbackground, colframe=cellborder]
\prompt{In}{incolor}{5}{\boxspacing}
\begin{Verbatim}[commandchars=\\\{\}]
\PY{c+c1}{\PYZsh{}Calculate ratios R/R\PYZus{}mbb with varied LECs}
\PY{n}{ratios\PYZus{}vary} \PY{o}{=} \PY{n}{model}\PY{o}{.}\PY{n}{ratios}\PY{p}{(}\PY{n}{reference\PYZus{}isotope}\PY{p}{,} \PY{n}{vary\PYZus{}LECs} \PY{o}{=} \PY{k+kc}{True}\PY{p}{)}
\PY{n}{ratios\PYZus{}vary}
\end{Verbatim}
\end{tcolorbox}

            \begin{tcolorbox}[breakable, size=fbox, boxrule=.5pt, pad at break*=1mm, opacityfill=0]
\prompt{Out}{outcolor}{5}{\boxspacing}
\begin{Verbatim}[commandchars=\\\{\}]
        48Ca  76Ge      82Se     124Sn     130Te     136Xe
0   0.188201   1.0  0.466779  0.704380  0.562058  0.607119
1   0.184358   1.0  0.490324  0.691977  0.551269  0.600332
2   0.293439   1.0  0.584830  0.768171  0.655088  0.691217
3   0.346657   1.0  0.791642  0.748110  0.655176  0.705475
4   0.209962   1.0  0.495675  0.707698  0.573777  0.616179
..       {\ldots}   {\ldots}       {\ldots}       {\ldots}       {\ldots}       {\ldots}
95  0.208918   1.0  0.462122  0.721153  0.581308  0.620530
96  0.268060   1.0  0.587650  0.739712  0.622569  0.664289
97  0.235483   1.0  0.510691  0.762411  0.630302  0.671020
98  0.153871   1.0  0.513988  0.642975  0.502340  0.559727
99  0.343899   1.0  0.771410  0.817641  0.727843  0.772934

[100 rows x 6 columns]
\end{Verbatim}
\end{tcolorbox}
        
    \begin{tcolorbox}[breakable, size=fbox, boxrule=1pt, pad at break*=1mm,colback=cellbackground, colframe=cellborder]
\prompt{In}{incolor}{6}{\boxspacing}
\begin{Verbatim}[commandchars=\\\{\}]
\PY{c+c1}{\PYZsh{}Get central ratio values of the LEC variation}
\PY{n}{ratios\PYZus{}vary}\PY{o}{.}\PY{n}{median}\PY{p}{(}\PY{p}{)}
\end{Verbatim}
\end{tcolorbox}

            \begin{tcolorbox}[breakable, size=fbox, boxrule=.5pt, pad at break*=1mm, opacityfill=0]
\prompt{Out}{outcolor}{6}{\boxspacing}
\begin{Verbatim}[commandchars=\\\{\}]
48Ca     0.229494
76Ge     1.000000
82Se     0.526431
124Sn    0.725552
130Te    0.594985
136Xe    0.637997
dtype: float64
\end{Verbatim}
\end{tcolorbox}
        
    \begin{tcolorbox}[breakable, size=fbox, boxrule=1pt, pad at break*=1mm,colback=cellbackground, colframe=cellborder]
\prompt{In}{incolor}{7}{\boxspacing}
\begin{Verbatim}[commandchars=\\\{\}]
\PY{c+c1}{\PYZsh{}Plot Ratios}
\PY{n}{model}\PY{o}{.}\PY{n}{plot\PYZus{}ratios}\PY{p}{(}\PY{l+s+s2}{\PYZdq{}}\PY{l+s+s2}{76Ge}\PY{l+s+s2}{\PYZdq{}}\PY{p}{,} \PY{n}{vary\PYZus{}LECs} \PY{o}{=} \PY{k+kc}{True}\PY{p}{,} \PY{n}{n\PYZus{}points} \PY{o}{=} \PY{l+m+mi}{1000}\PY{p}{,} \PY{n}{alpha} \PY{o}{=} \PY{l+m+mf}{0.1}\PY{p}{,} 
                  \PY{n}{show\PYZus{}central} \PY{o}{=} \PY{k+kc}{True}\PY{p}{,} \PY{n}{savefig} \PY{o}{=} \PY{k+kc}{True}\PY{p}{,} 
                  \PY{n}{file} \PY{o}{=} \PY{l+s+s2}{\PYZdq{}}\PY{l+s+s2}{CVR6\PYZus{}ratios.png}\PY{l+s+s2}{\PYZdq{}}\PY{p}{,} \PY{n}{dpi} \PY{o}{=} \PY{l+m+mi}{300}\PY{p}{)}
\end{Verbatim}
\end{tcolorbox}

    \begin{center}
    \adjustimage{max size={0.75\linewidth}{0.75\paperheight}}{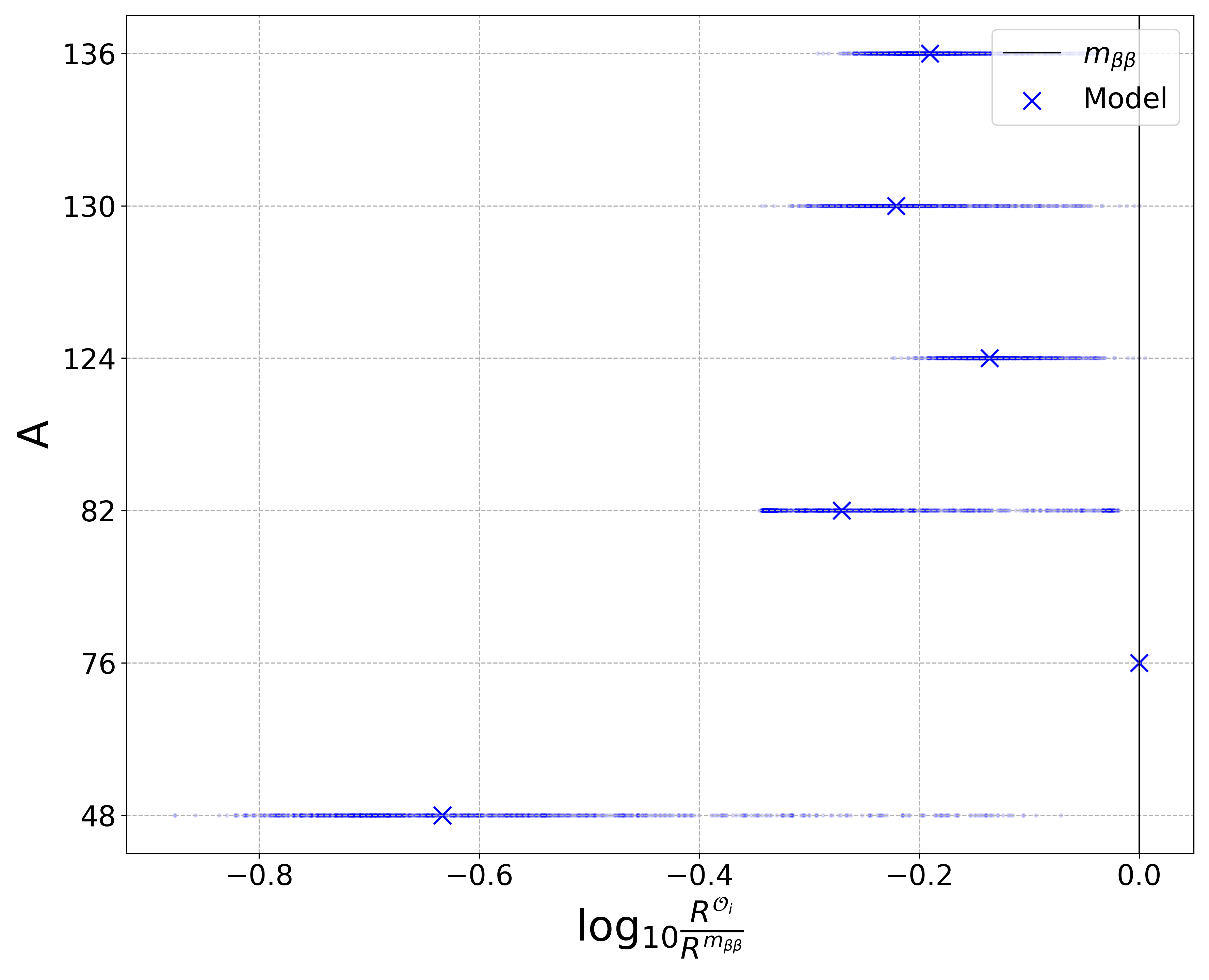}
    \end{center}
    { \hspace*{\fill} \\}

Here, the x-axis shows the range of ratio values while the y-axis gives the isotope mass number $A$.
%
%
\subsection{Phase-Space Observables}
The code can generate both  the single electron spectra $\mathrm{d}\Gamma/\mathrm{d}\epsilon$ defined in~\eqref{eq:spectrum}
as well as the energy dependant angular correlation coefficients $a_1/a_0$ defined in~\eqref{eq:angular_corr} for any given model. To calculate the numerical values for a given normalized electron energy
\begin{align}
    \overline{\epsilon} = \frac{\epsilon-m_e}{Q}
\end{align}
you can use the functions\\\\
\begin{tabularx}{\textwidth}{llX}
    \hline\hline\\
    \multicolumn{3}{l}{\hypertarget{model.spectrum}{\PY{n}{model}\PY{o}{.}\PY{n}{spectrum}\PY{p}{(}\PY{n}{Ebar}\PY{p}{,} \PY{n}{isotope} \PY{o}{=} \PY{l+s+s2}{\PYZdq{}}\PY{l+s+s2}{76Ge}\PY{l+s+s2}{\PYZdq{}}\PY{p}{,} \PY{n}{WC} \PY{o}{=} \PY{k+kc}{None}\PY{p}{,} \PY{n}{method} \PY{o}{=} \PY{k+kc}{None}\PY{p}{)}}}
    \\\\
    \multicolumn{3}{l}{\hypertarget{model.angular_corr}{\PY{n}{model}\PY{o}{.}\PY{n}{angular\PYZus{}corr}\PY{p}{(}\PY{n}{Ebar}\PY{p}{,} \PY{n}{isotope} \PY{o}{=} \PY{l+s+s2}{\PYZdq{}}\PY{l+s+s2}{76Ge}\PY{l+s+s2}{\PYZdq{}}\PY{p}{,} \PY{n}{WC} \PY{o}{=} \PY{k+kc}{None}\PY{p}{,} \PY{n}{method} \PY{o}{=} \PY{k+kc}{None}\PY{p}{)}}}
    \\\\
    \hline\\
    \textbf{Parameter} & \textbf{Type} & \textbf{Description}\\\\\hline\\
    \textbf{Ebar} & float & Normalized electron energy $0<\mathrm{Ebar}<1$. Either float or array of floats.
    \\\\
    \textbf{isotope} & string & Optional - Defines the isotope to be studied.
    \\\\
    \textbf{WC} & dictionary & Optional - Defines the Wilson coefficients as \{\str{WCname1} : WCvalue1\, ..., \str{WCnameN} : WCvalueN\}. If \bool{None} the models WCs will be used.
    \\\\
    \textbf{method} & string & Optional - Sets the NME calculation method. You can choose from \str{IBM2}, \str{SM} and \str{QRPA}. If \bool{None} the models method will be used.
    \\\\\hline\hline
\end{tabularx}
\\\\\\
Plots are then generated through\\\\
\begin{tabularx}{\textwidth}{llX}
    \hline\hline\\
    \multicolumn{3}{l}{\hypertarget{model.plot_spec}{\PY{n}{model}\PY{o}{.}\PY{n}{plot\PYZus{}spec}\PY{p}{(}\PY{n}{isotope}\PY{o}{=}\PY{l+s+s2}{\PYZdq{}}\PY{l+s+s2}{76Ge}\PY{l+s+s2}{\PYZdq{}}\PY{p}{,} \PY{n}{WC} \PY{o}{=} \PY{k+kc}{None}\PY{p}{,} \PY{n}{method}\PY{o}{=}\PY{k+kc}{None}\PY{p}{,} 
    \PY{n}{print\PYZus{}title} \PY{o}{=} \PY{k+kc}{False}\PY{p}{,}}}
    \\
    \multicolumn{3}{l}{\qquad\quad\PY{n}{addgrid} \PY{o}{=} \PY{k+kc}{True}\PY{p}{,} \PY{n}{show\PYZus{}mbb} \PY{o}{=} \PY{k+kc}{True}\PY{p}{,} \PY{n}{normalize\PYZus{}x} \PY{o}{=} \PY{k+kc}{True}\PY{p}{,} \PY{n}{savefig}\PY{o}{=}\PY{k+kc}{False}\PY{p}{,}}
    \\
    \multicolumn{3}{l}{\qquad\quad\PY{n}{file} \PY{o}{=} \PY{l+s+s2}{\PYZdq{}}\PY{l+s+s2}{spec.png}\PY{l+s+s2}{\PYZdq{}}\PY{p}{,} \PY{n}{dpi} \PY{o}{=} \PY{l+m+mf}{300}\PY{p}{)}}
    \\\\
    \multicolumn{3}{l}{\hypertarget{model.plot_corr}{\PY{n}{model}\PY{o}{.}\PY{n}{plot\PYZus{}corr}\PY{p}{(}\PY{n}{isotope}\PY{o}{=}\PY{l+s+s2}{\PYZdq{}}\PY{l+s+s2}{76Ge}\PY{l+s+s2}{\PYZdq{}}\PY{p}{,} \PY{n}{WC} \PY{o}{=} \PY{k+kc}{None}\PY{p}{,} \PY{n}{method}\PY{o}{=}\PY{k+kc}{None}\PY{p}{,} \PY{n}{print\PYZus{}title} \PY{o}{=} \PY{k+kc}{False}\PY{p}{,}}}
    \\
    \multicolumn{3}{l}{\qquad\quad\PY{n}{addgrid} \PY{o}{=} \PY{k+kc}{True}\PY{p}{,} \PY{n}{show\PYZus{}mbb} \PY{o}{=} \PY{k+kc}{True}\PY{p}{,} \PY{n}{normalize\PYZus{}x} \PY{o}{=} \PY{k+kc}{True}\PY{p}{,} \PY{n}{savefig}\PY{o}{=}\PY{k+kc}{False}\PY{p}{,}}
    \\
    \multicolumn{3}{l}{\qquad\quad\PY{n}{file} \PY{o}{=} \PY{l+s+s2}{\PYZdq{}}\PY{l+s+s2}{angular\_corr.png}\PY{l+s+s2}{\PYZdq{}}\PY{p}{,} \PY{n}{dpi} \PY{o}{=} \PY{l+m+mf}{300}\PY{p}{)}}
    \\\\
    \hline\\
    \textbf{Parameter} & \textbf{Type} & \textbf{Description}\\\\\hline\\
    \textbf{isotope} & string & Optional - Defines the isotope to be studied.
\end{tabularx}
\begin{tabularx}{\textwidth}{llX}
    \textbf{WC} & dictionary & Optional - Defines the Wilson coefficients as \{\str{WCname1} : WCvalue1\, ..., \str{WCnameN} : WCvalueN\}. If \bool{None} the models WCs will be used.
    \\\\
    \textbf{method} & string & Optional - Sets the NME calculation method. You can choose from \str{IBM2}, \str{SM} and \str{QRPA}. If \bool{None} the models method will be used.
    \\\\
    \textbf{print\_title} & string & Optional - Put a title into the plot.
    \\\\
    \textbf{addgrid} & bool & Optional - If \bool{True} a grid is plotted.
    \\\\
    \textbf{show\_mbb} & bool & Optional - If \bool{True} plot mass mechanism for comparison.
    \\\\
    \textbf{normalize\_x} & bool & Optional - If \bool{True} normalize x-axis to $\overline{\epsilon} \in [0,1]$. If \bool{False} show $T=\epsilon-m_e$ on the x-axis.x
    \\\\
    \textbf{savefig} & bool & Optional - If \bool{True} save figure as file
    \\\\
    \textbf{file} & string & Optional - Filename to save figure to.
    \\\\
    \textbf{dpi} & float & Optional - sets the resolution in dots per inch when saving the figure.
    \\\\\hline\hline
\end{tabularx}
\\\\\\
\textit{\textbf{Example: The light-neutrino-exchange mechanism with an additional LNV right-handed current}}\\
Let's look at our previous example again which combines a lepton number violating right-handed vector current with the standard mass mechanism. This time we choose $m_{\beta\beta} = 0.1\,$eV and $C_\mathrm{VR}^{(6)} = 3\times10^{-7}$. We initiate the model through
    
    \begin{tcolorbox}[breakable, size=fbox, boxrule=1pt, pad at break*=1mm,colback=cellbackground, colframe=cellborder]
\prompt{In}{incolor}{1}{\boxspacing}
\begin{Verbatim}[commandchars=\\\{\}]
\PY{k+kn}{import} \PY{n+nn}{numpy} \PY{k}{as} \PY{n+nn}{np}
\PY{k+kn}{from} \PY{n+nn}{nudobe} \PY{k+kn}{import} \PY{n}{EFT}
\PY{k+kn}{from} \PY{n+nn}{EFT} \PY{k+kn}{import} \PY{n}{LEFT}
\end{Verbatim}
\end{tcolorbox}

    \begin{tcolorbox}[breakable, size=fbox, boxrule=1pt, pad at break*=1mm,colback=cellbackground, colframe=cellborder]
\prompt{In}{incolor}{2}{\boxspacing}
\begin{Verbatim}[commandchars=\\\{\}]
\PY{n}{WC} \PY{o}{=} \PY{p}{\PYZob{}}\PY{l+s+s2}{\PYZdq{}}\PY{l+s+s2}{m\PYZus{}bb}\PY{l+s+s2}{\PYZdq{}} \PY{p}{:} \PY{l+m+mf}{0.1e\PYZhy{}9}\PY{p}{,} 
      \PY{l+s+s2}{\PYZdq{}}\PY{l+s+s2}{VR(6)}\PY{l+s+s2}{\PYZdq{}} \PY{p}{:} \PY{l+m+mf}{3e\PYZhy{}7}
     \PY{p}{\PYZcb{}}

\PY{n}{model} \PY{o}{=} \PY{n}{LEFT}\PY{p}{(}\PY{n}{WC}\PY{p}{,} \PY{n}{method} \PY{o}{=} \PY{l+s+s2}{\PYZdq{}}\PY{l+s+s2}{SM}\PY{l+s+s2}{\PYZdq{}}\PY{p}{,} 
             \PY{n}{name} \PY{o}{=} \PY{l+s+sa}{r}\PY{l+s+s2}{\PYZdq{}}\PY{l+s+s2}{\PYZdl{}m\PYZus{}}\PY{l+s+s2}{\PYZob{}}\PY{l+s+s2}{\PYZbs{}}\PY{l+s+s2}{beta}\PY{l+s+s2}{\PYZbs{}}\PY{l+s+s2}{beta\PYZcb{} + C\PYZus{}}\PY{l+s+s2}{\PYZbs{}}\PY{l+s+s2}{mathrm}\PY{l+s+si}{\PYZob{}VR\PYZcb{}}\PY{l+s+s2}{\PYZca{}}\PY{l+s+s2}{\PYZob{}}\PY{l+s+s2}{(6)\PYZcb{}\PYZdl{}}\PY{l+s+s2}{\PYZdq{}}\PY{p}{)}
\end{Verbatim}
\end{tcolorbox}
Then we define an array \texttt{Ebar} that represents the normalized kinetic energy of the outgoing electrons via $\overline{E} = (E_e-m_e)/Q$. Additionally, we need to define a parameter \texttt{e} which helps us to avoid poles in the calculation of phase-space quantities $g_{0k}$ and $h_{0k}$ (see Appendix~\ref{app:PSFs})
    \begin{tcolorbox}[breakable, size=fbox, boxrule=1pt, pad at break*=1mm,colback=cellbackground, colframe=cellborder]
\prompt{In}{incolor}{4}{\boxspacing}
\begin{Verbatim}[commandchars=\\\{\}]
\PY{n}{e} \PY{o}{=} \PY{l+m+mf}{1e\PYZhy{}5} \PY{c+c1}{\PYZsh{}avoid poles}

\PY{n}{Ebar} \PY{o}{=} \PY{n}{np}\PY{o}{.}\PY{n}{linspace}\PY{p}{(}\PY{l+m+mi}{0}\PY{o}{+}\PY{n}{e}\PY{p}{,} \PY{l+m+mi}{1}\PY{o}{\PYZhy{}}\PY{n}{e}\PY{p}{,} \PY{l+m+mi}{10}\PY{p}{)}
\end{Verbatim}
\end{tcolorbox}
We can then calculate the single electron spectrum
    \begin{tcolorbox}[breakable, size=fbox, boxrule=1pt, pad at break*=1mm,colback=cellbackground, colframe=cellborder]
\prompt{In}{incolor}{5}{\boxspacing}
\begin{Verbatim}[commandchars=\\\{\}]
\PY{n}{model}\PY{o}{.}\PY{n}{spectrum}\PY{p}{(}\PY{n}{Ebar}\PY{p}{)}
\end{Verbatim}
\end{tcolorbox}

            \begin{tcolorbox}[breakable, size=fbox, boxrule=.5pt, pad at break*=1mm, opacityfill=0]
\prompt{Out}{outcolor}{5}{\boxspacing}
\begin{Verbatim}[commandchars=\\\{\}]
array([4.35812489e-27, 5.46776633e-27, 5.60235186e-27, 5.22051437e-27,
       4.87200447e-27, 4.87200447e-27, 5.22051437e-27, 5.60235186e-27,
       5.46776633e-27, 4.35812489e-27])
\end{Verbatim}
\end{tcolorbox}
and the angular correlation coefficient
    \begin{tcolorbox}[breakable, size=fbox, boxrule=1pt, pad at break*=1mm,colback=cellbackground, colframe=cellborder]
\prompt{In}{incolor}{6}{\boxspacing}
\begin{Verbatim}[commandchars=\\\{\}]
\PY{n}{model}\PY{o}{.}\PY{n}{angular\PYZus{}corr}\PY{p}{(}\PY{n}{Ebar}\PY{p}{)}
\end{Verbatim}
\end{tcolorbox}

            \begin{tcolorbox}[breakable, size=fbox, boxrule=.5pt, pad at break*=1mm, opacityfill=0]
\prompt{Out}{outcolor}{6}{\boxspacing}
\begin{Verbatim}[commandchars=\\\{\}]
array([ 5.24575925e-04, -8.91742707e-02, -3.31779066e-01, -6.33496492e-01,
       -8.64038720e-01, -8.64038720e-01, -6.33496492e-01, -3.31779066e-01,
       -8.91742707e-02,  5.24575925e-04])
\end{Verbatim}
\end{tcolorbox}

Plots can be generated via
    \begin{tcolorbox}[breakable, size=fbox, boxrule=1pt, pad at break*=1mm,colback=cellbackground, colframe=cellborder]
\prompt{In}{incolor}{7}{\boxspacing}
\begin{Verbatim}[commandchars=\\\{\}]
\PY{n}{model}\PY{o}{.}\PY{n}{plot\PYZus{}spec}\PY{p}{(}\PY{n}{linewidth} \PY{o}{=} \PY{l+m+mi}{4}\PY{p}{,} \PY{n}{normalize\PYZus{}x} \PY{o}{=} \PY{k+kc}{True}\PY{p}{,} \PY{n}{savefig} \PY{o}{=} \PY{k+kc}{True}\PY{p}{,} 
                \PY{n}{file} \PY{o}{=} \PY{l+s+s2}{\PYZdq{}}\PY{l+s+s2}{spec.png}\PY{l+s+s2}{\PYZdq{}}\PY{p}{)}
\end{Verbatim}
\end{tcolorbox}

    \begin{center}
    \adjustimage{max size={0.75\linewidth}{0.75\paperheight}}{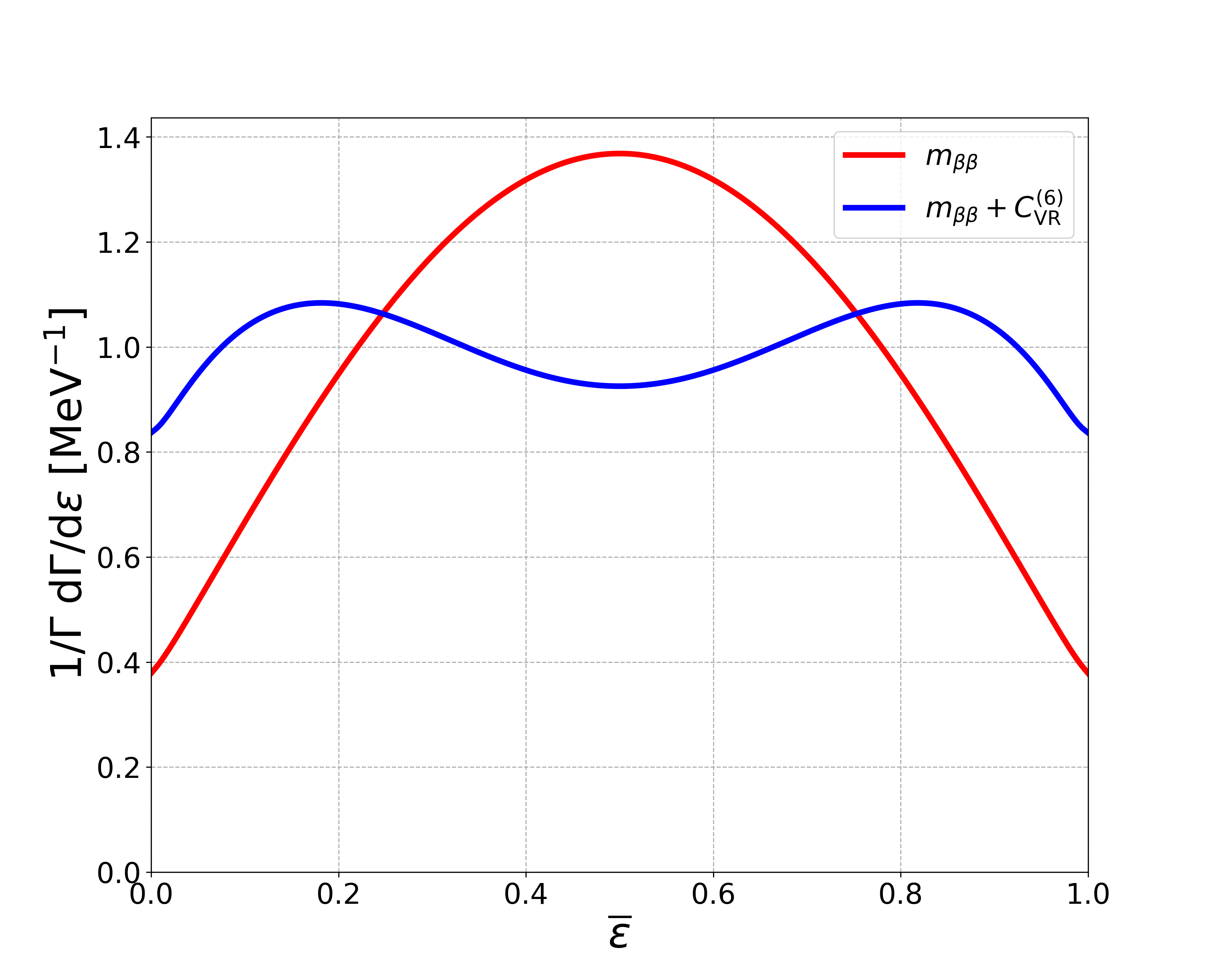}
    \end{center}
    { \hspace*{\fill} \\}

 for the single electron energy spectrum and via

    \begin{tcolorbox}[breakable, size=fbox, boxrule=1pt, pad at break*=1mm,colback=cellbackground, colframe=cellborder]
\prompt{In}{incolor}{8}{\boxspacing}
\begin{Verbatim}[commandchars=\\\{\}]
\PY{n}{model}\PY{o}{.}\PY{n}{plot\PYZus{}corr}\PY{p}{(}\PY{n}{linewidth} \PY{o}{=} \PY{l+m+mi}{4}\PY{p}{,} \PY{n}{normalize\PYZus{}x} \PY{o}{=} \PY{k+kc}{False}\PY{p}{,} \PY{n}{savefig} \PY{o}{=} \PY{k+kc}{True}\PY{p}{,} 
                \PY{n}{file} \PY{o}{=} \PY{l+s+s2}{\PYZdq{}}\PY{l+s+s2}{corr.png}\PY{l+s+s2}{\PYZdq{}}\PY{p}{)}
\end{Verbatim}
\end{tcolorbox}

    \begin{center}
    \adjustimage{max size={0.75\linewidth}{0.75\paperheight}}{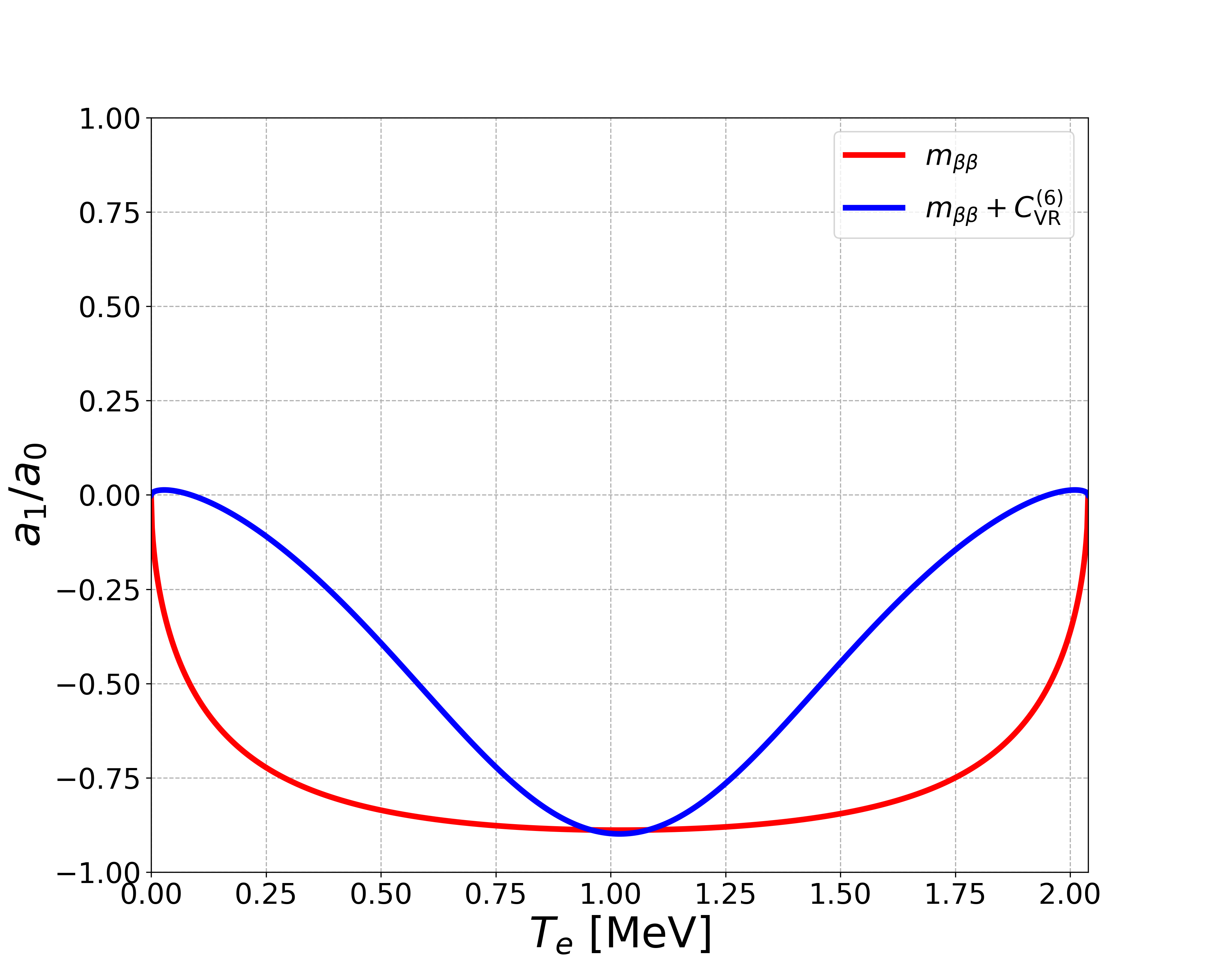}
    \end{center}
    { \hspace*{\fill} \\}
for the angular correlation. In the latter case we refrained from normalizing the x-axis to present both options.

While in case of the L$\nu$EM, the energy distribution for a single electron peaks, as expected, at $Q/2$, the exotic contribution proportional to $C_{\text{VR}}^{(6)}$ features a dip around that value (see e.g.~Ref.~\cite{Graf:2022lhj}), and therefore modifies the combined spectrum substantially, resulting in two equally high peaks on the sides of the distribution, i.e. for small and large energies of the electron, while there is a local minimum at $Q/2$. The negative angular correlation of the electrons obtained for the mass mechanism corresponds to the fact that they are preferably emitted back-to-back. The additional contribution from $C_{\text{VR}}^{(6)}$ distorts the shape, such that there are small ranges of energies, for which the correlation becomes positive and the dip around the middle of the distribution is more pronounced. This is caused, again, by a fairly distinctive shape of the correlation for this higher dimensional \0 mechanism, see e.g.~Ref.~\cite{Graf:2022lhj}.

Measuring the spectra of the individual electrons as well as the angular correlation between the emitted electrons can help to identify BSM scenarios with non-standard vector interactions~\citep{Graf:2022lhj}. Experiments like NEMO-3~\citep{NEMO:2008kpp,NEMO-3:2009fxe} and its next-generation upgrade SuperNEMO~\citep{arnold2010probing} are equipped to resolve the individual electron kinematics such that both the spectra and the angular correlation can be studied. In fact, these observables are available for the ordinary double beta decay observed by NEMO-3 and also in this case they can be used to probe new physics~\cite{Deppisch:2020mxv}.
 
%
%
\subsection{Limits on Wilson coefficients}
\subsubsection{Single operator dominance}\label{sec:opertor_limits}
Given experimental limits on the half-life of the $0\nu\beta\beta$-decay one can translate these into limits on the different Wilson coefficients assuming the contributions from a single operator dominates the \0 amplitude. Under this assumption and using the LEFT operators, the decay rates scale as
\begin{align}
    t_{1/2}^{-1}(C_i) \propto \left|C_i\right|^2\,.
\end{align}
 We can calculate limits on the different Wilson coefficients from a given half-life limit,
 $t_{1/2}^{\mathrm{lim}}(^AZ)$, through
\begin{align}
    \left|C_i\right| = \sqrt{\frac{t_{1/2}(C_i = 1, ^AZ) }{t_{1/2}^{\mathrm{lim}}(^AZ)}}\;\;.
\end{align}
Additionally, one can translate limits on the different Wilson coefficients into limits on the expected new-physics scale $\Lambda$. Assuming that LEFT operators of dimension $d$ scale as
\begin{align}
    C_i\simeq\left(\frac{v}{\Lambda}\right)^{(d-4)}
\end{align}
we can estimate the scale of new physics as
\begin{align}
    \Lambda \simeq v\left(C_i\right)^{1/(4-d)}
\end{align}
Because of operator mixing induced by the running of the Wilson coefficients from $m_W\rightarrow\Lambda_\chi$ it is important to clearly define the scale at which one wants to apply the assumption of single operator dominance. 

Similarly, when considering SMEFT the limits on the WCs can be obtained directly at the matching scale $m_W$. Here, limits are taken with respect to the dimensionful WC 
\begin{align}
    C_i \equiv \Tilde{C}_i/\Lambda^{(d-4)}
\end{align}
and correspondingly the new physics scale can be estimated assuming $\Tilde{C}_i=1$ as
\begin{align}
    \Lambda\simeq \left(C_i\right)^{1/(4-d)}
\end{align}

Both the limits on the different Wilson $C_i$ coefficients as well as the corresponding new physics scales can be accessed from the model classes or directly via\\\\
\begin{tabularx}{\textwidth}{llX}
    \hline\hline\\
    \multicolumn{3}{l}{\hypertarget{model.get_limits}{
    \PY{n}{LEFT\PYZus{}model}\PY{o}{.}\PY{n}{get\PYZus{}limits}\PY{p}{(}\PY{n}{half\PYZus{}life}\PY{p}{,} \PY{n}{isotope} \PY{o}{=} \PY{l+s+s2}{\PYZdq{}}\PY{l+s+s2}{76Ge}\PY{l+s+s2}{\PYZdq{}}\PY{p}{,} \PY{n}{method} \PY{o}{=} \PY{k+kc}{None}\PY{p}{,}}}\\
    \multicolumn{3}{l}{\qquad\quad\PY{n}{groups} \PY{o}{=} \PY{k+kc}{False}\PY{p}{,} \PY{n}{basis} \PY{o}{=} \PY{k+kc}{None}\PY{p}{,} \PY{n}{scale} \PY{o}{=} \PY{l+s+s2}{\PYZdq{}}\PY{l+s+s2}{up}\PY{l+s+s2}{\PYZdq{}}\PY{p}{)}}
    \\\\
    \multicolumn{3}{l}{\hypertarget{functions.limits_LEFT}{\PY{n}{functions}\PY{o}{.}\PY{n}{get\PYZus{}limits\PYZus{}LEFT\PY{p}{(}... , \PY{n}{unknown\PYZus{}LECs} \PY{o}{=} \PY{k+kc}{False}\PY{p}{,} \PY{n}{PSF\PYZus{}scheme} \PY{o}{=}\PY{l+s+s2}{\PYZdq{}}\PY{l+s+s2}{A}\PY{l+s+s2}{\PYZdq{}}\PY{p}{)}}}}
    \\\\
    \multicolumn{3}{l}{\PY{n}{SMEFT\PYZus{}model}\PY{o}{.}\PY{n}{get\PYZus{}limits}\PY{p}{(}...\PY{p}{)}}
    \\\\
    \multicolumn{3}{l}{\hypertarget{functions.limits_SMEFT}{\PY{n}{functions}\PY{o}{.}\PY{n}{get\PYZus{}limits\PYZus{}SMEFT\PY{p}{(}... , \PY{n}{unknown\PYZus{}LECs} \PY{o}{=} \PY{k+kc}{False}\PY{p}{,} \PY{n}{PSF\PYZus{}scheme} \PY{o}{=}\PY{l+s+s2}{\PYZdq{}}\PY{l+s+s2}{A}\PY{l+s+s2}{\PYZdq{}}\PY{p}{)}}}}\\\\
    \hline\\
    \textbf{Parameter} & \textbf{Type} & \textbf{Description}\\\\\hline\\
    \textbf{half\_life} & float & Experimental half-life limits
    \\\\
    \textbf{isotope} & string & Optional - Defines the corresponding isotope the half-life limit is obtained from.
    \\\\
    \textbf{method} & string & Optional - Sets the NME calculation method. You can choose from \str{IBM2}, \str{SM} and \str{QRPA}. If called via the \texttt{functions} module the preset value is \str{IBM2}. If called via a \texttt{model} class he models method will be used if \bool{None}.
    \\\\
    \textbf{groups} & bool & Optional - If set to \bool{True} operators with equal limits will be put into groups. Limits are then only shown for these operator groups. If \bool{False} all operator limits will be shown.
    \\\\
    \textbf{scale} & string & Optional - Only for LEFT models. Use \str{up} to get limits at $m_W$  or \str{down} to get limits at $\Lambda_\chi$. 
    \\\\
    \textbf{unknown\_LECs} & bool & Optional - If set to \bool{True} the unknown LECs will be set to their NDA estimates (see Table~\ref{Tab:LECs}). If set to \bool{False} the unknown LECs will be turned off i.e. set to 0.
    \\\\
    \textbf{PSF\_scheme} & string & Optional - Choose PSFs and electron wave functions - \str{A}: approximate solution to a uniform charge distribution. \str{B}: exact solution to a point-like charge
    \\\\\hline\hline
\end{tabularx}
\\\\\\
Plots of either the operator limits or the corresponding scales can be generated via 
\\\\
\begin{tabularx}{\textwidth}{llX}
    \hline\hline\\
    \multicolumn{3}{l}{\hypertarget{plots.limits_LEFT}{\PY{n}{plots}\PY{o}{.}\PY{n}{limits\PYZus{}LEFT}\PY{p}{(}\PY{n}{experiments}\PY{p}{,} \PY{n}{method} \PY{o}{=} \PY{k+kc}{None}\PY{p}{,} \PY{n}{unknown\PYZus{}LECs} \PY{o}{=} \PY{k+kc}{False}\PY{p}{,}}}
    \\
    \multicolumn{3}{l}{\qquad\quad\PY{n}{PSF\PYZus{}scheme} \PY{o}{=} \PY{l+s+s2}{\PYZdq{}}\PY{l+s+s2}{A}\PY{l+s+s2}{\PYZdq{}}\PY{n}{groups} \PY{o}{=} \PY{k+kc}{False}\PY{p}{,} \PY{n}{plottype} \PY{o}{=} \PY{l+s+s2}{\PYZdq{}}\PY{l+s+s2}{scales}\PY{l+s+s2}{\PYZdq{}}\PY{p}{,} \PY{n}{savefig} \PY{o}{=} \PY{k+kc}{True}\PY{p}{,}}
    \\
    \multicolumn{3}{l}{\qquad\quad\PY{n}{file} \PY{o}{=} \PY{l+s+s2}{\PYZdq{}}\PY{l+s+s2}{limits.png}\PY{l+s+s2}{\PYZdq{}}\PY{p}{,} \PY{n}{dpi} \PY{o}{=} \PY{l+m+mf}{300}\PY{p}{)}}
    \\\\
    \multicolumn{3}{l}{\hypertarget{plots.limits_SMEFT}{\PY{n}{plots}\PY{o}{.}\PY{n}{limits\PYZus{}SMEFT}\PY{p}{(}...\PY{p}{)}}}
    \\\\
    \hline\\
    \textbf{Parameter} & \textbf{Type} & \textbf{Description}\\\\\hline
    \\
    \textbf{experiments} & dict & Experimental half-life limits. Dictionary of the form \{name : \{ \str{half-life} : x, \str{isotope} : x, \str{label}: x\}\}
    \\\\
    \textbf{method} & string & Optional - Sets the NME calculation method. You can choose from \str{IBM2}, \str{SM} and \str{QRPA}. The preset value is \str{IBM2}.
    \\\\
    \textbf{unknown\_LECs} & bool & Optional - If set to \bool{True} the unknown LECs will be set to their NDA estimates (see Table~\ref{Tab:LECs}). If set to \bool{False} the unknown LECs will be turned off i.e. set to 0.
    \\\\
    \textbf{PSF\_scheme} & string & Optional - Choose PSFs and electron wave functions - \str{A}: approximate solution to a uniform charge distribution. \str{B}: exact solution to a point-like charge
    \\\\
    \textbf{groups} & bool & Optional - If set to \bool{True} operators with equal limits will be put into groups. Limits are then only shown for these operator groups. If \bool{False} all operator limits will be shown.
    \\\\
    \textbf{plottype} & string & Optional - Either \str{scales} or \str{limits} to plot on the y-axis\\\\
    \textbf{savefig} & bool & Optional - If \bool{True} save figure as file
    \\\\
    \textbf{file} & string & Optional - Filename to save figure to.
    \\\\
    \textbf{dpi} & float & Optional - sets the resolution in dots per inch when saving the figure.
    \\\\\hline\hline
\end{tabularx}
\\\\\\\\\\
\textbf{Example: Operator Limits obtained from the recent KamLAND-Zen result}  
    \begin{tcolorbox}[breakable, size=fbox, boxrule=1pt, pad at break*=1mm,colback=cellbackground, colframe=cellborder]
\prompt{In}{incolor}{1}{\boxspacing}
\begin{Verbatim}[commandchars=\\\{\}]
\PY{k+kn}{import} \PY{n+nn}{nudobe}
\end{Verbatim}
\end{tcolorbox}

    \begin{tcolorbox}[breakable, size=fbox, boxrule=1pt, pad at break*=1mm,colback=cellbackground, colframe=cellborder]
\prompt{In}{incolor}{2}{\boxspacing}
\begin{Verbatim}[commandchars=\\\{\}]
\PY{n}{LEFT\PYZus{}lims} \PY{o}{=} \PY{n}{nudobe}\PY{o}{.}\PY{n}{functions}\PY{o}{.}\PY{n}{get\PYZus{}limits\PYZus{}LEFT}\PY{p}{(}\PY{l+m+mf}{2.3e+26}\PY{p}{,} \PY{l+s+s2}{\PYZdq{}}\PY{l+s+s2}{136Xe}\PY{l+s+s2}{\PYZdq{}}\PY{p}{,} 
                                             \PY{n}{groups} \PY{o}{=} \PY{k+kc}{True}\PY{p}{)}
\PY{n}{LEFT\PYZus{}lims}
\end{Verbatim}
\end{tcolorbox}

            \begin{tcolorbox}[breakable, size=fbox, boxrule=.5pt, pad at break*=1mm, opacityfill=0]
\prompt{Out}{outcolor}{2}{\boxspacing}
\begin{Verbatim}[commandchars=\\\{\}]
                 Limits  Scales [GeV]
m\_bb       3.023437e-11  2.001563e+15
VL(6)      1.556091e-09  6.236166e+06
VR(6)      1.124938e-07  7.334505e+05
T(6)       8.488927e-10  8.443232e+06
S(6)       3.655193e-10  1.286708e+07
V(7)       1.873385e-05  9.262454e+03
S1(9)      2.637535e-05  2.026261e+03
S2(9)      8.871995e-08  6.328938e+03
S3(9)      3.182765e-07  4.902010e+03
S4(9)      5.466169e-08  6.972660e+03
S5(9)      1.609433e-08  8.904315e+03
V(9)       2.697231e-06  3.197065e+03
Vtilde(9)  5.138768e-06  2.810369e+03
\end{Verbatim}
\end{tcolorbox}
        
    \begin{tcolorbox}[breakable, size=fbox, boxrule=1pt, pad at break*=1mm,colback=cellbackground, colframe=cellborder]
\prompt{In}{incolor}{3}{\boxspacing}
\begin{Verbatim}[commandchars=\\\{\}]
\PY{n}{SMEFT\PYZus{}lims} \PY{o}{=} \PY{n}{nudobe}\PY{o}{.}\PY{n}{functions}\PY{o}{.}\PY{n}{get\PYZus{}limits\PYZus{}SMEFT}\PY{p}{(}\PY{l+m+mf}{2.3e+26}\PY{p}{,} \PY{l+s+s2}{\PYZdq{}}\PY{l+s+s2}{136Xe}\PY{l+s+s2}{\PYZdq{}}\PY{p}{,} 
                                               \PY{n}{method} \PY{o}{=} \PY{l+s+s2}{\PYZdq{}}\PY{l+s+s2}{QRPA}\PY{l+s+s2}{\PYZdq{}}\PY{p}{)}
\PY{n}{SMEFT\PYZus{}lims}
\end{Verbatim}
\end{tcolorbox}

            \begin{tcolorbox}[breakable, size=fbox, boxrule=.5pt, pad at break*=1mm, opacityfill=0]
\prompt{Out}{outcolor}{3}{\boxspacing}
\begin{Verbatim}[commandchars=\\\{\}]
             Limits [GeV\$\^{}\{4-d\}\$]  Scales [GeV]
LH(5)                3.923345e-16  2.548845e+15
LH(7)                1.296631e-20  4.256584e+06
LHD1(7)              2.964838e-13  1.499684e+04
LHD2(7)              2.441054e-12  7.426899e+03
LHDe(7)              7.199730e-17  2.403779e+05
LHW(7)               7.370338e-14  2.385087e+04
LLduD1(7)            1.041539e-15  9.865252e+04
LLQdH1(7)            3.052282e-17  3.199817e+05
LLQdH2(7)            1.530423e-16  1.869517e+05
LLQuH(7)             1.387754e-17  4.161301e+05
LeudH(7)             9.166365e-15  4.778236e+04
ddueue(9)            2.362703e-17  2.115038e+03
dQdueL1(9)           5.435754e-17  1.790392e+03
dQdueL2(9)           3.930148e-17  1.910373e+03
QudueL1(9)           5.435754e-17  1.790392e+03
QudueL2(9)           3.930148e-17  1.910373e+03
dQQuLL1(9)           1.978746e-20  8.724127e+03
dQQuLL2(9)           6.720479e-20  6.831561e+03
QuQuLL1(9)           5.453917e-20  7.122924e+03
QuQuLL2(9)           1.956554e-19  5.516983e+03
dQdQLL1(9)           5.453917e-20  7.122924e+03
dQdQLL2(9)           1.956554e-19  5.516983e+03
LLH4W1(9)            2.435831e-18  3.331736e+03
deueH2D(9)           6.898672e-20  6.795899e+03
dLuLH2D2(9)          6.898672e-20  6.795899e+03
duLLH2D(9)           3.440686e-20  7.810359e+03
dQLeH2D2(9)          5.579882e-17  1.781045e+03
dLQeH2D1(9)          2.341447e-17  2.118864e+03
deQLH2D(9)           2.567379e-17  2.080185e+03
QueLH2D2(9)          5.579882e-17  1.781045e+03
QeuLH2D2(9)          4.034355e-17  1.900401e+03
QLQLH2D2(9)          1.212675e-17  2.416858e+03
QLQLH2D5(9)          1.212675e-17  2.416858e+03
QQLLH2D2(9)          1.086507e-17  2.470549e+03
eeH4D2(9)            6.224144e-18  2.761748e+03
LLH4D23(9)           3.303900e-18  3.134690e+03
LLH4D24(9)           1.115316e-17  2.457652e+03
\end{Verbatim}
\end{tcolorbox}

    \begin{tcolorbox}[breakable, size=fbox, boxrule=1pt, pad at break*=1mm,colback=cellbackground, colframe=cellborder]
\prompt{In}{incolor}{4}{\boxspacing}
\begin{Verbatim}[commandchars=\\\{\}]
\PY{n}{nudobe}\PY{o}{.}\PY{n}{plots}\PY{o}{.}\PY{n}{limits\PYZus{}LEFT}\PY{p}{(}\PY{n}{experiments} \PY{o}{=} \PY{p}{\PYZob{}}\PY{l+s+s2}{\PYZdq{}}\PY{l+s+s2}{KamLAND}\PY{l+s+s2}{\PYZdq{}} \PY{p}{:} \PY{p}{\PYZob{}}\PY{l+s+s2}{\PYZdq{}}\PY{l+s+s2}{half\PYZhy{}life}\PY{l+s+s2}{\PYZdq{}}\PY{p}{:}\PY{l+m+mf}{2.3e+26}\PY{p}{,} 
                                                     \PY{l+s+s2}{\PYZdq{}}\PY{l+s+s2}{isotope}\PY{l+s+s2}{\PYZdq{}} \PY{p}{:} \PY{l+s+s2}{\PYZdq{}}\PY{l+s+s2}{136Xe}\PY{l+s+s2}{\PYZdq{}}\PY{p}{,} 
                                                     \PY{l+s+s2}{\PYZdq{}}\PY{l+s+s2}{label}\PY{l+s+s2}{\PYZdq{}}\PY{p}{:}\PY{l+s+s2}{\PYZdq{}}\PY{l+s+s2}{KamLAND\PYZhy{}Zen}\PY{l+s+s2}{\PYZdq{}}
                                                    \PY{p}{\PYZcb{}}\PY{p}{,} 
                                        \PY{l+s+s2}{\PYZdq{}}\PY{l+s+s2}{GERDA}\PY{l+s+s2}{\PYZdq{}}  \PY{p}{:} \PY{p}{\PYZob{}}\PY{l+s+s2}{\PYZdq{}}\PY{l+s+s2}{half\PYZhy{}life}\PY{l+s+s2}{\PYZdq{}} \PY{p}{:} \PY{l+m+mf}{1.8e+26}\PY{p}{,} 
                                                    \PY{l+s+s2}{\PYZdq{}}\PY{l+s+s2}{isotope}\PY{l+s+s2}{\PYZdq{}} \PY{p}{:} \PY{l+s+s2}{\PYZdq{}}\PY{l+s+s2}{76Ge}\PY{l+s+s2}{\PYZdq{}}\PY{p}{,} 
                                                    \PY{l+s+s2}{\PYZdq{}}\PY{l+s+s2}{label}\PY{l+s+s2}{\PYZdq{}} \PY{p}{:} \PY{l+s+s2}{\PYZdq{}}\PY{l+s+s2}{GERDA}\PY{l+s+s2}{\PYZdq{}}
                                                  \PY{p}{\PYZcb{}}
                                     \PY{p}{\PYZcb{}}\PY{p}{,}
                       \PY{n}{savefig} \PY{o}{=} \PY{k+kc}{True}\PY{p}{,} 
                       \PY{n}{file} \PY{o}{=} \PY{l+s+s2}{\PYZdq{}}\PY{l+s+s2}{LEFT\PYZus{}limits.png}\PY{l+s+s2}{\PYZdq{}}
                      \PY{p}{)}
\end{Verbatim}
\end{tcolorbox}

    \begin{center}
    \adjustimage{max size={1\linewidth}{1\paperheight}}{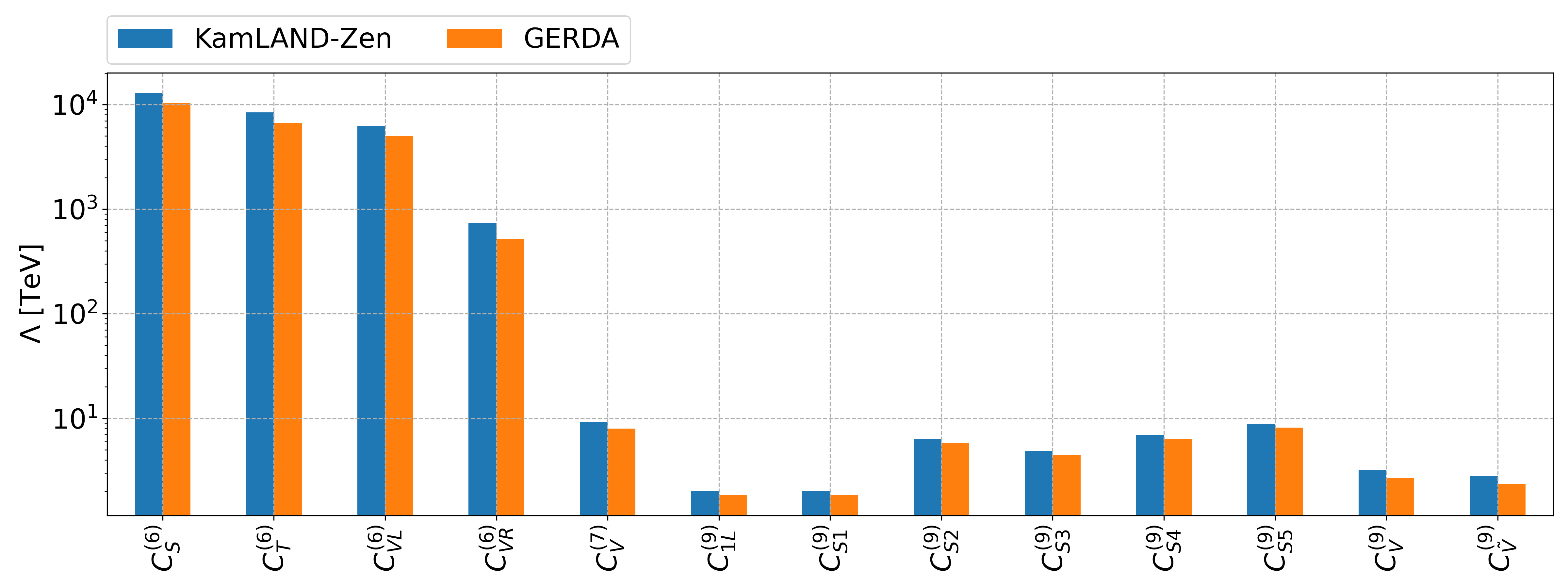}
    \end{center}
    { \hspace*{\fill} \\}

%
%
\subsubsection{Two-operator scenario}
If we drop the assumption of a single LNV operator the analysis becomes a bit more complicated. Assuming two non-vanishing real WCs $C_{x,y}$ with a relative phase $\phi$ the half-life is given as
\begin{align}
    T_{1/2}^{-1} = C_x^2M_{xx} + C_y^2M_{yy} + 2\mathrm{Re}[C_xC_y\exp{(i\phi)}]M_{xy}
\end{align}
Now, instead of calculating the half life in dependency on $C_{x,y}, M_{xx,xy,yy}$ and $\phi$ we can also solve for $C_y$
\begin{align}
    C_y = -C_x\frac{\cos(\phi)M_{xy}}{M_{yy}} \pm \sqrt{C_x^2\frac{\cos(\phi)M_{xy}^2-M_{xx}M_{yy}}{M_{yy}^2} + \frac{1}{T_{1/2} M_{yy}}}
\end{align}
Given an experimental limit on the half-life, we can use this relation to find the allowed parameter space. Within \texttt{nudobe} this is done with the function\\\\
\begin{tabularx}{\textwidth}{llX}
    \hline\hline\\
    \multicolumn{3}{l}{\hypertarget{functions.get_contours}{\PY{n}{functions}\PY{o}{.}\PY{n}{get\PYZus{}contours}\PY{p}{(}\PY{n}{WCx}\PY{p}{,} \PY{n}{WCy}\PY{p}{,} \PY{n}{half\PYZus{}life}\PY{p}{,} \PY{n}{isotope}\PY{p}{,} \PY{n}{method} \PY{o}{=} \PY{l+s+s2}{\PYZdq{}}\PY{l+s+s2}{IBM2}\PY{l+s+s2}{\PYZdq{}}\PY{p}{,}}}
    \\
    \multicolumn{3}{l}{\qquad\quad\PY{n}{phase} \PY{o}{=} \PY{l+m+mi}{3}\PY{o}{/}\PY{l+m+mi}{4}\PY{o}{*}\PY{n}{np}\PY{o}{.}\PY{n}{pi}\PY{p}{,} \PY{n}{n\PYZus{}points} \PY{o}{=} \PY{l+m+mi}{5000}\PY{p}{,} \PY{n}{x\PYZus{}min} \PY{o}{=} \PY{k+kc}{None}\PY{p}{,} \PY{n}{x\PYZus{}max} \PY{o}{=} \PY{k+kc}{None},}
    \\
    \multicolumn{3}{l}{\qquad\quad\PY{n}{unknown\PYZus{}LECs} \PY{o}{=} \PY{k+kc}{False}\PY{p}{,} \PY{n}{PSF\PYZus{}scheme} \PY{o}{=}\PY{l+s+s2}{\PYZdq{}}\PY{l+s+s2}{A}\PY{l+s+s2}{\PYZdq{}}\PY{p}{)}}
    \\\\
    \hline\\
    \textbf{Parameter} & \textbf{Type} & \textbf{Description}\\\\\hline\\
    \textbf{WCx} & string & Independent Wilson coefficient. Can be from SMEFT or LEFT.
    \\\\
    \textbf{WCy} & string & Dependent Wilson coefficient to be calculated from WCx and the half-life limit. Should be from the same EFT as WCx.
    \\\\
    \textbf{half\_life} & float & Experimental half-life limits
\end{tabularx}
\begin{tabularx}{\textwidth}{llX}
    \textbf{isotope} & string & Defines the corresponding isotope the half-life limit is obtained from.
    \\\\
    \textbf{method} & string & Optional - Sets the NME calculation method. You can choose from \str{IBM2}, \str{SM} and \str{QRPA}. The preset value is \str{IBM2}.
    \\\\
    \textbf{n\_points} & float & Optional - Sets the number of points to be calculated.
    \\\\
    \textbf{phase} & float & Optional - Sets the relative phase $\phi$.
    \\\\
    \textbf{x\_min} & float & Optional - Sets the minimum value of WCx to be studied.
    \\\\
    \textbf{x\_max} & float & Optional - Sets the maximum value of WCx to be studied. 
    \\\\
    \textbf{unknown\_LECs} & bool & Optional - If set to \bool{True} the unknown LECs will be set to their NDA estimates (see Table~\ref{Tab:LECs}). If set to \bool{False} the unknown LECs will be turned off i.e. set to 0.
    \\\\
    \textbf{PSF\_scheme} & string & Optional - Choose PSFs and electron wave functions - \str{A}: approximate solution to a uniform charge distribution. \str{B}: exact solution to a point-like charge
    \\\\\hline\hline
\end{tabularx}
\\\\\\
The resulting contours can be plotted via\\\\
\begin{tabularx}{\textwidth}{llX}
    \hline\hline\\
    \multicolumn{3}{l}{\hypertarget{plots.contours}{\PY{n}{plots}\PY{o}{.}\PY{n}{contours}\PY{p}{(}\PY{n}{WCx}\PY{p}{,} \PY{n}{WCy}\PY{p}{,} }}
    \\
    \multicolumn{3}{l}{\qquad\quad\PY{n}{limits} \PY{o}{=} \PY{p}{\PYZob{}}\PY{l+s+s2}{\PYZdq{}}\PY{l+s+s2}{KamLAND\PYZhy{}Zen}\PY{l+s+s2}{\PYZdq{}}\PY{p}{:} \PY{p}{\PYZob{}}\PY{l+s+s2}{\PYZdq{}}\PY{l+s+s2}{half\PYZhy{}life}\PY{l+s+s2}{\PYZdq{}} \PY{p}{:} \PY{l+m+mf}{2.3e+26}\PY{p}{,} \PY{l+s+s2}{\PYZdq{}}\PY{l+s+s2}{isotope}\PY{l+s+s2}{\PYZdq{}}   \PY{p}{:} \PY{l+s+s2}{\PYZdq{}}\PY{l+s+s2}{136Xe}\PY{l+s+s2}{\PYZdq{}}\PY{p}{,}}
    \\
    \multicolumn{3}{l}{\qquad\qquad\qquad\quad\;\,\PY{l+s+s2}{\PYZdq{}}\PY{l+s+s2}{color}\PY{l+s+s2}{\PYZdq{}}     \PY{p}{:} \PY{l+s+s2}{\PYZdq{}}\PY{l+s+s2}{b}\PY{l+s+s2}{\PYZdq{}}\PY{p}{,} \PY{l+s+s2}{\PYZdq{}}\PY{l+s+s2}{label}\PY{l+s+s2}{\PYZdq{}}     \PY{p}{:} \PY{l+s+s2}{\PYZdq{}}\PY{l+s+s2}{KamLAND\PYZhy{}Zen}\PY{l+s+s2}{\PYZdq{}}\PY{p}{,} \PY{l+s+s2}{\PYZdq{}}\PY{l+s+s2}{linewidth}\PY{l+s+s2}{\PYZdq{}} \PY{p}{:} \PY{l+m+mf}{1}\PY{p}{,} }
    \\
    \multicolumn{3}{l}{\qquad\qquad\qquad\quad\;\,\PY{l+s+s2}{\PYZdq{}}\PY{l+s+s2}{linealpha}\PY{l+s+s2}{\PYZdq{}} \PY{p}{:} \PY{l+m+mf}{1}\PY{p}{,} \PY{l+s+s2}{\PYZdq{}}\PY{l+s+s2}{linestyle}\PY{l+s+s2}{\PYZdq{}}     \PY{p}{:} \PY{l+s+s2}{\PYZdq{}}\PY{l+s+s2}{-}\PY{l+s+s2}{\PYZdq{}}\PY{p}{,} \PY{l+s+s2}{\PYZdq{}}\PY{l+s+s2}{alpha}\PY{l+s+s2}{\PYZdq{}} \PY{p}{:} \PY{k+kc}{None}\PY{p}{\PYZcb{}} \PY{p}{\PYZcb{}}\PY{p}{,} 
    }
    \\
    \multicolumn{3}{l}{\qquad\quad\PY{n}{method} \PY{o}{=} \PY{l+s+s2}{\PYZdq{}}\PY{l+s+s2}{IBM2}\PY{l+s+s2}{\PYZdq{}}\PY{p}{,} \PY{n}{unknown\PYZus{}LECs} \PY{o}{=} \PY{k+kc}{False}\PY{p}{,} \PY{n}{PSF\PYZus{}scheme} \PY{o}{=}\PY{l+s+s2}{\PYZdq{}}\PY{l+s+s2}{A}\PY{l+s+s2}{\PYZdq{}}\PY{p}{,}}\\
    \multicolumn{3}{l}{\qquad\quad\PY{n}{n\PYZus{}points} \PY{o}{=} \PY{l+m+mi}{5000}\PY{p}{,} \PY{n}{phase} \PY{o}{=} \PY{l+m+mi}{0}\PY{p}{,} \PY{n}{varyphases} \PY{o}{=} \PY{k+kc}{False}\PY{p}{,} \PY{n}{n\PYZus{}vary} \PY{o}{=} \PY{l+m+mi}{5}\PY{p}{,} \PY{n}{x\PYZus{}min} \PY{o}{=} \PY{k+kc}{None}\PY{p}{,}}\\
    \multicolumn{3}{l}{\qquad\quad\PY{n}{x\PYZus{}max} \PY{o}{=} \PY{k+kc}{None}\PY{p}{,} \PY{n}{savefig} \PY{o}{=} \PY{k+kc}{False}\PY{p}{,} \PY{n}{file} \PY{o}{=} \PY{l+s+s2}{\PYZdq{}}\PY{l+s+s2}{contour\PYZus{}limits.png}\PY{l+s+s2}{\PYZdq{}}\PY{p}{,}  \PY{n}{dpi}              \PY{o}{=} \PY{l+m+mi}{300}
            \PY{p}{)} }
    \\\\
    \hline\\
    \textbf{Parameter} & \textbf{Type} & \textbf{Description}\\\\\hline\\
    \textbf{WCx} & string & Independent Wilson coefficient. Can be from SMEFT or LEFT.
\end{tabularx}
\begin{tabularx}{\textwidth}{llX}
    \textbf{WCy} & string & Dependent Wilson coefficient to be calculated from WCx and the half-life limit. Should be from the same EFT as WCx.
    \\\\
    \textbf{limits} & dictionary & Experimental half-life limits. Given as a dictionary of the type \{Name : [half-life limit, isotope, color, label, linewidth, linealpha, linestyle, alpha]\}. The color will decide the color of the plot while the label sets the label for the legend. Linealpha sets the alpha channel for the outer line of the contour. If it is not defined it will be set equal to alpha which defines the alpha channel of the filled area. If alpha is not set or set to \bool{None} it is set automatically.
    \\\\
    \textbf{method} & string & Optional - Sets the NME calculation method. You can choose from \str{IBM2}, \str{SM} and \str{QRPA}. The preset value is \str{IBM2}.
    \\\\
    \textbf{unknown\_LECs} & bool & Optional - If set to \bool{True} the unknown LECs will be set to their NDA estimates (see Table~\ref{Tab:LECs}). If set to \bool{False} the unknown LECs will be turned off i.e. set to 0.
    \\\\
    \textbf{PSF\_scheme} & string & Optional - Choose PSFs and electron wave functions - \str{A}: approximate solution to a uniform charge distribution. \str{B}: exact solution to a point-like charge
    \\\\
    \textbf{n\_points} & float & Optional - Sets the number of points to be calculated.
    \\\\
    \textbf{phase} & float & Optional - Sets the relative phase $\phi$.
    \\\\
    \textbf{varyphases} & bool & Optional - If \bool{True} the relative complex phase will be varied and multiple plots will be generated.
    \\\\
    \textbf{n\_vary} & integer & Optional - Number of variations of the complex phase.
    \\\\
    \textbf{x\_min} & float & Optional - Sets the minimum value of WCx to be studied.
    \\\\
    \textbf{x\_max} & float & Optional - Sets the maximum value of WCx to be studied. 
    \\\\
    \textbf{savefig} & bool & Optional - If \bool{True} save figure as file
    \\\\
    \textbf{file} & string & Optional - Filename to save figure to.
\end{tabularx}
\begin{tabularx}{\textwidth}{llX}
    \textbf{dpi} & float & Optional - sets the resolution in dots per inch when saving the figure. 
    \\\\\hline\hline
\end{tabularx}
\\\\\\
\textbf{Example: The light-neutrino-exchange mechanism with additional lepton number violating right-handed current}\\
Again, we take a look at our standard example and want to study the combined limits on $m_{\beta\beta}$ and $C_{VR}^{(6)}$. We can get the lower and upper limits on $C_{VR}^{(6)}$ for different values of $m_{\beta\beta}$ via

        \begin{tcolorbox}[breakable, size=fbox, boxrule=1pt, pad at break*=1mm,colback=cellbackground, colframe=cellborder]
\prompt{In}{incolor}{1}{\boxspacing}
\begin{Verbatim}[commandchars=\\\{\}]
\PY{k+kn}{import} \PY{n+nn}{nudobe}
\end{Verbatim}
\end{tcolorbox}

    \begin{tcolorbox}[breakable, size=fbox, boxrule=1pt, pad at break*=1mm,colback=cellbackground, colframe=cellborder]
\prompt{In}{incolor}{2}{\boxspacing}
\begin{Verbatim}[commandchars=\\\{\}]
\PY{n}{contours} \PY{o}{=} \PY{n}{nudobe}\PY{o}{.}\PY{n}{functions}\PY{o}{.}\PY{n}{get\PYZus{}contours}\PY{p}{(}\PY{l+s+s2}{\PYZdq{}}\PY{l+s+s2}{m\PYZus{}bb}\PY{l+s+s2}{\PYZdq{}}\PY{p}{,} \PY{l+s+s2}{\PYZdq{}}\PY{l+s+s2}{VR(6)}\PY{l+s+s2}{\PYZdq{}}\PY{p}{,} \PY{l+m+mf}{2.3e+26}\PY{p}{,} 
                                         \PY{l+s+s2}{\PYZdq{}}\PY{l+s+s2}{136Xe}\PY{l+s+s2}{\PYZdq{}}\PY{p}{,} \PY{n}{phase} \PY{o}{=} \PY{l+m+mi}{0}\PY{p}{)}
\PY{n}{contours}
\end{Verbatim}
\end{tcolorbox}

            \begin{tcolorbox}[breakable, size=fbox, boxrule=.5pt, pad at break*=1mm, opacityfill=0]
\prompt{Out}{outcolor}{2}{\boxspacing}
\begin{Verbatim}[commandchars=\\\{\}]
              m\_bb     VR(6) min     VR(6) max
0    -3.096316e-11  2.304978e-08  2.678719e-08
1    -3.094502e-11  2.062428e-08  2.918349e-08
2    -3.092687e-11  1.913349e-08  3.064508e-08
3    -3.090873e-11  1.795119e-08  3.179817e-08
4    -3.089058e-11  1.693975e-08  3.278041e-08
{\ldots}            {\ldots}           {\ldots}           {\ldots}
3409  3.089058e-11 -3.278041e-08 -1.693975e-08
3410  3.090873e-11 -3.179817e-08 -1.795119e-08
3411  3.092687e-11 -3.064508e-08 -1.913349e-08
3412  3.094502e-11 -2.918349e-08 -2.062428e-08
3413  3.096316e-11 -2.678719e-08 -2.304978e-08

[3414 rows x 3 columns]
\end{Verbatim}
\end{tcolorbox}
And a plot showing the contours of the allowed parameter space resulting from the recent KamLAND-Zen result is generated via

    \begin{tcolorbox}[breakable, size=fbox, boxrule=1pt, pad at break*=1mm,colback=cellbackground, colframe=cellborder]
\prompt{In}{incolor}{3}{\boxspacing}
\begin{Verbatim}[commandchars=\\\{\}]
\PY{n}{nudobe}\PY{o}{.}\PY{n}{plots}\PY{o}{.}\PY{n}{contours}\PY{p}{(}\PY{l+s+s2}{\PYZdq{}}\PY{l+s+s2}{m\PYZus{}bb}\PY{l+s+s2}{\PYZdq{}}\PY{p}{,} \PY{l+s+s2}{\PYZdq{}}\PY{l+s+s2}{VR(6)}\PY{l+s+s2}{\PYZdq{}}\PY{p}{,} \PY{n}{phase} \PY{o}{=} \PY{l+m+mi}{0}\PY{p}{,} 
                      \PY{n}{limits} \PY{o}{=} \PY{p}{\PYZob{}}\PY{l+s+s2}{\PYZdq{}}\PY{l+s+s2}{KamLAND\PYZhy{}Zen}\PY{l+s+s2}{\PYZdq{}} \PY{p}{:} \PY{p}{\PYZob{}}\PY{l+s+s2}{\PYZdq{}}\PY{l+s+s2}{half\PYZhy{}life}\PY{l+s+s2}{\PYZdq{}} \PY{p}{:} \PY{l+m+mf}{2.3e+26}\PY{p}{,} 
                                \PY{l+s+s2}{\PYZdq{}}\PY{l+s+s2}{isotope}\PY{l+s+s2}{\PYZdq{}}   \PY{p}{:} \PY{l+s+s2}{\PYZdq{}}\PY{l+s+s2}{136Xe}\PY{l+s+s2}{\PYZdq{}}\PY{p}{,}
                                \PY{l+s+s2}{\PYZdq{}}\PY{l+s+s2}{color}\PY{l+s+s2}{\PYZdq{}}     \PY{p}{:} \PY{l+s+s2}{\PYZdq{}}\PY{l+s+s2}{b}\PY{l+s+s2}{\PYZdq{}}\PY{p}{,} 
                                \PY{l+s+s2}{\PYZdq{}}\PY{l+s+s2}{label}\PY{l+s+s2}{\PYZdq{}}     \PY{p}{:}\PY{l+s+s2}{\PYZdq{}}\PY{l+s+s2}{KamLAND\PYZhy{}Zen}\PY{l+s+s2}{\PYZdq{}}\PY{p}{\PYZcb{}}\PY{p}{\PYZcb{}}\PY{p}{)}
\end{Verbatim}
\end{tcolorbox}

    \begin{center}
    \adjustimage{max size={0.75\linewidth}{0.75\paperheight}}{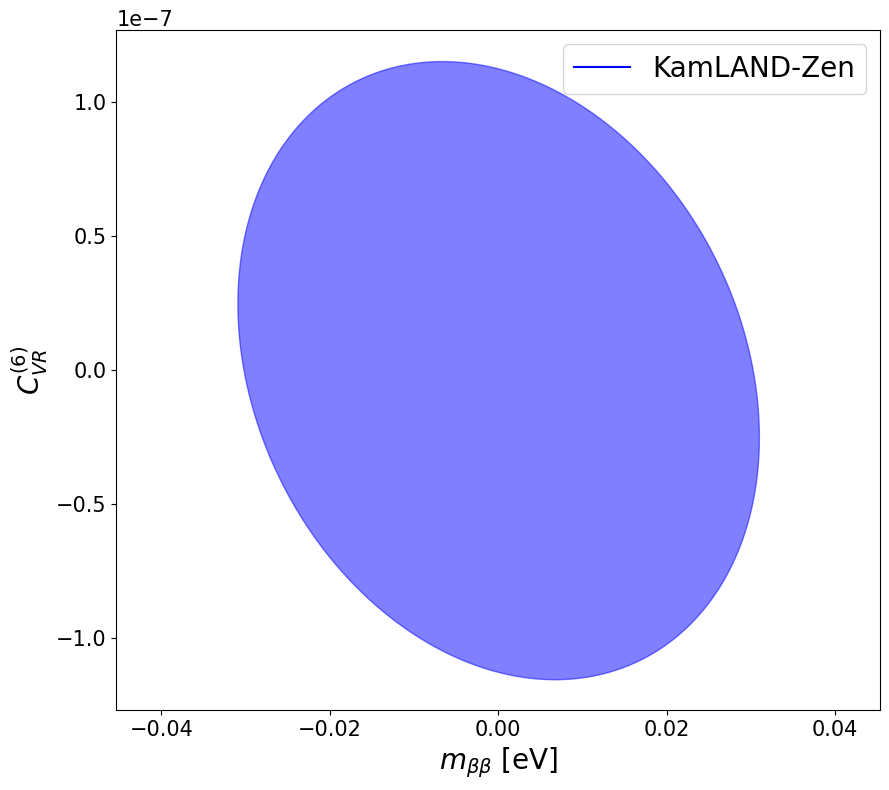}
    \end{center}
    { \hspace*{\fill} \\}

With variations of $\phi$ the resulting allowed parameter space can be plotted via
    \begin{tcolorbox}[breakable, size=fbox, boxrule=1pt, pad at break*=1mm,colback=cellbackground, colframe=cellborder]
\prompt{In}{incolor}{4}{\boxspacing}
\begin{Verbatim}[commandchars=\\\{\}]
\PY{n}{nudobe}\PY{o}{.}\PY{n}{plots}\PY{o}{.}\PY{n}{contours}\PY{p}{(}\PY{l+s+s2}{\PYZdq{}}\PY{l+s+s2}{m\PYZus{}bb}\PY{l+s+s2}{\PYZdq{}}\PY{p}{,} \PY{l+s+s2}{\PYZdq{}}\PY{l+s+s2}{VR(6)}\PY{l+s+s2}{\PYZdq{}}\PY{p}{,} 
                      \PY{n}{limits} \PY{o}{=} \PY{p}{\PYZob{}}\PY{l+s+s2}{\PYZdq{}}\PY{l+s+s2}{KamLAND\PYZhy{}Zen}\PY{l+s+s2}{\PYZdq{}} \PY{p}{:} \PY{p}{\PYZob{}}\PY{l+s+s2}{\PYZdq{}}\PY{l+s+s2}{half\PYZhy{}life}\PY{l+s+s2}{\PYZdq{}} \PY{p}{:} \PY{l+m+mf}{2.3e+26}\PY{p}{,} 
                                \PY{l+s+s2}{\PYZdq{}}\PY{l+s+s2}{isotope}\PY{l+s+s2}{\PYZdq{}}   \PY{p}{:} \PY{l+s+s2}{\PYZdq{}}\PY{l+s+s2}{136Xe}\PY{l+s+s2}{\PYZdq{}}\PY{p}{\PYZcb{}}
                               \PY{p}{\PYZcb{}}\PY{p}{,} 
                      \PY{n}{varyphases} \PY{o}{=} \PY{k+kc}{True}\PY{p}{)}
\end{Verbatim}
\end{tcolorbox}

    \begin{center}
    \adjustimage{max size={0.75\linewidth}{0.75\paperheight}}{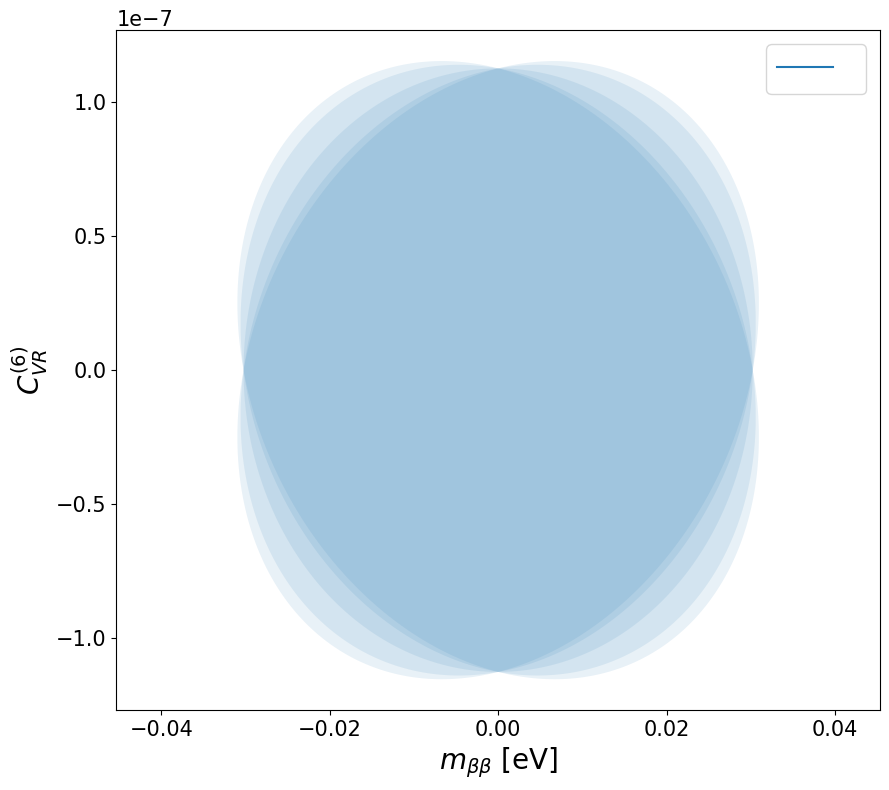}
    \end{center}
    { \hspace*{\fill} \\}
Now, instead of one area, the plot shows 5 different contour areas each corresponding to a different choice of the complex relative phase $\Phi$.

And limits from multiple different experiments in different isotopes can be plotted via
    \begin{tcolorbox}[breakable, size=fbox, boxrule=1pt, pad at break*=1mm,colback=cellbackground, colframe=cellborder]
\prompt{In}{incolor}{5}{\boxspacing}
\begin{Verbatim}[commandchars=\\\{\}]
\PY{n}{nudobe}\PY{o}{.}\PY{n}{plots}\PY{o}{.}\PY{n}{contours}\PY{p}{(}\PY{l+s+s2}{\PYZdq{}}\PY{l+s+s2}{m\PYZus{}bb}\PY{l+s+s2}{\PYZdq{}}\PY{p}{,} \PY{l+s+s2}{\PYZdq{}}\PY{l+s+s2}{VR(6)}\PY{l+s+s2}{\PYZdq{}}\PY{p}{,}
\PY{n}{phase} \PY{o}{=} \PY{l+m+mi}{0}\PY{p}{,}
\PY{n}{limits} \PY{o}{=} \PY{p}{\PYZob{}}\PY{l+s+s2}{\PYZdq{}}\PY{l+s+s2}{KamLAND\PYZhy{}Zen}\PY{l+s+s2}{\PYZdq{}} \PY{p}{:} \PY{p}{\PYZob{}}\PY{l+s+s2}{\PYZdq{}}\PY{l+s+s2}{half\PYZhy{}life}\PY{l+s+s2}{\PYZdq{}} \PY{p}{:} \PY{l+m+mf}{2.3e+26}\PY{p}{,} 
                           \PY{l+s+s2}{\PYZdq{}}\PY{l+s+s2}{isotope}\PY{l+s+s2}{\PYZdq{}}   \PY{p}{:} \PY{l+s+s2}{\PYZdq{}}\PY{l+s+s2}{136Xe}\PY{l+s+s2}{\PYZdq{}}\PY{p}{,}
                           \PY{l+s+s2}{\PYZdq{}}\PY{l+s+s2}{color}\PY{l+s+s2}{\PYZdq{}}     \PY{p}{:} \PY{l+s+s2}{\PYZdq{}}\PY{l+s+s2}{b}\PY{l+s+s2}{\PYZdq{}}\PY{p}{,} 
                           \PY{l+s+s2}{\PYZdq{}}\PY{l+s+s2}{label}\PY{l+s+s2}{\PYZdq{}}     \PY{p}{:}\PY{l+s+s2}{\PYZdq{}}\PY{l+s+s2}{KamLAND\PYZhy{}Zen}\PY{l+s+s2}{\PYZdq{}}\PY{p}{\PYZcb{}}\PY{p}{,}
          \PY{l+s+s2}{\PYZdq{}}\PY{l+s+s2}{GERDA}\PY{l+s+s2}{\PYZdq{}}       \PY{p}{:} \PY{p}{\PYZob{}}\PY{l+s+s2}{\PYZdq{}}\PY{l+s+s2}{half\PYZhy{}life}\PY{l+s+s2}{\PYZdq{}} \PY{p}{:} \PY{l+m+mf}{1.8e+26}\PY{p}{,} 
                           \PY{l+s+s2}{\PYZdq{}}\PY{l+s+s2}{isotope}\PY{l+s+s2}{\PYZdq{}}   \PY{p}{:} \PY{l+s+s2}{\PYZdq{}}\PY{l+s+s2}{76Ge}\PY{l+s+s2}{\PYZdq{}}\PY{p}{,} 
                           \PY{l+s+s2}{\PYZdq{}}\PY{l+s+s2}{color}\PY{l+s+s2}{\PYZdq{}}     \PY{p}{:} \PY{l+s+s2}{\PYZdq{}}\PY{l+s+s2}{r}\PY{l+s+s2}{\PYZdq{}}\PY{p}{,} 
                           \PY{l+s+s2}{\PYZdq{}}\PY{l+s+s2}{label}\PY{l+s+s2}{\PYZdq{}}     \PY{p}{:} \PY{l+s+s2}{\PYZdq{}}\PY{l+s+s2}{GERDA}\PY{l+s+s2}{\PYZdq{}}\PY{p}{\PYZcb{}}\PY{p}{,}
          \PY{l+s+s2}{\PYZdq{}}\PY{l+s+s2}{Cupid\PYZhy{}Mo}\PY{l+s+s2}{\PYZdq{}}    \PY{p}{:} \PY{p}{\PYZob{}}\PY{l+s+s2}{\PYZdq{}}\PY{l+s+s2}{half\PYZhy{}life}\PY{l+s+s2}{\PYZdq{}} \PY{p}{:} \PY{l+m+mf}{1.5e+24}\PY{p}{,} 
                           \PY{l+s+s2}{\PYZdq{}}\PY{l+s+s2}{isotope}\PY{l+s+s2}{\PYZdq{}}   \PY{p}{:} \PY{l+s+s2}{\PYZdq{}}\PY{l+s+s2}{100Mo}\PY{l+s+s2}{\PYZdq{}}\PY{p}{,} 
                           \PY{l+s+s2}{\PYZdq{}}\PY{l+s+s2}{color}\PY{l+s+s2}{\PYZdq{}}     \PY{p}{:} \PY{l+s+s2}{\PYZdq{}}\PY{l+s+s2}{g}\PY{l+s+s2}{\PYZdq{}}\PY{p}{,} 
                           \PY{l+s+s2}{\PYZdq{}}\PY{l+s+s2}{label}\PY{l+s+s2}{\PYZdq{}}     \PY{p}{:} \PY{l+s+s2}{\PYZdq{}}\PY{l+s+s2}{Cupid\PYZhy{}Mo}\PY{l+s+s2}{\PYZdq{}}\PY{p}{\PYZcb{}}\PY{p}{\PYZcb{}}\PY{p}{)}
\end{Verbatim}
\end{tcolorbox}

    \begin{center}
    \adjustimage{max size={0.75\linewidth}{0.75\paperheight}}{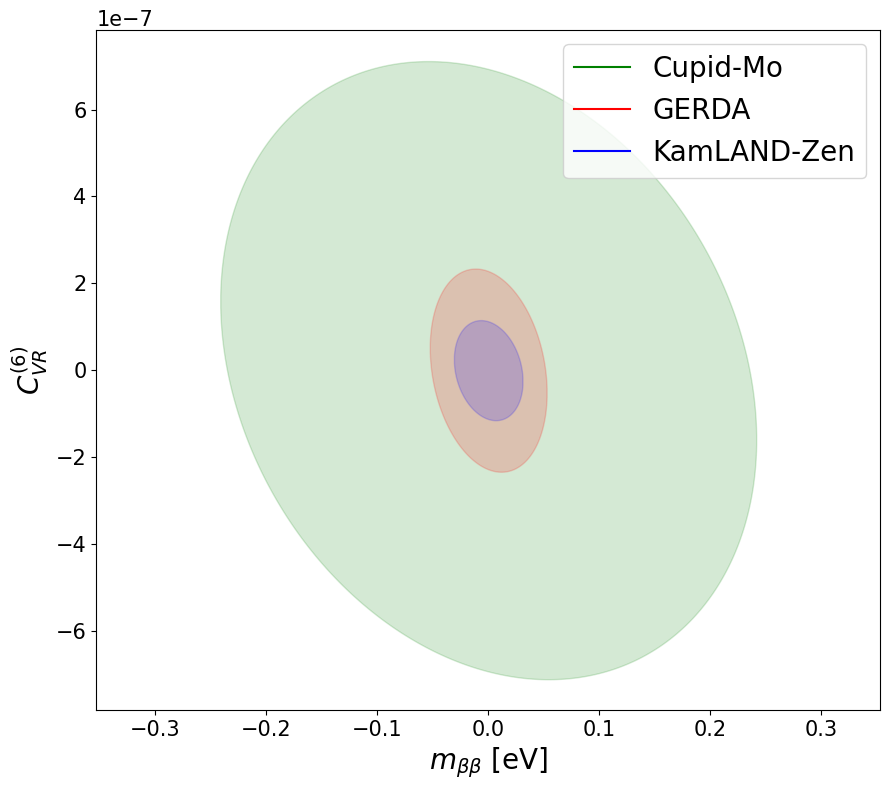}
    \end{center}
    { \hspace*{\fill} \\}
Because of different NMEs related to the different isotopes, the contours not only have different radii but also different orientations and shapes.

%
%


\newpage
\appendix
\section{List of Operators}\label{app:operator_list}
In total, \nudobe contains 32 different LEFT operators and 36 different SMEFT operators that contribute to \0. The relevant LEFT operators up to dimension 9 have been obtained in \citep{Cirigliano:2017djv,Cirigliano_2018} and are summarized in Table~\ref{tab:LEFT_operators}. The general LEFT Lagrangian can be written as
\begin{align}
    \mathcal{L}_\mathrm{LEFT} = -\frac{1}{2}m_{\beta\beta}\overline{\nu_L^C}\nu_L 
                                + \frac{1}{v^2}\sum_i C_i^{(6)}\mathcal{O}_i^{(6)}
                                + \frac{1}{v^3}\sum_j C_j^{(7)}\mathcal{O}_j^{(7)}
                                + \frac{1}{v^5}\sum_k C_k^{(9)}\mathcal{O}_k^{(9)}
\end{align}
In the SMEFT we consider all LNV operators up to dimension 7 and all dimension 9 operators that match onto LEFT dimension 9~\citep{Liao:2020jmn}. The relevant SMEFT operators are summarized in Table~\ref{tab:SMEFT_operators_SR} and the SMEFT Lagrangian can be written as
\begin{align}
    \mathcal{L}_\mathrm{SMEFT} = C_{LH}^{(5)}\mathcal{O}_{LH}^{(5)} + \sum_i C_i^{(7)}\mathcal{O}_i^{(7)} + \sum_j C_j^{(9)} \mathcal{O}_j^{(9)}
\end{align}
Therefore, we define the Wilson coefficients of LEFT as dimensionless numbers while the Wilson coefficients of the SMEFT are dimensionful, i.e., $C_{i,\mathrm{SMEFT}}^{d}\propto \Lambda^{4-d}$. We choose this somewhat awkward convention because it is done in most of the $0\nu\beta\beta$ literature.
\\\\
In defining the operator basis, we use the following conventions: 
The partial derivative acting to the left and right is defined as
\begin{align}
    \overline{\Psi_1}\overset{\leftrightarrow}{\partial}_\mu \Psi_2 = \overline{\Psi_1}\left(\partial_\mu \Psi_2\right) - \left(\partial_\mu\overline{\Psi_1}\right) \Psi_2
\end{align}
The covariant derivative used to define the SMEFT operators (ignoring $SU(3)$) is given by
\begin{align}
    D_\mu = \left(\partial_\mu-i\frac{g_2}{2}\tau^IW^I_\mu-ig_1Y B_\mu\right)\;,
\end{align}
with the Pauli matrices labeled via $\tau^I, I=1,2,3$. The generators of $SU(3)$, that are present in the definition of several LEFT operators, are denoted by
\begin{align}
    t_a = \frac{\lambda_a}{2}\;,
\end{align}
where $\lambda_a, a=1,...,8$ represent the Gell-Mann matrices.
\newpage
\begin{table}[h!]
\centering
\scalebox{0.71}{
    \centering
    \begin{tabularx}{1.246\textwidth}{| l | l | l | r | r | r |}
        \hline\textbf{Class} & \multicolumn{1}{c|}{\textbf{Op. Name}} & \multicolumn{1}{c|}{\textbf{Code Label}} & \multicolumn{1}{c|}{\textbf{Op. Structure}} & \multicolumn{1}{c|}{\textbf{Half-Life Group}} & \multicolumn{1}{c|}{\textbf{SMEFT Dim.}}\\\hline
        \multicolumn{4}{c}{\quad}\\
        \multicolumn{4}{c}{\textbf{Dim 3}}\\\hline
        $\Psi^2$ & $\mathcal{O}_{m_{\beta\beta}}$ & m\_bb & $-\frac{1}{2}m_{ee}\overline{\nu_{L,e}^C}\nu_{L,e}$ & $m_{\beta\beta}$ & 5
        \\\hline
        \multicolumn{6}{c}{\quad}\\
        \multicolumn{6}{c}{\textbf{Dim 6}}\\\hline
        \multirow{5}{*}{$\Psi^4$} & $\mathcal{O}_{SL}^{(6)}$ & SL(6) & $\big[\overline{u_R} d_L\big]\big[\overline{e_L}\nu_L^C\big]$ & $C_S^{(6)}$ & 7
        \\
         & $\mathcal{O}_{SR}^{(6)}$ & SR(6) & $\big[\overline{u_L} d_R\big]\big[\overline{e_L}\nu_L^C\big]$ & $C_S^{(6)}$ & 7
         \\
         & $\mathcal{O}_{VL}^{(6)}$ & VL(6) & $\big[\overline{u_L}\gamma^\mu d_L\big]\big[\overline{e_R}\gamma_\mu\nu_L^C\big]$ & $C_{VL}^{(6)}$ & 7
         \\
         & $\mathcal{O}_{VR}^{(6)}$ & VR(6) & $\big[\overline{u_R}\gamma^\mu d_R\big]\big[\overline{e_R}\gamma_\mu\nu_L^C\big]$  & $C_{VR}^{(6)}$ & 7
         \\
         & $\mathcal{O}_{T}^{(6)}$ & T(6) & $\big[\overline{u_L}\sigma^{\mu\nu} d_R\big]\big[\overline{e_L}\sigma_{\mu\nu}\nu_L^C\big]$  & $C_{T}^{(6)}$ & 7
         \\\hline
        \multicolumn{6}{c}{\quad}\\
        \multicolumn{6}{c}{\textbf{Dim 7}}\\\hline
        \multirow{2}{*}{$\Psi^4\partial$} & VL(7) & $\mathcal{O}_{VL}^{(7)}$ & $\big[\overline{u_L}\gamma^\mu d_L\big]\big[\overline{e_L}\overset{\leftrightarrow}{\partial}_\mu\nu_L^C\big]$ & $C_{V}^{(7)}$ & 7
        \\
         & $\mathcal{O}_{VR}^{(7)}$ & VR(7) & $\big[\overline{u_R}\gamma^\mu d_R\big]\big[\overline{e_L}\overset{\leftrightarrow}{\partial}_\mu\nu_L^C\big]$ & $C_{V}^{(7)}$ & 7
         \\\hline
        \multicolumn{6}{c}{\quad}\\
        \multicolumn{6}{c}{\textbf{Dim 9}}\\\hline
        \multirow{24}{*}{$\Psi^6$} & $\mathcal{O}_{1L}^{(9)}$ & 1L(9)) & $\big[\overline{u_L}\gamma_\mu d_L\big]\big[\overline{u_L}\gamma^\mu d_L\big]\big[\overline{e_{L}}e_{L}^C\big]$ & $C_{S1}^{(9)}$ & 7
        \\
         & ${\mathcal{O}_{1R}^{(9)}}$ & 1R(9) & $\big[\overline{u_R}\gamma_\mu d_R\big]\big[\overline{u_R}\gamma^\mu d_R\big]\big[\overline{e_{R}}e_{R}^C\big]$ & $C_{S1}^{(9)}$ & 9
         \\
         & ${\mathcal{O}_{1L}^{(9)}}'$ & 1L(9)prime & $\big[\overline{u_R}\gamma_\mu d_R\big]\big[\overline{u_R}\gamma^\mu d_R\big]\big[\overline{e_{L}}e_{L}^C\big]$ & $C_{S1}^{(9)}$ & 11
         \\
         & ${\mathcal{O}_{1R}^{(9)}}'$ & 1R(9)prime & $\big[\overline{u_R}\gamma_\mu d_R\big]\big[\overline{u_R}\gamma^\mu d_R\big]\big[\overline{e_{R}}e_{R}^C\big]$ & $C_{S1}^{(9)}$ & 9
         \\
         & $\mathcal{O}_{2L}^{(9)}$ & 2L(9) & $\big[\overline{u_R} d_L\big]\big[\overline{u_R} d_L\big]\big[\overline{e_{L}}e_{R}^C\big]$ & $C_{S2}^{(9)}$ & 9
         \\
         & $\mathcal{O}_{2R}^{(9)}$ & 2R(9) & $\big[\overline{u_R} d_L\big]\big[\overline{u_R} d_L\big]\big[\overline{e_{R}}e_{R}^C\big]$ & $C_{S2}^{(9)}$ & 11
         \\
         & ${\mathcal{O}_{2L}^{(9)}}'$ & 2L(9)prime & $\big[\overline{u_L} d_R\big]\big[\overline{u_L} d_R\big]\big[\overline{e_{L}}e_{L}^C\big]$ & $C_{S2}^{(9)}$ & 9
         \\
         & ${\mathcal{O}_{2R}^{(9)}}'$ & 2R(9)prime & $\big[\overline{u_L} d_R\big]\big[\overline{u_L} d_R\big]\big[\overline{e_{R}}e_{R}^C\big]$ & $C_{S2}^{(9)}$ & 11
         \\
         & $\mathcal{O}_{3L}^{(9)}$ & 3L(9) & $\big[\overline{u_R}^\alpha d_L^\beta\big]\big[\overline{u_R}^\beta d_L^\alpha\big]\big[\overline{e_{L}}e_{L}^C\big]$ & $C_{S3}^{(9)}$ & 9
         \\
         & $\mathcal{O}_{3R}^{(9)}$ & 3R(9) & $\big[\overline{u_R}^\alpha d_L^\beta\big]\big[\overline{u_R}^\beta d_L^\alpha\big]\big[\overline{e_{R}}e_{R}^C\big]$ & $C_{S3}^{(9)}$ & 11
         \\
         & ${\mathcal{O}_{3L}^{(9)}}'$ & 3L(9)prime & $\big[\overline{u_L}^\alpha d_R^\beta\big]\big[\overline{u_L}^\beta d_R^\alpha\big]\big[\overline{e_{L}}e_{L}^C\big]$ & $C_{S3}^{(9)}$ & 9
         \\
         & ${\mathcal{O}_{3R}^{(9)}}'$ & 3R(9)prime & $\big[\overline{u_L}^\alpha d_R^\beta\big]\big[\overline{u_L}^\beta d_R^\alpha\big]\big[\overline{e_{R}}e_{R}^C\big]$ & $C_{S3}^{(9)}$ & 11
         \\
         & ${\mathcal{O}_{4L}^{(9)}}$ & 4L(9) & $\big[\overline{u_L}\gamma^\mu d_L\big]\big[\overline{u_R}\gamma_\mu d_R\big]\big[\overline{e_{L}}e_{L}^C\big]$ & $C_{S4}^{(9)}$ & 7
         \\
         & ${\mathcal{O}_{4R}^{(9)}}$ & 4R(9) & $\big[\overline{u_L}\gamma^\mu d_L\big]\big[\overline{u_R}\gamma_\mu d_R\big]\big[\overline{e_{R}}e_{R}^C\big]$ & $C_{S4}^{(9)}$ & 9
         \\
         & ${\mathcal{O}_{5L}^{(9)}}$ & 5L(9) & $\big[\overline{u_L}^\alpha\gamma^\mu d_L^\beta\big]\big[\overline{u_R}^\beta\gamma_\mu d_R^\alpha\big]\big[\overline{e_{L}}e_{L}^C\big]$ & $C_{S5}^{(9)}$ & 9
         \\
         & ${\mathcal{O}_{5R}^{(9)}}$ & 5R(9) & $\big[\overline{u_L}^\alpha\gamma^\mu d_L^\beta\big]\big[\overline{u_R}^\beta\gamma_\mu d_R^\alpha\big]\big[\overline{e_{R}}e_{R}^C\big]$ & $C_{S5}^{(9)}$ & 11
         \\
         & ${\mathcal{O}_{6}^{(9)}}$ & 6(9) & $\big[\overline{u_L}\gamma_\mu d_L\big]\big[\overline{u_L} d_R\big]\big[\overline{e}\gamma^\mu\gamma_5 e^C\big]$ & $C_{V}^{(9)}$ & 9
         \\
         & ${\mathcal{O}_{6}^{(9)}}'$ & 6(9)prime & $\big[\overline{u_R}\gamma_\mu d_R\big]\big[\overline{u_R} d_L\big]\big[\overline{e}\gamma^\mu\gamma_5 e^C\big]$ & $C_{V}^{(9)}$ & 9
         \\
         & ${\mathcal{O}_{7}^{(9)}}$ & 7(9) & $\big[\overline{u_L}t^A\gamma_\mu d_L\big]\big[\overline{u_L}t^A d_R\big]\big[\overline{e}\gamma^\mu\gamma_5 e^C\big]$ & $C_{\Tilde{V}}^{(9)}$ & 9
         \\
         & ${\mathcal{O}_{7}^{(9)}}'$ & 7(9)prime & $\big[\overline{u_R}t^A\gamma_\mu d_R\big]\big[\overline{u_R}t^A d_L\big]\big[\overline{e}\gamma^\mu\gamma_5 e^C\big]$ & $C_{\Tilde{V}}^{(9)}$ & 9
         \\
         & ${\mathcal{O}_{8}^{(9)}}$ & 8(9) & $\big[\overline{u_L}\gamma_\mu d_L\big]\big[\overline{u_R} d_L\big]\big[\overline{e}\gamma^\mu\gamma_5 e^C\big]$ & $C_{V}^{(9)}$ & 9
         \\
         & ${\mathcal{O}_{8}^{(9)}}'$ & 8(9)prime & $\big[\overline{u_R}\gamma_\mu d_R\big]\big[\overline{u_L} d_R\big]\big[\overline{e}\gamma^\mu\gamma_5 e^C\big]$ & $C_{V}^{(9)}$ & 9
         \\
         & ${\mathcal{O}_{9}^{(9)}}$ & 9(9) & $\big[\overline{u_L}t^A\gamma_\mu d_L\big]\big[\overline{u_R}t^A d_L\big]\big[\overline{e}\gamma^\mu\gamma_5 e^C\big]$ & $C_{\Tilde{V}}^{(9)}$ & 9
         \\
         & ${\mathcal{O}_{9}^{(9)}}'$ & 9(9)prime & $\big[\overline{u_R}t^A\gamma_\mu d_R\big]\big[\overline{u_L}t^A d_R\big]\big[\overline{e}\gamma^\mu\gamma_5 e^C\big]$ & $C_{\Tilde{V}}^{(9)}$ & 9
        \\\hline
    \end{tabularx}}
    \caption{LEFT operators contributing to $0\nu\beta\beta$-decay at tree level. The primed dimension 9 operators represent a parity flip in the quark currents. The ``Code Label'' column shows the name of the operators used in \nudobe. Additionally, the ``Half-Life Group'' column identifies operators that result in the same half-life. Finally, the ``SMEFT Dim.'' column shows the lowest SMEFT dimension an operator can originate from at the SMEFT$\rightarrow$LEFT matching scale $m_W$. When evolving the operators down to the chiral scale $\Lambda_\chi$, mixing between the different LEFT operators can occur such that, e.g., the operators $\mathcal{O}_{5L,R}^{(9)}$ receive contributions from SMEFT dim 7 and 9 at $\Lambda_\chi$.}
    \label{tab:LEFT_operators}
    
\end{table}

\newpage
\begin{table}[h!]
\centering
\scalebox{0.71}{
    \centering
    \begin{tabularx}{1.2313\textwidth}{|l | l | l | r | r|}
        \hline\textbf{Class} & \multicolumn{1}{c|}{\textbf{Op. Name}} & \multicolumn{1}{c|}{\textbf{Code Label}} & \multicolumn{1}{c|}{\textbf{Op. Structure}} & \multicolumn{1}{c|}{\textbf{LEFT Matching}}\\\hline
        \multicolumn{5}{c}{\quad}\\
        \multicolumn{5}{c}{\textbf{Dim 5}}\\\hline
        $\Psi^2H^2$ & $\mathcal{O}_{LH}^{(5)}$ & LH(5) & $\epsilon_{ij}\epsilon_{kl}\big[\overline{L^C_i}L_k\big] \big[H_jH_n\big]$ & $m_{\beta\beta}$\\\hline
        \multicolumn{5}{c}{\quad} \\
        \multicolumn{5}{c}{\textbf{Dim 7}}
        \\\hline
        $\Psi^2 H^4$ & $\mathcal{O}_{LH}^{(7)}$ & LH(7) & $\epsilon_{ij}\epsilon_{kl}\big[\overline{L^C_i}L_k\big] \big[H_jH_n\big]\big[ H^\dagger H\big]$ & $m_{\beta\beta}$
        \\\hline
        $\Psi^2 H^3D$ & $\mathcal{O}_{LHDe}^{(7)}$ & LHDe(7) & $\epsilon_{ij}\epsilon_{kl}\big[\overline{L^C_i}\gamma^\mu e\big]\big[H_jH_k\big]\big[D_\mu H\big]_l$ & $\mathcal{O}_{VL}^{(6)}$
        \\\hline
        \multirow{2}{*}{$\Psi^2 H^2D^2$}
        & $\mathcal{O}_{LHD1}^{(7)}$ & LHD1(7) & $\epsilon_{ij}\epsilon_{kl}\big[\overline{L^C_i}\big(D_\mu L\big)_j\big]\big[H_k\big(D_\mu H\big)_l\big]$ & $\mathcal{O}_{VL}^{(7)},\mathcal{O}_{1L}^{(9)}$
        \\
         & $\mathcal{O}_{LHD2}^{(7)}$ & LHD2(7) & $\epsilon_{ik}\epsilon_{jl}\big[\overline{L^C_i}\big(D_\mu L\big)_j\big]\big[H_k\big(D_\mu H\big)_l\big]$ & $\mathcal{O}_{SL}^{(6)}, \mathcal{O}_{SR}^{(6)}, \mathcal{O}_{VL}^{(7)}$
         \\\hline
         $\Psi^2 H^2W$ & $\mathcal{O}_{LHW}^{(7)}$ & LHW(7) & $\epsilon_{ij}\epsilon_{km}\tau^I_{ml}g_2\big[\overline{L^C_i}\sigma^{\mu\nu}L_k\big]\big[H_jH_l\big]W^I_{\mu\nu}$ &$\mathcal{O}_{VL}^{(6)}, \mathcal{O}_{VL}^{(7)}, \mathcal{O}_{1L}^{(9)}$ 
         \\\hline
         $\Psi^4 D$ & $\mathcal{O}_{LLduD1}^{(7)}$ & LLduD1(7) & $\epsilon_{ij}\big[\overline{d}\gamma^\mu u\big]\big[\overline{L^C_i}\big(D_\mu L\big)_j\big]$ & $\mathcal{O}_{VR}^{(7)}, \mathcal{O}_{4L}^{(9)}$
         \\\hline
         \multirow{4}{*}{$\Psi^4 H$}
          & $\mathcal{O}_{LLQdH1}^{(7)}$ & LLQdH1(7) & $\epsilon_{ik}\epsilon_{jl}\big[\overline{d}L_i\big]\big[\overline{Q^C}_j L_k\big]H_l$ & $\mathcal{O}_{SR}^{(6)}, \mathcal{O}_{T}^{(6)}$
          \\
          & $\mathcal{O}_{LLQdH2}^{(7)}$ & LLQdH2(7) & $\epsilon_{ij}\epsilon_{kl}\big[\overline{d}L_i\big]\big[\overline{Q^C}_j L_k\big]H_l$ & $\mathcal{O}_{T}^{(6)}$
          \\
          & $\mathcal{O}_{LLQuH}^{(7)}$ & LLQuH(7) & $\epsilon_{ij}\big[\overline{Q}_k u\big]\big[\overline{L^C}_kL_i\big]H_j$ & $\mathcal{O}_{SL}^{(6)}$
          \\
          & $\mathcal{O}_{LeudH}^{(7)}$ & LeudH(7) & $\epsilon_{ij}\big[\overline{L^C}_i\gamma^\mu e\big]\big[\overline{d}\gamma_\mu u\big]H_j$ & $\mathcal{O}_{VR}^{(6)}$\\\hline\multicolumn{4}{c}{\quad}
          \\
        \multicolumn{5}{c}{\quad}\\
        \multicolumn{5}{c}{\textbf{Dim 9}}\\\hline
         \multirow{11}{*}{$\Psi^6$} & 
         $\mathcal{O}_{ddueue}^{(9)}$ & ddueue(9) & $\big[\overline{d^\alpha}d^{C\beta}\big]\big[\overline{u^{C\alpha}}e\big]\big[\overline{u^{C\beta}}e\big]$ & ${\mathcal{O}_{1R}^{(9)}}'$
         \\
          & $\mathcal{O}_{dQdueL1}^{(9)}$ & dQdueL1(9) & $\epsilon_{ij} \big[\overline{d}Q_i\big]\big[\overline{d}\gamma^\mu u\big]\big[\overline{e^C}\gamma_\mu L_j\big]$ & ${\mathcal{O}_{8}^{(9)}}'$
          \\
          & $\mathcal{O}_{dQdueL2}^{(9)}$ & dQdueL2(9) & $\epsilon_{ij} \big[\overline{d^\alpha}Q_i^\beta\big]\big[\overline{d^\beta}\gamma^\mu u^\alpha\big]\big[\overline{e^C}\gamma_\mu L_j\big]$ & ${\mathcal{O}_{8}^{(9)}}',{\mathcal{O}_{9}^{(9)}}'$
          \\
          & $\mathcal{O}_{QudueL1}^{(9)}$ & QudueL1(9) & $ \big[\overline{Q}u\big]\big[\overline{d}\gamma^\mu u\big]\big[\overline{e^C}\gamma_\mu L_j\big]$ & ${\mathcal{O}_{6}^{(9)}}'$
          \\
          & $\mathcal{O}_{QudueL2}^{(9)}$ & QudueL2(9) & $\big[\overline{Q^\alpha}u^\beta\big]\big[\overline{d^\beta}\gamma^\mu u^\alpha\big]\big[\overline{e^C}\gamma_\mu L_j\big]$ & ${\mathcal{O}_{6}^{(9)}}',{\mathcal{O}_{7}^{(9)}}'$
          \\
          & $\mathcal{O}_{dQdQLL1}^{(9)}$ & dQdQLL1(9) & $\epsilon_{ik}\epsilon_{jl} \big[\overline{d}Q_i\big]\big[\overline{d}\gamma^\mu Q_j\big]\big[\overline{L^C_k}\gamma_\mu L_l\big]$ & ${\mathcal{O}_{2L}^{(9)}}'$
          \\
          & $\mathcal{O}_{dQdQLL2}^{(9)}$ & dQdQLL2(9) & $\epsilon_{ik}\epsilon_{jl} \big[\overline{d^\alpha}Q_i^\beta\big]\big[\overline{d^\beta}\gamma^\mu Q_j^\alpha\big]\big[\overline{L^C_k}\gamma_\mu L_l\big]$ & ${\mathcal{O}_{3L}^{(9)}}'$
          \\
          & $\mathcal{O}_{dQQuLL1}^{(9)}$ & dQQuLL1(9) & $\epsilon_{ij} \big[\overline{d}Q_i\big]\big[\overline{Q} u\big]\big[\overline{L^C} L_j\big]$ & $\mathcal{O}_{5L}^{(9)}$
          \\
          & $\mathcal{O}_{dQQuLL2}^{(9)}$ & dQQuLL2(9) & $\epsilon_{ij} \big[\overline{d^\alpha}Q_i^\beta\big]\big[\overline{Q^\beta} u^\alpha\big]\big[\overline{L^C} L_j\big]$ & $\mathcal{O}_{4L}^{(9)}$
          \\
          & $\mathcal{O}_{QuQuLL1}^{(9)}$ & QuQuLL1(9) & $ \big[\overline{Q_i}u\big]\big[\overline{Q_j} u\big]\big[\overline{L^C_i} L_j\big]$ & $\mathcal{O}_{2L}^{(9)}$
          \\
          & $\mathcal{O}_{QuQuLL2}^{(9)}$ & QuQuLL2(9) & $ \big[\overline{Q_i^\alpha}u^\beta\big]\big[\overline{Q_j^\beta} u^\alpha\big]\big[\overline{L^C_i} L_j\big]$ & $\mathcal{O}_{3L}^{(9)}$
          \\\hline
          $\Psi^2H^4W$ & $\mathcal{O}_{LLH^4W1}^{(9)}$ & LLH4W1(9) & $\epsilon_{ij}\epsilon_{km}\tau^I_{ml}g_2\big[\overline{L^C_i}\sigma^{\mu\nu}L_k\big]\big[H_jH_l\big]W^I_{\mu\nu}\big[H^\dagger H\big]$ & $\mathcal{O}_{VL}^{(6)}, \mathcal{O}_{VL}^{(7)}, \mathcal{O}_{1L}^{(9)}$
          \\\hline
          \multirow{3}{*}{$D^2\Psi^2H^4$} & $\mathcal{O}_{eeH^4D^2}^{(9)}$ & eeH4D2(9) & $\epsilon_{ij}\epsilon_{kl}\big[\overline{e^C}e\big]\big[H_i\big(D_\mu H\big)_j\big]\big[H_k\big(D^\mu H\big)_l\big]$ & $\mathcal{O}_{1R}^{(9)}$
          \\
           & $\mathcal{O}_{LLH^4D^23}^{(9)}$ & LLH4D23(9) & $\epsilon_{ik}\epsilon_{jl}\big[\big(\overline{D_\mu L^C}\big)_i\big(D^\mu L\big)_j\big]\big[H_k H_l\big]\big[H^\dagger H\big]$ & $\mathcal{O}_{SL}^{(6)}, \mathcal{O}_{SR}^{(6)}, \mathcal{O}_{VL}^{(7)}, \mathcal{O}_{1L}^{(9)}$
           \\
           & $\mathcal{O}_{LLH^4D^24}^{(9)}$ & LLH4D24(9) & $\epsilon_{ik}\epsilon_{jl}\big[\overline{L^C}_i\big(D^\mu L\big)_j\big]\big[\big(D_\mu H\big)_k H_l\big]\big[H^\dagger H\big]$ & $\mathcal{O}_{SL}^{(6)}, \mathcal{O}_{SR}^{(6)}, \mathcal{O}_{VL}^{(7)}, \mathcal{O}_{1L}^{(9)}$
           \\\hline
           \multirow{9}{*}{$D\Psi^4H^2$} & $\mathcal{O}_{deueH^2D}^{(9)}$ & deueH2D(9) & $\epsilon_{ij}\big[\overline{d}\gamma^\mu e\big]\big[\overline{u^C}e\big]\big[H_i \big(iD_\mu H\big)_j\big]$ & $\mathcal{O}_{4R}^{(9)}$
           \\
           & $\mathcal{O}_{dQLeH^2D2}^{(9)}$ & dQLeH2D2(9) & $\epsilon_{ik}\epsilon_{jl}\big[\overline{d}Q_i\big]\big[\overline{L^C_j}\gamma^\mu e\big]\big[H_k \big(iD_\mu H\big)_l\big]$ & $\mathcal{O}_{6}^{(9)}$
           \\
           & $\mathcal{O}_{dLQeH^2D1}^{(9)}$ & dLQeH2D1(9) & $\epsilon_{ik}\epsilon_{jl}\big[\overline{d}L_i\big]\big[\overline{Q^C_j}\gamma^\mu e\big]\big[\big(iD_\mu H\big)_k H_l\big]$ & $\mathcal{O}_{6}^{(9)}, \mathcal{O}_{7}^{(9)}$
           \\
           & $\mathcal{O}_{dLuLH^2D2}^{(9)}$ & dLuLH2D2(9) & $\epsilon_{ik}\epsilon_{jl}\big[\overline{d}L_i\big]\big[\overline{u^C}\gamma_\mu L_j\big]\big[\Tilde{H}_k \big(iD^\mu H\big)_l\big]$ & $\mathcal{O}_{4L}^{(9)}$
           \\
           & $\mathcal{O}_{duLLH^2D}^{(9)}$ & duLLH2D(9) & $\epsilon_{ik}\epsilon_{jl}\big[\overline{d}\gamma_\mu u\big]\big[\overline{L^C}_i \big(i D^\mu L\big)_j\big]\big[\Tilde{H}_k H_l\big]$ & $\mathcal{O}_{SL}^{(6)},\mathcal{O}_{SR}^{(6)},\mathcal{O}_{VR}^{(7)},\mathcal{O}_{4L}^{(9)}$
           \\
           & $\mathcal{O}_{deQLH^2D}^{(9)}$ & deQLH2D(9) & $\epsilon_{ik}\epsilon_{jl}\big[\overline{d}\gamma^\mu e\big]\big[\overline{Q^C_i}\big(iD_\mu L\big)_j\big]\big[H_k H_l\big]$ & $
                \begin{matrix}
                   \mathcal{O}_{SR}^{(6)}, \mathcal{O}_{VL}^{(6)} \mathcal{O}_{VR}^{(6)}, \mathcal{O}_{T}^{(6)}, 
                   \\
                   \mathcal{O}_{6}^{(9)}, \mathcal{O}_{7}^{(9)}, ({\mathcal{O}_{VR}^{(7)}}')
                \end{matrix}$
           \\
           & $\mathcal{O}_{QueLH^2D2}^{(9)}$ & QueLH2D2(9) & $\epsilon_{jk}\big[\overline{Q_i}u\big]\big[\overline{u^C}\gamma^\mu L_j\big]\big[H_i \big(iD_\mu H\big)_k\big]$ & $\mathcal{O}_{8}^{(9)}$
           \\
           & $\mathcal{O}_{QeuLH^2D2}^{(9)}$ & QeuLH2D2(9) & $\delta_{ik}\epsilon_{jl}\big[\overline{Q_i}e\big]\big[\overline{u^C}\gamma^\mu L_j\big]\big[H_k \big(iD_\mu H\big)_l\big]$ & $\mathcal{O}_{8}^{(9)}, \mathcal{O}_{9}^{(9)}$
           \\
           & $\mathcal{O}_{QLQLH^2D2}^{(9)}$ & QLQLH2D2(9) & $\epsilon_{ik}\epsilon_{jl}\big[\overline{Q}\gamma^\mu L\big]\big[\overline{Q^C_i}\gamma^\mu L_j\big]\big[H_k \big(iD_\mu H\big)_l\big]$ & $\mathcal{O}_{1L}^{(9)}$
           \\
           & $\mathcal{O}_{QLQLH^2D5}^{(9)}$ & QLQLH2D5(9) & $\epsilon_{ik}\epsilon_{jl}\big[\overline{Q}\gamma^\mu L_i\big]\big[\overline{Q^C_j}\gamma^\mu L\big]\big[\big(iD_\mu H\big)_k H_l \big]$ & $\mathcal{O}_{1L}^{(9)}$
           \\
           & $\mathcal{O}_{QQLLH^2D2}^{(9)}$ & QQLLH2D2(9) & $\epsilon_{ik}\epsilon_{jl}\big[\overline{Q}\gamma^\mu Q_i\big]\big[\overline{L^C}\big(iD_\mu L\big)_j\big]\big[H_k H_l \big]$ & $\mathcal{O}_{SL}^{(6)}, \mathcal{O}_{SR}^{(6)}, \mathcal{O}_{VL}^{(7)}, \mathcal{O}_{1L}^{(9)}$
           \\
           \hline
    \end{tabularx}}
    \caption{Here we show the different SMEFT dimension 9 operators that we considered. Again, the ``Code Label'' column shows the name of the operators as used in \nudobe. The ``LEFT Matching'' column provides a list of LEFT operators each SMEFT operator matches onto with the precise matching given in Appendix~\ref{app:operator_matching}.}
    \label{tab:SMEFT_operators_SR}
\end{table}
\newpage
\section{Operator Matching}\label{app:operator_matching}
The transition amplitudes and correspondingly all $0\nu\beta\beta$ observables calculated by \nudobe are evaluated at LEFT level. Hence, models which are generated at SMEFT level first have to be matched down onto LEFT. At SMEFT level lepton number violation by two units without baryon number violation occurs only at odd dimensions~\citep{Kobach:2016ami}. 
Here we provide the explicit matching conditions at the matching scale $E=m_W$
\begin{align}
    \begin{split}
        m_{\beta\beta} & = -v^2C_{LH}^{(5)} - \frac{v^4}{2}C_{LH}^{(7)}\;,
        \\
        C_{SL}^{(6)}   & = v^3\bigg(\frac{1}{\sqrt{2}}C_{LLQuH1}^{(7)*} 
                                   +\frac{m_u}{v}\frac{V_{ud}}{2}C_{LHD2}^{(7)*}\bigg) \\
                         &+ v^4\bigg(+m_u\frac{V_{ud}}{2}C_{LLH^4D^23}^{(9)*}
                                     -m_u\frac{V_{ud}}{4} C_{LLH^4D^24}^{(9)*}\\
                         &\qquad\quad-m_u\frac{1}{4} C_{QQLLH^2D2}^{(9)*}
                                     -m_d\frac{1}{4}C_{duLLH^2D}^{(9)*}\bigg)\;,
        \\
        C_{SR}^{(6)} &= v^3\bigg(\frac{1}{2\sqrt{2}}C_{LLQdH1}^{(7)*}
                               - \frac{V_{ud}}{2} \frac{m_d}{v}C_{LHD2}^{(7)*}\bigg) \\
                     &+ v^4\bigg(-m_d\frac{V_{ud}}{2}C_{LLH^4D^23}^{(9)*} 
                                 + m_d\frac{V_{ud}}{4}C_{LLH^4D^24}^{(9)*} 
                                 + \frac{m_d}{4}C_{QQLLH^2D2}^{(9)*}\\
                     &\qquad\quad+ \frac{m_u}{4} C_{duLLH^2D}^{(9)}
                                 + \frac{m_e}{8} C_{deQLH^2D}^{(9)*}\bigg)\;,
        \\
        C_{VL}^{(6)} &= v^3\bigg(-\frac{i}{\sqrt{2}}V_{ud}C_{LHDe}^{(7)*} 
                                 +4\frac{m_e}{v}C_{LHW}^{(7)*}
                                 \bigg)\\
                      &+ v^4\bigg(2m_eV_{ud}C_{LLH^4W1}^{(9)*}
                                 -\frac{m_d}{4} C_{deQLH^2D}^{(9)*}
                                 \bigg)\;,
        \\
        C_{VR}^{(6)} &= v^3\frac{1}{\sqrt{2}}C_{LeudH}^{(7)*} 
                        -v^4 \frac{m_u}{4} C_{deQLH^2D}^{(9)*}\;,
        \\
        C_T^{(6)}     &= v^3\bigg(\frac{1}{8\sqrt{2}}C_{LLQdH1}^{(7)*} 
                               +\frac{1}{4\sqrt{2}}C_{LLQdH2}^{(7)*}
                               \bigg)
                        +v^4\frac{m_e}{16}C_{deQLH^2D}^{(9)*}\;,
        \\
        C_{VL}^{(7)} &= v^3\bigg(V_{ud}C_{LHD1}^{(7)*}
                                 - \frac{V_{ud}}{2}C_{LHD2}^{(7)*} 
                                 +4 V_{ud}C_{LHW}^{(7)*}
                            \bigg)\\ 
                     &+ v^5\bigg(2V_{ud}C_{LLH^4W1}^{(9)*} 
                                 + \frac{V_{ud}}{2}C_{LLH^4D^23}^{(9)*}
                                 - \frac{V_{ud}}{4}C_{LLH^4D^24}^{(9)*} 
                                 - \frac{1}{4}C_{QQLLH^2D2}^{(9)*}
                            \bigg)\;,
        \\
        C_{VR}^{(7)} &= v^3\bigg(-iC_{LLduD1}^{(7)*}\bigg) 
                      + v^5\bigg(\frac{1}{4}C_{duLLH^2D}^{(9)*}\bigg)\;,
    \end{split}
\end{align}
for the mass-mechanism and the long-range operators and
\begin{align}
    \begin{split}
        C_{1L}^{(9)} & = v^3\bigg(2V_{ud}^2C_{LHD1}^{(7)*}
                                  +8V_{ud}^2C_{LHW}^{(7)*}
                            \bigg)\\ 
                     &  + v^5\bigg(4V_{ud}^2C_{LLH^4W1}^{(9)*}
                                   -V_{ud}^2 C_{LLH^4D^23}^{(9)*}
                                   - V_{ud}^2 C_{LLH^4D^24}^{(9)*} - V_{ud}C_{QQLLH^2D2}^{(9)*} 
                                   \\&\qquad\quad
                                   - \frac{V_{ud}}{2}C_{QLQLH^2D2}^{(9)*} 
                                   - \frac{V_{ud}}{2}C_{QLQLH^2D5}^{(9)*}\bigg)\;,
        \\
        C_{1R}^{(9)} &  = -v^5V_{ud}^2C_{eeH^4D^2}^{(9)*}\;,\quad
        {C_{1R}^{(9)}}' = \frac{v^5}{4}C_{ddueue}^{(9)*}\;,\quad
        \\
        C_{2L}^{(9)}   &= -v^5C_{QuQuLL1}^{(9)*}\;,\quad
        {C_{2L}^{(9)}}' = -v^5C_{dQdQLL1}^{(9)*}\;,
        \\
        C_{3L}^{(9)} &  = -v^5C_{QuQuLL2}^{(9)*}\;,\quad
        {C_{3L}^{(9)}}' = -v^5C_{dQdQLL2}^{(9)*}\;,\quad
        \\
        {C_{4L}^{(9)}} &= -v^3 i2V_{ud}C_{LLduD1}^{(7)*}
                          +v^5\bigg(V_{ud}C_{duLLH^2D}^{(9)*}
                                    - \frac{V_{ud}}{2}C_{dLuLH^2D2}^{(9)*} 
                                    - \frac{1}{2} C_{dQQuLL2}^{(9)*}
                              \bigg)\;,
        \\
        {C_{4R}^{(9)}} &= -v^5\frac{V_{ud}}{2} C_{deueH^2D}^{(9)*}\;,
        \\
        {C_{5L}^{(9)}} &= -\frac{1}{2}v^5C_{dQQuLL1}^{(9)*}\;,\quad
        \\
        {C_{6}^{(9)}}  &= v^5\bigg(-\frac{2}{3}V_{ud}C_{dLQeH^2D1}^{(9)*} 
                                   +\frac{V_{ud}}{2}C_{dQLeH^2D2}^{(9)*} 
                                   -\frac{5}{12}V_{ud}C_{deQLH^2D}^{(9)*}\bigg)\;,
        \\
        {C_{6}^{(9)}}' &= v^5\bigg(\frac{1}{6}C_{QudueL2}^{(9)*}
                                  +\frac{1}{2}C_{QudueL1}^{(9)*}\bigg)\;,\quad
        \\
        {C_{7}^{(9)}}  &= v^5\bigg(-V_{ud} C_{dLQeH^2D1}^{(9)} -V_{ud}C_{deQLH^2D}^{(9)*}\bigg)\;,\quad
        {C_{7}^{(9)}}'  = v^5 C_{QudueL2}^{(9)*}\;,
        \\
        {C_{8}^{(9)}}  &= v^5\bigg(-\frac{V_{ud}}{2}C_{QueLH^2D2}^{(9)*} 
                                   +\frac{V_{ud}}{6}C_{QeuLH^2D2}^{(9)*}\bigg)\;,\quad
        \\
        {C_{8}^{(9)}}' &= v^5\bigg(\frac{1}{6}C_{dQdueL2}^{(9)*} 
                                  +\frac{1}{2}C_{dQdueL1}^{(9)*}\bigg)\;,
        \\
        {C_{9}^{(9)}} & = v^5V_{ud}C_{QeuLH^2D2}^{(9)*}\;,\quad
        {C_{9}^{(9)}}'  = v^5 C_{dQdueL2}^{(9)*}\;,
    \end{split}
\end{align}
for the short-range LEFT operators. The dim-9 SMEFT operator $\mathcal{O}_{deQLH^2D}$, additionally, matches onto the low-energy $\Delta_L=2$ operator ${\mathcal{O}_{{VR}}^{(7)}}'=\big[\overline{e_R}\gamma^\mu \nu_L^C\big]\big[\overline{u_L}i\overset{\leftrightarrow}{\partial}_\mu d_R\big]$. To study the relevance of ${\mathcal{O}_{{VR}}^{(7)}}'$ we can be express it in terms of $\mathcal{O}_{VL}^{(6)}, \mathcal{O}_{VR}^{(6)}$ and an additional tensor contribution
\begin{align}
\begin{split}
    {\mathcal{O}_{VR}^{(7)}}' = &\frac{2G_F}{\sqrt{2}v}
     {C_{VR}^{(7)}}' \Big[\overline{e_R}\gamma^\mu \nu_L^C\Big]\Big[\overline{u_L}i\overset{\leftrightarrow}{\partial}_\mu d_R\Big]\\
    =&\frac{2G_F}{\sqrt{2}v}{C_{VR}^{(7)}}'\Big[\overline{e_R}\gamma^\mu \nu_L^C\Big]\bigg(m_d\Big[\overline{u_L}\gamma_\mu d_L\Big] + m_u\Big[\overline{u_R}\gamma_\mu d_R\Big]\bigg) \\+&\frac{2G_F}{\sqrt{2}v} {C_{VR}^{(7)}}'\bigg(\Big[\partial_\nu\overline{e_R}\gamma_\mu \nu_L^C\Big] + \Big[\overline{e_R}\gamma_\mu \partial_\nu\nu_L^C\Big]\bigg)\Big[\overline{u_L}\sigma^{\mu\nu}d_R\Big]
\end{split}
\end{align}
With the chiral power counting $\epsilon_\chi = m_\pi/\Lambda_\chi$ and $\Lambda_\chi \sim m_N \sim 1\,\mathrm{GeV}$ we find that the tensor operator's $\bigg(\Big[\partial_\nu\overline{e_R}\gamma_\mu \nu_L^C\Big] + \Big[\overline{e_R}\gamma_\mu \partial_\nu\nu_L^C\Big]\bigg)\Big[\overline{u_L}\sigma^{\mu\nu}d_R\Big]$leading contribution to the decay amplitude is at $\mathcal{O}(\epsilon_\chi^5)$ as it is proportional to the lepton momenta $k\sim\epsilon_{\chi}^3\Lambda_\chi$, the neutrino momentum $q\sim\epsilon_\chi\Lambda_\chi$ and requires the inclusion of NLO nuclear currents $~\sim \epsilon_\chi$. We can therefore savely ignore it. With $m_d\sim m_u\sim\epsilon^2_\chi\Lambda_\chi$ and the leading order contributions from $C_{VL}^{(6)}~\sim \Lambda_\chi\epsilon_\chi^2$ and $C_{VR}^{(6)}~\sim \Lambda_\chi\epsilon_\chi^3$, the leading order contribution from the remaining parts of ${\mathcal{O}_{VR}^{(7)}}'$ is at $~\sim\frac{\Lambda_\chi^2}{v}\epsilon_\chi^4$ from ${C_{VR}^{(7)}}'m_d\Big[\overline{e_R}\gamma^\mu \nu_L^C\Big]\Big[\overline{u_L}\gamma_\mu d_L\Big]$. Therefore, compared to the other dim-7 and dim-9 operators~\citep{Cirigliano_2018} ${\mathcal{O}_{VR}^{(7)}}'$ is suppressed by an additional factor of $\epsilon_\chi^2$ and we ignore it. 

\section{Phase Space Factors}\label{app:PSFs}
\subsection{Calculation of PSFs}
The different phase-space factors $G_{0k}$ that enter the half-life formula are defined as
\begin{align}
\begin{split}
    G_{0k} = C_k\frac{G_F^4m_e^2}{64\pi^5\ln{2}R^2}&\int \delta\bigg(\epsilon_1 + \epsilon_2 +E_f-E_i\bigg) \\
    &\times \bigg(h_{0k}(\epsilon_1, \epsilon_2, R)\cos{\theta} + g_{0k}(\epsilon_1, \epsilon_2, R)\bigg) \\
    &\times p_1 p_2 \epsilon_1 \epsilon_2 \,\text{d}\epsilon_1\,\text{d}\epsilon_2\,\text{d}(\cos{\theta}),\label{eq:PSF_definition}
\end{split}
\end{align}
where $\epsilon_i, p_i$ denote the energy and momentum of the $i-$th. single electron, $\theta$ is the angle between the emitted electrons, we defined
\begin{align}
    C_k = \left\{
          \begin{matrix}
              9/2         & ,k = 4\\
              m_e R/2     & ,k = 6\\
              (m_e R/2)^2 & ,k = 9\\
              1           & ,\mathrm{else\;\;\;}
          \end{matrix}
          \right.,
\end{align}
and $h_{0k}, g_{0k}$ were introduced in Ref.~\citep{Stefanik:2015twa}. The constants $C_k$ take care of the different conventions used in Refs.~\citep{Stefanik:2015twa} and~\citep{Cirigliano_2018}. \nudobe takes care of this re-scaling internally, such that no action is required from the user and $C_k=1$ can be chosen when calculating PSFs to use in \nudobe. The quark-mixing parameter $V_{ud}$ is not part of $G_{0k}$ in contrast to the usual literature convention. This is to account for the fact that higher dimensional typically do not exchange two $W$ bosons. Instead, factors of $V_{ud}$ enter the sub-amplitudes $\mathcal{A}_k$ (see Ref.~\citep{Cirigliano_2018}).
\nudobe includes two approximation schemes for the calculation of PSFs and electron wave functions, i.e., an approximate solution to a uniform charge distribution (Scheme A) as well as an exact solution to a point-like nuclear potential (Scheme B).
\subsection{Related Observables}
Apart from the total decay rate, we are also interested in the differential rates, i.e., the energy and angular electron spectra \citep{Stefanik:2015twa}. The normalized single electron spectrum can be determined through
\begin{align}
    \frac{\mathrm{d}\Gamma}{\mathrm{d}\epsilon_1} \left(\left\{C_i\right\}, \Tilde{\epsilon}\right) \propto \sum_k g_{0k}\left(\epsilon, \Delta M - \epsilon, R\right)\left|A_{k}(\{C_i\})\right|^2p_1 p_2 \epsilon\left(\Delta M-\epsilon\right)\;,\label{eq:spectrum}
\end{align}
Obviously, the electron spectra depend on the precise EFT model, i.e., the chosen set of Wilson coefficients $\left\{C_i\right\}$. This dependency is represented in the sub-amplitudes $\mathcal{A}_k(\left\{C_i\right\})$. We defined the mass difference of the mother and daughter isotope $\Delta M = E_i-E_f$ and the
normalized electron energy $\Tilde{\epsilon}_i = (\epsilon_i-m_e)/Q$ in terms of the $Q$-value $Q = \Delta M - 2 m_e$. The nuclear radius $R$ is determined from the mass number $A$
\begin{align}
    R = 1.2\,\mathrm{fm}\times A^{1/3}\,.
\end{align}
Similarly, the energy-dependent angular correlation coefficient $a_1/a_0$ is given by
\begin{align}
    \frac{\mathrm{d}\Gamma}{\mathrm{d}\cos\theta\mathrm{d}\Tilde{\epsilon}_1} = a_0\left(1+\frac{a_1}{a_0}\cos\theta\right)\;,\qquad\frac{a_1}{a_0}\left(\{C_i\}, \Tilde{\epsilon}\right) = \frac{\sum_j h_{0j}(\epsilon, \Delta M-\epsilon, R)\left|A_j(\{C_i\})\right|^2}{\sum_k g_{0k}(\epsilon, \Delta M-\epsilon, R) \left|A_k(\{C_i\})\right|^2}\;.\label{eq:angular_corr}
\end{align}

\section{Nuclear Matrix Elements}\label{app:NMEs}
The different NMEs that enter the calculation of the sub-amplitudes $\mathcal{A}_k$ are defined via
\begin{align}
    h^{ij}_K(r) = \frac{2}{\pi}R_A\int_0^\infty d|\vec{q}| h_K^{ij}(\vec{q}^2)j_\lambda(|\vec{q}|r), \quad 
    h^{ij}_{K,sd}(r) = C_{sd}\frac{2}{\pi}\frac{R_A}{m_e m_p}\int_0^\infty d|\vec{q}|\vec{q}^2 h_K^{ij}(\vec{q}^2)j_\lambda(|\vec{q}|r)
\end{align}
with
\begin{align}
\begin{split}
    &h_{GT,T}^{AA}(\vec{q}^2) = \frac{g_A^2(\vec{q}^2)}{g_A^2}, \quad h_{GT}^{AP}(\vec{q}^2) = \frac{g_P(\vec{q}^2)g_A(\vec{q}^2)}{g_A^2} \frac{\vec{q}^2}{3m_N}, \quad h_{GT}^{PP}(\vec{q}^2) = \frac{g_P^2(\vec{q}^2)}{g_A^2}\frac{\vec{q}^4}{12 m_N^2}
    \\
    &h_{GT}^{MM}(\vec{q}^2) = g_M^2(\vec{q}^2)\frac{\vec{q}^2}{6g_A^2m_N^2}, \quad h_F(\vec{q}^2) = g_V(\vec{q}^2)
\end{split}
\end{align}
and
\begin{align}
\begin{split}
    &g_V(\vec{q}^2) = \left(1+\frac{\vec{q}^2}{\Lambda_V^2}\right)^{-2}, \quad
    g_A(\vec{q}^2) = g_A \left(1+\frac{\vec{q}^2}{\Lambda_A^2}\right)^{-2},
    \\ 
    &g_M(\vec{q}^2) = (1+\kappa_1)g_V(\vec{q}^2),\quad
    g_P(\vec{q}^2) = -\frac{2m_Ng_A(\vec{q}^2)}{\vec{q}^2+m_\pi^2}
\end{split}
\end{align}
as
\begin{align}
    \begin{split}
        M_{F,(sd)}        &= \left<0^+\right|\sum_{m,n}h_{F,(sd)}(r)\tau^{+(m)}\tau^{+(n)}\left|0^+\right>,\\
        M_{GT,(sd)}^{ij}  &= \left<0^+\right|\sum_{m,n}h_{GT,(sd)}^{ij}(r)\vec{\sigma}^{(m)}\cdot\vec{\sigma}^{(n)}\tau^{+(m)}\tau^{+(n)}\left|0^+\right>,\\
        M_{T,(sd)}^{ij}   &= \left<0^+\right|\sum_{m,n}h_{T,(sd)}^{ij}(r)S^{mn}(\hat{\vec{r}})\tau^{+(m)}\tau^{+(n)}\left|0^+\right>,
    \end{split}
\end{align}
with the position-space tensor
\begin{align}
    S^{mn}(\hat{\vec{r}}) = 3(\vec{\sigma}^{(m)}\cdot\hat{\vec{r}})(\vec{\sigma}^{(n)}\cdot\hat{\vec{r}}) - \vec{\sigma}^{(m)}\cdot\vec{\sigma}^{(n)}
\end{align}
Again, the re-scaling factor
\begin{align}
    C_{sd} = \frac{m_e m_p}{m_\pi^2}
\end{align}
takes care of different conventions used in the common literature and~\citep{Cirigliano_2018}. \nudobe takes care of this re-scaling internally when importing NMEs from the corresponding \texttt{.csv} files. That is, the NMEs defined in the \texttt{.csv} files are multiplied by $C_{sd}$ when initiating a model class such that, similar to the case of $C_k$ in the PSFs, users can choose $C_{sd}=1$ when calculating NMEs to use in \nudobe.
\bibliographystyle{utphys}
\bibliography{references.bib}
\end{document}

%% file: python_preamble.tex
    \usepackage[breakable]{tcolorbox}
    \usepackage{parskip} 
    
    \usepackage{iftex}
    \ifPDFTeX
    	\usepackage[T1]{fontenc}
    \else
    	\usepackage{fontspec}
    \fi

    \usepackage{graphicx}
    
    \usepackage[Export]{adjustbox} 
    \adjustboxset{max size={0.9\linewidth}{0.9\paperheight}}
    \usepackage{float}
    \usepackage{xcolor} 
    \usepackage{enumerate} 
    \usepackage{geometry} 
    \usepackage{amsmath} 
    \usepackage{amssymb} 
    \usepackage{textcomp} 
    \AtBeginDocument{%
    }
    \usepackage{upquote} 
    \usepackage{eurosym} 
    \usepackage[mathletters]{ucs} 
    \usepackage{fancyvrb} 
    \usepackage{grffile} 
    \makeatletter 
    \def\Gread@@xetex#1{%
      \IfFileExists{"\Gin@base".bb}%
      {\Gread@eps{\Gin@base.bb}}%
      {\Gread@@xetex@aux#1}%
    }
    \makeatother

    \usepackage{titling}
    \usepackage{longtable} 
    \usepackage{booktabs}  
    \usepackage[inline]{enumitem} 
    \usepackage[normalem]{ulem} 
    \usepackage{mathrsfs}

    \definecolor{urlcolor}{rgb}{0,.145,.698}
    \definecolor{linkcolor}{rgb}{.71,0.21,0.01}
    \definecolor{citecolor}{rgb}{.12,.54,.11}

    \definecolor{ansi-black}{HTML}{3E424D}
    \definecolor{ansi-black-intense}{HTML}{282C36}
    \definecolor{ansi-red}{HTML}{E75C58}
    \definecolor{ansi-red-intense}{HTML}{B22B31}
    \definecolor{ansi-green}{HTML}{00A250}
    \definecolor{ansi-green-intense}{HTML}{007427}
    \definecolor{ansi-yellow}{HTML}{DDB62B}
    \definecolor{ansi-yellow-intense}{HTML}{B27D12}
    \definecolor{ansi-blue}{HTML}{208FFB}
    \definecolor{ansi-blue-intense}{HTML}{0065CA}
    \definecolor{ansi-magenta}{HTML}{D160C4}
    \definecolor{ansi-magenta-intense}{HTML}{A03196}
    \definecolor{ansi-cyan}{HTML}{60C6C8}
    \definecolor{ansi-cyan-intense}{HTML}{258F8F}
    \definecolor{ansi-white}{HTML}{C5C1B4}
    \definecolor{ansi-white-intense}{HTML}{A1A6B2}
    \definecolor{ansi-default-inverse-fg}{HTML}{FFFFFF}
    \definecolor{ansi-default-inverse-bg}{HTML}{000000}

    \DefineVerbatimEnvironment{Highlighting}{Verbatim}{commandchars=\\\{\}}

    \let\Oldtex\TeX
    \let\Oldlatex\LaTeX
    \renewcommand{\TeX}{\textrm{\Oldtex}}
    \renewcommand{\LaTeX}{\textrm{\Oldlatex}}
    \title{Showcase}

\makeatletter
\def\PY@reset{\let\PY@it=\relax \let\PY@bf=\relax%
    \let\PY@ul=\relax \let\PY@tc=\relax%
    \let\PY@bc=\relax \let\PY@ff=\relax}
\def\PY@tok#1{\csname PY@tok@#1\endcsname}
\def\PY@toks#1+{\ifx\relax#1\empty\else%
    \PY@tok{#1}\expandafter\PY@toks\fi}
\def\PY@do#1{\PY@bc{\PY@tc{\PY@ul{%
    \PY@it{\PY@bf{\PY@ff{#1}}}}}}}
\def\PY#1#2{\PY@reset\PY@toks#1+\relax+\PY@do{#2}}

\expandafter\def\csname PY@tok@w\endcsname{\def\PY@tc##1{\textcolor[rgb]{0.73,0.73,0.73}{##1}}}
\expandafter\def\csname PY@tok@c\endcsname{\let\PY@it=\textit\def\PY@tc##1{\textcolor[rgb]{0.25,0.50,0.50}{##1}}}
\expandafter\def\csname PY@tok@cp\endcsname{\def\PY@tc##1{\textcolor[rgb]{0.74,0.48,0.00}{##1}}}
\expandafter\def\csname PY@tok@k\endcsname{\let\PY@bf=\textbf\def\PY@tc##1{\textcolor[rgb]{0.00,0.50,0.00}{##1}}}
\expandafter\def\csname PY@tok@kp\endcsname{\def\PY@tc##1{\textcolor[rgb]{0.00,0.50,0.00}{##1}}}
\expandafter\def\csname PY@tok@kt\endcsname{\def\PY@tc##1{\textcolor[rgb]{0.69,0.00,0.25}{##1}}}
\expandafter\def\csname PY@tok@o\endcsname{\def\PY@tc##1{\textcolor[rgb]{0.40,0.40,0.40}{##1}}}
\expandafter\def\csname PY@tok@ow\endcsname{\let\PY@bf=\textbf\def\PY@tc##1{\textcolor[rgb]{0.67,0.13,1.00}{##1}}}
\expandafter\def\csname PY@tok@nb\endcsname{\def\PY@tc##1{\textcolor[rgb]{0.00,0.50,0.00}{##1}}}
\expandafter\def\csname PY@tok@nf\endcsname{\def\PY@tc##1{\textcolor[rgb]{0.00,0.00,1.00}{##1}}}
\expandafter\def\csname PY@tok@nc\endcsname{\let\PY@bf=\textbf\def\PY@tc##1{\textcolor[rgb]{0.00,0.00,1.00}{##1}}}
\expandafter\def\csname PY@tok@nn\endcsname{\let\PY@bf=\textbf\def\PY@tc##1{\textcolor[rgb]{0.00,0.00,1.00}{##1}}}
\expandafter\def\csname PY@tok@ne\endcsname{\let\PY@bf=\textbf\def\PY@tc##1{\textcolor[rgb]{0.82,0.25,0.23}{##1}}}
\expandafter\def\csname PY@tok@nv\endcsname{\def\PY@tc##1{\textcolor[rgb]{0.10,0.09,0.49}{##1}}}
\expandafter\def\csname PY@tok@no\endcsname{\def\PY@tc##1{\textcolor[rgb]{0.53,0.00,0.00}{##1}}}
\expandafter\def\csname PY@tok@nl\endcsname{\def\PY@tc##1{\textcolor[rgb]{0.63,0.63,0.00}{##1}}}
\expandafter\def\csname PY@tok@ni\endcsname{\let\PY@bf=\textbf\def\PY@tc##1{\textcolor[rgb]{0.60,0.60,0.60}{##1}}}
\expandafter\def\csname PY@tok@na\endcsname{\def\PY@tc##1{\textcolor[rgb]{0.49,0.56,0.16}{##1}}}
\expandafter\def\csname PY@tok@nt\endcsname{\let\PY@bf=\textbf\def\PY@tc##1{\textcolor[rgb]{0.00,0.50,0.00}{##1}}}
\expandafter\def\csname PY@tok@nd\endcsname{\def\PY@tc##1{\textcolor[rgb]{0.67,0.13,1.00}{##1}}}
\expandafter\def\csname PY@tok@s\endcsname{\def\PY@tc##1{\textcolor[rgb]{0.73,0.13,0.13}{##1}}}
\expandafter\def\csname PY@tok@sd\endcsname{\let\PY@it=\textit\def\PY@tc##1{\textcolor[rgb]{0.73,0.13,0.13}{##1}}}
\expandafter\def\csname PY@tok@si\endcsname{\let\PY@bf=\textbf\def\PY@tc##1{\textcolor[rgb]{0.73,0.40,0.53}{##1}}}
\expandafter\def\csname PY@tok@se\endcsname{\let\PY@bf=\textbf\def\PY@tc##1{\textcolor[rgb]{0.73,0.40,0.13}{##1}}}
\expandafter\def\csname PY@tok@sr\endcsname{\def\PY@tc##1{\textcolor[rgb]{0.73,0.40,0.53}{##1}}}
\expandafter\def\csname PY@tok@ss\endcsname{\def\PY@tc##1{\textcolor[rgb]{0.10,0.09,0.49}{##1}}}
\expandafter\def\csname PY@tok@sx\endcsname{\def\PY@tc##1{\textcolor[rgb]{0.00,0.50,0.00}{##1}}}
\expandafter\def\csname PY@tok@m\endcsname{\def\PY@tc##1{\textcolor[rgb]{0.40,0.40,0.40}{##1}}}
\expandafter\def\csname PY@tok@gh\endcsname{\let\PY@bf=\textbf\def\PY@tc##1{\textcolor[rgb]{0.00,0.00,0.50}{##1}}}
\expandafter\def\csname PY@tok@gu\endcsname{\let\PY@bf=\textbf\def\PY@tc##1{\textcolor[rgb]{0.50,0.00,0.50}{##1}}}
\expandafter\def\csname PY@tok@gd\endcsname{\def\PY@tc##1{\textcolor[rgb]{0.63,0.00,0.00}{##1}}}
\expandafter\def\csname PY@tok@gi\endcsname{\def\PY@tc##1{\textcolor[rgb]{0.00,0.63,0.00}{##1}}}
\expandafter\def\csname PY@tok@gr\endcsname{\def\PY@tc##1{\textcolor[rgb]{1.00,0.00,0.00}{##1}}}
\expandafter\def\csname PY@tok@ge\endcsname{\let\PY@it=\textit}
\expandafter\def\csname PY@tok@gs\endcsname{\let\PY@bf=\textbf}
\expandafter\def\csname PY@tok@gp\endcsname{\let\PY@bf=\textbf\def\PY@tc##1{\textcolor[rgb]{0.00,0.00,0.50}{##1}}}
\expandafter\def\csname PY@tok@go\endcsname{\def\PY@tc##1{\textcolor[rgb]{0.53,0.53,0.53}{##1}}}
\expandafter\def\csname PY@tok@gt\endcsname{\def\PY@tc##1{\textcolor[rgb]{0.00,0.27,0.87}{##1}}}
\expandafter\def\csname PY@tok@err\endcsname{\def\PY@bc##1{\setlength{\fboxsep}{0pt}\fcolorbox[rgb]{1.00,0.00,0.00}{1,1,1}{\strut ##1}}}
\expandafter\def\csname PY@tok@kc\endcsname{\let\PY@bf=\textbf\def\PY@tc##1{\textcolor[rgb]{0.00,0.50,0.00}{##1}}}
\expandafter\def\csname PY@tok@kd\endcsname{\let\PY@bf=\textbf\def\PY@tc##1{\textcolor[rgb]{0.00,0.50,0.00}{##1}}}
\expandafter\def\csname PY@tok@kn\endcsname{\let\PY@bf=\textbf\def\PY@tc##1{\textcolor[rgb]{0.00,0.50,0.00}{##1}}}
\expandafter\def\csname PY@tok@kr\endcsname{\let\PY@bf=\textbf\def\PY@tc##1{\textcolor[rgb]{0.00,0.50,0.00}{##1}}}
\expandafter\def\csname PY@tok@bp\endcsname{\def\PY@tc##1{\textcolor[rgb]{0.00,0.50,0.00}{##1}}}
\expandafter\def\csname PY@tok@fm\endcsname{\def\PY@tc##1{\textcolor[rgb]{0.00,0.00,1.00}{##1}}}
\expandafter\def\csname PY@tok@vc\endcsname{\def\PY@tc##1{\textcolor[rgb]{0.10,0.09,0.49}{##1}}}
\expandafter\def\csname PY@tok@vg\endcsname{\def\PY@tc##1{\textcolor[rgb]{0.10,0.09,0.49}{##1}}}
\expandafter\def\csname PY@tok@vi\endcsname{\def\PY@tc##1{\textcolor[rgb]{0.10,0.09,0.49}{##1}}}
\expandafter\def\csname PY@tok@vm\endcsname{\def\PY@tc##1{\textcolor[rgb]{0.10,0.09,0.49}{##1}}}
\expandafter\def\csname PY@tok@sa\endcsname{\def\PY@tc##1{\textcolor[rgb]{0.73,0.13,0.13}{##1}}}
\expandafter\def\csname PY@tok@sb\endcsname{\def\PY@tc##1{\textcolor[rgb]{0.73,0.13,0.13}{##1}}}
\expandafter\def\csname PY@tok@sc\endcsname{\def\PY@tc##1{\textcolor[rgb]{0.73,0.13,0.13}{##1}}}
\expandafter\def\csname PY@tok@dl\endcsname{\def\PY@tc##1{\textcolor[rgb]{0.73,0.13,0.13}{##1}}}
\expandafter\def\csname PY@tok@s2\endcsname{\def\PY@tc##1{\textcolor[rgb]{0.73,0.13,0.13}{##1}}}
\expandafter\def\csname PY@tok@sh\endcsname{\def\PY@tc##1{\textcolor[rgb]{0.73,0.13,0.13}{##1}}}
\expandafter\def\csname PY@tok@s1\endcsname{\def\PY@tc##1{\textcolor[rgb]{0.73,0.13,0.13}{##1}}}
\expandafter\def\csname PY@tok@mb\endcsname{\def\PY@tc##1{\textcolor[rgb]{0.40,0.40,0.40}{##1}}}
\expandafter\def\csname PY@tok@mf\endcsname{\def\PY@tc##1{\textcolor[rgb]{0.40,0.40,0.40}{##1}}}
\expandafter\def\csname PY@tok@mh\endcsname{\def\PY@tc##1{\textcolor[rgb]{0.40,0.40,0.40}{##1}}}
\expandafter\def\csname PY@tok@mi\endcsname{\def\PY@tc##1{\textcolor[rgb]{0.40,0.40,0.40}{##1}}}
\expandafter\def\csname PY@tok@il\endcsname{\def\PY@tc##1{\textcolor[rgb]{0.40,0.40,0.40}{##1}}}
\expandafter\def\csname PY@tok@mo\endcsname{\def\PY@tc##1{\textcolor[rgb]{0.40,0.40,0.40}{##1}}}
\expandafter\def\csname PY@tok@ch\endcsname{\let\PY@it=\textit\def\PY@tc##1{\textcolor[rgb]{0.25,0.50,0.50}{##1}}}
\expandafter\def\csname PY@tok@cm\endcsname{\let\PY@it=\textit\def\PY@tc##1{\textcolor[rgb]{0.25,0.50,0.50}{##1}}}
\expandafter\def\csname PY@tok@cpf\endcsname{\let\PY@it=\textit\def\PY@tc##1{\textcolor[rgb]{0.25,0.50,0.50}{##1}}}
\expandafter\def\csname PY@tok@c1\endcsname{\let\PY@it=\textit\def\PY@tc##1{\textcolor[rgb]{0.25,0.50,0.50}{##1}}}
\expandafter\def\csname PY@tok@cs\endcsname{\let\PY@it=\textit\def\PY@tc##1{\textcolor[rgb]{0.25,0.50,0.50}{##1}}}

\def\PYZbs{\char`\\}
\def\PYZus{\char`\_}
\def\PYZob{\char`\{}
\def\PYZcb{\char`\}}
\def\PYZca{\char`\^}

\def\PYZsh{\char`\#}

\def\PYZdl{\char`\$}
\def\PYZhy{\char`\-}

\def\PYZdq{\char`\"}

\makeatother

    \makeatletter
        \newbox\Wrappedcontinuationbox 
        \newbox\Wrappedvisiblespacebox 
        \newcommand*\Wrappedvisiblespace {\textcolor{red}{\textvisiblespace}} 
        \newcommand*\Wrappedcontinuationsymbol {\textcolor{red}{\llap{\tiny$\m@th\hookrightarrow$}}} 
        \newcommand*\Wrappedcontinuationindent {3ex } 
        \newcommand*\Wrappedafterbreak {\kern\Wrappedcontinuationindent\copy\Wrappedcontinuationbox} 
        \newcommand*\Wrappedbreaksatspecials {%
            \def\PYGZus{\discretionary{\char`\_}{\Wrappedafterbreak}{\char`\_}}%
            \def\PYGZob{\discretionary{}{\Wrappedafterbreak\char`\{}{\char`\{}}%
            \def\PYGZcb{\discretionary{\char`\}}{\Wrappedafterbreak}{\char`\}}}%
            \def\PYGZca{\discretionary{\char`\^}{\Wrappedafterbreak}{\char`\^}}%
            \def\PYGZam{\discretionary{\char`\&}{\Wrappedafterbreak}{\char`\&}}%
            \def\PYGZlt{\discretionary{}{\Wrappedafterbreak\char`\<}{\char`\<}}%
            \def\PYGZgt{\discretionary{\char`\>}{\Wrappedafterbreak}{\char`\>}}%
            \def\PYGZsh{\discretionary{}{\Wrappedafterbreak\char`\#}{\char`\#}}%
            \def\PYGZpc{\discretionary{}{\Wrappedafterbreak\char`\%}{\char`\%}}%
            \def\PYGZdl{\discretionary{}{\Wrappedafterbreak\char`\$}{\char`\$}}%
            \def\PYGZhy{\discretionary{\char`\-}{\Wrappedafterbreak}{\char`\-}}%
            \def\PYGZsq{\discretionary{}{\Wrappedafterbreak\textquotesingle}{\textquotesingle}}%
            \def\PYGZdq{\discretionary{}{\Wrappedafterbreak\char`\"}{\char`\"}}%
            \def\PYGZti{\discretionary{\char`\~}{\Wrappedafterbreak}{\char`\~}}%
        } 
        \newcommand*\Wrappedbreaksatpunct {%
            \lccode`\~`\.\lowercase{\def~}{\discretionary{\hbox{\char`\.}}{\Wrappedafterbreak}{\hbox{\char`\.}}}%
            \lccode`\~`\,\lowercase{\def~}{\discretionary{\hbox{\char`\,}}{\Wrappedafterbreak}{\hbox{\char`\,}}}%
            \lccode`\~`\;\lowercase{\def~}{\discretionary{\hbox{\char`\;}}{\Wrappedafterbreak}{\hbox{\char`\;}}}%
            \lccode`\~`\:\lowercase{\def~}{\discretionary{\hbox{\char`\:}}{\Wrappedafterbreak}{\hbox{\char`\:}}}%
            \lccode`\~`\?\lowercase{\def~}{\discretionary{\hbox{\char`\?}}{\Wrappedafterbreak}{\hbox{\char`\?}}}%
            \lccode`\~`\!\lowercase{\def~}{\discretionary{\hbox{\char`\!}}{\Wrappedafterbreak}{\hbox{\char`\!}}}%
            \lccode`\~`\/\lowercase{\def~}{\discretionary{\hbox{\char`\/}}{\Wrappedafterbreak}{\hbox{\char`\/}}}%
            \catcode`\.\active
            \catcode`\,\active 
            \catcode`\;\active
            \catcode`\:\active
            \catcode`\?\active
            \catcode`\!\active
            \catcode`\/\active 
            \lccode`\~`\~ 	
        }
    \makeatother

    \let\OriginalVerbatim=\Verbatim
    \makeatletter
    \renewcommand{\Verbatim}[1][1]{%
        \sbox\Wrappedcontinuationbox {\Wrappedcontinuationsymbol}%
        \sbox\Wrappedvisiblespacebox {\FV@SetupFont\Wrappedvisiblespace}%
        \def\FancyVerbFormatLine ##1{\hsize\linewidth
            \vtop{\raggedright\hyphenpenalty\z@\exhyphenpenalty\z@
                \doublehyphendemerits\z@\finalhyphendemerits\z@
                \strut ##1\strut}%
        }%
        \def\FV@Space {%
            \nobreak\hskip\z@ plus\fontdimen3\font minus\fontdimen4\font
            \discretionary{\copy\Wrappedvisiblespacebox}{\Wrappedafterbreak}
            {\kern\fontdimen2\font}%
        }%
        
        \Wrappedbreaksatspecials
        \OriginalVerbatim[#1,codes*=\Wrappedbreaksatpunct]%
    }
    \makeatother

    \definecolor{incolor}{HTML}{303F9F}
    \definecolor{outcolor}{HTML}{D84315}
    \definecolor{cellborder}{HTML}{CFCFCF}
    \definecolor{cellbackground}{HTML}{F7F7F7}
    
    \makeatletter
    \newcommand{\boxspacing}{\kern\kvtcb@left@rule\kern\kvtcb@boxsep}
    \makeatother
    \newcommand{\prompt}[4]{
        \ttfamily\llap{{\color{#2}[#3]:\hspace{3pt}#4}}\vspace{-\baselineskip}
    }

%% file: main.bbl
\providecommand{\href}[2]{#2}\begingroup\raggedright\begin{thebibliography}{10}

\bibitem{Umehara:2008ru}
S.~Umehara {\em et~al.}, ``{Neutrino-less double-beta decay of Ca-48 studied by
  Ca F(2)(Eu) scintillators},''
  \href{http://dx.doi.org/10.1103/PhysRevC.78.058501}{{\em Phys. Rev. C}
  {\bfseries 78} (2008) 058501},
  \href{http://arxiv.org/abs/0810.4746}{{\ttfamily arXiv:0810.4746 [nucl-ex]}}.

\bibitem{GERDA:2020xhi}
{\bfseries GERDA} Collaboration, M.~Agostini {\em et~al.}, ``{Final Results of
  GERDA on the Search for Neutrinoless Double-$\beta$ Decay},''
  \href{http://dx.doi.org/10.1103/PhysRevLett.125.252502}{{\em Phys. Rev.
  Lett.} {\bfseries 125} no.~25, (2020) 252502},
  \href{http://arxiv.org/abs/2009.06079}{{\ttfamily arXiv:2009.06079
  [nucl-ex]}}.

\bibitem{CUPID-0:2018rcs}
{\bfseries CUPID-0} Collaboration, O.~Azzolini {\em et~al.}, ``{First Result on
  the Neutrinoless Double-$\beta$ Decay of $^{82}Se$ with CUPID-0},''
  \href{http://dx.doi.org/10.1103/PhysRevLett.120.232502}{{\em Phys. Rev.
  Lett.} {\bfseries 120} no.~23, (2018) 232502},
  \href{http://arxiv.org/abs/1802.07791}{{\ttfamily arXiv:1802.07791
  [nucl-ex]}}.

\bibitem{NEMO-3:2009fxe}
{\bfseries NEMO-3} Collaboration, J.~Argyriades {\em et~al.}, ``{Measurement of
  the two neutrino double beta decay half-life of Zr-96 with the NEMO-3
  detector},'' \href{http://dx.doi.org/10.1016/j.nuclphysa.2010.07.009}{{\em
  Nucl. Phys. A} {\bfseries 847} (2010) 168--179},
  \href{http://arxiv.org/abs/0906.2694}{{\ttfamily arXiv:0906.2694 [nucl-ex]}}.

\bibitem{CUPID:2020aow}
{\bfseries CUPID} Collaboration, E.~Armengaud {\em et~al.}, ``{New Limit for
  Neutrinoless Double-Beta Decay of $^{100}$Mo from the CUPID-Mo Experiment},''
  \href{http://dx.doi.org/10.1103/PhysRevLett.126.181802}{{\em Phys. Rev.
  Lett.} {\bfseries 126} no.~18, (2021) 181802},
  \href{http://arxiv.org/abs/2011.13243}{{\ttfamily arXiv:2011.13243
  [nucl-ex]}}.

\bibitem{Danevich:2016eot}
F.~A. Danevich {\em et~al.}, ``{Search for double beta decay of $^{116}$Cd with
  enriched $^{116}$CdWO$_4$ crystal scintillators (Aurora experiment)},''
  \href{http://dx.doi.org/10.1088/1742-6596/718/6/062009}{{\em J. Phys. Conf.
  Ser.} {\bfseries 718} no.~6, (2016) 062009},
  \href{http://arxiv.org/abs/1601.05578}{{\ttfamily arXiv:1601.05578
  [nucl-ex]}}.

\bibitem{Arnaboldi:2002te}
C.~Arnaboldi {\em et~al.}, ``{A Calorimetric search on double beta decay of
  Te-130},'' \href{http://dx.doi.org/10.1016/S0370-2693(03)00212-0}{{\em Phys.
  Lett. B} {\bfseries 557} (2003) 167--175},
  \href{http://arxiv.org/abs/hep-ex/0211071}{{\ttfamily arXiv:hep-ex/0211071}}.

\bibitem{CUORE:2019yfd}
{\bfseries CUORE} Collaboration, D.~Q. Adams {\em et~al.}, ``{Improved Limit on
  Neutrinoless Double-Beta Decay in $^{130}$Te with CUORE},''
  \href{http://dx.doi.org/10.1103/PhysRevLett.124.122501}{{\em Phys. Rev.
  Lett.} {\bfseries 124} no.~12, (2020) 122501},
  \href{http://arxiv.org/abs/1912.10966}{{\ttfamily arXiv:1912.10966
  [nucl-ex]}}.

\bibitem{EXO-200:2017vqi}
{\bfseries EXO-200} Collaboration, J.~B. Albert {\em et~al.}, ``{Searches for
  double beta decay of $^{134}$Xe with EXO-200},''
  \href{http://dx.doi.org/10.1103/PhysRevD.96.092001}{{\em Phys. Rev. D}
  {\bfseries 96} no.~9, (2017) 092001},
  \href{http://arxiv.org/abs/1704.05042}{{\ttfamily arXiv:1704.05042
  [hep-ex]}}.

\bibitem{KamLAND-Zen:2022tow}
{\bfseries KamLAND-Zen} Collaboration, S.~Abe {\em et~al.}, ``{First Search for
  the Majorana Nature of Neutrinos in the Inverted Mass Ordering Region with
  KamLAND-Zen},'' \href{http://arxiv.org/abs/2203.02139}{{\ttfamily
  arXiv:2203.02139 [hep-ex]}}.

\bibitem{NEMO:2008kpp}
{\bfseries NEMO} Collaboration, J.~Argyriades {\em et~al.}, ``{Measurement of
  the Double Beta Decay Half-life of Nd-150 and Search for Neutrinoless Decay
  Modes with the NEMO-3 Detector},''
  \href{http://dx.doi.org/10.1103/PhysRevC.80.032501}{{\em Phys. Rev. C}
  {\bfseries 80} (2009) 032501},
  \href{http://arxiv.org/abs/0810.0248}{{\ttfamily arXiv:0810.0248 [hep-ex]}}.

\bibitem{2020CupidProspects}
E.~Armengaud {\em et~al.}, ``The cupid-mo experiment for neutrinoless
  double-beta decay: performance and prospects,''
  \href{http://dx.doi.org/10.1140/epjc/s10052-019-7578-6}{{\em The European
  Physical Journal C} {\bfseries 80} no.~1, (Jan, 2020) }.
  \url{http://dx.doi.org/10.1140/epjc/s10052-019-7578-6}.

\bibitem{2017LEGEND200}
{\bfseries LEGEND} Collaboration, N.~Abgrall {\em et~al.}, ``{The Large
  Enriched Germanium Experiment for Neutrinoless Double Beta Decay (LEGEND)},''
  \href{http://dx.doi.org/10.1063/1.5007652}{{\em AIP Conf. Proc.} {\bfseries
  1894} no.~1, (2017) 020027},
  \href{http://arxiv.org/abs/1709.01980}{{\ttfamily arXiv:1709.01980
  [physics.ins-det]}}.

\bibitem{legendcollaboration2021legend1000}
{\bfseries LEGEND} Collaboration, N.~Abgrall {\em et~al.}, ``{The Large
  Enriched Germanium Experiment for Neutrinoless $\beta\beta$ Decay}:
  {LEGEND-1000 Preconceptual Design Report},''
  \href{http://arxiv.org/abs/2107.11462}{{\ttfamily arXiv:2107.11462
  [physics.ins-det]}}.

\bibitem{nEXO:2021ujk}
{\bfseries nEXO} Collaboration, G.~Adhikari {\em et~al.}, ``{nEXO: neutrinoless
  double beta decay search beyond 10$^{28}$ year half-life sensitivity},''
  \href{http://dx.doi.org/10.1088/1361-6471/ac3631}{{\em J. Phys. G} {\bfseries
  49} no.~1, (2022) 015104}, \href{http://arxiv.org/abs/2106.16243}{{\ttfamily
  arXiv:2106.16243 [nucl-ex]}}.

\bibitem{SNO:2021xpa}
{\bfseries SNO+} Collaboration, V.~Albanese {\em et~al.}, ``{The SNO+
  experiment},'' \href{http://dx.doi.org/10.1088/1748-0221/16/08/P08059}{{\em
  JINST} {\bfseries 16} no.~08, (2021) P08059},
  \href{http://arxiv.org/abs/2104.11687}{{\ttfamily arXiv:2104.11687
  [physics.ins-det]}}.

\bibitem{PhysRevD.25.2951}
J.~Schechter and J.~W.~F. Valle, ``Neutrinoless double-$\ensuremath{\beta}$
  decay in su(2)$\ifmmode\times\else\texttimes\fi{}$u(1) theories,''
  \href{http://dx.doi.org/10.1103/PhysRevD.25.2951}{{\em Phys. Rev. D}
  {\bfseries 25} (Jun, 1982) 2951--2954}.
  \url{https://link.aps.org/doi/10.1103/PhysRevD.25.2951}.

\bibitem{2019review_Werner}
M.~J. Dolinski, A.~W. Poon, and W.~Rodejohann, ``Neutrinoless double-beta
  decay: Status and prospects,''
  \href{http://dx.doi.org/10.1146/annurev-nucl-101918-023407}{{\em Annual
  Review of Nuclear and Particle Science} {\bfseries 69} no.~1, (Oct, 2019)
  219–251}. \url{http://dx.doi.org/10.1146/annurev-nucl-101918-023407}.

\bibitem{Agostini:2022zub}
M.~Agostini, G.~Benato, J.~A. Detwiler, J.~Men\'endez, and F.~Vissani,
  ``{Toward the discovery of matter creation with neutrinoless double-beta
  decay},'' \href{http://arxiv.org/abs/2202.01787}{{\ttfamily arXiv:2202.01787
  [hep-ex]}}.

\bibitem{Cirigliano:2022oqy}
V.~Cirigliano {\em et~al.}, ``{Neutrinoless Double-Beta Decay: A Roadmap for
  Matching Theory to Experiment},''
  \href{http://arxiv.org/abs/2203.12169}{{\ttfamily arXiv:2203.12169
  [hep-ph]}}.

\bibitem{Deppisch:2012nb}
F.~F. Deppisch, M.~Hirsch, and H.~Päs, ``{Neutrinoless Double Beta Decay and
  Physics Beyond the Standard Model},''
  \href{http://dx.doi.org/10.1088/0954-3899/39/12/124007}{{\em J. Phys. G}
  {\bfseries 39} (2012) 124007},
  \href{http://arxiv.org/abs/1208.0727}{{\ttfamily arXiv:1208.0727 [hep-ph]}}.

\bibitem{Li:2020flq}
G.~Li, M.~Ramsey-Musolf, and J.~C. Vasquez, ``{Left-Right Symmetry and Leading
  Contributions to Neutrinoless Double Beta Decay},''
  \href{http://dx.doi.org/10.1103/PhysRevLett.126.151801}{{\em Phys. Rev.
  Lett.} {\bfseries 126} no.~15, (2021) 151801},
  \href{http://arxiv.org/abs/2009.01257}{{\ttfamily arXiv:2009.01257
  [hep-ph]}}.

\bibitem{Deppisch:2006hb}
F.~Deppisch and H.~Pas, ``{Pinning down the mechanism of neutrinoless double
  beta decay with measurements in different nuclei},''
  \href{http://dx.doi.org/10.1103/PhysRevLett.98.232501}{{\em Phys. Rev. Lett.}
  {\bfseries 98} (2007) 232501},
  \href{http://arxiv.org/abs/hep-ph/0612165}{{\ttfamily arXiv:hep-ph/0612165}}.

\bibitem{Gehman:2007qg}
V.~M. Gehman and S.~R. Elliott, ``{Multiple-Isotope Comparison for Determining
  0 nu beta beta Mechanisms},''
  \href{http://dx.doi.org/10.1088/0954-3899/34/4/006}{{\em J. Phys. G}
  {\bfseries 34} (2007) 667--678},
  \href{http://arxiv.org/abs/hep-ph/0701099}{{\ttfamily arXiv:hep-ph/0701099}}.
  [Erratum: J.Phys.G 35, 029701 (2008)].

\bibitem{Graf:2022lhj}
L.~Gr\'af, M.~Lindner, and O.~Scholer, ``{Unraveling the
  0\ensuremath{\nu}\ensuremath{\beta}\ensuremath{\beta} decay mechanisms},''
  \href{http://dx.doi.org/10.1103/PhysRevD.106.035022}{{\em Phys. Rev. D}
  {\bfseries 106} no.~3, (2022) 035022},
  \href{http://arxiv.org/abs/2204.10845}{{\ttfamily arXiv:2204.10845
  [hep-ph]}}.

\bibitem{Agostini:2022bjh}
M.~Agostini, F.~F. Deppisch, and G.~Van~Goffrier, ``{Probing the Mechanism of
  Neutrinoless Double-Beta Decay in Multiple Isotopes},''
  \href{http://arxiv.org/abs/2212.00045}{{\ttfamily arXiv:2212.00045
  [hep-ph]}}.

\bibitem{Pas:1999fc}
H.~Pas, M.~Hirsch, H.~Klapdor-Kleingrothaus, and S.~Kovalenko, ``{Towards a
  superformula for neutrinoless double beta decay},''
  \href{http://dx.doi.org/10.1016/S0370-2693(99)00330-5}{{\em Phys. Lett. B}
  {\bfseries 453} (1999) 194--198}.

\bibitem{Pas:2000vn}
H.~Pas, M.~Hirsch, H.~V. Klapdor-Kleingrothaus, and S.~G. Kovalenko, ``{A
  Superformula for neutrinoless double beta decay. 2. The Short range part},''
  \href{http://dx.doi.org/10.1016/S0370-2693(00)01359-9}{{\em Phys. Lett. B}
  {\bfseries 498} (2001) 35--39},
  \href{http://arxiv.org/abs/hep-ph/0008182}{{\ttfamily arXiv:hep-ph/0008182}}.

\bibitem{Prezeau:2003xn}
G.~Prezeau, M.~Ramsey-Musolf, and P.~Vogel, ``{Neutrinoless double beta decay
  and effective field theory},''
  \href{http://dx.doi.org/10.1103/PhysRevD.68.034016}{{\em Phys. Rev. D}
  {\bfseries 68} (2003) 034016},
  \href{http://arxiv.org/abs/hep-ph/0303205}{{\ttfamily arXiv:hep-ph/0303205}}.

\bibitem{Graf:2018ozy}
L.~Graf, F.~F. Deppisch, F.~Iachello, and J.~Kotila, ``{Short-Range
  Neutrinoless Double Beta Decay Mechanisms},''
  \href{http://dx.doi.org/10.1103/PhysRevD.98.095023}{{\em Phys. Rev. D}
  {\bfseries 98} no.~9, (2018) 095023},
  \href{http://arxiv.org/abs/1806.06058}{{\ttfamily arXiv:1806.06058
  [hep-ph]}}.

\bibitem{Cirigliano:2017djv}
V.~Cirigliano, W.~Dekens, J.~de~Vries, M.~L. Graesser, and E.~Mereghetti,
  ``{Neutrinoless double beta decay in chiral effective field theory: lepton
  number violation at dimension seven},''
  \href{http://dx.doi.org/10.1007/JHEP12(2017)082}{{\em JHEP} {\bfseries 12}
  (2017) 082}, \href{http://arxiv.org/abs/1708.09390}{{\ttfamily
  arXiv:1708.09390 [hep-ph]}}.

\bibitem{Cirigliano_2018}
V.~Cirigliano, W.~Dekens, J.~de~Vries, M.~L. Graesser, and E.~Mereghetti, ``A
  neutrinoless double beta decay master formula from effective field theory,''
  \href{http://dx.doi.org/10.1007/jhep12(2018)097}{{\em Journal of High Energy
  Physics} {\bfseries 2018} no.~12, (Dec, 2018) }.
  \url{http://dx.doi.org/10.1007/JHEP12(2018)097}.

\bibitem{Dekens:2020ttz}
W.~Dekens, J.~de~Vries, K.~Fuyuto, E.~Mereghetti, and G.~Zhou, ``{Sterile
  neutrinos and neutrinoless double beta decay in effective field theory},''
  \href{http://dx.doi.org/10.1007/JHEP06(2020)097}{{\em JHEP} {\bfseries 06}
  (2020) 097}, \href{http://arxiv.org/abs/2002.07182}{{\ttfamily
  arXiv:2002.07182 [hep-ph]}}.

\bibitem{Kobach:2016ami}
A.~Kobach, ``{Baryon Number, Lepton Number, and Operator Dimension in the
  Standard Model},''
  \href{http://dx.doi.org/10.1016/j.physletb.2016.05.050}{{\em Phys. Lett. B}
  {\bfseries 758} (2016) 455--457},
  \href{http://arxiv.org/abs/1604.05726}{{\ttfamily arXiv:1604.05726
  [hep-ph]}}.

\bibitem{Weinberg:1979sa}
S.~Weinberg, ``{Baryon and Lepton Nonconserving Processes},''
  \href{http://dx.doi.org/10.1103/PhysRevLett.43.1566}{{\em Phys. Rev. Lett.}
  {\bfseries 43} (1979) 1566--1570}.

\bibitem{Lehman:2014jma}
L.~Lehman, ``{Extending the Standard Model Effective Field Theory with the
  Complete Set of Dimension-7 Operators},''
  \href{http://dx.doi.org/10.1103/PhysRevD.90.125023}{{\em Phys. Rev. D}
  {\bfseries 90} no.~12, (2014) 125023},
  \href{http://arxiv.org/abs/1410.4193}{{\ttfamily arXiv:1410.4193 [hep-ph]}}.

\bibitem{Liao:2020jmn}
Y.~Liao and X.-D. Ma, ``{An explicit construction of the dimension-9 operator
  basis in the standard model effective field theory},''
  \href{http://dx.doi.org/10.1007/JHEP11(2020)152}{{\em JHEP} {\bfseries 11}
  (2020) 152}, \href{http://arxiv.org/abs/2007.08125}{{\ttfamily
  arXiv:2007.08125 [hep-ph]}}.

\bibitem{Cirigliano:2018hja}
V.~Cirigliano, W.~Dekens, J.~De~Vries, M.~L. Graesser, E.~Mereghetti,
  S.~Pastore, and U.~Van~Kolck, ``{New Leading Contribution to Neutrinoless
  Double-\ensuremath{\beta} Decay},''
  \href{http://dx.doi.org/10.1103/PhysRevLett.120.202001}{{\em Phys. Rev.
  Lett.} {\bfseries 120} no.~20, (2018) 202001},
  \href{http://arxiv.org/abs/1802.10097}{{\ttfamily arXiv:1802.10097
  [hep-ph]}}.

\bibitem{Cirigliano:2019vdj}
V.~Cirigliano, W.~Dekens, J.~De~Vries, M.~L. Graesser, E.~Mereghetti,
  S.~Pastore, M.~Piarulli, U.~Van~Kolck, and R.~B. Wiringa, ``{Renormalized
  approach to neutrinoless double- $\beta$ decay},''
  \href{http://dx.doi.org/10.1103/PhysRevC.100.055504}{{\em Phys. Rev. C}
  {\bfseries 100} no.~5, (2019) 055504},
  \href{http://arxiv.org/abs/1907.11254}{{\ttfamily arXiv:1907.11254
  [nucl-th]}}.

\bibitem{Grzadkowski:2010es}
B.~Grzadkowski, M.~Iskrzynski, M.~Misiak, and J.~Rosiek, ``{Dimension-Six Terms
  in the Standard Model Lagrangian},''
  \href{http://dx.doi.org/10.1007/JHEP10(2010)085}{{\em JHEP} {\bfseries 10}
  (2010) 085}, \href{http://arxiv.org/abs/1008.4884}{{\ttfamily arXiv:1008.4884
  [hep-ph]}}.

\bibitem{Henning:2015alf}
B.~Henning, X.~Lu, T.~Melia, and H.~Murayama, ``{2, 84, 30, 993, 560, 15456,
  11962, 261485, ...: Higher dimension operators in the SM EFT},''
  \href{http://dx.doi.org/10.1007/JHEP08(2017)016}{{\em JHEP} {\bfseries 08}
  (2017) 016}, \href{http://arxiv.org/abs/1512.03433}{{\ttfamily
  arXiv:1512.03433 [hep-ph]}}. [Erratum: JHEP 09, 019 (2019)].

\bibitem{Liao:2019tep}
Y.~Liao and X.-D. Ma, ``{Renormalization Group Evolution of Dimension-seven
  Operators in Standard Model Effective Field Theory and Relevant
  Phenomenology},'' \href{http://dx.doi.org/10.1007/JHEP03(2019)179}{{\em JHEP}
  {\bfseries 03} (2019) 179}, \href{http://arxiv.org/abs/1901.10302}{{\ttfamily
  arXiv:1901.10302 [hep-ph]}}.

\bibitem{Li:2020gnx}
H.-L. Li, Z.~Ren, J.~Shu, M.-L. Xiao, J.-H. Yu, and Y.-H. Zheng, ``{Complete
  set of dimension-eight operators in the standard model effective field
  theory},'' \href{http://dx.doi.org/10.1103/PhysRevD.104.015026}{{\em Phys.
  Rev. D} {\bfseries 104} no.~1, (2021) 015026},
  \href{http://arxiv.org/abs/2005.00008}{{\ttfamily arXiv:2005.00008
  [hep-ph]}}.

\bibitem{Murphy:2020rsh}
C.~W. Murphy, ``{Dimension-8 operators in the Standard Model Eective Field
  Theory},'' \href{http://dx.doi.org/10.1007/JHEP10(2020)174}{{\em JHEP}
  {\bfseries 10} (2020) 174}, \href{http://arxiv.org/abs/2005.00059}{{\ttfamily
  arXiv:2005.00059 [hep-ph]}}.

\bibitem{Li:2020xlh}
H.-L. Li, Z.~Ren, M.-L. Xiao, J.-H. Yu, and Y.-H. Zheng, ``{Complete set of
  dimension-nine operators in the standard model effective field theory},''
  \href{http://dx.doi.org/10.1103/PhysRevD.104.015025}{{\em Phys. Rev. D}
  {\bfseries 104} no.~1, (2021) 015025},
  \href{http://arxiv.org/abs/2007.07899}{{\ttfamily arXiv:2007.07899
  [hep-ph]}}.

\bibitem{Jenkins:2017jig}
E.~E. Jenkins, A.~V. Manohar, and P.~Stoffer, ``{Low-Energy Effective Field
  Theory below the Electroweak Scale: Operators and Matching},''
  \href{http://dx.doi.org/10.1007/JHEP03(2018)016}{{\em JHEP} {\bfseries 03}
  (2018) 016}, \href{http://arxiv.org/abs/1709.04486}{{\ttfamily
  arXiv:1709.04486 [hep-ph]}}.

\bibitem{Jenkins:2017dyc}
E.~E. Jenkins, A.~V. Manohar, and P.~Stoffer, ``{Low-Energy Effective Field
  Theory below the Electroweak Scale: Anomalous Dimensions},''
  \href{http://dx.doi.org/10.1007/JHEP01(2018)084}{{\em JHEP} {\bfseries 01}
  (2018) 084}, \href{http://arxiv.org/abs/1711.05270}{{\ttfamily
  arXiv:1711.05270 [hep-ph]}}.

\bibitem{Dekens:2019ept}
W.~Dekens and P.~Stoffer, ``{Low-energy effective field theory below the
  electroweak scale: matching at one loop},''
  \href{http://dx.doi.org/10.1007/JHEP10(2019)197}{{\em JHEP} {\bfseries 10}
  (2019) 197}, \href{http://arxiv.org/abs/1908.05295}{{\ttfamily
  arXiv:1908.05295 [hep-ph]}}. [Erratum: JHEP 11, 148 (2022)].

\bibitem{Liao:2020zyx}
Y.~Liao, X.-D. Ma, and Q.-Y. Wang, ``{Extending low energy effective field
  theory with a complete set of dimension-7 operators},''
  \href{http://dx.doi.org/10.1007/JHEP08(2020)162}{{\em JHEP} {\bfseries 08}
  (2020) 162}, \href{http://arxiv.org/abs/2005.08013}{{\ttfamily
  arXiv:2005.08013 [hep-ph]}}.

\bibitem{Li:2020tsi}
H.-L. Li, Z.~Ren, M.-L. Xiao, J.-H. Yu, and Y.-H. Zheng, ``{Low energy
  effective field theory operator basis at d \ensuremath{\leq} 9},''
  \href{http://dx.doi.org/10.1007/JHEP06(2021)138}{{\em JHEP} {\bfseries 06}
  (2021) 138}, \href{http://arxiv.org/abs/2012.09188}{{\ttfamily
  arXiv:2012.09188 [hep-ph]}}.

\bibitem{Hyvarinen:2015bda}
J.~Hyv\"arinen and J.~Suhonen, ``{Nuclear matrix elements for $0\nu\beta\beta$
  decays with light or heavy Majorana-neutrino exchange},''
  \href{http://dx.doi.org/10.1103/PhysRevC.91.024613}{{\em Phys. Rev. C}
  {\bfseries 91} no.~2, (2015) 024613}.

\bibitem{Terasaki:2020ndc}
J.~Terasaki, ``{Strength of the isoscalar pairing interaction determined by a
  relation between double-charge change and double-pair transfer for double-
  $\beta$ decay},'' \href{http://dx.doi.org/10.1103/PhysRevC.102.044303}{{\em
  Phys. Rev. C} {\bfseries 102} no.~4, (2020) 044303},
  \href{http://arxiv.org/abs/2003.03542}{{\ttfamily arXiv:2003.03542
  [nucl-th]}}.

\bibitem{Simkovic:2013qiy}
F.~\v{S}imkovic, V.~Rodin, A.~Faessler, and P.~Vogel,
  ``{0\ensuremath{\nu}\ensuremath{\beta}\ensuremath{\beta} and
  2\ensuremath{\nu}\ensuremath{\beta}\ensuremath{\beta} nuclear matrix
  elements, quasiparticle random-phase approximation, and isospin symmetry
  restoration},'' \href{http://dx.doi.org/10.1103/PhysRevC.87.045501}{{\em
  Phys. Rev. C} {\bfseries 87} no.~4, (2013) 045501},
  \href{http://arxiv.org/abs/1302.1509}{{\ttfamily arXiv:1302.1509 [nucl-th]}}.

\bibitem{Mustonen:2013zu}
M.~T. Mustonen and J.~Engel, ``{Large-scale calculations of the
  double-\ensuremath{\beta} decay of $^{76}Ge,^{130}Te,^{136}Xe$, and
  $^{150}Nd$ in the deformed self-consistent Skyrme quasiparticle random-phase
  approximation},'' \href{http://dx.doi.org/10.1103/PhysRevC.87.064302}{{\em
  Phys. Rev. C} {\bfseries 87} no.~6, (2013) 064302},
  \href{http://arxiv.org/abs/1301.6997}{{\ttfamily arXiv:1301.6997 [nucl-th]}}.

\bibitem{Fang:2018tui}
D.-L. Fang, A.~Faessler, and F.~Simkovic, ``{0\ensuremath{\nu}$\beta\beta$
  -decay nuclear matrix element for light and heavy neutrino mass mechanisms
  from deformed quasiparticle random-phase approximation calculations for
  $^{76}$Ge, $^{82}$Se, $^{130}$Te, $^{136}$Xe , and $^{150}$Nd with isospin
  restoration},'' \href{http://dx.doi.org/10.1103/PhysRevC.97.045503}{{\em
  Phys. Rev. C} {\bfseries 97} no.~4, (2018) 045503},
  \href{http://arxiv.org/abs/1803.09195}{{\ttfamily arXiv:1803.09195
  [nucl-th]}}.

\bibitem{LopezVaquero:2013yji}
N.~L\'opez~Vaquero, T.~R. Rodr\'\i{}guez, and J.~L. Egido, ``{Shape and pairing
  fluctuations effects on neutrinoless double beta decay nuclear matrix
  elements},'' \href{http://dx.doi.org/10.1103/PhysRevLett.111.142501}{{\em
  Phys. Rev. Lett.} {\bfseries 111} no.~14, (2013) 142501},
  \href{http://arxiv.org/abs/1401.0650}{{\ttfamily arXiv:1401.0650 [nucl-th]}}.

\bibitem{Yao:2014uta}
J.~M. Yao, L.~S. Song, K.~Hagino, P.~Ring, and J.~Meng, ``{Systematic study of
  nuclear matrix elements in neutrinoless double-$\beta$ decay with a
  beyond-mean-field covariant density functional theory},''
  \href{http://dx.doi.org/10.1103/PhysRevC.91.024316}{{\em Phys. Rev. C}
  {\bfseries 91} no.~2, (2015) 024316},
  \href{http://arxiv.org/abs/1410.6326}{{\ttfamily arXiv:1410.6326 [nucl-th]}}.

\bibitem{Rodriguez:2010mn}
T.~R. Rodriguez and G.~Martinez-Pinedo, ``{Energy density functional study of
  nuclear matrix elements for neutrinoless $\beta\beta$ decay},''
  \href{http://dx.doi.org/10.1103/PhysRevLett.105.252503}{{\em Phys. Rev.
  Lett.} {\bfseries 105} (2010) 252503},
  \href{http://arxiv.org/abs/1008.5260}{{\ttfamily arXiv:1008.5260 [nucl-th]}}.

\bibitem{Deppisch:2020ztt}
F.~F. Deppisch, L.~Graf, F.~Iachello, and J.~Kotila, ``{Analysis of light
  neutrino exchange and short-range mechanisms in $0\nu\beta\beta$ decay},''
  \href{http://dx.doi.org/10.1103/PhysRevD.102.095016}{{\em Phys. Rev. D}
  {\bfseries 102} no.~9, (2020) 095016},
  \href{http://arxiv.org/abs/2009.10119}{{\ttfamily arXiv:2009.10119
  [hep-ph]}}.

\bibitem{Barea:2015kwa}
J.~Barea, J.~Kotila, and F.~Iachello, ``{$0\nu\beta\beta$ and $2\nu\beta\beta$
  nuclear matrix elements in the interacting boson model with isospin
  restoration},'' \href{http://dx.doi.org/10.1103/PhysRevC.91.034304}{{\em
  Phys. Rev. C} {\bfseries 91} no.~3, (2015) 034304},
  \href{http://arxiv.org/abs/1506.08530}{{\ttfamily arXiv:1506.08530
  [nucl-th]}}.

\bibitem{Coraggio:2020hwx}
L.~Coraggio, A.~Gargano, N.~Itaco, R.~Mancino, and F.~Nowacki, ``{Calculation
  of the neutrinoless double-$\beta$ decay matrix element within the realistic
  shell model},'' \href{http://dx.doi.org/10.1103/PhysRevC.101.044315}{{\em
  Phys. Rev. C} {\bfseries 101} no.~4, (2020) 044315},
  \href{http://arxiv.org/abs/2001.00890}{{\ttfamily arXiv:2001.00890
  [nucl-th]}}.

\bibitem{Neacsu:2014bia}
A.~Neacsu and M.~Horoi, ``{Shell model studies of the $^{130}Te$ neutrinoless
  double-beta decay},''
  \href{http://dx.doi.org/10.1103/PhysRevC.91.024309}{{\em Phys. Rev. C}
  {\bfseries 91} (2015) 024309},
  \href{http://arxiv.org/abs/1411.4313}{{\ttfamily arXiv:1411.4313 [nucl-th]}}.

\bibitem{Menendez:2008jp}
J.~Menendez, A.~Poves, E.~Caurier, and F.~Nowacki, ``{Disassembling the Nuclear
  Matrix Elements of the Neutrinoless beta beta Decay},''
  \href{http://dx.doi.org/10.1016/j.nuclphysa.2008.12.005}{{\em Nucl. Phys. A}
  {\bfseries 818} (2009) 139--151},
  \href{http://arxiv.org/abs/0801.3760}{{\ttfamily arXiv:0801.3760 [nucl-th]}}.

\bibitem{Menendez:2017fdf}
J.~Men\'endez, ``{Neutrinoless $\beta\beta$ decay mediated by the exchange of
  light and heavy neutrinos: The role of nuclear structure correlations},''
  \href{http://dx.doi.org/10.1088/1361-6471/aa9bd4}{{\em J. Phys. G} {\bfseries
  45} no.~1, (2018) 014003}, \href{http://arxiv.org/abs/1804.02105}{{\ttfamily
  arXiv:1804.02105 [nucl-th]}}.

\bibitem{Workman:2022ynf}
{\bfseries Particle Data Group} Collaboration, R.~L. Workman and Others,
  ``{Review of Particle Physics},''
  \href{http://dx.doi.org/10.1093/ptep/ptac097}{{\em PTEP} {\bfseries 2022}
  (2022) 083C01}.

\bibitem{Cirigliano:2020dmx}
V.~Cirigliano, W.~Dekens, J.~de~Vries, M.~Hoferichter, and E.~Mereghetti,
  ``{Toward Complete Leading-Order Predictions for Neutrinoless Double $\beta$
  Decay},'' \href{http://dx.doi.org/10.1103/PhysRevLett.126.172002}{{\em Phys.
  Rev. Lett.} {\bfseries 126} no.~17, (2021) 172002},
  \href{http://arxiv.org/abs/2012.11602}{{\ttfamily arXiv:2012.11602
  [nucl-th]}}.

\bibitem{Wirth:2021pij}
R.~Wirth, J.~M. Yao, and H.~Hergert, ``{Ab~Initio Calculation of the Contact
  Operator Contribution in the Standard Mechanism for Neutrinoless Double Beta
  Decay},'' \href{http://dx.doi.org/10.1103/PhysRevLett.127.242502}{{\em Phys.
  Rev. Lett.} {\bfseries 127} no.~24, (2021) 242502},
  \href{http://arxiv.org/abs/2105.05415}{{\ttfamily arXiv:2105.05415
  [nucl-th]}}.

\bibitem{PhysRevD.10.275}
J.~C. Pati and A.~Salam, ``Lepton number as the fourth "color",''
  \href{http://dx.doi.org/10.1103/PhysRevD.10.275}{{\em Phys. Rev. D}
  {\bfseries 10} (Jul, 1974) 275--289}.
  \url{https://link.aps.org/doi/10.1103/PhysRevD.10.275}.

\bibitem{PhysRevD.11.2558}
R.~N. Mohapatra and J.~C. Pati, ``"natural" left-right symmetry,''
  \href{http://dx.doi.org/10.1103/PhysRevD.11.2558}{{\em Phys. Rev. D}
  {\bfseries 11} (May, 1975) 2558--2561}.
  \url{https://link.aps.org/doi/10.1103/PhysRevD.11.2558}.

\bibitem{PhysRevD.12.1502}
G.~Senjanovic and R.~N. Mohapatra, ``Exact left-right symmetry and spontaneous
  violation of parity,'' \href{http://dx.doi.org/10.1103/PhysRevD.12.1502}{{\em
  Phys. Rev. D} {\bfseries 12} (Sep, 1975) 1502--1505}.
  \url{https://link.aps.org/doi/10.1103/PhysRevD.12.1502}.

\bibitem{Duka_2000}
P.~Duka, J.~Gluza, and M.~Zrałek, ``Quantization and renormalization of the
  manifest left–right symmetric model of electroweak interactions,''
  \href{http://dx.doi.org/10.1006/aphy.1999.5988}{{\em Annals of Physics}
  {\bfseries 280} no.~2, (Mar, 2000) 336–408}.
  \url{http://dx.doi.org/10.1006/aphy.1999.5988}.

\bibitem{Hirsch:1996ye}
M.~Hirsch, H.~V. Klapdor-Kleingrothaus, and S.~G. Kovalenko, ``{New leptoquark
  mechanism of neutrinoless double beta decay},''
  \href{http://dx.doi.org/10.1103/PhysRevD.54.R4207}{{\em Phys. Rev. D}
  {\bfseries 54} (1996) R4207--R4210},
  \href{http://arxiv.org/abs/hep-ph/9603213}{{\ttfamily arXiv:hep-ph/9603213}}.

\bibitem{Blennow:2010th}
M.~Blennow, E.~Fernandez-Martinez, J.~Lopez-Pavon, and J.~Menendez,
  ``{Neutrinoless double beta decay in seesaw models},''
  \href{http://dx.doi.org/10.1007/JHEP07(2010)096}{{\em JHEP} {\bfseries 07}
  (2010) 096}, \href{http://arxiv.org/abs/1005.3240}{{\ttfamily arXiv:1005.3240
  [hep-ph]}}.

\bibitem{Barea:2015zfa}
J.~Barea, J.~Kotila, and F.~Iachello, ``{Limits on sterile neutrino
  contributions to neutrinoless double beta decay},''
  \href{http://dx.doi.org/10.1103/PhysRevD.92.093001}{{\em Phys. Rev. D}
  {\bfseries 92} (2015) 093001},
  \href{http://arxiv.org/abs/1509.01925}{{\ttfamily arXiv:1509.01925
  [hep-ph]}}.

\bibitem{Asaka:2016zib}
T.~Asaka, S.~Eijima, and H.~Ishida, ``{On neutrinoless double beta decay in the
  $\nu$MSM},'' \href{http://dx.doi.org/10.1016/j.physletb.2016.09.044}{{\em
  Phys. Lett. B} {\bfseries 762} (2016) 371--375},
  \href{http://arxiv.org/abs/1606.06686}{{\ttfamily arXiv:1606.06686
  [hep-ph]}}.

\bibitem{deVries:2022nyh}
J.~de~Vries, G.~Li, M.~J. Ramsey-Musolf, and J.~C. Vasquez, ``{Light sterile
  neutrinos, left-right symmetry, and
  0\ensuremath{\nu}\ensuremath{\beta}\ensuremath{\beta} decay},''
  \href{http://dx.doi.org/10.1007/JHEP11(2022)056}{{\em JHEP} {\bfseries 11}
  (2022) 056}, \href{http://arxiv.org/abs/2209.03031}{{\ttfamily
  arXiv:2209.03031 [hep-ph]}}.

\bibitem{Dekens:2023iyc}
W.~Dekens, J.~de~Vries, E.~Mereghetti, J.~Men\'endez, P.~Soriano, and G.~Zhou,
  ``{Neutrinoless double-beta decay in the neutrino-extended Standard Model},''
  \href{http://arxiv.org/abs/2303.04168}{{\ttfamily arXiv:2303.04168
  [hep-ph]}}.

\bibitem{Kotila:2012zza}
J.~Kotila and F.~Iachello, ``{Phase space factors for double-$\beta$ decay},''
  \href{http://dx.doi.org/10.1103/PhysRevC.85.034316}{{\em Phys. Rev. C}
  {\bfseries 85} (2012) 034316},
  \href{http://arxiv.org/abs/1209.5722}{{\ttfamily arXiv:1209.5722 [nucl-th]}}.

\bibitem{Stoica:2013lka}
S.~Stoica and M.~Mirea, ``{New calculations for phase space factors involved in
  double-$\beta$ decay},''
  \href{http://dx.doi.org/10.1103/PhysRevC.88.037303}{{\em Phys. Rev. C}
  {\bfseries 88} no.~3, (2013) 037303},
  \href{http://arxiv.org/abs/1307.0290}{{\ttfamily arXiv:1307.0290 [nucl-th]}}.

\bibitem{numpy}
C.~R.~H. others, ``Array programming with {NumPy},''
  \href{http://dx.doi.org/10.1038/s41586-020-2649-2}{{\em Nature} {\bfseries
  585} no.~7825, (Sept., 2020) 357--362}.
  \url{https://doi.org/10.1038/s41586-020-2649-2}.

\bibitem{pandas_software}
T.~pandas~development team, ``pandas-dev/pandas: Pandas 1.1.3,'' Oct., 2020.
\newblock \url{https://doi.org/10.5281/zenodo.4067057}.

\bibitem{pandas_paper}
{W}es {M}c{K}inney,
  \href{http://dx.doi.org/10.25080/Majora-92bf1922-00a}{``{D}ata {S}tructures
  for {S}tatistical {C}omputing in {P}ython,''} in {\em {P}roceedings of the
  9th {P}ython in {S}cience {C}onference}, {S}t\'efan van~der {W}alt and
  {J}arrod {M}illman, eds., pp.~56 -- 61.
\newblock 2010.

\bibitem{matplotlib}
J.~D. Hunter, ``Matplotlib: A 2d graphics environment,''
  \href{http://dx.doi.org/10.1109/MCSE.2007.55}{{\em Computing in Science \&
  Engineering} {\bfseries 9} no.~3, (2007) 90--95}.

\bibitem{scipy}
P.~Virtanen {\em et~al.}, ``{SciPy} 1.0: Fundamental algorithms for scientific
  computing in python,''
  \href{http://dx.doi.org/10.1038/s41592-019-0686-2}{{\em Nature Methods}
  {\bfseries 17} (2020) 261--272}.

\bibitem{mpmath}
F.~Johansson {\em et~al.}, {\em mpmath: a {P}ython library for
  arbitrary-precision floating-point arithmetic (version 1.1.0)}, December,
  2013.
\newblock {\tt http://mpmath.org/}.

\bibitem{Stefanik:2015twa}
D.~Stefanik, R.~Dvornicky, F.~Simkovic, and P.~Vogel, ``{Reexamining the light
  neutrino exchange mechanism of the $0\nu\beta\beta$ decay with left- and
  right-handed leptonic and hadronic currents},''
  \href{http://dx.doi.org/10.1103/PhysRevC.92.055502}{{\em Phys. Rev. C}
  {\bfseries 92} no.~5, (2015) 055502},
  \href{http://arxiv.org/abs/1506.07145}{{\ttfamily arXiv:1506.07145
  [hep-ph]}}.

\bibitem{ParticleDataGroup:2016lqr}
{\bfseries Particle Data Group} Collaboration, C.~Patrignani {\em et~al.},
  ``{Review of Particle Physics},''
  \href{http://dx.doi.org/10.1088/1674-1137/40/10/100001}{{\em Chin. Phys. C}
  {\bfseries 40} no.~10, (2016) 100001}.

\bibitem{Nicholson:2018mwc}
A.~Nicholson {\em et~al.}, ``{Heavy physics contributions to neutrinoless
  double beta decay from QCD},''
  \href{http://dx.doi.org/10.1103/PhysRevLett.121.172501}{{\em Phys. Rev.
  Lett.} {\bfseries 121} no.~17, (2018) 172501},
  \href{http://arxiv.org/abs/1805.02634}{{\ttfamily arXiv:1805.02634
  [nucl-th]}}.

\bibitem{Bhattacharya:2016zcn}
T.~Bhattacharya, V.~Cirigliano, S.~Cohen, R.~Gupta, H.-W. Lin, and B.~Yoon,
  ``{Axial, Scalar and Tensor Charges of the Nucleon from 2+1+1-flavor Lattice
  QCD},'' \href{http://dx.doi.org/10.1103/PhysRevD.94.054508}{{\em Phys. Rev.
  D} {\bfseries 94} no.~5, (2016) 054508},
  \href{http://arxiv.org/abs/1606.07049}{{\ttfamily arXiv:1606.07049
  [hep-lat]}}.

\bibitem{Cirigliano:2021qko}
V.~Cirigliano, W.~Dekens, J.~de~Vries, M.~Hoferichter, and E.~Mereghetti,
  ``{Determining the leading-order contact term in neutrinoless double $\beta$
  decay},'' \href{http://dx.doi.org/10.1007/JHEP05(2021)289}{{\em JHEP}
  {\bfseries 05} (2021) 289}, \href{http://arxiv.org/abs/2102.03371}{{\ttfamily
  arXiv:2102.03371 [nucl-th]}}.

\bibitem{arnold2010probing}
R.~Arnold, C.~Augier, J.~Baker, A.~Barabash, A.~Basharina-Freshville,
  M.~Bongrand, V.~Brudanin, A.~Caffrey, S.~Cebri{\'a}n, A.~Chapon, {\em
  et~al.}, ``Probing new physics models of neutrinoless double beta decay with
  supernemo,'' {\em The European Physical Journal C} {\bfseries 70} no.~4,
  (2010) 927--943.

\bibitem{Deppisch:2020mxv}
F.~F. Deppisch, L.~Graf, and F.~\v{S}imkovic, ``{Searching for New Physics in
  Two-Neutrino Double Beta Decay},''
  \href{http://dx.doi.org/10.1103/PhysRevLett.125.171801}{{\em Phys. Rev.
  Lett.} {\bfseries 125} no.~17, (2020) 171801},
  \href{http://arxiv.org/abs/2003.11836}{{\ttfamily arXiv:2003.11836
  [hep-ph]}}.

\end{thebibliography}\endgroup
